\documentclass[12pt]{article}
\usepackage{amsmath,amsfonts,amssymb}

\usepackage{amssymb,amscd,amsmath,amsxtra}
\begin{document}
\thispagestyle{empty}

\def\theequation{\arabic{section}.\arabic{equation}}
\def\a{\alpha}
\def\al{\alpha}
\def\b{\beta}
\def\g{\gamma}
\def\d{\delta}
\def\D{\Delta}
\def\dd{\rm d}
\def\e{\epsilon}
\def\ve{\varepsilon}
\def\z{\zeta}
\def\B{\mbox{\bf B}}
\def\B{\mbox{\bf B}}\def\cp{\mathbb {CP}^3}
\newcommand{\h}{\hspace{0.5cm}}

\newcommand{\vf}{\varphi}
\newcommand{\ul}{\underline}
\newcommand{\p}{\partial}
\newcommand{\s}{\sigma}
\newcommand{\uz}{\underline z}
\newcommand{\us}{\underline\sigma}
\newcommand{\da}{\delta^p(\us_1 - \us_2)}
\newcommand{\la}{\lambda}
\newcommand{\La}{\Lambda}
\newcommand{\Di}{\left(\p_0-\la^{i}\p_i\right)}
\newcommand{\Dj}{\left(\p_0-\la^{j}\p_j\right)}
\newcommand{\ct}{\cal{t}}
\newcommand{\cf}{\cal{\varphi}}

\def\CA{{\cal A}}       \def\CB{{\cal B}}       \def\CC{{\cal C}}
\def\CD{{\cal D}}       \def\CE{{\cal E}}       \def\CF{{\cal F}}
\def\CG{{\cal G}}       \def\CH{{\cal H}}       \def\CI{{\cal J}}
\def\CJ{{\cal J}}       \def\CK{{\cal K}}       \def\CL{{\cal L}}
\def\CM{{\cal M}}       \def\CN{{\cal N}}       \def\CO{{\cal O}}
\def\CP{{\cal P}}       \def\CQ{{\cal Q}}       \def\CR{{\cal R}}
\def\CS{{\cal S}}       \def\CT{{\cal T}}       \def\CU{{\cal U}}
\def\CV{{\cal V}}       \def\CW{{\cal W}}       \def\CX{{\cal X}}
\def\CY{{\cal Y}}       \def\CZ{{\cal Z}}

\def\a{\alpha}
\def\b{\beta}
\def\g{\gamma}
\def\ga{\gamma}
\def\d{\delta}
\def\e{\epsilon}
\def\ve{\varepsilon}
\def\th{\theta}
\def\vt{\vartheta}
\def\k{\kappa}
\def\l{\lambda}
\def\m{\mu}
\def\n{\nu}
\def\x{\xi}
\def\r{\rho}
\def\vr{\varrho}
\def\s{\sigma}
\def\t{\tau}
\def\th{\theta}
\def\z{\zeta }
\def\vp{\varphi}
\def\G{\Gamma}
\def\Ga{\Gamma}
\def\D{\Delta}
\def\T{\Theta}
\def\X{{\bf X}}
\def\P{\Pi}
\def\S{\Sigma}
\def\L{\lambda_{_{^L}} }
\def\M{ {\bf M}}
\def\O{\Omega}
\def\oo{\hat \omega   }
\def\ov{\over}
\def\o{\omega }
\def\hf{ {1\over 2} }
\def\bz{\bar z}
\def\tphi{\tilde\phi}

\def\rb{$\bullet\,$}

\def\half{{1\over{2}}}
\def\pd{\partial}
\def\del{\bf\nabla}
\def\ol#1{{\overline#1}}
\newcommand{\beq}{\begin{equation}}
\newcommand{\eeq}{\end{equation}}
\newcommand{\beaq}{\begin{eqnarray}}
\newcommand{\eeaq}{\end{eqnarray}}
\def\val{\alpha}
\def\al{\alpha}
\def\ga{\gamma}
\def\ep{\epsilon}
\def\ze{\zeta}
\def\et{\eta}
\def\io{\iota}
\def\ka{\kappa}
\def\la{\lambda}
\def\th{\theta}
\def\rh{\rho}
\def\si{\sigma}
\def\up{\upsilon}
\def\ph{\phi}
\def\ch{\chi}
\def\om{\omega}
\def\ve{\varepsilon}
\def\vp{\varphi}
\def\vt{\vartheta}
\def\phit{{\tilde\varphi}}
\def\Ga{\Gamma}
\def\De{\Delta}
\def\Th{\Theta}
\def\La{\Lambda}
\def\Si{\Sigma}
\def\Up{\Upsilon}
\def\Ph{\Phi}
\def\Om{\Omega}
\def\half{{1\over{2}}}
\def\pd{\partial}
\def\bp{\bar\partial}

\def\dbar{{\overline\partial}}

\def\hfb{ \frac{b}{2}}
\def\bhf{\frac{1}{2b}}
\def\Pll{P^{_{L}}}
\def\Pmm{P^{_{M}}}
\def\hC{\hat{C}}
\def\tq{\tilde q}
\def\gpl{G^+}
\def\gmi{G^-}
\def\gph{G^+_{-\half}}
\def\gmh{G^-_{-\half}}

\def\bps{\bar\psi}
\def\ra{\rangle}


\def\[{\left[}
\def\]{\right]}
\def\({\left(}
\def\){\right)}
\def\tphi{\tilde\phi}
\def\hs{\hat{su}(2)}
\def\no{\noindent}

\def\bpsi {\bar\psi}
\def\der {\partial}
\def\bder {\bar\partial}
\def\lch #1#2{L_{#1}^{(#2)}}
\def\blch #1#2{\bar L_{#1}^{(#2)}}
\def\fr #1#2{{#1\over #2}}
\def\bz {\bar z}
\def\lo {\bz\bder-z\der}
\def\lkts {L_{-k}^{(2s)}}
\def\blkts {\bar L_{-k}^{(2s)}}
\def\loi {\bz_i\bder_i-z_i\der_i}
\def\cL {{\cal L}}
\newcommand{\sfrac}[2]{{\textstyle\frac{#1}{#2}}}

\newcommand{\bL}{\mbox{\bf L}}
\newcommand{\bM}{\mbox{\bf M}}
\newcommand{\bR}{\mbox{\bf R}}
\newcommand{\bT}{\mbox{\bf T}}
\newcommand{\cP}{{\cal P}}
\newcommand{\buno}{\mbox{\bf 1}}

\newcommand{\eq}{\begin{equation}}
\newcommand{\en}{\end{equation}}

\newcommand{\spz}{\hspace{0.7cm}}
\newcommand{\virg}{\spz,\spz}
\newcommand{\de}{\partial_u}

\newcommand{\beqa}{\begin{eqnarray}}
\newcommand{\eeqa}{\end{eqnarray}}


\newcommand{\vph}{{\varphi}}
\newcommand{\vrh}{\boldsymbol{\rho}}
\newcommand{\vbe}{\boldsymbol{\beta}}
\newcommand{\vaa}{\boldsymbol{a}}
\newcommand{\va}{{\bf a}}
\newcommand{\vP}{{\bf P}}
\newcommand{\ovz}{{\overline z}}
\newcommand{\vQ}{{\bf Q}}
\newcommand{\bfe}{{\bf e}}
\newcommand{\mbar}{{\overline m}}
\newcommand{\tid}{\tilde{\Delta}}
\newcommand{\rap}{\over{2\pi}{2\mu_\pi}}
\newcommand{\fleche}[1]{\raisebox{-0.4cm}{~\shortstack{${=}$\\${\em{#1}}$}}}
\newcommand{\Eqr}[1]{(\ref{#1})}
\renewcommand{\theequation}{\thesection.\arabic{equation}}


\newcommand{\ba}{\begin{eqnarray}}
\newcommand{\ea}{\end{eqnarray}}

\newcommand{\virb}{\overline{\bf Vir}}
\newcommand{\ext}{{\bf Ext}}
\newcommand{\Db}{\bar{\Delta}}
\newcommand{\ih}{\hat{\imath}}
\newcommand{\jh}{\hat{\jmath}}
\newcommand{\dx}{\partial_x}
\newcommand{\dt}{\partial_t}
\newcommand{\tp}{\otimes}
\newcommand{\lt}{\left(}
\newcommand{\rt}{\right)}
\newcommand{\lqu}{\left[}
\newcommand{\rqu}{\right]}
\newcommand{\lgr}{ \left\{ }
\newcommand{\rgr}{ \right\} }
\newcommand{\dla}{\partial_{\lambda}}

\def\hepth#1{{arXiv:hep-th/}#1}
\def\cqg#1#2#3{{ Class. Quantum Grav.} {\bf #1} (#2) #3}
\def\np#1#2#3{{Nucl. Phys.} {\bf B#1} (#2) #3}
\def\pl#1#2#3{{Phys. Lett. }{\bf B#1} (#2) #3}
\def\prl#1#2#3{{Phys. Rev. Lett.}{\bf #1} (#2) #3}
\def\physrev#1#2#3{{Phys. Rev.} {\bf D#1} (#2) #3}
\def\ap#1#2#3{{Ann. Phys.} {\bf #1} (#2) #3}
\def\prep#1#2#3{{Phys. Rep.} {\bf #1} (#2) #3}
\def\rmp#1#2#3{{Rev. Mod. Phys. }{\bf #1} (#2) #3}
\def\rmatp#1#2#3{{Rev. Math. Phys. }{\bf #1} (#2) #3}
\def\cmp#1#2#3{{Comm. Math. Phys.} {\bf #1} (#2) #3}
\def\mpl#1#2#3{{Mod. Phys. Lett. }{\bf #1} (#2) #3}
\def\ijmp#1#2#3{{Int. J. Mod. Phys.} {\bf #1} (#2) #3}
\def\lmp#1#2#3{{Lett. Math. Phys.} {\bf #1} (#2) #3}
\def\tmatp#1#2#3{{Theor. Math. Phys.} {\bf #1} (#2) #3}
\def\jhep#1#2#3{{JHEP} {\bf #1} (#2) #3}


\begin{titlepage}
\vspace*{1.cm}
\renewcommand{\thefootnote}{\fnsymbol{footnote}}
\begin{center}
\LARGE{BULGARIAN ACADEMY OF SCIENCES \\ Institute for Nuclear
Research and Nuclear Energy}
\vspace*{4cm}

{\Huge \bf Symmetries in Two Dimensional Conformal Field Theories and Related Integrable Models}
\end{center}

 \vskip 20mm

\baselineskip 18pt

\begin{center}

\centerline{{\huge \bf Marian Stanishkov}}
 \vskip 0.6cm
{\huge {\bf Thesis}}

\Large{\bf presented for obtaining the scientific degree\\ "Doctor
of sciences"}

\vspace*{3cm}

\Large{\bf SOFIA, 2017}
\end{center}

\end{titlepage}

\newpage

\def\nn{\nonumber}
\def\tr{{\rm tr}\,}
\def\p{\partial}
\def\ov{\over}
\newcommand{\bea}{\begin{eqnarray}}
\newcommand{\eea}{\end{eqnarray}}
\renewcommand{\thefootnote}{\fnsymbol{footnote}}
\newcommand{\be}{\begin{equation}}
\newcommand{\ee}{\end{equation}}
\def\rb{$\bullet\,$}
\renewcommand{\thefootnote}{\arabic{footnote}}
\setcounter{footnote}{0}

\tableofcontents

\newpage

\setcounter{equation}{0}
\section{Introduction}

The two-dimensional (2D) conformal field theories (CFT's) have a wide application in the description of the scaling and universality behaviour of the 2D statistical systems at the second order phase transition points. They play also a basic role in the description of the string theories providing the symmetry of the worldsheet. As was shown in \cite{bpz} the specific properties of the reducible representations of the Virasoro algebra (describing the so called minimal models) allows one to solve the conformal bootstrap and to find the the exact critical exponents, the explicit multipoint correlation functions of the fields and finally to obtain the full structure of their associative operator product expansion (OPE) algebra. The different conformal models, characterized by the value of the central charge $c$ and their 2D operator content, describe different universal behaviours of the statistical systems near the critical point.
In this sense the classification of the universality classes in two dimensions is equivalent to exhausting the minimal models of all possible extensions of the Virasoro algebra (e.g. supersymmetry, $W$-algebras etc.). The solution of this problem in the context of string models will provide us with all the possible classical string backgrounds.
Another example of 2D CFT is the Lioville field theory (LFT). It has been studied actively for its relevance for non-critical string theories and 2D quantum gravity. This theory is also interesting on its own as an example  of irrational CFT. Most of the CFT formalism developed for rational CFT's do not apply to this class of theories mainly because they have continuously infinite number of primary fields. Various methods have been proposed to derive the structure constants and the reflection amplitudes, which are basic building blocks to complete the conformal bootstrap \cite{ZZtri}.

As mentioned above a natural generalization of the 2D CFT is its $N=1$ supersymmetric extension. Two kinds of fields appear in this theory belonging to the so called Neveu-Swartz (NS) and Ramond (R) sectors respectively. An important technical problem is their full description in the supersymmetric minimal models. It turns out that the construction of the NS sector of these models requires a slight modification of the Virasoro algebra methods only \cite{bpz}. The difficulties with the fusion rules and multipoint functions of the Ramond fields are connected with the double valuedness of the supercurrent $G(z)$: $G(e^{2\pi i}z)=e^{i k\pi}G(z), k=0,1$ in the presence of the R fields.  Because of that the direct generalization of of the NS sector null vector's method to the R sector seems to be ineffective. A more constructive and certainly more powerful approach to the minimal models is the Coulomb gas representation proposed in \cite{df} for the Virasoro algebra models. In the case of $N=1$ superconformal minimal models the Coulomb gas method was developed in \cite{bkt} and \cite{myn12}.
One could further consider $N=1$ minimal models defined on hyperelliptic surfaces. It turns out that they also can be described by a generalized Coulomb gas representation. The partition function of the models on such $Z_2$ surfaces are then constructed in terms of the correlation functions of fields from the twisted sector of the corresponding branched sphere models. It is interesting to investigate also the renormalization group properties of the supersymmetric minimal models $SM_p$ perturbed by the least relevant field. The first order corrections were obtained in \cite{pogsc1}.
It was argued that there exists an infrared (IR) fixed point of the renormalization group (RG) flow which coincides with the minimal superconformal model $SM_{p-2}$. It is interesting to check this result in the second order of the perturbation theory. Calculation up to the second order is always a challenge even in two dimensions. The problem is that one needs the corresponding four-point function which is not known exactly even in two dimensions.
Fortunately, in the scheme proposed in \cite{pogsc2} one needs the value of this function up to the zeroth order in the small parameter $\epsilon$ describing the dimension of the perturbing field. In addition to the minimal superconformal models one can consider the example of an irrational CFT - the $N=1$ supersymmetric LFT (SLFT). This model has some motivations. It is applicable to the superstring theories and the 2D supergravity with fermionic matter fields. One can also understand the role of the extended conformal symmetry in the irrational CFT's by studying this model. The methods used for the bulk LFT (in obtaining the 3-point functions for example) could be applied successfully to SLFT although the latter becomes algebraically more complicated. It is interesting to extend this formalism to the CFT defined in the two-dimensional space-time geometry with a boundary condition (BC) which preserves the conformal symmetry. It is known \cite{cardy} that the conformally invariant BC's can be associated with the primary fields for the case of rational CFT's. It has been an issue whether this could be extended to the irrational theories. Another motivation is to understand open string theories in various nontrivial background space-time geometries.

The Vrasoro and the $N=1$ superconformal minimal models are just the first two members of a more general infinite series of the so called coset models. More precisely, let us consider a model $M(k,l)$ based on the symmetric coset $\hat{su}(2)_k\times \hat{su}(2)_l/\hat{su}(2)_{k+l}$, $k$ and $l$ are integers \cite{gko}. It is written in terms of the $\hat{su}(2)_k$
WZNW models of level $k$. The WZNW model is a conformal theory with a stress-energy tensor given by the Sugawara construction:\\ $T_k(z)={1\ov k+2}J^2(z)$ where $J^a(z)$ is the
$\hat{su}(2)$ current. The central charge of the corresponding Virasoro algebra is $c_k={3k\ov k+2}$. The coset theory $M(k,l)$ is then also a conformal field theory with a stress tensor $T=T_k+T_l-T_{k+l}$. The resulting central charge can be read from this construction and is labeled by the two integers $k$ and $l$. The dimensions of the primary fields
$\phi_{m,n}(l,k)$ of the minimal coset models are known \cite{kmq} and are labelled by the integers $m,n$ and an additional integer number $s$. It is known \cite{kmq,rava,argy} that the coset theory $M(k,l)$
possesses a larger parafermionic-like symmetry. The simplest (lowest dimensional) current $A(z)$ for example has a dimension $\D_A={l+4\ov l+2}$. For $l=2$ it coincides with the supercurrent $G(z)$ of the superconformal theory, for $l=4$ the corresponding current generates the well known $4/3$-parafermionic series of models. In general, under this symmetry
the primary fields are divided in sectors labelled by the integer $s$. The problem of the description of the coset theory in terms of the representations of the parafermionic algebra lies in the difficulty with its nonlocal nature. Because of that it is not well understood. We will present in Section 4 another recursive construction of $M(k,l)$ based on the product of lower level models. Another interesting problem is the investigation of the renormalization group properties of the coset models perturbed by the least relevant field, in the lines of what was done for the Virasoro and $N=1$ superconformal theories.

The conformal theories with $N=2$ supersymmetry are natural generalizations of the $N=0$ (Virasoro) and $N=1$ superconformal theories. The extended $N=2$ superconformal symmetry in two dimensions \cite{adem} plays an important role in the classification and the construction of the classical superstring vacua, giving rise to $N=1$ SUSY models in four dimensions \cite{bdfm}. The different ways to realize explicitly such vacua and the corresponding low-energy effective action \cite{dhvw} are mainly based on the properties of appropriate $N=2$ superconformal models. In principle all the properties of the effective four-dimensional theory we wish to have can be encoded as specific properties of the 2D $N=2$ superconformal models realizing the corresponding superstring compactification. For example the fusion rules (FR's) and the structure constants determine the Yukawa couplings of the massless low-energy particles and the N-point functions of the 2D fields (representing the vertices of these particles) are the main ingredients in the construction of the 4D scattering amplitudes. All these issues make it interesting the classification of the $N=2$ superconformal models with $c\le 9$ and the systematic study of their properties. An important point in the realization of this program is the full description of the minimal models or the $N=2$ unitary discrete series \cite{bfk,vpz} and of the specific tensor products of these models. According to Gepner's compactification scheme \cite{gep5} they are related to certain Calabi-Yau non-linear sigma models. Different elements, needed for the explicit construction of the $N=2$ minimal models can be found in \cite{gepqiu}. The remaining open problems are mainly related  to the appropriate description of the Ramond (R) and twisted (T) sectors of the minimal models, to the computation of the 4-point functions of all the fields in NS, R and T sectors, the corresponding structure constants and the FR's.

The $N=2$ SLFT is instead an example of an irrational CFT which has continuously infinite number of primary fields. In spite of the extended symmetry, it turns out that exact correlation functions of the $N=2$ SLFT are much more difficult to derive than the $N=0$ and $N=1$ cases. The main reason is that the $N=2$ SLFT has no strong-weak coupling duality. The invariance of the LFT and the $N=1$ SLFT under $b\to 1/b$ is realized when the background charge changes to $b+1/b$ from its classical value $1/b$ after quantum corrections \cite{ctorn}.
All the physical quantities like the correlation functions depend on the coupling constant through this combination. This duality as well as the functional relations based on the conformal bootstrap methods are essential ingredients to obtain exact correlation functions uniquely for $N=0$ \cite{ZZtri} and $N=1$ LFT \cite{my.no3} and their boundary extensions. Differently, the $N=2$ SLFT is not renormalized and no duality appears. This non-renormalization is a general aspect of the $N=2$ superconformal field theories in two dimensions. Without the self-duality the functional relations satisfied by the correlation functions cannot be solved uniquely. In \cite{my.no1}, an $N=2$ super-CFT has been proposed as a dual theory to the $N=2$ SLFT under the transformation $b\to 1/b$.

It is of special interest to study the $N=2$ SLFT in the presence of a (super) conformally invariant boundary. Computing the corresponding one-point functions is more complicated than in the case of $N=0$ and $N=1$ SLFT. The standard approach for the computation of the one-point functions is the conformal bootstrap method \cite{boot1,boot2,my.no2} which can generate functional relations using the conformally invariant boundary actions as boundary screening operators. The boundary action of the $N=2$ SLFT has been derived in \cite{ay}. But the $N=2$ SLFT with this  boundary action is not self-dual either and one needs to know the boundary action of the dual $N=2$ theory. Without this one cannot solve the functional relations uniquely. Due to the non-locality of the bulk action of the dual $N=2$ theory the method used in the $N=2$ SLFT \cite{ay} seems not be applicable. We need a different approach. One possibility is to use the so called modular bootstrap. The modular bootstrap method is a generalization of the Cardy formulation for the conformal BC's to the irrational CFT's. We can first derive the one point functions from the modular transformation properties. Then we relate them to the bulk and boundary actions of the $N=2$ SLFT and its dual theory by the conformal bootstrap method.

It is known \cite{eqm} that in the LFT an infinite set of relations hold for the quantum operators. These equations relate different basic Liouville primary fields $V_\a(z)$ (represented by the vertex operators $\exp(\a\phi)$. They are parameterized by a pair of positive integers $(m,n)$ and are called "higher equations of motion" (HEM), because the first one $(1,1)$ coincides with the usual Liouville equation of motion. Higher equations turn out to be useful in practical calculations. In particular, they were used to derive a general 4-point correlation function in the minimal Liouville gravity. Similar operator valued relations have been found also for the $N=1$ SLFT \cite{n1eqm}. It is intriguing to understand if such relations appear also for the extended $N=2$ SLFT.

The two dimensional integrable system is a classical or quantum field theory with the property of having an infinite number of local integrals of motion in involution (LIM). This kind of symmetry does not allow the determination of the most interesting features of the system because of its Abelian character. Instead, the presence of an infinite dimensional non-Abelian algebra could complete this Abelian algebra giving rise to the possibility of building its representations, i.e. the spectrum of local fields. One could call this non-commutative algebra a spectrum generating algebra. In different models the presence of this spectrum generating symmetry is often connected to the Abelian one. This is the case of the simplest integrable quantum theories--the 2D CFT's--their common crucial property being their covariance under the infinite dimensional Virasoro symmetry. Indeed, the highest weight representations of this algebra classify all the local fields in the 2D CFT's and turn out to be reducible because of occurrence of vectors of null Hermitian product with all the other vectors, the so called null-vector. The factorization by the modules generated over the null-vectors leads to a number of important algebraic-geometrical properties such as fusion algebras, differential equations for the correlation functions etc.
Unfortunately, this beautiful picture collapses when one pushes the system away from criticality by perturbing the original CFT with some relevant local field. From the infinite dimensional Virasoro symmetry only the Poincare subalgebra survives the perturbation. Consequently, one of the most important open problems in 2D integrable quantum field theories (IQFT's) is the construction of the spectrum of local fields and the computation of their correlation functions. Actually, the CFT possesses a bigger $W$-like symmetry and in particular it is invariant under an infinite dimensional Abelian subalgebra of the latter \cite{syff}. With suitable deformations, this Abelian subalgebra survives the perturbation resulting in the so called LIM. As we noticed, being Abelian this symmetry does not carry sufficient information and in particular one cannot build the spectrum of an integrable theory by means of the LIM alone. It has been conjectured in \cite{bbs} that one could add to the LIM $I_{2m+1}$ non-commuting charges $J_{2m}$ in such a way that the resulting algebra would be sufficient to create all the states of a particular class of perturbed theory, the so called restricted sine-Gordon theory. Therein it was also discovered that a sort of null-vector condition appears in the above procedure leading to certain equations for the form-factors.

One of the simplest integrable field theories is the 2D sine-Gordon (SG) model. It possesses an infinite number of conserved charges $I_{2n+1},n\in Z$ in involution. It is known \cite{blec} that the SG theory possesses also an infinite dimensional symmetry provided by the $\hat{sl}(2)_q$ algebra. However, this symmetry connects the correlation functions of the fields in the same multiplet without giving a sufficient information about the functions themselves. It is known also to some extent that there should be another kind of symmetry present in the SG theory. Actually, it could be obtained as a particular scaling limit of the so called $XYZ$-spin chain \cite{lut}. The latter is known to possess an infinite symmetry obeying the so called Deformed Virasoro algebra (DVA) \cite{luk}. It is natural to suppose that in the scaling limit, represented by SG, there should be some infinite dimensional symmetry, a particular limit of DVA. What remains unclear is how this symmetry is realized in SG theory, for example what is the action of the corresponding generators on the exponential fields, what kind of restrictions it imposes on their correlation functions etc.

In a 2D integrable QFT which can be realized as a CFT perturbed by some relevant operator it is well known that any correlation function of local fields $\CO_a(x)$ in the short distance limit can be reduced down to the one-point functions $<\CO_a'(x)>$ by successive application of the operator product expansion \cite{alz}.
These vacuum expectation values (VEV's) contain important information about the IR environment. Important progress has been made concerning the evaluation of some VEV's in different integrable QFT - for example the VEV's of exponential fields in the sinh-Gordon and SG models \cite{lukz} and in the Bullough-Dodd (BD) model with real and imaginary coupling \cite{flzz}.
However, the higher order corrections to the short distance expansion of the correlation functions involve the VEV's of the descendent fields. The knowledge of these quantities improves the analytical prediction for the short distance expansion of the correlation functions which can be better compared with the results obtained from the numerical study of the model.

\setcounter{equation}{0}
\section{$N=1$ superconformal theories}

In this Section we present a detailed discussion of the Ramond sector of the superconformal models and the explicit structure of the OPE algebras of the fields of these models. It appears that all the elements of our construction: vertices, screening operators, 3- and 4-point functions etc. can be written in terms of of the Ising model fields $\s(z),\psi(z)$ and a free scalar field $\phi(z)$. These constructions allow us to find the fusion rules (FR's) for the R and NS fields. The main advantage of the Ramond Coulomb gas method lies in the calculation of the multipoint correlation functions. An important step in the construction of the 2D models (combining both the left and right chiral ones) is the requirement of the crossing symmetry of the 4-point functions. The solution of this problem allows us to calculate explicitly the operator content of the 2D $N=1$ superconformal models and the exact structure constants of the OPE algebra of the fields of the model.
Next, we describe also the superconformal minimal models on a restricted class of surfaces which can be represented as a double covering of the branched sphere. Such surfaces are known as hyperelliptic surfaces. The strategy we use is to reduce the genus $g$ problem to the corresponding $g=0$ problem. The partition function is computed using a generalized Coulomb gas representation.

We are interested also in the renormalization group properties of the superconformal minimal models $SM_{p}$ perturbed by the last component of the superfield $\Phi_{1,3}$ in the second order of the perturbation theory. We present the computation of the conformal blocks in the NS sector and the mixed conformal blocks of NS and R fields. The computation of the beta-function and the IR fixed point confirms that it coincides up to second order with the model $SM_{p-2}$. The matrix of anomalous dimensions for certain NS and R fields are also computed. The results are in perfect agreement with the conjectured RG flow.

Finally, we consider the $N=1$ supersymmetric LFT which is an example of irrational CFT. We propose exact expressions for the 3-point correlation functions in NS and R sectors of the theory. Using the reflection properties of the Liouville vertex operators we introduce the so called reflection amplitudes for the NS and R fields.
We then extend our considerations to a supersymmetric LFT defined in the 2D space-time geometry with a boundary condition (BC) which preserves the (super-) conformal symmetry. We use the functional relation method for the boundary SLFT with super-conformal boundary action \cite{boot1} to derive the one-point function of a bulk operator and correlation functions of two boundary operators for a given conformal BC. Here the conformal BC is denoted by a continuous parameter related to the coupling constant in the boundary action. Another development is to generalize this method to the boundary SLFT defined in the Lobachevskiy plane, or the pseudosphere. We show that in both cases the results are consistent with the Cardy formalism \cite{cardy}. We also show that the boundary 2-point functions of the (NS) boundary operators satisfy the same relation as those of the LFT.

The results of this Section have been published in \cite{myn12,my.no3,my.no2}, \cite{myne1}-\cite{myne5}, (1.-8.).

\subsection{$N=1$ minimal models}

The infinite superconformal algebra in two dimensions (2D) splits into a direct sum of two algebras: a left one generated by the stress-energy tensor $T(z)$ (of dimension 2) and its fermionic superpartner $G(z)$ (of dimension ${3\ov 2}$) \cite{bkt}:
\bea\label{opea} T(z_1)T(z_2) &=&{c\ov 2z_{12}^4}+{2T(z_2)\ov z_{12}^2}+{1\ov z_{12}}\p T(z_2)+\ldots,\\
\nn T(z_1)G(z_2) &=&{3\ov 2z_{12}^2}G(z_2)+{1\ov z_{12}}\p G(z_2)+\ldots,\\
\nn G(z_1)G(z_2) &=&{2c\ov 3z_{12}^3}+{2\ov z_{12}} T(z_2)+\ldots \eea
and a right one defined by the corresponding singular terms in the OPE of $\bar T(\bar z)$ and $\bar G(\bar z)$. We shall restrict our discussion in the following to the "chiral", one-dimensional part, leaving the 2D construction for the end of this Section.

The two different boundary conditions for the supercurrent $G(e^{2\pi i}z)=\pm G(z)$ imply two different Laurent mode expansions:
\be
\nn G^{(p)}(z)=\sum_{n\in Z}{G^{(p)}_{n+1/2}\ov z^{n+2}},\qquad G^{(a)}(z)=\sum_{n\in Z}{G^{(a)}_{n}\ov z^{n+3/2}}
\ee
while the stress-energy tensor has the usual mode expansion:
\be
\nn T(z)=\sum_{n\in Z}{L_{n}\ov z^{n+2}}
\ee
Then the OPE's (\ref{opea}) take the well known form of the Neveu-Schwarz (NS) and Ramond (R) algebras:
\bea\label{alg} [L_n,L_m]&=&(n-m)L_{n+m}+{c\ov 12} n(n^2-1)\d_{n+m,0},\\
\nn [L_n,G_\a]&=&({n\ov 2}-\a)G_{n+\a},\\
\nn \{G_\a,G_\b\}&=&2L_{\a+\b}+{c\ov 3}(\a^2-{1\ov4})\d_{\a+\b,0}
\eea
where $\a,\b\in Z+1/2$ for the NS sector and $\a,\b\in Z$ for the R sector. According to \cite{bkt} we can define the primary states $|\D>$ corresponding to the lowest weight representations of the superalgebra (\ref{alg}) requiring:
\be
L_n|\D>=G_\a|\D>=0, \quad (n,\a>0),\quad L_0|\D>=\D|\D>.
\ee
Since in the Ramond sector a $G(z)$ zero mode appears:
\be
\nn [G_0,L_0]=0,\qquad G_0^2=L_0-{c\ov 24}
\ee
the lowest energy Ramond state is doubly degenerate (for $\D\neq {c\ov 24}$), i.e. both states $|\D,\pm>$ defined as:
\be\label{ramst}
\nn G_0|\D,+>=|\D,->,\qquad G_0|\D,->=(\D-{c\ov 24})|\D,+>
\ee
correspond to the same eigenvalue $\D$ of $L_0$. Introducing the invariant vacuum state $|0>$:
\be
\nn L_n|0>=G_\a|0>=0,\qquad n\ge -1,\a\ge -{1\ov 2}
\ee
we can represent the NS state $|\D>$ in terms of the NS primary superfields\\ $\Phi_\D(z,\theta)=\phi_\D(z)+\theta\psi_{\D+1/2}(z)$:
\be
\nn |\D>=\phi(0)|0>,\qquad G_{-1/2}|\D>=\psi(0)|0>
\ee
In the OPE language these algebraic properties of the primary fields have the form:
\bea\label{pri}
T(z_1)\phi(z_2) &=& {\D\ov z_{12}^2}\phi(z_2)+{1\ov z_{12}}\p\phi(z_2)+\ldots,\\
\nn G(z_1)\phi(z_2) &=& {1\ov z_{12}}\psi(z_2)+\ldots,\\
\nn G(z_1)\psi(z_2) &=& {2\D\ov z_{12}^2}\phi(z_2)+{1\ov z_{12}}\p\phi(z_2)+\ldots
\eea
Using the mode expansion of $T(z)$ and $G(z)$ one obtains from (\ref{pri}) the Ward identities:
\bea\label{wi}
[L_n,\Phi(z,\theta)] &=&\left(z^{n+1}\p +(n+1)z^n(\D+{1\ov 2}\theta{\p\ov\p\theta})\right)\Phi(z,\theta),\\
\nn [G_{n+{1\ov 2}},\Phi(z,\theta)] &=& z^n\left(z({\p\ov\p\theta}-\theta{\p\ov\p z})-2\D(n+1)\theta\right)\Phi(z,\theta).
\eea

The $R$-primary states are created from the $NS$ vacuum by the corresponding Ramond primary fields $R_\D^\pm(z)$:
\be
|\D,\pm>=R_\D^\pm(0)|0>
\ee
The latter have the following OPE with the supercurrent:
\be\label{oper}
G(z_1)R_\D^\pm(z_2) = {a^\pm(\D)\ov z_{12}^{3/2}}R_\D^\mp(z_2)+\ldots,
\ee
where:
\be
a^+(\D)=1,\qquad a^-(\D)=\D-{c\ov 24}
\ee
and the OPE with the stress-tensor is the same as for the $NS$ fields. It is clear from the OPE's (\ref{pri}) and (\ref{oper}) that the fields from the NS and R sectors realize different analytic behaviour of the supercurrent - periodic and antiperiodic respectively, which is a manifestation of the hidden $Z_2$ symmetry of of the $N=1$ superconformal theories. Correspondingly, NS fields describe the even and R fields the odd sectors of such theories with respect to this discrete symmetry. We shall also use a diagonal basis for the Ramond fields:
\be\nn
G(z_1)\tilde R_\D(z_2) = \mp {1\ov z_{12}^{3/2}}\sqrt{\D-{c\ov 24}}\tilde R_\D(z_2)+\ldots
\ee
(for $\D\ne {c\ov 24}$) and we defined:
\be\nn
\tilde R_\D=\sqrt{\D-{c\ov 24}}R^+\pm R^-.
\ee

It is well known \cite{bkt} that the (reducible) unitary representations of the $N=1$ superconformal algebras given by:
\bea\label{nonereps}
c&=&{3\ov 2} -{12\ov p(p+2)}\\
\nn \D_{n,m}&=&{((p+2)n-p m)^2-4\ov 8p(p+2)}+{1\ov 32}(1-(-1)^{n-m})
\eea
($n-m\in 2Z$ for the NS sector and $n-m\in 2Z+1$ for the R sector) determine an infinite series of exactly solvable minimal models. The basic property of the superconformal
representations (or superconformal families) $[\phi_{\D_{n,m}}]$ is that at level ${1\ov 2}nm$ there exist descendent fields $\phi_{\D_{n,m}+{1\ov 2}nm}$ which are again primary fields. Then the covariant condition $\phi_{\D_{n,m}+{1\ov 2}nm}=0$ separates the irreducible part of the representations $[\phi_{\D_{n,m}}]$. In the NS sector these null vector conditions, together with the Ward identities (\ref{wi}) lead to differential equations for the $n$-point functions of the fields $\phi_{\D_{n,m}}$. These equations allow one to find explicitly for example the corresponding fusion rules. This was explained in details in \cite{qiu,myne1}.

The difficulties with the application of this method for the Ramond fields come from the branch cut singularity in the OPE (\ref{oper}). In \cite{myn12} it was described a modification of the null vector's method based on the specific analytic properties of the Ramond fields. It consists basically in defining an auxiliary correlation function in which the branch cut is cancelled by multiplying the original function with a suitable power of the coordinates. Then, the OPE (\ref{oper}) and the null vector condition lead to an equation for the original correlation function. We omit here the explicit calculations since the Coulomb gas construction that we present below turns out to be more powerful. We shall only consider in more details the computation of the Ising model 4-point functions, which is performed using a similar null vector technics, since they are important elements of the Ramond Coulomb gas construction \cite{myne2}.

Let us consider the semi-direct sum of the Virasoro algebra and the algebra of the Laurent coefficients $\psi_n$ of the antiperiodic Majorana field:
\be\nn
\psi(z)=\sum_{n\in Z}{\psi_n\ov z^{n+1/2}},\qquad \psi(e^{2\pi i}z)=-\psi(z),
\ee
\bea\label{algis}
[L_n,\psi_m]&=&-({n\ov 2}+m)\psi_{n+m},\\
\nn \{\psi_n,\psi_m\}&=&\d_{n+m,0}
\eea
which is an analog of the Ramond sector for the case of the Ising model. Because of the zero mode of the fermion:
\be\nn
[L_0,\psi_0]=0,\qquad \psi_0^2={1\ov 2}
\ee
the lowest energy state of this algebra, $|\s^\pm>$, is doubly degenerate and has dimension $\qquad$ $\D^\pm ={1\ov 16}$. The "spin fields" corresponding to this "Ramond states"
$|\s^\pm>=\s^\pm(0)|0>$ produce a branch cut singularity of the antiperiodic fermionic field:
\be\nn
\psi(z)\s^\pm(w)=\sqrt{{1\ov 2(z-w)}}\s^\mp(w)+\ldots .
\ee
We shall also use a diagonal basis $\tilde\s={1\ov \sqrt{2}}(\s^+\mp\s^-)$ with the following OPE:
\be\label{tilis}
\psi(z)\tilde\s(w)=\mp \sqrt{{1\ov 2(z-w)}}\tilde\s(w)+\ldots.
\ee

The $SL(2,R)$ invariance allows to write the 4-point function of the field $\s(z)$ in the form:
\be\nn
\CF(z)=\lim_{w\rightarrow\infty}w^{1/8}<\s(w)\s(z)\s(1)\s(0)>=z^{1/8}(1-z)^{-1/8}f(z).
\ee

The "null vector method" used in the case of Ramond fields in \cite{myne2} can be applied also for the above function in the Ising model. Namely, the singular part of the OPE (\ref{tilis})
and the first level null vector of the Ising algebra:
\be\nn
\left( L_{-1}-{1\ov 2}\psi_{-1}\psi_0\right)|\s>=0
\ee
allow us to find a first order differential equation for the unknown function $f(z)$:
\be\nn
4z{df\ov dz}-{1\ov \sqrt{1-{1\ov z}}}\left(1-\sqrt{1-{1\ov z}}\right)f(z)=0.
\ee
The final solution reads:
\be
\CF(z)=\sqrt{{1\ov 2}}z^{1/8}(1-z)^{-1/8}\sqrt{1+\sqrt{1-{1\ov z}}}.
\ee
The normalization is fixed by the OPE (\ref{tilis}) and by normalizing the two-point function of the spin field to one.

Repeating the same procedure for the 4-point function:
\be\nn
\tilde\CF(z)=<\tilde\s(\infty)\s(z)\tilde\s(1)\s(0)>
\ee
we obtain a similar result:
\be
\tilde\CF(z)=\sqrt{{1\ov 2}}z^{1/8}(1-z)^{-1/8}\sqrt{1-\sqrt{1-{1\ov z}}}.
\ee
This method, together with the OPE:
\be\nn
\psi(z_1)\psi(z_2)={1\ov z_{12}}+2 z_{12}T(z_2)+\ldots
\ee
leads to a recursive equation for the following correlation function:
\be
G^N(z,v_i)=<\s(\infty)\s(z)\s(1)\prod_{i=1}^N\psi(v_i)\s(0)>.
\ee
The solution of this equation can be written in the form of the Wick-like theorem:
\bea\label{wic}
G^N(z,v_i)&=&\big(\prod_{i=1}^Nf_i(v_i,N-i+1)+\sum_{all<ij>}\prod_{l\ne i,j}^{N-2}f_l g_{ij}+\\
\nn &+& \sum_{all<ij>,<k,l>}\prod_{p\ne ijkl}^{N-4}f_p g_{ij}g_{kl}+\ldots\big)\CF(z)
\eea
where:
\bea\nn
f_i(v_i,N)&=&\sqrt{{1\ov 2}}\left({\sqrt{z-1}\ov 1-v_i}+(-1)^{n+1}{\sqrt{z}\ov v_i}\right)\sqrt{{(1-v_i)v_i\ov z-v_i}},\\
\nn g_{ij}&=&{(-1)^{i+j-1}\ov v_{ij}}\sqrt{{v_i(1-v_i)(z-v_i)\ov v_j(1-v_j)(z-v_j)}}.
\eea
As we will se below these functions are an important ingredient in the calculation of the 4-point functions of the Ramond fields.

The natural language for the description of the two-dimensional minimal models is the so called Coulomb gas construction. Its generalization to the case of of $N=1$ superconformal models is based on the free scalar (chiral) superfields $S(z,\theta)=\phi(z)+\theta\psi(z)$ (and antichiral $\bar S(\bar z,\bar\theta)$) with the action:
\be\label{ac}
A(S,\bar S)={2\ov \pi}\int dzd\bar z({1\ov 2}\p\phi\bar\p\bar\phi-\psi\p\bar\psi).
\ee
It follows from this action that the propagator is given by:
\bea
<S(z_1,\theta_1)&S&(z_2,\theta_2)>=-\ln{\hat{z}_{12}\ov R},\\
\nn \hat{z}_{12}&=&z_1-z_2-\theta_1\theta_2
\eea
($R$ is the infrared cut-off). In this language the superfields of the conformal grid are constructed in terms of the so called NS vertices:
\be
V_\a(z,\theta)=:e^{i\a S(z,\theta)}:,\qquad \a\in R.
\ee
where $::$ defines certain normal ordering.

Actually, the action (\ref{ac}) leads to a free theory with central charge $c={3\ov 2}$. The construction which leads to the anomalous central charge:
\be\label{anc}
c={3\ov 2} -{12\ov p(p+2)}
\ee
is generated by the modified action:
\be\label{mac}
\CA(S,\bar S)={2\ov \pi}\int dzd\bar zd\theta d\bar\theta({1\ov 2}D S \bar D\bar S -2i\a_0 \hat R(S+\bar S)
\ee
where $\hat R$ is the curvature superfield. The latter can be chosen such that its inclusion in (\ref{mac}) is equivalent to an introduction of a vertex operator
$\exp(-2i\a_0 S)$ at infinity. Then the correlation function:
\be
<\prod_{i=1}^N V_{\a_i}(z_i,\theta_i)>=\int \CD S\CD \bar S\prod_{i=1}^N e^{i\a_i S(z_i,\theta_i)}e^{-\CA(S,\bar S)}
\ee
calculated with the modified action should satisfy the neutrality condition:
\be\label{neutr}
\sum_{i=1}^N \a_i=2\a_0
\ee
(in order to cancel the cutoff dependence). For instance, the only non-zero two-point function is:
\be\nn
<V_\a(z_1,\theta_1)V_{2\a_0-\a}(z_2,\theta_2)>=\hat z_{12}^{-\a(\a-2\a_0)}
\ee
and therefore the vertices $V_\a$ and $V_{2\a_0-\a}$ represent fields with the same dimension:
\be\label{dimal}
\D(\a)=\D(2\a_0-\a)={1\ov 2}\a(\a-2\a_0).
\ee
From the action (\ref{mac}) we can derive the expressions for the stress energy tensor and the supercurrent:
\bea\label{curs}
T(z)&=&-{1\ov 2}((\p\phi)^2-\psi\p\psi)+i\a_0\p^2\phi,\\
\nn G(z)&=&i\psi\p\phi+2\a_0\p\psi
\eea
and the central charge is a function of the charge at infinity $\a_0$:
\be\label{cazero}
c={3\ov 2}-12\a_0^2.
\ee
Thus the different superconformal minimal models are parameterized by their charges at infinity:
\be\label{alzer}
\a_0^2={1\ov p(p+2)}.
\ee

In the Coulomb gas construction the 4-point functions of the NS fields (of the same dimension) are defined as \cite{bkt}:
\bea\label{nscoul}
&<\prod_{k=1}^4 \phi_\D(z_k,\theta_k)>=\oint_{C_{i}}\prod_{i=1}^{n-1}d\zeta_i d v_i \oint_{C_{j}}\prod_{j=1}^{m-1}d\eta_j d w_j\\
\nn &<V_\a(z_1,\theta_1)V_\a(z_2,\theta_2)V_{2\a_0-\a}(z_3,\theta_3)V_\a(z_4,\theta_4)\prod_{i=1}^{n-1}<V_{\a_+}(v_i,\zeta_i)\prod_{j=1}^{m-1}V_{\a_-}(w_j,\eta_j)>.
\eea

The superinvariant dimensionless screening operators:
\be
J_\pm=\oint_{C_\pm}d\theta d z V_{\a_\pm}(z,\theta)\sim \oint_{C_\pm} d z \psi(z)e^{i\a_\pm\phi(z)}
\ee
with charges and dimensions:
\be
\a_\pm=\a_0\pm \sqrt{\a_0^2+1},\qquad \D(\a_\pm)={1\ov 2}\a_\pm(\a_\pm-2\a_0)={1\ov 2}
\ee
are introduced in (\ref{nscoul}) in order to screen the extra charge $2\a$. They generate non trivial solutions of the neutrality condition (\ref{neutr}):
\be\label{chg}
\a_{n,m}={1\ov 2}\left( (1-n)\a_+ +(1-m)\a_-\right).
\ee

This quantization of the charges of the superfields leads to  the well known quantization of the dimensions of the minimal models:
\be\nn
\D_{n,m}={1\ov 8}\left((n\a_+ +m\a_-)^2-(\a_+ +\a_-)^2\right)
\ee
which exactly coincides with the Kac formula (\ref{nonereps}) if $n-m\in 2Z$.

The Ramond fields of the minimal models should have the same stress-energy tensor $T$ and the supercurrent $G$ as the NS fields. The only difference is that in this case $G(z)$ has to be an antiperiodic field and therefore we have to impose antiperiodic boundary conditions on the free Majorana field $\psi(z)$. Because of that the scalar field $\phi(z)$ and $\psi(z)$ cannot be combined in a superfield multiplet.

As it was explained in details in \cite{myn12}, the spin fields $\s$ and $\tilde\s$ (corresponding to the lowest energy states of $\psi(z)$) play an important role in the construction of the Ramond vertices. Namely,  one can define the latter as follows:
\be\label{rveron}
\tilde R_\a(z)=\tilde\s(z):e^{i\a\phi(z)}:.
\ee
The direct inspection based on the expression for the super stress-energy tensor (\ref{curs}), eq. (\ref{tilis}) and the Wick theorem for the free fields shows that the Ramond vertices satisfy (\ref{ramst}), (\ref{oper}) with dimensions given by:
\be\label{dram}
\D_R(\a)={1\ov 16}+{1\ov 2}\a(\a-2\a_0).
\ee
In fact, we have:
\be\nn
G(z_1)R_\a(z_2)={\a-\a_0\ov \sqrt{2}(z_1-z_2)^{3/2}}R_\a(z_2)+\ldots,
\ee
and a simple algebra gives:
\be\nn
\sqrt{1\ov 2}(\a-\a_0)=\pm\sqrt{\D_R(\a)-{c\ov 24}}.
\ee
Therefore the vertex (\ref{rveron}) form a correct representation of the Ramond algebra.

Accepting that the screening operators $J_\pm$ are the same as for the NS sector:
\be
J_\pm= \oint_{C_\pm} d z \psi(z)e^{i\a_\pm\phi(z)}
\ee
with antiperiodic $\psi(z)$, we can construct the correlation function of four Ramond fields (of equal dimensions) modifying the NS average procedure (\ref{nscoul}):
\bea\label{rcoul}
&<\prod_{k=1}^4 R_\D(z_k)>=\oint_{C_{i}}\prod_{i=1}^{n-1} d v_i \oint_{C_{j}}\prod_{j=1}^{m-1}d w_j\\
\nn &<R_\a(z_1)R_\a(z_2)R_{2\a_0-\a}(z_3)R_\a(z_4)\prod_{i=1}^{n-1} <\psi e^{i\a_-\phi}(v_i)\prod_{j=1}^{m-1}\psi e^{i\a_+\phi}(w_j)>.
\eea

Since the neutrality condition implies that one has again the same charge quantization (\ref{chg}) the dimensions (\ref{dram}) are quantized in accordance with the Kac formula (\ref{nonereps}):
\be
\D_{n,m}^R={1\ov 16}+{1\ov 8}\left((n\a_+ +m\a_-)^2-(\a_+ +\a_-)^2\right)
\ee
where now $n-m\in 2Z+1$. This screening procedure works well also in the case of mixed R-NS correlation functions and in the general case of multi-point functions.

The screening procedure (\ref{nscoul}) and the neutrality condition (\ref{neutr}) applied to the three-point functions generate the fusion rules for the fields of a given minimal model. In fact, the primary field $\phi_{x,y}$ which enters the OPE of two given fields $\phi_{n_1,m_1}$ and $\phi_{n_2,m_2}$ should have a non-zero 3-point function:
\be\nn
<\phi_{n_1,m_1}(z_1)\phi_{n_2,m_2}(z_2)\phi_{x,y}(z_3)>.
\ee
The $Z_2$ charge conservation implies the following qualitative description of the $N=1$ supersymmetric OPE algebra of fields:
\be\nn
[R][R]\sim [NS],\qquad [R][NS]\sim [R],\qquad [NS][NS]\sim [NS].
\ee

We begin with the fusion rules in the NS sector considering the correlation functions of three superfields. It is known that there exist two different structures in it, an even part and an odd one:
\bea\label{odev}
<&N&(z_1,\theta_1)N(z_2,\theta_2)N(z_3,\theta_3)>=\\
\nn &=&(\hat{z}_{12})^{\D_3-\D_1-\D_2}(\hat{z}_{13})^{\D_2-\D_1-\D_3}(\hat{z}_{23})^{\D_1-\D_2-\D_3}(a_1+a_2\eta),\\
\nn &\eta&=(\hat{z}_{12}\hat{z}_{13}\hat{z}_{23})^{-{1/2}}(\theta_1\hat{z}_{23}+\theta_2\hat{z}_{13}+\theta_3\hat{z}_{12}+\theta_1\theta_2\theta_3)
\eea
($a_1$ and $a_2$ are some constants). This structure gives rise to different fusion rules-odd and even-generated by the corresponding odd and even parts of the function
(\ref{odev}). This is explained in details in ref. \cite{myne1}.

In the Coulomb gas picture there exist three different ways to construct the 3-point function of the fields $N_{n_1,m_1}$, $N_{n_2,m_2}$, $N_{n_3,m_3}$ depending on which vertex has conjugate charge. In each case there exists a number of screening operators which assure the neutrality condition (\ref{neutr}). This screening procedure leads to a chain of equations for the unknown charge $\a_{x,y}$. We have to take the common solution of these equations which is:
\bea\nn
x&=&|n_1-n_2|+1, |n_1-n_2|+3,\ldots,n_1+n_2-1,\\
\nn y&=&|m_1-m_2|+1, |m_1-m_2|+3,\ldots,m_1+m_2-1.
\eea
This gives the fusion rules for the fields $N_{n_1,m_1}$, $N_{n_2,m_2}$ \cite{myne1}:
\be
[N_{n_1,m_1}][N_{n_2,m_2}]=\sum_{x=|n_1-n_2|+1}^{n_1+n_2-1}\sum_{y=|m_1-m_2|+1}^{m_1+m_2-1}[N_{x,y}]
\ee
($[]$ stays for the conformal family of the corresponding field). The even fusion rules are recovered when there is an even number of screening operators, i.e. $x+y\in 2Z_+$ while the odd ones correspond to an odd number of the latter $x+y\in 2Z_++1$ \cite{myne1}.

In order to find the fusion rules of two Ramond fields $R_{n_1,m_1}$ and $R_{n_2,m_2}$ we have to look at the non-zero 3-point functions with the NS superfield $N_{x,y}$:
$<R_{n_1,m_1}R_{n_2,m_2}N_{x,y}>$. The same procedure as the one described above leads to the following FR's of two R-fields:
\be
[R_{n_1,m_1}][R_{n_2,m_2}]=\sum_{x=|n_1-n_2|+1}^{n_1+n_2-1}\sum_{y=|m_1-m_2|+1}^{m_1+m_2-1}[N_{x,y}].
\ee
This result was first achieved in \cite{myn12}. The corresponding mixed FR's $[R][NS]\sim [R]$ are direct consequences of the FR's we already found. The set of these FR's completes the structure of the associative OPE algebra of fields of the corresponding supersymmetric minimal models.

Let us now turn to the computation of the four point functions of the Ramond fields. Using the vertex representation and the screening procedure we can express them in the form:
\bea\label{rf}
&<\prod_{k=1}^4 R_{n_k,m_k}(z_k)>=\oint_{C_{i}}\prod_{i=1}^{n-1} d v_i \oint_{C_{j}}\prod_{j=1}^{m-1}d w_j\\
\nn &<\s(z_1)\s(z_2)\s(z_3)\prod_{i=1}^{n-1} \psi(v_i)\prod_{j=1}^{m-1}\psi(w_j)\s(z_4)>\times\\
\nn &\times <e^{i\a_{n_1,m_1}\phi(z_1)}e^{i\a_{n_2,m_2}\phi(z_2)}e^{i\a_{n_3,m_3}\phi(z_3)}e^{i(2\a_0-\a_{n_4,m_4})\phi(z_4)}\\
\nn &\prod_{i=1}^{n-1} e^{i\a_-\phi(v_i)}\prod_{j=1}^{m-1} e^{i\a_+\phi(w_j)}>
\eea
The second factor in the integrand is the well-known multi-point function of the modified Coulomb system:
\be\nn
<\prod_{k=1}^N e^{i\a_k\phi(z_k)}>=\prod_{l< n=1}^N (z_{ln})^{\a_n\a_l},\qquad \sum_{i=1}^N \a_i=2\a_0,
\ee
while the first factor is given by the solution to the recursive equation for the Ising model (\ref{wic}) (called below $G^{(n,m)}(z,v_i,w_j)$). Then the four point function of the Ramond fields (of same dimension and charge $\a=\a_{n,m}$) takes the form:
\bea
&<&R_{n,m}(\infty)R_{n,m}(z)R_{n,m}(1)R_{n,m}(0)>=\\
\nn &=&z^{\a^2}(1-z)^{\a(2\a_0-\a)}\oint_{C_{i}}\prod_{i=1}^{n-1} d v_i \oint_{C_{j}}\prod_{j=1}^{m-1}d w_j G^{(n,m)}(z,v_i,w_j)\times\\
\nn &\times&\prod_{l<k=2}^{n-1}v_{lk}^{\a_-^2}\prod_{s<t=2}^{m-1}w_{st}^{\a_+^2}\prod_{i=1}^{n-1}\prod_{j=1}^{m-1}((v_i-z)v_i)^{\a_-\a}(v_i-1)^{\a_-(2\a_0-\a)}\times\\
\nn &\times&((w_j-z)w_j)^{\a_+\a}(w_j-1)^{\a_+(2\a_0-\a)}(v_i-w_j)^{\a_-\a_+}.
\eea
The integration contours $C_i$ are fixed by the branch cut singularities of the integrand. Thus, for the general expression of the four point function of the Ramond fields we should take a linear combination of all four point functions corresponding to the possible independent choices of contours $C_i$. In the simplest example of one screening operator, say $J_+$, there are two independent contours, one from $0$ to $z$ and the other from 1 to $\infty$, and the corresponding integrals are expressed in terms of hypergeometric function. As a result we get the following result for the 4-point function of the corresponding field $R_{1,2}$ (after taking the sum of $\qquad$ $<RRRR>$ and $<\tilde R R\tilde R R>$):
\be
G^p_{12}(z)=z^{-(p+12)/8p}(1-z)^{(p+4)/8p}\sum_{i=1}^4 A^i W_i(z)
\ee
where (using the notation $h=1/p$ and $B(x,y)={\Gamma(x)\Gamma(y)\ov \Gamma(x+y)}$):
\bea\label{ww}
W_1(z)&=&\sqrt{1+\sqrt{1-{1\ov z}}}\big(B(1-h,h)\sqrt{z}F(1+h,-h,1-2h;z)-\\
\nn &-&B(-h,-h)\sqrt{1-z}F(1+h,-h,-2h;z)\big),\\
\nn W_2(z)&=&\sqrt{1+\sqrt{1-{1\ov z}}}\big(B(1+3h,-h)\sqrt{z}z^{2h}F(h,1+h,1+2h;z)+\\
\nn &+&B(2+3h,-h)\sqrt{1-z}z^{-2-2h}F(1+h,2+3h,2+2h;z)\big),\\
\nn W_3(z)&=&\sqrt{1-\sqrt{1-{1\ov z}}}\big(B(1-h,h)\sqrt{z}F(1+h,-h,1-2h;z)+\\
\nn &+&B(-h,-h)\sqrt{1-z}F(1+h,-h,-2h;z)\big),\\
\nn W_4(z)&=&\sqrt{1-\sqrt{1-{1\ov z}}}\big(B(1+3h,-h)\sqrt{z}z^{2h}F(h,1+h,1+2h;z)-\\
\nn &-&B(2+3h,-h)\sqrt{1-z}z^{-2-2h}F(1+h,2+3h,2+2h;z)\big).
\eea

In the same way we can calculate the 4-point function of the Ramond fields $R_{2,1}$ since it needs only the insertion of one $J_-$ screening operator. From the branch cut analysis of the integrand one can see that there are two independent integration contours for each $p$. Denoting $a=1/(p+2)$ we have four independent solutions:
\be
G^p_{21}(z)=z^{-(p-10)/8(p+2)}(1-z)^{(p-2)/8(p+2)}\sum_{i=1}^4 A^i Y_i(z)
\ee
where:
\bea\label{yy}
Y_1(z)&=&\sqrt{1+\sqrt{1-{1\ov z}}}\big(B(a,1+a)\sqrt{z}F(1-a,a,1+2a;z)-\\
\nn &-&B(a,a)\sqrt{1-z}F(1-a,a,2a;z)\big),\\
\nn Y_2(z)&=&\sqrt{1+\sqrt{1-{1\ov z}}}\big(B(1-3a,a)\sqrt{z}z^{-2a}F(-a,1-3a,1-2a;z)+\\
\nn &+&B(2-3a,a)\sqrt{1-z}z^{1-2a}F(1-a,2-3a,2-2a;z)\big),\\
\nn Y_3(z)&=&\sqrt{1-\sqrt{1-{1\ov z}}}\big(B(a,1+a)\sqrt{z}F(1-a,a,1+2a;z)+\\
\nn &+&B(a,a)\sqrt{1-z}F(1-a,a,2a;z)\big),\\
\nn Y_4(z)&=&\sqrt{1-\sqrt{1-{1\ov z}}}\big(B(1-3a,a)\sqrt{z}z^{-2a}F(-a,1-3a,1-2a;z)-\\
\nn &-&B(2-3a,a)\sqrt{1-z}z^{1-2a}F(1-a,2-3a,2-2a;z)\big).
\eea

Up to now we dealt with the one-dimensional, or chiral, fields. In the 2D minimal models the real two-dimensional fields are constructed as a product of left and right chiral fields
$\phi(z,\bar z)=\phi_\D(z)\bar\phi_{\bar \D}(\bar z)$. Then the true 2D four-point correlation function should obey the crossing symmetry relations. For scalar fields, with $\D=\bar\D$, they have the following form:
\be
\nn G_{nm}^{lk}(z,\bar z)=G_{nl}^{mk}(1-z,1-\bar z)=z^{-2\D_n}\bar z^{-2\bar\D_n}G_{nk}^{lm}({1\ov z},{1\ov\bar z})
\ee
where $G_{nm}^{lk}(z,\bar z)=<\phi_k(\infty)\phi_l(1)\phi_n(z,\bar z)\phi_m(0)>$.

It is known \cite{df} that the crossing symmetry is equivalent to the monodromy invariance of the 4-point functions. Denoting again by $\{W_i(z)\}$ the set of functions corresponding to the possible independent contours (for the field $R_{1,2}$) one can write the 2D 4-point function in the form:
\be
\nn G(z,\bar z)=\sum_{i,j}I_{i,j}W_i(z)\bar W_j(\bar z),
\ee
where $I_{i,j}$ are unknown coefficients. Since the functions $\{W_i(z)\}$ have branch cut singularities in the points $0,1,\infty$ they transform nontrivially along closed curves enclosing the singular points:
\be
\nn W_i(z)\rightarrow (g_l)_{ik}W_k(z),\qquad l=0,1,\infty.
\ee
The matrices $g_l$ generate the monodromy group of the functions $W_i(z)$. The correlation functions of the scalar fields should be uniquely defined in the 2D space, i.e. they have to be monodromy invariant \cite{df}:
\bea
\nn G(z,\bar z)&=&\sum_{i,j}I_{i,j}W_i\bar W_j=\sum_{i,j}\sum_{k,l}I_{i,j}(g_l)_{ik}W_k(\bar g_l)_{jp}\bar W_p=\\
\nn &=&\sum_{k,l}\big(\sum_{i,j}(g_l^t)_{ki}I_{ij}(\bar g_l)_{jl}\big)W_k\bar W_l.
\eea
Thus we obtain the following equation for the unknown coefficients $I_{ij}$:
\be
\nn I_{kp}=\sum_{ij}(g_l^t)_{ki}I_{ij}(\bar g_l)_{jp}.
\ee
These equations determine $I_{ij}$ up to an overall factor related to the normalization of the two-point functions. Solving the latter one obtains for example the 2D correlation function of four Ramond fields $R_{1,2}$ for any value of the central charge $c_p$:
\bea\label{rot}
G^p_{1,2}(z,\bar z)&=&\la_{1,2}(p)|z|^{-(p+12)/4p}|1-z|^{(p+4)/4p}\times\\
\nn &\times&\big[ W_1\bar W_1+ W_3\bar W_3+\big( 4\cos^2({\pi\ov p})-1\big)(W_2\bar W_2+ W_4\bar W_4)\big].
\eea
where $W_i(z)$ are those defined in (\ref{ww}). In the case of the Ramond fields $R_{2,1}$ similar calculations lead to the following 4-point function:
\bea\label{rto}
G^p_{2,1}(z,\bar z)&=&\la_{2,1}(p)|z|^{-(10-p)/4(p+2)}|1-z|^{(p-2)/4(p+2)}\times\\
\nn &\times&\big[ Y_1\bar Y_1+ Y_3\bar Y_3+\big( 4\cos^2({\pi\ov p+2})-1\big)(Y_2\bar Y_2+ Y_4\bar Y_4)\big].
\eea
and $Y_i$ were defined in (\ref{yy}).

Analogously, the monodromy invariant expression for the correlator involving the Ramond fields $R_{1,2}$ and $R_{2,1}$
\be
\nn G^p_{12,21}(z,\bar z)=<R_{1,2}(\infty)R_{1,2}(z,\bar z)R_{2,1}(1)R_{2,1}(0)>
\ee
is
\bea
G^p_{12,21}(z,\bar z)&=&{1\ov 8}|z|^{-3/4}|1-z|^{-3/4}(N_1\bar N_1+N_2\bar N_2),\\
\nn N_1(z)&=&\sqrt{1+\sqrt{1-{1\ov z}}}\big( {1-2z+2\sqrt{z(1-z)}\ov \sqrt{z}}\big),\\
\nn N_1(z)&=&\sqrt{1-\sqrt{1-{1\ov z}}}\big( {1-2z-2\sqrt{z(1-z)}\ov \sqrt{z}}\big)
\eea
with the usual normalization of the 2-point functions to one.

Looking at the singularities of the 4-point functions in the limit $z\rightarrow 1$ we can recover the OPE:
\bea
R_{21}R_{21}&\sim& 1+c_1(p)\Phi_{31}\\
\nn R_{12}R_{12}&\sim& 1+c_2(p)\Phi_{13}\\
\nn R_{12}R_{21}&\sim& c_3(p)\Phi_{22}
\eea
and the structure constants $c_1, c_2, c_3$. In fact, using the OPE:
\be\nn
\phi_{\D_1}(z,\bar z)\phi_{\D_2}(1)=\sum_k {c_k\ov |1-z|^{2(\D_1+\D_2-\D_k)}}\phi_k(1)
\ee
with a normalization fixed by the identity, i.e. $c_0=1$, we have:
\bea
\nn G(z,\bar z)&=&<R_{\D}(\infty)R_\D(z,\bar z)R_\D(1)R_\D(0)>\sim\\
\nn &\sim&\sum_k {c_k\ov |1-z|^{2(2\D-\D_k)}}<R_{\D}(\infty)\Phi_k(1)R_\D(0)>\sim\\
\nn &\sim&\sum_k {c_k^2\ov |1-z|^{2(2\D-\D_k)}}.
\eea
Hence, the contribution of the different conformal families is identified by the power singularities.

In this way we obtain the normalization and the structure constants. Explicitly we get:
\bea
\nn \la_{2,1}(a)&=&{1\ov 32 \cos^2(\pi a)}\big|{\Gamma(-a)\ov \Gamma(a)\Gamma(-2a)}\big|^2,\\
\nn c_1(a)&=&{1\ov2}{\Gamma(2a)\Gamma((p-1)a)\ov \Gamma(-2a)\Gamma((p+3)a)}\sqrt{4\cos^2(\pi a)-1},\\
\nn \la_{1,2}(h)&=&{1\ov 32 \cos^2(\pi h)}\big|{\Gamma(h)\ov \Gamma(-h)\Gamma(2h)}\big|^2,\\
\nn c_2(h)&=&{3\ov2}{\Gamma(3h)\Gamma(-2h)\ov \Gamma(2h)\Gamma(-h)}\sqrt{4\cos^2(\pi h)-1}.
\eea
The mixed structure constant is simply $c_3(p)={1\ov 2}$.

Using the mixed R-NS 4-point functions we can also extract the structure constants of the NS field $\Phi_{31}$ with itself. For that purpose we compute the correlation function:
\be
G^p_{31}(z)=<\Phi_{3,1}(\infty)\Phi_{3,1}(1)R_{2,1}(z)R_{2,1}(0)>.
\ee
In terms of the Coulomb gas construction it is given by an integral with a single screening operator insertion:
\bea
\nn G^p_{31}(z)&=&\sum_i\oint_{C_i}dv<V_{2\a_0-\a_{31}}(\infty)V_{\a_{31}}(1)V_{\a_{21}}(z)V_{\a_{21}}(0)V_{\a_-}(v)>=\\
\nn &=&z^{\a_{21}^2}(1-z)^{\a_{31}\a_{21}}\sum_i\oint_{C_i}dv v^{\a_{-}\a_{21}}(1-v)^{\a_{31}\a_{-}}(z-v)^{\a_{-}\a_{21}}<\s(z)\psi(v)\s(0)>=\\
\nn &=&z^{\a_{21}^2+3/8}(1-z)^{\a_{31}\a_{21}}\sum_i\oint_{C_i}dv v^{\a_{-}\a_{21}-1/2}(1-v)^{\a_{31}\a_{-}}(z-v)^{\a_{-}\a_{21}-1/2}
\eea
and is therefore expressed in terms of hypergeometric functions. The monodromy invariant solution with the usual normalization is:
\bea\label{fito}
G^p_{31}(z,\bar z)&=&|z|^{(5p+6)/4(p+2)}|1-z|^{p/(p+2)}\big\{|z^{-2pa} F(pa,a,2a,z)|^2 +\\
\nn &+&{s(2a)s(4a)\ov s^2(a)}\big({\Gamma^2(2a)\Gamma(2pa)\ov \Gamma^2(a)\Gamma((2p+2)a)}\big)^2 |F((p+1)a,2pa,(2p+2)a,z)|^2\big\}
\eea
($s(x)=\sin(\pi x)$). In the limit $z\rightarrow 0$ the first term in (\ref{fito}) gives the contribution of the identity family while the second one - the contribution of the $\Phi_{31}$ operator. In this way we get:
\bea
\nn G^p_{31}&\sim&c_1(p)|z|^{-2(2\D_{21}-\D_{31})}<\Phi_{31}(\infty)\Phi_{31}(1)\Phi_{31}(0)>\sim\\
\nn &\sim&c_1(p)\big(a_1+a_2|\eta|^2\big)|z|^{-2(2\D_{21}-\D_{31})}.
\eea
Since the value of $c_1(p)$ is known we can extract the structure constants $a_1$ and $a_2$ of the even and odd part of the field $\Phi_{31}$:
\bea
\nn a_1&=&0\\
\nn a_2&=&{\Gamma^3(2a)\Gamma^2(2-4a)\ov \Gamma(4a)\Gamma(1-4a)\Gamma(2a-1)\Gamma^2(2-2a)}\sqrt{{\Gamma(1-a)\Gamma(3a)\ov \Gamma^3(a)\Gamma(1-3a)}}.
\eea

At the end of this Section we consider the case of the $N=1$ superconformal minimal models on $Z_2$ hyperelliptic supersurfaces $SX_g^{(2)}$ of genus $g$. The problem is to find the appropriate minimal models on the branched supersphere which describe the $N=1$ minimal models on $SX_g^{(2)}$. We consider the case of a split $SX_g^{(2)}$ which can be defined as a double cover of the supersphere branched over $2g+2$ points $a_i$. The split super hyperelliptic map is given by:
\be\label{hymap}
w_i(z,\theta)=\sqrt{z-a_i},\qquad \chi_i(z,\theta)={\theta\ov \sqrt{2}(z-a_i)^{1/4}}
\ee
whose monodromy group is $Z_4$.

Under (\ref{hymap}) the stress tensor $T(z)$ and the supercurrent $G(z)$ on $SX_g^{(2)}$ get mapped onto $T^{(k)}(z)$ and $G^{(k)}(z)$ ($k=0,1$) defined on the corresponding sheets. For a given $k$ they satisfy the usual OPE's and for $k\ne m$ the OPE contains only regular terms. Consequently, the central charge is twice that of the sphere one $\tilde c=2c$.
It is more useful to define diagonal currents through :
\bea\label{diapi}
T&=&T^{(0)}+T^{(1)},\qquad T^\dagger=T^{(0)}-T^{(1)},\\
\nn G&=&G^{(0)}-iG^{(1)},\qquad G^\dagger=G^{(0)}+iG^{(1)}.
\eea
The $Z_4$ charge of $T$, $G$, $T^\dagger$ and $G^\dagger$ is $0,1,2$ and $3$ respectively.

The primary states and fields are divided into four sectors  $V_{[l]}(z)$ ($l=0,1,2,3$) according to the $Z_4$ boundary conditions for the generators. The primary fields $V_{[l]}(z)$ realize these boundary conditions through their OPE's with the generators, defining in this way the modes of the latter in the various sectors. The direct comparison with the parafermionic algebras \cite{zfpf} shows that they represent the $Z_4$-disorder sectors of the generalized parafermionic $Z_4^{p=2}$ algebra. But for $SX_g^{(2)}$ the $Z_4$ monodromies do not exhaust all the discrete symmetries. In fact, the sheet-interchanging $Z_2$ symmetry acts as a charge conjugation for the generators, introducing in such way a C-disorder sector and leading to the $D_4^{p=2}$ parafermionic symmetry. As a consequence, we have to introduce, in addition to $V_{[l]}(z)$, new fields $W_l(z)$ representing this C-disorder sector.
In conclusion, we are led to the statement that the $N=1$ superconformal algebra on split $SX_g^{(2)}$ maps into the $D_4^{p=2}$ parafermionic algebra \cite{zfpf1} on the branched supersphere.
Therefore one can construct the superconformal models on $SX_g^{(2)}$ in terms of the minimal models of the  parafermionic algebra. The unitary degenerate representations of this algebra can be obtained by the GKO method \cite{gko} which results in the quantization of the central charge:
$\tilde c=3-{24/p(p+2)}=2c,\quad p=3,4,\ldots .$
The easiest way to construct these representations is by taking the usual (supersymmetric) Coulomb gas realization of $T^{(k)}$ and $G^{(k)}$ in terms of free Majorana fermions $\psi^{(k)}$ and free scalar fields $\phi^{(k)}$. It is actually more convenient to use again a diagonal basis $\phi^{(\dagger)}$, $\psi^{(\dagger)}$, analogous to that of (\ref{diapi}).

We note that with respect to the Virasoro subalgebra generated by $T$ the Coulomb gas system splits into a sphere model with $c_{sp}=1-24/p(p+2)$ described by $\phi$, a $Z_2$-orbifold with $c_{orb}=1$ corresponding to $\phi^\dagger$ and the $X_g^{(2)}$ Ising model with $c_{Is}=1$. This suggests that the vertex operators representing the different
$D_4^{p=2}$ sectors can be constructed as products of $\exp(\a\phi)$, $\exp(\b\phi^\dagger)$, the $Z_2$ twist fields $\s_\epsilon$ and certain Ising model fields. For the fields
$V_{[l]}$ from the order sector we have for example:

\no NS and R sectors:
\be\label{nsnb}
V_{[0]}=\exp(a_0\phi+b_0\phi^\dagger),\quad V_{[2]}=V_{[2]}^\psi \exp(a_2\phi+b_2\phi^\dagger),
\ee
NS branched sector:
\be\label{nsbr}
V_{[1]}=V_{[1]}^\psi \s_0 \exp(\a\phi),\quad  V_{[3]}=V_{[3]}^\psi \s_1 \exp(\a\phi)
\ee
and similarly for the C-disorder fields. Here we denoted by $V_{[1]}^\psi$ etc. the fields from the Ising model. They can be constructed in the same way as the one we discuss here specified to the simpler $N=0$ non-supersymmetric theory \cite{myne3}.

We are going now to construct the screening operators. As usual, these operators are particular NS vertices whose contour integrals  are invariant under the action of the generators. In our case these requirements are satisfied by:
\bea\nn
\hat Q_+^\pm&=&\oint dz(\psi+\psi^\dagger)\exp(\hat a_\pm(\phi+\phi^\dagger)),\quad \hat Q_-^\pm=\oint dz(\psi-\psi^\dagger)\exp(\hat a_\pm(\phi-\phi^\dagger)),\\
\nn \hat a_\pm^2&-&\a_0\hat a_\pm={1\ov 4},\quad \hat a_\pm=\half (\a_0\pm\sqrt{\a_0^2+1}),\quad \hat a_++\hat a_-=\a_0,\quad \hat a_+\hat a_-=-{1\ov 4}.
\eea
The vertices representing the primary fields in the minimal models have the form (\ref{nsnb}), (\ref{nsbr}) but with quantized charges. Of special interest for us here are the fields from the NS branched sectors $l=1$ and $l=3$, in which case we have:
\bea\nn
\a_{n,m}&=&\half(2-n)\hat a_++\half(2-m)\hat a_-,\quad n-m\in 2Z,\\
\nn \D_{n,m}^{[1]}&=&\D_{n,m}^{[3]}=\a_{n,m}^2-2\a_0\a_{n,m}+{3\ov 32}={\left(n(p+2)-mp\right)^2-16\ov 16p(p+2)}+{3\ov 32},\\
\nn 1\le &n&\le p-1,\qquad \nn 1\le m\le p+1.
\eea

Our final goal will be the construction of the partition function of $N=1$ minimal models on split $SX_g^{(2)}$. It turns out that for this reason we need the lowest energy NS branching operators $V_{1,1}^{(1)}$ and $V_{1,1}^{(3)}$ only. In fact we have to calculate the vacuum expectation value $<I>_g$ of the identity $I$, $\D_I=0$ on $SX_g^{(2)}$,
with no other marked points on $SX_g^{(2)}$, i.e. each point on $SX_g^{(2)}$ represents the NS vacua $V_{1,1}^g$, $\D^g=0$. By doing the map (\ref{hymap})
$w^2(z)=z-a_i, i=1,\ldots, 2g+2$ we produce $2g+2$ branching operators $V_{1,1}^{\epsilon_i}(a_i)\leftarrow I$ and therefore:
\be\label{parti}
Z_g(a_i)\equiv <I>_g=\langle \prod_{i=1}^{2g+2}V_{1,1}^{\epsilon_i}(a_i)V_{1,1}^{\epsilon_i}(\bar a_i)\rangle_{g=0},\quad \sum\epsilon_i=0 \mod 4.
\ee
In this way the calculation of the partition function $Z_g$ reduces to the problem of the construction of the $(2g+2)$-point function of the primary fields $V_{1,1}^{\epsilon_i}(a_i)$.

We restrict ourselves to the case of $g=2$ split supersurface. The standard screening procedure leads to the following expression for the "chiral" part of the $g=2$ partition function:
\bea
Y^{g=2}_{\CC^\pm,p}&=&\langle\prod_{i=1}^6 V_{1,1}^{\epsilon_i}(a_i)(\tilde Q_\pm^-)^r(\hat Q_\pm^-\hat Q_\pm^+)^{rp/2-1}\rangle=\\
\nn &=&\prod_{l=1}^L\prod_{m=1}^M\oint_{\CC_l^-}du_l\oint_{\CC_m^+}dv_m\langle\prod_{i=1}^6\exp(\half \a_0\phi(a_i))\exp(\hat a_-\phi(u_l))\exp(\hat a_+\phi(v_m))\rangle\times\\
\nn &\times&\langle \prod_{i=1}^6 \s_{\epsilon_i}(a_i)\exp(\pm\hat a_-\phi^\dagger(u_l))\exp(\pm\hat a_+\phi^\dagger(v_m))\rangle
          \langle \prod_{i=1}^6 V_{\epsilon_i}^\psi(a_i)\psi^{(k_l)}(u_l)\psi^{(k_m)}(v_m)\rangle
\eea
where $L=\half r(p+2)-1,M=\half rp-1,r=1$ for even $p$ and $r=2$ for odd $p$, $\sum\epsilon_i=0 \mod 4$ and:
\be\nn
\tilde Q_\pm^-=\oint_\CC dx(\psi(x)\pm\psi^\dagger(x))\exp(-\half p\a_0(\phi(x)\pm\phi^\dagger(x))),\quad 2\hat a_-=-p\a_0,\quad k_i=0,1,
\ee
$\psi^{(0)}$, $\psi^{(1)}$ are Ising fermions on $X_{g=2}^{(2)}$.
It is clear that the integrand splits into a product of three correlation functions. The first one, involving the exponential fields is trivial. The correlation function
$\tilde G(p_a,a_i,x_k)$ of $2g+2=6$ twist fields $\s_{\epsilon_i}$ and an arbitrary number of untwisted vertices $\exp(q\phi^\dagger)$ can be computed using the methods explained in \cite{alztw,myne3}. It is expressed, up to power-like prefactors, in terms of abelian differentials of first and third kind. It remains to construct the multipoint Ising fermion correlation function on $X_{g=2}^{(2)}$. As noticed above, this function is computable using the technics we used here for the case of $N=0$ non-supersymmetric case with $c=1$. Combining all these ingredients and satisfying the conditions for monodromy invariance of the six-point function (\ref{parti}) we obtain an integral representation for the $g=2$ partition functions of the $N=1$ superconformal minimal models $Z_{g=2}$.

\subsection{RG flow in $N=1$ minimal models}

In this section we consider a minimal superconformal theory $SM_p$ perturbed by the least relevant field.
Let us remind that the fields in the NS sector are organized in 2D superfields:
\be
\Phi(z,\bar z,\th,\bar\th)=\phi+\th\psi+\bar\th\bar\psi+\th\bar\th\tilde\phi.
\ee
 The first (and the last) component of a spinless superfield of dimensions $\D=\bar\D$ ($\D+\hf=\bar\D+\hf$) is expressed as a product of ``chiral fields" depending on $z$ and $\bar z$, respectively. We use the same notations $\phi$ and $\tilde\phi$ below for these chiral components. If we fix the two-point function of the first component $\phi$ to one, that of the second components is $(2\D)^2$ by supersymmetry. Since it is assumed that these functions are all equal to one in the renormalization procedure, we have to normalize the second component $\tilde\phi\rightarrow {1\ov 2\D}\tilde\phi$.

We will consider the superminimal model $SM_p$ with $p\rightarrow\infty$  perturbed by the least relevant field $\tilde\phi=\tilde\phi_{1,3}$ of dimension $\D=\D_{1,3}+\hf=1-\e$, $\e={2\ov p+2}\rightarrow 0$:
\be
\nn {\cal L}(x)={\cal L}_0(x)+\l \tilde\phi(x).
\ee
It is obvious that this theory is also supersymmetric, since the perturbation can be written as a covariant super-integral over the superfield $\Phi_{1,3}$.

The two-point function of arbitrary fields up to the second order is then given by:
\bea\label{secex}
<\phi_1(x)\phi_2(0)>&=&<\phi_1(x)\phi_2(0)>_0-\l\int <\phi_1(x)\phi_2(0)\tilde\phi(y)>_0 d^2y+\\
\nn &+&{\l^2\ov 2}\int <\phi_1(x)\phi_2(0)\tilde\phi(x_1)\tilde\phi(x_2)>_0 d^2x_1 d^2x_2 +\ldots
\eea
where $\phi_1$, $\phi_2$ can be the first or the last components of a superfield or Ramond fields of dimensions $\D_1$, $\D_2$.
Since the first order corrections were considered in \cite{pogsc1}, we will focus on the second order.

One can use the conformal transformation properties of the fields to bring the double integral to the form:
\bea\label{bint}
&\int&<\phi_1(x)\phi_2(0)\tilde\phi(x_1)\tilde\phi(x_2)>_0 d^2x_1d^2x_2 =\\
\nn &=&(x\bar x)^{2-\D_1-\D_2-2\D}\int I(x_1) <\tilde\phi(x_1)\phi_1(1)\phi_2(0)\tilde\phi(\infty)>_0 d^2x_1
\eea
where:
\be
\nn I(x)=\int |y|^{2(a-1)}|1-y|^{2(b-1)}|x-y|^{2c} d^2y
\ee
and $a=2\e+\D_2-\D_1$,$b=2\e+\D_1-\D_2$, $c=-2\e$. It is well known that the integral for $I(x)$ can be expressed in terms of hypergeometric functions:
\bea\label{ix}
I(x)&=&{\pi\g(b)\g(a+c)\ov \g(a+b+c)}|F(1-a-b-c,-c,1-a-c,x)|^2+\\
\nn &+&{\pi\g(1+c)\g(a)\ov \g(1+a+c)}|x^{a+c}F(a,1-b,1+a+c,x)|^2
\eea

This form is useful for evaluating $I(x)$ near $x=0$.
Using the transformation properties of the hypergeometric functions, (\ref{ix}) can be rewritten as a function of $1-x$ and ${1\ov x}$ which is suitable for the investigation of $I(x)$ around the points $1$ and $\infty$, respectively.

It is clear that the integral (\ref{bint}) is singular. We follow the
regularization procedure proposed in \cite{pogsc2} . It consists basically
in cutting discs in the two-dimensional surface of radius $l$
(${1\ov l}$) around singular points $0$, $1$ ($\infty$):
$D_{l,0}=\{x\in C,|x|<l\}$, $D_{l,1}=\{x\in C,|x-1|<l\}$,
$D_{l,\infty}=\{x\in C,|x|>1/l\}$ with $0<l_0\ll l\ll 1$ where $l_0$ is the ultraviolet cut-off.
Clearly $l$ should be canceled in the calculations and should not appear in the
final result. We call the region outside these discs as $\O_{l,l_0}$ where the integration is well-defined.
 It is useful to do this integration in
radial coordinates. Since the correlation function exhibits poles
only at the points $0$ and $1$, the phase integration can be
performed by using residue theorem and the resulting rational
integral in the radial direction is straightforward. Near the
singular points one can use the OPE. In doing that it turns out that
we count twice two lens-like regions around the point $1$ so we have
to subtract those integrals. Explicit formulas as well as a more
detailed explanation can be found in \cite{pogsc2}.

Let us start with the correlation function that enters in the
integral (\ref{bint}) for the case of NS fields. As we explained in the previous section this could be done using the Coulomb gas construction. We will need however an explicit expression that could be integrated. So we will adopt here another strategy.

The basic ingredients for the computation of the
four-point correlation functions are the conformal blocks. These are
quite complicated objects in general and closed formula were not
known. It was argued that they coincide (up to factors)
with the instanton partition function of certain $N=2$ YM theories
on ALE spaces. This was proved by a recurrence
relation satisfied by the conformal blocks \cite{bbnz,vbrec} which we will use here. We need the expressions for
the first few levels conformal blocks in order to have a guess for
the limit $\e\rightarrow 0$.

The chiral components of the fields obey the OPEs:
\bea\label{bops}
\phi_1(x)\phi_2(0)&=&x^{\D-\D_1-\D_2}\sum_{N=0}^\infty x^N
C_N \phi_\D(0)\\
\nn \tilde\phi_1(x)\phi_2(0)&=&x^{\D-\D_1-\D_2-1/2}\sum_{N=0}^\infty x^N \tilde
C_N \phi_\D(0)\\
\nn \phi_1(x)\tilde
\phi_2(0)&=&x^{\D-\D_1-\D_2-1/2}\sum_{N=0}^\infty x^N \tilde C'_N
\phi_\D(0)\\
\nn \tilde \phi_1(x)\tilde
\phi_2(0)&=&x^{\D-\D_1-\D_2-1}\sum_{N=0}^\infty x^N C'_N \phi_\D(0)
\eea
where $C_N$'s are polynomials of order $N$ in the generators of the
superconformal algebra $L_{-k}$ and $G_{-\a}$ ($k,\a>0$) with coefficients depending on the
dimensions $\D$, $\D_1$, $\D_2$, which we omitted,
usually called chain vectors. Here $N$ runs over all
nonnegative integers or half-integers depending on
the fusion rules of $SM_p$.

Acting by positive mode generators on the both sides of these OPEs and using the super-conformal transformation properties gives the chain equations for $L$'s:
\be
L_k C_N=(\D+k\D_1-\D_2+N-k)C_{N-k}
\ee
(here $C$ is any of of the chain vectors with the corresponding
dimensions of the fields) and for $G$'s:
\bea\label{chr}
G_k C_N&=&\tilde C_{N-k}\\
\nn G_k\tilde C_N &=&(\D+2k\D_1-\D_2+N-k)C_{N-k}\\
\nn G_k\tilde C'_N &=&C'_{N-k}\\
\nn G_k C'_N&=&(\D+2k\D_1-\D_2+N-k-\hf)\tilde C'_{N-k}
\eea
for $k>\hf$, and:
\bea
\nn G_\hf \tilde C'_N&=&2\D_2C_{N-\hf}+C'_{N-\hf}\\
\nn G_\hf  C'_N&=&-2\D_2\tilde C_{N-\hf}+(\D+\D_1-\D_2+N-1)\tilde C'_{N-\hf}.
\eea

There are two independent constants at the zeroth level in the OPEs (\ref{bops}), the other two are expressible through them:
\be
\tilde C'_0=-\tilde C_0,\hskip1cm C'_0=(\D-\D_1-\D_2)C_0.
\ee
The above chain relations could be solved order by order. As
mentioned before, in \cite{bbnz,vbrec} a recursion relation for the chain
vectors can be also found.

The conformal blocks are readily obtained by the chain vectors.  Presented as
vectors in the basis of $L$'s and $G$'s, the conformal block can be
expressed as:
\be
F(\D,\D_{i})=\sum_{N=0}^\infty x^N F_N =\sum_{N=0}^\infty x^N C_N(\D,\D_3,\D_4)S_N^{-1}C_N(\D,\D_1,\D_2)
\ee
where $S_N$ is the Shapovalov matrix at level $N$. What of $C_N$'s appear depends on the external fields involved.

The conformal blocks are in general quite complicated objects.
Fortunately, in view of the renormalization scheme and the
regularization of the integrals, we need to compute them here only
up to the zero-th order in $\e$. This simplifies significantly the
problem.

Once the conformal blocks are known, the correlation function of spinless fields for our $SM_p$ models is written as:
\be\nn
\sum_n C_n|F(\D_n,\D_{i})|^2
\ee
where the range of $n$ depends on the fusion rules and $C_n$ is the
corresponding structure constant. In what follows we compute the conformal blocks up to sufficiently high
level and then check also the crossing symmetry and the behavior near the singular points $1$ and $\infty$.

We start with the computation of the $\b$-function and the fixed point.
For the computation of the $\b$-function in the second order, we need
the four-point function of the perturbing field. Here we consider a
more general function:
\be\nn
<\tphi(x)\tphi(0)\tphi_{n,n+2}(1)\tphi_{n,n+2}(\infty)>.
\ee
There are three "channels" (or intermediate fields) in the corresponding
conformal block: two even, corresponding to the identity and
$\phi_{1,5}$ and one odd - to $\tilde\phi$ itself. From the
procedure we explained above, we get the following expression for
this correlation function:
\bea\label{npt}
&<&\tilde\phi(x)\tilde\phi(0)\tilde\phi_{n,n+2}(1)\tilde\phi_{n,n+2}(\infty)>\\
\nn &=& \left|{(1 - 2 x + 7/3 x^2 - 4/3 x^3 + 1/3 x^4)\ov x^2 (1 - x)^2}\right|^2+{2(n + 3)\ov 3 (n + 1)}\left| {(1 - 3/2 x + 3/2 x^2 - 1/2 x^3)\ov x (1 - x)^2}\right|^2\\
\nn &+& {( n+3) ( n+4)\ov 18  n ( n+1)}\left| {(1 - x + x^2)\ov (1 - x)^2}\right|^2.
\eea
We checked explicitly the crossing symmetry and the $x\rightarrow 1$ limit of this function.
The function that enters the integral is obtained by the
conformal transformation $x\rightarrow 1/x$ (explicit formula is
presented below).

The integration over the safe region far from the singularities
yields ($I(x)\sim {\pi\ov \epsilon}$):
\bea
\nn &\int_{\Omega_{l,l_0}}& I(x)<\tilde\phi(x)\tilde\phi(0)\tilde\phi(1)\tilde\phi(\infty)>d^2x\\
\nn  &=&-{35 \pi^2\ov 24 \e} + {2 \pi^2\ov\e l^2} + {\pi^2\ov 2 \e l_0^2} -
{ 16 \pi^2 \log l\ov 3 \e} - {8 \pi^2 \log 2 l_0\ov 3 \e}
\eea
and we omitted the terms of order $l$ or $l_0/l$.

We have to subtract the integrals over the lens-like regions
since they  would be accounted twice. Here is the result of that integration:
\be\nn
{\pi^2 \ov \e}
\left(-{1\ov l^2}+{1\ov 2 l_0^2} +{61\ov 24}-{8\ov 3}\log{l\ov 2l_0}\right).
\ee

Next we have to compute the integrals near the singular points $0,1$ and $\infty$. For that purpose we can use the OPE of the
fields and take the appropriate limit of $I(x)$.
Near the point $0$ the relevant OPE is:
\be
\tilde\phi(x)\tilde\phi(0)=(x\bar x)^{-2(\Delta_{1,3}+\hf)}(1+\ldots)
+ \hat C_{(1,3)(1,3)}^{(1,3)}(x\bar x)^{-(\Delta_{1,3}+\hf)}(\tilde\phi(0)+\ldots).
\ee
The channel $\phi_{1,5}$ gives after integration a term proportional
to $l/l_0$ which is negligible. The structure constants we need here and in the calculations below were computed in \cite{myne33}.
In the present case we have:
\be
\hat C_{(1,3)(1,3)}^{(1,3)}={2\ov \sqrt 3} - 2 \sqrt 3 \e
\ee
to the first order in $\e$. The value of $I(x)$ near $0$ can be found
by taking the limit in (\ref{ix}) written in terms of $1/x$.
Finally one gets:
\be\nn
\int_{D_{l,0}\backslash D_{l_0,0}} I(x)<\tilde\phi(x)\tilde\phi(0)\tilde\phi(1)\tilde\phi(\infty)>d^2x
=-{\pi^2\ov l^2 \e} + {8 \pi^2\ov
 3 \e^2} -{16 \pi^2\ov\e} + {8\ov3} {\pi^2 \log l\ov\e}.
\ee
Since the integral near $1$ gives obviously the same result, we just need to add the above result twice.
To compute the integral near infinity, we use a relation
\be\label{infi}
<\phi_1(x)\phi_2(0)\phi_3(1)\phi_4(\infty)>=(x\bar x)^{-2\D_1}<\phi_1(1/x)\phi_4(0)\phi_3(1)\phi_2(\infty)>
\ee
and $I(x)\sim {\pi\ov\e}(x\bar x)^{-2\e}$.
This gives
\be
\nn \int_{D_{l,\infty}\backslash D_{l_0,\infty}} I(x)<\tphi(x)\tphi(0)\tphi(1)\tphi(\infty)>d^2x
=-{\pi^2\ov l^2 \e} + {4 \pi^2\ov 3 \e^2}  - {8 \pi^2\ov\e} + {8 \pi^2 \log l\ov3\e}.
\ee

Putting all together, we obtain the finite part of the integral:
\be
{20\pi^2\ov 3 \e^2}-{44\pi^2\ov\e}.
\ee

Taking into account also the first order term (proportional to
the above structure constant), we get the final
result (up to the second order) for the two-point function of the
perturbing field:
\bea\label{twoptf}
G(x,\l)&=&<\tphi (x)\tphi(0)>\\
\nn &=&(x\bar x)^{-2+2\e}\left[1-\l {4\pi\ov \sqrt 3}\left({1\ov\e}-3\right)(x\bar x)^\e+{\l^2\ov 2}\left({20\pi^2\ov 3 \e^2}-{44\pi^2\ov\e}\right)(x\bar x)^{2\e}
+\ldots\right].
\eea

Now we introduce a renormalized coupling constant and a renormalized field $\tphi^g=\p_g {\cal L}$ which is normalized by $<\tphi^g (1)\tphi^g(0)>=1$.
Under the scale transformation $x^{\mu}\to t x^{\mu}$, the Lagrangian transforms to the trace of the energy-momentum tensor $\Theta$:
\be\nn
\Theta(x)=\p_t {\cal L}=\beta(g)\p_g {\cal L}=\beta(g)\tphi^g.
\ee
Comparing these with the original bare Lagrangian where $\tphi=\p_{\l}{\cal L}$ and $\Theta=\e\l\tphi$ leads to
the $\b$-function given by:
\be
\beta(g)=\e\l{\p g \ov\p\l}=\e\l\sqrt{ G(1,\l)},
\ee
where $G(1,\l)$ is given by (\ref{twoptf}) with $x=1$.
One can invert this and compute the bare coupling constant and the $\beta$-function in terms of $g$:
\bea\label{bare}
\l&=&g+g^2{\pi\ov \sqrt 3}\left({1\ov\e}-3\right)+g^3{\pi^2\ov 3}\left({1\ov\e^2}-{5\ov\e}\right)+{\cal O}(g^4),\\
\beta(g)&=&\e g-g^2{\pi\ov\sqrt 3}(1-3\e)-{2\pi^2\ov 3}g^3+{\cal O}(g^4).
\eea
In this calculations, we keep only the relevant terms by assuming the
coupling constant $\l$ (and $g$) to be order of ${\cal O}(\e)$.

A non-trivial IR fixed point occurs at the zero of the $\beta$-function:
\be\label{fx}
g^*={\sqrt{3}\ov\pi}\e(1+\e).
\ee
It corresponds to the IR CFT  $SM_{p-2}$ as can be seen from the central charge difference:
\be
c^*-c=-8\pi^2\int_0^{g^*}\beta(g)d g=-4\e^3-12\e^4+{\cal O}(\e^5).
\ee
The anomalous dimension of the perturbing field becomes:
\be
\D^*=1-\p_g\beta(g)|_{g^*}=1+\e+2\e^2+{\cal O}(\e^3)
\ee
which matches with that of the second component of the superfield
$\Phi_{3,1}^{p-2}$ of $SM_{p-2}$.

Now we turn to the computation of the mixing coefficients of the super-fields in the NS sector.
Actually, using the second component of a super-field as a perturbing field guarantees the preservation of
super-symmetry along the RG flow.
The dimension, which is close to ($1/2,1/2$), and the fusion rules between the super-fields $\Phi_{n,n\pm
2}$ and $D\bar D\Phi_{n,n}$ where $D$ is the covariant super-derivative
suggest that the operators mix along the RG-trajectory.
We will compute the corresponding dilatation matrix for the anomalous dimensions of the second
components while  the mixing of the first ones is a consequence of the supersymmetry.
For this purpose we compute the two-point functions and the corresponding integrals. Actually, the computation of the integrals goes along the same lines of that of the perturbing field itself. So we will present just the final result of the integration.

\noindent\rb Function $<\tphi_{n,n+2}(1)\tphi_{n,n+2}(0)>$

The corresponding 4-point function in the second order of the perturbation was
already written above (\ref{npt}). The integration goes in the same way as before.
Combining all the integrals, we get:
\be\nn
-{2\pi^2 (-20 - 143 n - 121 n^2 - 33 n^3 - 3 n^4)\ov
 3 n (1 + n) (3 + n)^2 \e^2}-
{2\pi^2(5 + n) (8 + 151 n + 143 n^2 + 45 n^3 + 5 n^4) \ov 3 n (1 + n) (3 + n)^2 \e}.
\ee
We note that the final result is very similar to that of the Virasoro case \cite{pogsc2}.
This will be also the case with the next integrals.

\noindent\rb Function $<\tphi_{n,n+2}(1)\tphi_{n,n-2}(0)>$

The relevant four-point function in this case in the zeroth order of $\e$ is:
\be\nn
<\tphi(x)\tphi(0)\tphi_{n,n+2}(1)\tphi_{n,n-2}(\infty)>=
{1\ov 3} \sqrt{{(-4 + n^2)\ov n^2}}\left|{ 1\ov(1 - x)^2 }(1 - x + x^2)\right|^2
\ee
$\phi_{1,5}$ is only channel appearing here.
Collecting again all the integrals gives:
\be\nn
{80 (1 - 2 \e) \pi^2\ov 3 \e^2 n (-9 + n^2) \sqrt{-4 + n^2}}.
\ee

\noindent\rb Function $<\phi_{n,n}(1)\tphi_{n,n+2}(0)>$

The 4-point function computed in \cite{myne5} is equal to:
\be\nn
<\tphi(x)\tphi(0)\phi_{n,n}(1)\tphi_{n,n+2}(\infty)>={2\ov 3}\sqrt{{n+2\ov n}}|x|^{-2}
\ee
which leads after integration to:
\be\nn
-{4 (-1 + n)\pi^2\ov 3 (3 + n) (5 + n)}\sqrt{{n+2\ov n}}\left(11 + 3 n + \e (1 + n) (9 + 2 n)\right).
\ee

\vfill
\newpage

\noindent\rb Function $<\phi_{n,n}(1)\phi_{n,n}(0)>$

Finally, we need the function $<\tphi(x)\tphi(0)\phi_{n,n}(1)\phi_{n,n}(\infty)>$.
This function happens to coincide exactly with the same function in the Virasoro case. Moreover, as we shall show in Section 3, it is the same for all levels $l$ of the general $\hat{su}(2)$ coset models. We shall present the exact expression in that Section. Because of this almost all integrals are the same. The only difference comes from the corresponding structure constants. Taking this into account, our final result here is:
\be\nn
{(-1 + n^2)\pi^2 \ov 6}(1 + \e)
\ee

Since the dimension of the first component $\phi_{n,n}$ is close to zero, it doesn't mix with other fields.
Therefore, we need to compute only its anomalous dimension.
Taking into account also the first order
contribution, the final result for the two-point function is:
\bea\nn
G_n(x,\l)=<\phi_{n,n}(x)\phi_{n,n}(0)>&=&(x\bar
x)^{-2\D_{n,n}}\big[1 - \l \left({\sqrt 3 \pi\ov 6} (-1 + n^2) \e(1+3\e) \right)(x\bar x)^\e+\\
\nn &+& {\l^2\ov 2} \left({\pi^2\ov 6} (1 +  \e) (-1 + n^2)\right)(x\bar x)^{2\e}+...\big].
\eea

Computation of the anomalous dimension goes in exactly the same way as for the perturbing field:
\bea\nn
\D_{n,n}^g &=&\D_{n,n}-{\e\l\ov 2}\p_\l G_n(1,\l)=\\
\nn &=&\D_{n,n}+{\sqrt 3\pi g\ov 12} \e^2 (1 + 3 \e) (-1 + n^2)-{\pi^2 g^2\ov 12} \e^2 (-1 + n^2),
\eea
where we again kept the appropriate terms of order $\e\sim g$. Then, at the fixed point (\ref{fx}), this becomes:
\be\nn
\D_{n,n}^{g^*}=
{(-1 + n^2)( \e^2 + 3 \e^3 + 7 \e^4+...)\ov 8}
\ee
which coincides with the dimension of the first component of the superfield $\Phi_{n,n}^{(p-2)}$
of the model $SM_{p-2}$.

We are now in a position to compute the matrix of anomalous dimensions in the NS sector. The renormalization scheme we use is presented in \cite{pogsc2} and is essentially a variation of that originally proposed by Zamolodchikov \cite{zamrg}.
The renormalized fields are expressed through the bare ones by:
\be
\phi^g_\a=B_{\a\b}(\l)\phi_\b
\ee
(here $\phi$ could be the first or last component).
The two-point functions of the renormalized fields:
\be\label{norm}
G_{\a\b}^g(x)=<\phi_\a^g(x)\phi_\b^g(0)>,\quad G_{\a\b}^g(1)=\delta_{\a\b},
\ee
satisfy the Callan-Symanzic equation:
\be
(x\p_x-\b(g)\p_g)G_{\a\b}^g+\sum_{\rho=1}^2(\G_{\a\rho}G_{\rho\b}^g+\G_{\b\rho}G_{\a\rho}^g)=0
\ee
where the matrix of anomalous dimensions $\Gamma$ is given by:
\be\label{ano}
\G=B\hat\D B^{-1}-\e\l B\p_\l B^{-1}
\ee
where $\hat\D=diag(\D_1,\D_2)$ is a
diagonal matrix of the bare dimensions.
The matrix $B$ itself is
computed from the matrix of the bare two-point functions we computed
using the normalization condition (\ref{norm}) and requiring the matrix
$\G$ to be symmetric.

We computed above some of the entries of the $3\times 3$ matrix
of two-point functions in the second order. This matrix is obviously
symmetric. It turns out also that the remaining functions
$<\tphi_{n,n-2}(1)\tphi_{n,n-2}(0)>$ and
$<\phi_{n,n}(1)\tphi_{n,n-2}(0)>$ can be obtained from the computed
ones $<\tphi_{n,n+2}(1)\tphi_{n,n+2}(0)>$ and
$<\phi_{n,n}(1)\tphi_{n,n+2}(0)>$ by just taking $n\rightarrow -n$.
Let us denote for convenience the basis of fields:
\be
\nn \phi_1=\tphi_{n,n+2},\quad
\phi_2=(2\D_{n,n}(2\D_{n,n}+1))^{-1}\p\bar\p \phi_{n,n},\quad
\phi_3=\tphi_{n,n-2}
\ee
where we normalized the field $\phi_2$ so that its bare two-point function is $1$. It is straightforward to
modify the functions involving $\phi_2$ taking into account the derivatives and the normalization.

We can write the matrix of the two-point functions up to the second
order in the perturbation expansion as:
\be\label{tpfo}
G_{\a,\b}(x,\l)=<\phi_\a(x)\phi_\b(0)>=
(x\bar x)^{-\D_\a-\D_\b}\left[\delta_{\a,\b}-\l C^{(1)}_{\a,\b}(x\bar x)^{\e}+{\l^2\ov 2}C^{(2)}_{\a,\b}(x\bar x)^{2\e}+...\right].
\ee
The two-point functions in the first order are proportional to the
structure constants \cite{zamrg}:
\be\label{fir}
C^{(1)}_{\a,\b}=\hat C_{(1,3)(\a)(\b)}{\pi \g(\e+\D_\a-\D_\b)\g(\e-\D_\a+\D_\b)\ov\g(2\e)}
\ee
and are obviously symmetric. The second order entries $C^{(2)}_{\a,\b}$ are the result of integration of the corresponding 4-point functions and were presented above.

Now, following the renormalization procedure, sketched above, we can obtain the matrix of anomalous dimensions (\ref{ano}). The bare coupling constant $\l$ is
expressed through $g$ by (\ref{bare}) and the bare dimensions, up to order $\e^2$. The results are:
\bea
\nn \G_{1,1}&=&\D_1-{(3 + n) (-1 + \e (2 + n)) \pi g\ov \sqrt 3 (1 + n)}+{4 g^2 \pi^2(2 + n)\ov 3 (1 + n)}\\
\nn \G_{1,2}&=&\G_{2,1}=-{(-1 + \e) (-1 + n) \sqrt{{2 + n\ov 3n}} \pi g\ov (1 + n)}+{2 g^2 (-1 + n) \sqrt{{2 + n\ov n}} \pi^2\ov 3 (1 + n)}\\
\nn \G_{1,3}&=&\G_{3,1}=0\\
\nn \G_{2,2}&=&\D_2-{2\sqrt 3 \pi (-2 + \e  +  \e n^2 ) g\ov
 3 (-1 + n^2)}+{2 g^2 (3 + n^2) \pi^2\ov 3 (-1 + n^2)}\\
\nn \G_{2,3}&=&\G_{3,2}=-{(-1 + \e) \sqrt{{-2 + n\ov 3n}} (1 + n) \pi g\ov (-1 + n)}\\
 \G_{3,3}&=&\D_3+{(1 + \e (-2 + n)) (-3 + n) \pi g\ov \sqrt 3 (-1 + n)}+{4 g^2 \pi^2(-2  + n )\ov 3 (-1 + n)}
\eea
where:
\bea
\nn \D_1&=&1 -{n+1\ov 2} \e + {1\ov 8} (-1 + n^2) \e^2,\quad
\D_2=1+{1\ov 8} (-1 + n^2) \e^2,\\
\nn \D_3&=&1 +{n-1\ov 2} \e + {1\ov 8} (-1 + n^2) \e^2.
\eea
Evaluating this matrix at the fixed point (\ref{fx}), we get:
\bea
\nn \G_{1,1}^{g^*}&=&1 + {(20 - 4 n^2) \e\ov 8 (1 + n)} + {(39 - n - 7 n^2 + n^3) \e^2\ov
 8 (1 + n)}\\
\nn \G_{1,2}^{g^*}&=&\G_{2,1}^{g^*}={(-1 + n) \sqrt{{2 + n\ov n}} \e(1+2\e)\ov n+1}\\
\nn \G_{1,3}^{g^*}&=&\G_{3,1}^{g^*}=0\\
\nn \G_{2,2}^{g^*}&=&1 + {4 \e\ov -1 + n^2} + {(65 - 2 n^2 + n^4) \e^2\ov 8 (-1 + n^2)}\\
\nn \G_{2,3}^{g^*}&=&\G_{3,2}^{g^*}={\sqrt{{-2 + n\ov n}} (1 + n) \e(1+2\e)\ov n-1}\\
\nn \G_{3,3}^{g^*}&=&1 + {(-5 + n^2) \e\ov 2 (-1 + n)} + {(-39 - n + 7 n^2 + n^3) \e^2\ov
 8 (-1 + n)}
\eea
whose eigenvalues are (up to order $\e^2$):
\bea
\nn \D_1^{g^*}&=&1 +  {1 + n\ov 2} \e + {7 +8 n + n^2\ov 8} \e^2\\
\nn \D_2^{g^*}&=&1 + {-1 + n^2\ov 8} \e^2\\
\nn \D_3^{g^*}&=&1 + {1-n\ov 2} \e +  {7 - 8 n + n^2\ov 8} \e^2.
\eea
This result coincides with dimensions $\D_{n+2,n}^{(p-2)}+1/2$,$\D_{n,n}^{(p-2)}+1$ and $\D_{n-2,n}^{(p-2)}+1/2$ of the model $SM_{p-2}$ up to this order.
The corresponding normalized eigenvectors should be identified with the fields of $SM_{p-2}$:
\bea
\nn \tphi_{n+2,n}^{(p-2)}&=&{2 \ov n (1 + n)}\phi_1^{g^*} + {2
\sqrt{{2 + n\ov n}}\ov 1 + n}\phi_2^{g^*} + {\sqrt{-4 + n^2}\ov
n}\phi_3^{g^*}\\
\nn \phi_2^{(p-2)}&=&-{2 \sqrt{{2 + n\ov n}}\ov 1 +
n}\phi_1^{g^*} -{-5 +
  n^2\ov 1 + n^2}\phi_2^{g^*} +{2\sqrt{{n-2\ov n}}\ov n-1}\phi_3^{g^*}\\
\nn \tphi_{n-2,n}^{(p-2)}&=&{\sqrt{-4 + n^2}\ov n}\phi_1^{g^*}  - { 2
\sqrt{{-2 + n\ov n}}\ov n-1}\phi_2^{g^*} +{ 2\ov n(n-1)}\phi_3^{g^*}.
\eea
We used as before the notation $\tphi$ for the last component of the
corresponding superfield and:
\be\nn
\phi_2^{(p-2)}={1\ov 2\D_{n,n}^{p-2}(2\D_{n,n}^{p-2}+1)}\p\bar\p
\phi_{n,n}^{(p-2)}
\ee
is the normalized derivative of the corresponding first component. We notice that these eigenvectors are finite
as $\e\rightarrow 0$ with exactly the same entries as in the Virasoro minimal models. We will show in Section 3 that this is the case also for any level $l$ of the general $\hat{su}(2)$ coset model.

Let us now consider the mixing of the fields in the Ramond sector.

The computation of the conformal blocks in the Ramond sector is more involved.
A way of computing them was recently proposed in \cite{belmu} where
conformal blocks in the first few levels were shown to coincide with the
instanton partition function of certain $N=2$ YM theories in four
dimensions by a generalized AGT correspondence up to prefactors.

Following \cite{belmu} one can compute NS-R conformal blocks only for a
special choice of the points. After that we can get the function
necessary for the integration in the second order by using its
conformal transformation properties.

The difficulties arise because of the branch cut in the OPE of
Ramond fields with the supercurrent:
\be\label{gr}
G(z)R^\ve(0)={\b
R^{-\ve}(0)\ov z^{{3\ov 2}}}+{G_{-1}R^\ve(0)\ov z^\hf}
\ee
where $\b=\sqrt{\D-{\hat c\ov 16}}$, $\hat c={2\ov 3}c$, $\ve=\pm 1$.
Therefore one cannot obtain the usual commutation relations.
Here the Ramond field $R^\ve$ is doubly degenerate
because of the zero mode of $G$ in this sector.

The difficulty can be removed in the following way. Consider the OPE
between NS and Ramond fields:
\bea\label{rop}
\phi_1(x)R_2^\ve(0)&=&x^{\D-\D_1-\D_2}\sum_{N=0}^\infty x^N C_N^\ve
R_\D^\ve(0),\\
\nn \tilde\phi_1(x)R_2^\ve(0)&=&x^{\D-\D_1-\D_2-\hf}\sum_{N=0}^\infty x^N
\tilde C_N^\ve R_\D^{-\ve}(0).
\eea
Here $N$ runs over nonnegative
integers as $G$'s have integer valued modes in the Ramond sector.
Applying $G_0$ on both sides of (\ref{rop}) and taking into account (\ref{gr}),
we obtain:
\bea\label{gor}
G_0C_N^\e &=&\tilde
C_N^\ve+\b_2C_N^{-\ve},\\
\nn G_0\tilde C_N^\ve
&=&(\D-\D_2+N)C_N^\ve-\b_2\tilde C_N^{-\ve}.
\eea
From the consistency conditions, $\tilde C_0^\ve$ is given by:
\be
\nn \tilde C_0^\ve=\b C_0^\ve-\b_2C_0^{-\ve}.
\ee
Acting with $G_k$ with $k>0$ gives chain relations:
\bea\label{rcr}
G_k C_N^\e &=&\tilde C_{N-k}^\e\\
\nn G_k\tilde
C_N^\e &=&(\D+2k\D_1-\D_2+N-k) C_{N-k}^\e
\eea
and $L_k$ acts as usual
with the appropriate dimensions (see \cite{belmu} for the details).

One has to solve these chain relations order by order or to use the recursion formulae. Then the conformal block for the function $<N(x)R(0)N(1)R(\infty)>$ ($N$ here stays for the first or the last component of a NS field)
 is obtained in the same way as in the NS case:
\be
\nn F(x,\D,\D_i)=\sum_{N=0}^\infty x^N C_N(\D,\D_3,\D_4)S_N^{-1}C_N(\D,\D_1,\D_2)
\ee
where $C_N$ could be actually $C_N$ or $\tilde C_N$ depending on the function in consideration.
Finally the correlation function is constructed as:
\be
\nn <N(x)R(0)N(1)R(\infty)>=\sum_n C_n|F_n(x)|^2,
\ee
where
$C_n$'s are the structure constants and the range of $n$ is dictated by the fusion rules.
The function that enters the integral is then obtained by the conformal transformation.

As already mentioned, the conformal block in general is very
complicated. Fortunately, it is sufficient to compute the finite
term as $\e\rightarrow 0$. We did the computation for the functions
below up to high order and then check the behavior near the singular
points. It turns out also that the two-point function do not depend
on which of the fields $R^\ve$ are involved. So we drop the subscript $\ve$
from our notations in what follows.

\noindent\rb Function $<R_{n,n+1}(1)R_{n,n+1}(0)>$

Our calculation for the corresponding 4-point function gives:
\bea
\nn &<&\tphi(x)R_{n,n+1}(0)R_{n,n+1}(\infty)\tphi(1)>={n^2-1\ov
12n^2}\left|{1\ov x (1 - x)^2} (1 + {n\ov n + 1} x - {1\ov n + 1}
x^2)\right|^2+\\
\nn &+&{(2 + n)^2\ov 48 n^2}\left|{1\ov x (1 - x)^2} (1 +  {2n\ov
n + 2} x + {n - 2\ov n + 2} x^2)\right|^2+{n+3\ov 12(n+1)}\left|{1\ov
(1-x)^2}(1+x)\right|^2.
\eea
To obtain the function that enters the integral, we use the conformal
transformation properties. One can easily get:
\bea
\nn &<&\tphi(x)R_{n,n+1}(0)R_{n,n+1}(1)\tphi(\infty)>=(x\bar
x)^{-2\D_{1,3}-1}<\tphi({x-1\ov
x})R_{n,n+1}(0)R_{n,n+1}(\infty)\tphi(1)>\\
\nn &=&{n^2-1\ov
12n^2}\left|{(2x-1)(n x+1)\ov (n+1)x(x-1)}\right|^2+{(2 + n)^2\ov 48
n^2}\left|{(2x-1)(n(2x-1)+2)\ov (n+2)x(x-1)}\right|^2+{n+3\ov 12(n+1)}\left|{2x-1\ov
x}\right|^2.
\eea
As in the NS case we omit the detailed calculation of the integrals since it goes in the same way. The final result of the integration of this function is:
\be
\nn {\pi^2\left[44 + 64 n + 24 n^2 + 3 n^3 -
   8 \e (1 + n) (5 + 14 n + 7 n^2 + n^3)\right]\ov 12 \e^2 n (2 + n)^2}.
\ee

\noindent\rb Function $<R_{n,n-1}(1)R_{n,n+1}(0)>$

The calculation of the four-point function with the perturbing
fields can be done in the same way:
\be
\nn <\tphi(x)R_{n,n+1}(0)\tphi(1)R_{n,n-1}(\infty)>={\sqrt{n^2-1}\ov 12
n}\left|{1\ov x(1-x)}(1+x)\right|^2.
\ee
Performing the same transformation as in the previous case, the integrand becomes:
\be
\nn <\tphi(x)R_{n,n+1}(0)R_{n,n-1}(1)\tphi(\infty)>={\sqrt{n^2-1}\ov 12 n}\left|{2x-1\ov x(1-x)}\right|^2.
\ee
Here is the final result in the second order:
\be
\nn {4\pi^2 \sqrt{n^2 - 1} (44 - 5 n^2-2\e (20+n^2)) \ov 3 \e^2 n(n^2-16)(n^2-4)}.
\ee

The functions we computed above are enough for our computation since the other two functions
$<R_{n,n-1}(1)R_{n,n-1}(0)>$ and $<R_{n,n+1}(1)R_{n,n-1}(0)>$ can be
obtained from $\qquad$ $<R_{n,n+1}(1)R_{n,n+1}(0)>$ and
$<R_{n,n-1}(1)R_{n,n+1}(0)>$ by just changing $n\rightarrow -n$  as in the
case of NS fields.
Let us introduce again a basis: $R_1=R_{n,n+1}$, $R_2=R_{n,n-1}$. From
the general formula (\ref{fir}) and the bare dimensions of the fields
\bea
\nn \D_1&=&{3\ov 16}  -\left({n\ov4}+{1\ov8}\right) \e + {1\ov 8} (n^2-1) \e^2,\\
\nn \D_2&=&{3\ov 16}  +\left({n\ov4}-{1\ov8}\right) \e + {1\ov 8} (n^2-1) \e^2
\eea
we can get the $2\times 2$ matrix of two-point functions in the first order.
Together with our results in the second order presented above and following the same procedure as in the NS case, we get for the matrix of anomalous dimensions up to order
$\e^2\sim g^2$:
\bea
\nn \G_{1,1}&=&\D_1-{g (n+2) (-1 - \e + 2 \e n) \pi\ov 4 \sqrt 3 n}+{g^2 \pi^2\ov 4},\\
\nn \G_{1,2}&=&\G_{2,1}={(1 + \e) g \sqrt{n^2-1} \pi\ov 2 \sqrt 3 n},\\
\nn \G_{2,2}&=&\D_2+{g (n-2) (1 + \e + 2 \e n) \pi\ov 4 \sqrt 3 n}+{g^2 \pi^2\ov 4},
\eea
which, at the fixed point (\ref{fx}), becomes:
\bea
\nn \G_{1,1}^{g^*}&=&{3\ov 16} + {(4 + n - 2 n^2) \e\ov 8 n} + {(8 + n - 4 n^2 + n^3) \e^2\ov 8 n},\\
\nn \G_{1,2}^{g^*}&=&\G_{2,1}^{g^*}={\sqrt{n^2-1} \e(1+2\e)\ov 2 n},\\
\nn \G_{2,2}^{g^*}&=&{3\ov 16} + {(-4 + n + 2 n^2) \e\ov 8 n} + {(-8 + n + 4 n^2 + n^3) \e^2\ov 8 n}.
\eea
The eigenvalues of this matrix are:
\bea
\nn \D_1^{g^*}&=&{3\ov 16} + \left({1\ov 8} + {n\ov 4}\right) \e + {1\ov 8} (1 + 4 n + n^2) \e^2,\\
\nn \D_2^{g^*}&=&{3\ov 16} + \left({1\ov 8} - {n\ov 4}\right) \e + {1\ov 8} (1 - 4 n + n^2) \e^2.
\eea
As expected, they coincide with the dimensions of the Ramond fields $\D_{n+1,n}^{(p-2)}$ and $\D_{n-1,n}^{(p-2)}$ of the  $SM_{p-2}$. The corresponding fields are expressed as a  (normalized) linear combination:
\bea
\nn R_{n+1,n}^{(p-2)}&=&{1\ov n}R_1^{g^*}+{\sqrt{n^2-1}\ov n}R_2^{g^*},\\
\nn R_{n-1,n}^{(p-2)}&=&-{\sqrt{n^2-1}\ov n}R_1^{g^*}+{1\ov n}R_2^{g^*}.
\eea

\subsection{Three-point correlation functions in $N=1$ super-Liouville theory}

In this section we consider the super Liouville field
theory. We formulate the supersymmetric theory by the action:
\begin{equation}\label{acti}
S_{SL}=\frac {1}{4\pi} \int d^2zd^2\theta\left[\frac 12 D_\alpha
\Phi D^\alpha\Phi - Q\hat R\Phi + \mu e^{b\Phi}\right]
\end{equation}
where the real superfield $\Phi$ possesses the expansion:
$$
\Phi = \phi +\theta\psi +\bar\theta\bar\psi+\theta\bar\theta\tphi,
$$
(the component fields here depend on both $z$ and $\bar z$), $\hat R$ is the
scalar curvature corresponding to the background metric and $\mu$ in (\ref{acti})
is the cosmological constant. The classical equations of motion for (\ref{acti})
are\footnote{One can choose the background metric to be flat and therefore
$\hat R$ will not be essential in the sequal.}:
\begin{equation}
\nn D_\alpha D^\alpha\Phi = Q\hat R +\mu e^{b\Phi}.
\end{equation}
The superspace notations that we shall use in the case of super Liouville theory are:
\begin{align}
\nn & Z=(z,\theta)\\
\nn & Z_1 -Z_2 = z_1 -z_2 -\theta_1\theta_2.
\end{align}
The super stress tensor of the super Liouville theory is expressed in
terms of the real superfield $\Phi$:
\begin{equation}
\nn T_{SL}=-\frac 12 D\Phi\partial\Phi +\frac Q2 D\partial\Phi,
\end{equation}
the central charge of the super-Virasoro algebra with this normalization being given by:
\begin{equation}\label{cli}
\hat c=1+2Q^2.
\end{equation}

Similarly to the minimal models, the superconformal primary fields are divided in two sectors depending
on the boundary conditions of the supercurrent. In the Neveu-Schwarz
sector they are represented by the vertex operators:
\begin{equation}\label{nver}
V_\alpha=e^{\alpha\Phi(Z,\bar Z)}
\end{equation}
of dimension $\Delta_\alpha =\frac 12\alpha(Q-\alpha)$ and in the Ramond
sector by:
\begin{equation}\label{rversl}
R_\alpha^\epsilon =\sigma^\epsilon e^{\alpha\Phi(z,\bar z)}
\end{equation}
where $\sigma^\epsilon$ is the so called spin field and
$\Delta_\alpha=\frac 12\alpha(Q-\alpha)+\frac{1}{16}$ ($\epsilon =\pm$).
The requirement for the cosmological term in (\ref{acti}) to be (1/2,1/2) form in
order to be able to integrate over the surface, gives a connection between
$Q$ and $b$:
\begin{equation}
Q=b+\frac 1b.
\end{equation}
It is easy to see that the operators $V_\alpha=e^{\alpha\Phi}$ and
$V_{Q-\alpha}=e^{(Q-\alpha)\Phi}$ have equal dimensions and therefore they are
reflection image of each other (the same is true also for $R_\alpha^\epsilon$ and
$R_{Q-\alpha}^{\epsilon}$).

In order to compute the 3-point function in the NS sector we shall follow the approach of \cite{ZZtri}.
Consider first the three-point correlation function of Liouville vertex
operators from Neveu-Schwarz sector. The perturbative expansion in the
cosmological constant $\mu$ is given by:
\bea\label{expan}
&\langle& V_{\alpha_1}(Z_1)V_{\alpha_2}(Z_2)V_{\alpha_3}(Z_3)\rangle
=\int \CD\Phi e^{-S_{SL}}e^{\alpha_1\Phi (Z_1)}e^{\alpha_2\Phi (Z_2)}
e^{\alpha_3\Phi (Z_3)}\\
\nn &=&\sum\limits_{s=0}^\infty\left (\frac{\mu}{2\pi}\right)^s\frac{1}{s!}
\int \CD \Phi
e^{-S_{SL} '}\left (\int d^2zd^2\theta e^{b\Phi}\right)^s\prod\limits_{i=1}^{3}
e^{\alpha_i\Phi (Z_i)}.
\eea
The free superfield action $S_{SL} '$:
\be
S_{SL} '=\frac{1}{4\pi}\int d^2zd^2\theta\left[\frac 12 D_\alpha\Phi D^\alpha\Phi
-Q\hat R\Phi \right],
\ee
is used above to compute the resulting integral.

Specializing to the case of correlation functions on the sphere we
shall concenrate the curvature at infinity ($\infty$,0). Therefore,
we can use the super Coulomb gas formalism in order to evaluate the
correlation function on the r.h.s in (\ref{expan}). As it is well known, (\ref{expan}) is nonzero
only if:
\begin{equation}\label{con}
sb=Q-\sum\limits_{i=1}^3\alpha_i
\end{equation}
for any order $s$ of the pertubation series.
The result for the $s^{th}$ term in the expansion (\ref{expan}) is:
\be
\nn \langle V_{\alpha_1}(Z_1)V_{\alpha_2}(Z_2)V_{\alpha_3}(Z_3)\rangle_s
=\left (\frac{\mu}{2\pi}\right)^s\frac{1}{s!}
\prod\limits_{i<j}^{3}|Z_i -Z_j|^{-2\alpha_i\alpha_j}
\int\prod\limits_{j=1}^{s}D^2Y_j\prod\limits_{i=1}^{3}
|Z_i-Y_j|^{-2b\alpha_i}\prod\limits_{i<j}^{s}|Y_i-Y_j|^{-2b^2}
\ee
For $N=1$ case there exists a supersymmetric extension of the
Dotsenko-Fateev integrals and an analogous integral expression for the
structure constants can be extracted. Applied to our problem this
integral expression gives for the three-point function in the case of
integer number of screening charges the following result:
\begin{align}\label{fex}
& \langle V_{\alpha_1}(Z_1)V_{\alpha_2}(Z_2)V_{\alpha_3}(Z_3)\rangle_s \\ \notag
&=\left (\frac{\mu}{8}
\gamma\left(\frac{b^2}{2}+\frac 12\right)\right)^s
\prod\limits_{i<j}^{3}|Z_i -Z_j|^{-2\delta_{ij}}
\prod\limits_{j=1}^{s}\gamma\left (\frac j2 -\left[\frac j2\right]
-j\frac{b^2}{2}\right)\\ \notag
&\times\prod\limits_{j=0}^{s-1}\prod\limits_{i=1}^{3}
\gamma\left (1-\frac j2 +\left [\frac j2\right]-b\alpha_i -
j\frac{b^2}{2}\right)\times
\begin{cases} 1 \qquad\qquad s\in 2\Bbb N \\
       \frac{\eta\bar\eta}{b^2}\qquad s\in 2\Bbb N+1
\end{cases}
\end{align}
where:
\be\label{notg}
\nn \gamma(x)=\frac{\Gamma (x)}{\Gamma (1-x)}, \quad\delta_{ij}=\Delta_i +\Delta_j -\Delta_k, \quad i\neq j\neq k,
\quad\Delta_i = \alpha_i (Q-\alpha_i)
\ee
In the above formula we reminded the definition of $\eta$ as the $SL(2|1)$ odd invariant
for any given three
points ($Z_1,Z_2,Z_3$). In contrast to the bosonic case here the correlation
function is different for $s\in 2\Bbb N$ and $s\in 2\Bbb N+1$.

At this point we use the interpretation of (\ref{con}) along the lines
of \cite{pfku}. It was suggested to consider (\ref{con}) as a kind of "on-mass-shell"
condition for the exact correlation function. This condition means that
the exact correlation function should satisfy:
\begin{equation}\label{resi}
\underset {\sum_{i=1}^{3}\alpha_i=Q-sb}{res}
\langle V_{\alpha_1}V_{\alpha_2}V_{\alpha_3}\rangle =
\frac{\left(-\mu\right)^s}{s!}
\langle V_{\alpha_1}V_{\alpha_2}V_{\alpha_3}
\underbrace{\int D^2Z V_b\dots \int D^2Z V_b}_s\rangle_{\sum_{i=1}^{3}
\alpha_i=Q-sb}
\end{equation}
when (\ref{con}) holds for $s=0,1,2\dots$. In general (\ref{resi}) alone seems to be
unsufficient to determine $N$-point function, but for
three Liouville vertex operators the situation is simple: the
coordinate dependence on the left hand side and right hand side is as in the
three-point function (\ref{expan}). Therefore we have the following "on-mass-shell"
condition for the structure constants:
\begin{equation}
\nn \underset {sb=Q-\sum_i\alpha_i}{res}C^{even(odd)}
(\alpha_1,\alpha_2,\alpha_3)=I_s^{even(odd)}(\alpha_1,\alpha_2,\alpha_3)
\end{equation}
where we have denoted by $I_s^{even(odd)}(\alpha_1,\alpha_2,\alpha_3)$
the coordinate independent part of the $s^{th}$ term in the expansion (\ref{fex}).

Now we have to generalize the special function $\Upsilon (x)$ introduced
in \cite{ZZtri}. For both, bosonic and supersymmetric cases, we define the
function ($0<Re(x)<Q$):
\begin{align}\label{defrli}
 \log\mathcal R(x,a)= & \frac 12\int\limits_0^{\infty}\frac{dt}{t}
\left\{\left[\left(\frac Q2-x\right)^2+\left(\frac Q2-a\right)^2
\right]e^{-t}\right.\\ \notag
& \left. -2\frac{sh^2\left[\left(\frac Q2-x\right)+\left(\frac Q2-a\right)
\right]\frac t4+sh^2\left[\left(\frac Q2-x\right)-\left(\frac Q2-a\right)
\right]\frac t4}{sh\frac{t}{2b}sh\frac{bt}{2}}\right\}.
\end{align}
The simplest properties that are clear from (\ref{defrli}) are:
\be\label{prr}
 \mathcal R (\frac Q2,\frac Q2)=1,\quad
 \mathcal R (x,a)= \mathcal R (Q-x,a),\quad
 \mathcal R (x,a)= \mathcal R (a,x).
\ee
Let us define also:
\begin{equation}
\nn \mathcal R_0= \left.\frac{d \mathcal R (x,a)}{dx}\right|_{x=a=0}. \notag
\end{equation}
We propose the following expression
as an exact three-point function in the supersymmetric Liouville theory:
\begin{align}
& C^{even}(\alpha_1,\alpha_2,\alpha_3)=
\left[\frac{\mu}{8}\gamma\left(\frac{b^2}{2}+\frac 12\right)
b^{-1-b^2}\right]^{\frac{Q-\sum_i\alpha_i}{b}}\\ \notag
& \\ \notag
& \times\frac{\mathcal R_0 \mathcal R(2\alpha_1,0) \mathcal R(2\alpha_2,0)
\mathcal R(2\alpha_3,0)}{\mathcal R(\alpha_1+\alpha_2+\alpha_3 -Q,0)
\mathcal R(x_1,0) \mathcal R(x_2,0) \mathcal R(x_3,0)}
\end{align}

\begin{align}
& C^{odd}(\alpha_1,\alpha_2,\alpha_3)=
\left[\frac{\mu}{8}\gamma\left(\frac{b^2}{2}+\frac 12\right)
b^{-1-b^2}\right]^{\frac{Q-\sum_i\alpha_i}{b}}\\ \notag
& \\ \notag
& \times\frac{\mathcal R_0 \mathcal R(2\alpha_1,0) \mathcal R(2\alpha_2,0)
\mathcal R(2\alpha_3,0)}{\mathcal R(\alpha_1+\alpha_2+\alpha_3 -Q,b)
\mathcal R(x_1,b) \mathcal R(x_2,b) \mathcal R(x_3,b)}
\end{align}
where:
\begin{equation}\label{xii}
\nn x_i=\alpha_j+\alpha_k-\alpha_i; \qquad i\ne j\ne k
\end{equation}

Thus, we have in general:
\begin{equation}
\langle V_{\alpha_1}V_{\alpha_2}V_{\alpha_3}\rangle =
\left( C^{even}(\alpha_1,\alpha_2,\alpha_3) +
\eta\bar\eta C^{odd}(\alpha_1,\alpha_2,\alpha_3)\right)
\prod\limits_{i<j}|Z_i-Z_j|^{\delta_{ij}}.
\end{equation}
This expression for the exact three-point function is
based on the properties of the defined $\mathcal R(x,a)$ function described
below.

We now pass to the Ramond sector. According to the explicit form of the Ramond vertex operator (\ref{rversl}) its three-point
function has the following perturbative expansion:
\begin{align}\label{rfu}
&\langle R^{\epsilon_1}_{\alpha_1}(z_1)
R^{\epsilon_2}_{\alpha_2}(z_2)V_{\alpha_3}
(Z_3)\rangle = \sum\limits_{s=0}^\infty\left({\mu\ov 2\pi}\right)^s\int du_1\dots du_s\langle
\prod\limits_{i=1}^3 e^{\alpha_i\phi(z_i)}\prod\limits_{j=1}^s e^{b\phi(u_j)}
\rangle \\ \notag
&\times \langle\sigma^{\epsilon_1}(z_1)\sigma^{\epsilon_2}(z_2)
\left (1+\theta_3\bar\theta_3\psi(z_3)\right)
\prod\limits_{j=1}^s\psi(u_j)\rangle
\end{align}
As before, in order the free bosonic correlator to be nonzero we have to impose
the condition (\ref{con}). It can be interpreted again as a "on-mass-shell" condition
for the exact correlation function. The explicit expression for the integrals in
(\ref{rfu}) can be extracted from the corresponding formulae for the Ramond fields
\cite{myn12,myne2,pogzam}. The final result is:
\begin{align}\nn
& \langle R^{\epsilon_1}_{\alpha_1}(Z_1)R^{\epsilon_2}_{\alpha_2}(Z_2)
V_{\alpha_3}(Z_3)\rangle_s
=\left (\frac{\mu}{8}
\gamma\left(\frac{b^2}{2}+\frac 12\right)\right)^s
\prod\limits_{i<j}^{3}|Z_i -Z_j|^{-2\delta_{ij}}\\ \notag
& \times\prod\limits_{j=1}^{s}\gamma\left (\frac j2 -\left[\frac j2\right]
-j\frac{b^2}{2}\right)\prod\limits_{j=0}^{s-1}\prod\limits_{i=1}^{2}
\gamma\left (1+\frac j2 -\left [\frac j2\right]-b\alpha_i -
j\frac{b^2}{2}-\frac 12\right)\\ \notag
& \times\gamma\left (1-\frac j2 +\left [\frac j2\right]-b\alpha_3 -
j\frac{b^2}{2}\right)\times A_{\epsilon_1,\epsilon_2}
\end{align}
where:
\begin{align}\nn
& A_{\epsilon,\epsilon}=
\begin{cases} 1 ,\qquad \qquad\qquad\quad s= 2\Bbb N\\
      \theta_3\bar\theta_3\frac{|z_1-z_3||z_2-z_3|}{|z_1-z_2|},\quad s=2\Bbb N+1
\end{cases}
\end{align}
\begin{align}\nn
A_{\epsilon,-\epsilon}=
\begin{cases} 1 ,\qquad \qquad\qquad\quad s=2\Bbb N+1\\
      \theta_3\bar\theta_3\frac{|z_1-z_3||z_2-z_3|}{|z_1-z_2|};\quad s=2\Bbb N
\end{cases}
\end{align}
and $\gamma(x),\,\, \delta_{ij},\,\,\Delta_i$ as in (\ref{notg}).
Finally, we propose the following expression for the exact three-point function
in the Ramond sector:
\begin{equation}\nn
\langle R^{\epsilon_1}_{\alpha_1}(z_1)R^{\epsilon_2}_{\alpha_2}(z_2)
V_{\alpha_3}(Z_3)\rangle = \left(C^{\epsilon_1,\epsilon_2}+
\theta_3\bar\theta_3\frac{|z_1-z_3||z_2-z_3|}{|z_1-z_2|}
\tilde C^{\epsilon_1,\epsilon_2}\right)\prod\limits_{i<j}|z_i-z_j|^{\delta_{ij}}
\end{equation}
where:
\begin{align}\label{cee}
& C^{\epsilon,\epsilon}(\alpha_1,\alpha_2,\alpha_3)=
\left[\frac{\mu}{8}\gamma\left(\frac{b^2}{2}+\frac 12\right)
b^{-1-b^2}\right]^{\frac{Q-\sum_i\alpha_i}{b}}\\ \notag
& \\ \notag
& \times\frac{\mathcal R_0 \mathcal R(2\alpha_1,b) \mathcal R(2\alpha_2,b)
\mathcal R(2\alpha_3,0)}{\mathcal R(\alpha_1+\alpha_2+\alpha_3 -Q,0)
\mathcal R(x_1,b) \mathcal R(x_2,b) \mathcal R(x_3,0)}
\end{align}
\begin{align}\label{ceme}
& C^{\epsilon,-\epsilon}(\alpha_1,\alpha_2,\alpha_3)=
\left[\frac{\mu}{8}\gamma\left(\frac{b^2}{2}+\frac 12\right)
b^{-1-b^2}\right]^{\frac{Q-\sum_i\alpha_i}{b}}\\ \notag
& \\ \notag
& \times\frac{\mathcal R_0 \mathcal R(2\alpha_1,b) \mathcal R(2\alpha_2,b)
\mathcal R(2\alpha_3,0)}{\mathcal R(\alpha_1+\alpha_2+\alpha_3 -Q,b)
\mathcal R(x_1,0) \mathcal R(x_2,0) \mathcal R(x_3,b)}
\end{align}
and $\tilde C^{\epsilon_1,\epsilon_2}$ can be determined using the
supersymmetry ($x_i$ is as in (\ref{xii})).

Now we are going to discuss the pole structure of the
three-point correlation function and to define the so called reflection
amplitudes.
For this purpose we start with some transformation properties
and functional relations for $\mathcal R(x,a)$ defined in (\ref{prr}). Using the integral
representation of $\mathcal R(x,a)$ (\ref{defrli}) one can check that the following functional
relation holds:
\begin{equation}\nn
\mathcal R(x+b,a)=b^{-bx+ab}\gamma\left(\frac{bx-ba+1}{2}\right)\mathcal R(x,a+b).
\end{equation}
It is clear that due to the "self-duality" of $\mathcal R(x,a)$
(i.e. the invariance under $b\to 1/b$) one can conclude that:
\begin{equation}\nn
\mathcal R(x+1/b,a)=b^{\frac xb-\frac ab}\gamma\left(\frac{x-a+b}{2b}\right)
\mathcal R(x,a+1/b).
\end{equation}
It is easy to verify that, using the above properties,
$\Upsilon_1(x)=\mathcal R(x,0)$ and
$\Upsilon_2(x)=\mathcal R(x,b)$ are entire functions of $x$ with the following
zero-structure:
\begin{align}\nn
&\Upsilon_1(x)=0\quad for\quad x=-nb-\frac mb;\quad n-m=even\\ \notag
&\Upsilon_2(x)=0\quad for\quad x=-nb-\frac mb;\quad n-m=odd
\end{align}
and due to (\ref{prr}):
\begin{align}\nn
&\Upsilon_1(x)=0\quad for\quad x=(n+1)b+\frac{m+1}{b};\quad n-m=even\\ \notag
&\Upsilon_2(x)=0\quad for\quad x=(n+1)b+\frac{m+1}{b};\quad n-m=odd
\end{align}
($n,m$ are non-negative integers).

Using all the above properties of $\mathcal R(x,a)$ it is straightforward to
check that the proposed exact three-point functions satisfy the
"on-mass-shell" condition (\ref{con}).

As in the bosonic case \cite{ZZtri}, the proposed correlators as a function of
$\alpha=\sum_i^3\alpha_i$
have more poles than expected: at $\alpha=Q-n/b -mb$ and at
$\alpha=2Q+n/b +mb$. They appear when more general multiple integrals
are considered:
\begin{equation}\label{mut}
\underset{\sum_i\alpha_i=Q-\frac nb -mb}{res}
\langle V_{\alpha_1}V_{\alpha_2}V_{\alpha_3}\rangle=
\frac{(\tilde\mu)^n(\mu)^m}{n!m!}
\langle\prod\limits_{i=1}^{3}V_{\alpha_i}(Z_i)\prod\limits_{k=1}^{n}\int
V_{1/b}(X_k)\prod\limits_{l=1}^{m}\int V_{b}(Y_l)\rangle
\end{equation}
where:
\begin{equation}\nn
\frac{\tilde\mu}{8}\gamma\left(\frac{1}{2b^2}+\frac 12\right)=
\left(\frac{\mu}{8}\gamma\left(\frac{b^2}{2}+\frac
12\right)\right)^{\frac{1}{b^2}}.
\end{equation}
We note that the correlation function (\ref{mut}) is self-dual with respect
to $b\to\frac 1b, \mu\to\tilde\mu$.

As it was mentioned above the Liouville vertex operators $V_\alpha$
and $V_{Q-\alpha}$ have the same dimensions and are interpreted as reflection images of each other. We shall use this
property in order to define the so called reflection amplitudes in the
supersymmetric case.
We start with the reflection amplitude in the Neveu-Schwarz sector. We define the latter as:
\begin{equation}\nn
C^{even(odd)}(Q-\alpha_1,\alpha_2,\alpha_3)=
S^{NS}(\alpha_1)C^{even(odd)}(\alpha_1,\alpha_2,\alpha_3)
\end{equation}
where $S^{NS}(\alpha_1)$ is the reflection amplitude.
Using the functional relations of $\mathcal R(x,a)$
we find that it equals to:
\begin{equation}\label{rens}
S^{NS}(\alpha)=
\left[\frac{\mu}{8}
\gamma\left(\frac{b^2}{2}+\frac 12\right)
\right]^{\frac{2\alpha_1-Q}{b}}
b^{2+\frac{2(Q-2\alpha)}{b}}
\frac{\gamma\left(b\alpha -\frac{b^2}{2}+\frac 12\right)}
{\gamma\left(2-\frac{\alpha}{b}+\frac{1}{2b^2}-\frac 12\right)}.
\end{equation}

As in the bosonic case we associate the reflection amplitude with the
two-point correlation function \cite{ZZtri}.

Now we pass to the Ramond sector and consider the correlation functions
(\ref{cee}), (\ref{ceme}) together. In these functions we have two Ramond fields
($R^{\epsilon_1}_{\alpha_1},
R^{\epsilon_2}_{\alpha_2}$) and one Neveu-Schwarz field
($V_{\alpha_3}$). Therefore the
reflection of the first two fields will give us one reflection amplitude, but
the reflection of the third field will differs from the first one.
We find:
\begin{equation}\nn
C^{\epsilon,\pm\epsilon}(Q-\alpha_1,\alpha_2,\alpha_3)=
S^R(\alpha_1)C^{\epsilon,\mp\epsilon}(\alpha_1,\alpha_2,\alpha_3)
\end{equation}
where:
\begin{equation}\nn
S^R(\alpha_1)=
\left[\frac{\mu}{8}
\gamma\left(\frac{b^2}{2}+\frac 12\right)
\right]^{\frac{2\alpha_1-Q}{b}}
b^{\frac{2(Q-2\alpha)}{b}}
\frac{\gamma\left(b\alpha_1-\frac{b^2}{2}\right)}
{\gamma\left(1-\frac{\alpha_1}{b}+\frac{1}{2b^2}\right)}.
\end{equation}
For the reflection of the Neveu-Schwarz field we found:
\begin{equation}\nn
C^{\epsilon,\pm\epsilon}(\alpha_1,\alpha_2,Q-\alpha_3)=
S^{NS}(\alpha_3)C^{\epsilon,\pm\epsilon}(\alpha_1,\alpha_2,\alpha_3)
\end{equation}
where $S^{NS}(\alpha)$ is as in (\ref{rens}).

\subsection{One-point function of $N=1$ super-Liouville theory with boundary}

Let us remind that the $N=1$ SLFT describes a supermultiplet consisting of a bosonic
field and its fermionic partner interacting with exponential potential.
In this Section we prefer to use the component fields rather than the superfield language. In terms of the component fields, the Lagrangian can be expressed by:
\be
{\cal L}_{\rm SL}={1\over{8\pi}}(\partial_{a}\phi)^2
-\frac{1}{2\pi}(\bar\psi\partial\bar\psi + \psi\bar\partial\psi)
+i\mu b^2\psi \bar\psi e^{b\phi}
+{\pi\mu^2 b^2\over{2}}(:e^{b\phi}:)^2.
\label{lagrangian}
\ee
The central charge in this normalization reads:
\be
c_{SL}={3\over{2}}(1+2Q^2).
\ee
The NS and R primary fields are expressed again in terms of vertex operators (\ref{nver}) and (\ref{rversl}) respectively. Their dimensions are the same as in the previous section.
The physical states can be denoted by a real parameter $P$ defined by:
\be\nn
\alpha={Q\over{2}}+iP.
\ee

In the  first part of this Section we will be interested in the SLFT on a pseudosphere.
This is a generalization of \cite{boot2} where the LFT is studied in
the geometry of the infinite constant negative curvature surface,
the so-called Lobachevskiy plane, i.e. the pseudosphere.
The equations of motion for the component fields of the SLFT are given by:
\bea\nn
\partial\bar\partial\phi &=& 4\pi^2\mu b^2\left( \mu
e^{b\phi}-i\bar\psi\psi\right)e^{b\phi}\\
\nn \partial\bar\psi &=& -i\mu e^{b\phi}\psi,\qquad
\bar\partial\psi= i\mu e^{b\phi}\bar\psi.
\eea
We will assume that the fermion vanishes in the classical limit so that
the background metric is determined by the bosonic field satisfying:
\be\nn
e^{\varphi(z)}={4R^2\over{(1-|z|^2)^2}},
\ee
where $\varphi=2b\phi$ and $R^{-2}=4\pi^2\mu^2 b^3$.
The parameter $R$ is interpreted as the radius of the pseudosphere
in which the points at the circle $|z|=1$ are infinitely
far away from any internal point.
This circle can be interpreted as the ``boundary'' of the pseudosphere.
In the same way as the LFT, we can now use the Poincar\'e model of
the Lobachevskiy plane with complex coordinate $\xi$ in the
upper half plane.

We want to compute exact one-point functions of the (NS) and (R)
bulk operators $N_\alpha$ and $R_\alpha$.
Due to the superconformal invariance, these one-point functions are given by:
\be\nn
\langle N_\alpha(\xi)\rangle
= {U^N(\alpha)\over |\xi-\bar\xi|^{2\Delta^N_\alpha}},\qquad
\langle R_\alpha(\xi)\rangle
= {U^R(\alpha)\over |\xi-\bar\xi|^{2\Delta^R_\alpha}}.
\ee
We will simply refer to the coefficients $U^N(\alpha)$ and $U^R(\alpha)$ as
bulk one-point functions.
To derive the functional relations satisfied by these one-point
functions, we should consider the bulk degenerate fields
which are defined by some differential equations
with certain orders.

The degenerate fields in the (NS) sector are given by:
\be
\nn N_{\alpha_{m,n}}=e^{\alpha_{m,n}\phi},\quad
\alpha_{m,n}={1\over{2b}}(1-m)+{b\over{2}}(1-n),
\quad{\rm with}\quad m-n={\rm even}
\ee
and those in the (R) sector by:
\be
\nn R_{\alpha_{m,n}}=\sigma^{(\epsilon)} e^{\alpha_{m,n}\phi},\quad
{\rm with}\quad m-n={\rm odd}.
\ee
One of the essential features of these fields is that the operator
product expansion (OPE) of a degenerate field with any primary
field is given by a linear combination of only finite
number of primary fields and their decendents.
The simplest degenerate fields are $N_{-b}$ for the (NS) sector and
$R_{-b/2}$ for the (R) sector.

The OPE of $R_{-b/2}$ with a (NS) primary field is given by:
\begin{equation}
N_{\alpha}R_{-b/2}=C^{(N)}_{+}(\alpha)\left[R_{\alpha-b/2}\right]
+C^{(N)}_{-}(\alpha)\left[R_{\alpha+b/2}\right],
\label{nsope}
\end{equation}
where $[\ldots]$ stands for entire family of conformal decendents
corresponding to a primary field.
The structure constants can be computed
using Coulomb gas integrals.
One can set $C^{(N)}_{+}=1$ since no screening insertion is needed.
The other structure constant $C^{(N)}_{-}$ needs just one
insertion of the SLFT interaction and can be computed to be:
\be
\nn C^{(N)}_{-}(\alpha)=
{\pi\mu b^2\gamma\left(\alpha b-{b^2\over{2}}
-{1\over{2}}\right)\over{\gamma\left({1-b^2\over{2}}\right)\gamma(\alpha b)}}.
\ee

Similarly, the OPE with the (R) primary field is
\begin{equation}
R_{\alpha}R_{-b/2}=C^{(R)}_{+}(\alpha)\left[N_{\alpha-b/2}\right]
+C^{(R)}_{-}(\alpha)\left[N_{\alpha+b/2}\right]
\label{rope}
\end{equation}
where $C^{(R)}_{+}=1$ as before and
$C^{(R)}_{-}$ is given by:
\be\nn
C^{(R)}_{-}(\alpha)=
{\pi\mu b^2\gamma\left(\alpha b-{b^2\over{2}}\right)\over{
\gamma\left({1-b^2\over{2}}\right)\gamma\left(\alpha b+{1\over{2}}\right)}}.
\label{cminus}
\ee

Now we consider the bulk two-point functions of
the degenerate field $R_{-b/2}$ and a (NS) field $N_\alpha$,
\be\nn
G^{N}_{\alpha,-b/2}(\xi,\xi')=\langle N_{\alpha}(\xi')R_{-b/2}(\xi)\rangle.
\ee
It is straightforward from (\ref{nsope}) to show that
the two-point function satisfy:
\bea\label{gna}
G^{N}_{\alpha,-b/2}(\xi,\xi')
&=&C^{(N)}_{+}(\alpha)U^{R}\left(\alpha-{b\over{2}}\right)
{\cal G}_{+}(\xi,\xi')+\\
\nn &+&C^{(N)}_{-}(\alpha)U^{R}\left(\alpha+{b\over{2}}\right)
{\cal G}_{-}(\xi,\xi')
\eea
where ${\cal G}_{\pm}(\xi,\xi')$ are expressed in terms of some special
conformal blocks:
\be\nn
{\cal G}_{\pm}(\xi,\xi')={|\xi'-{\overline \xi'}|^{2\Delta^{N}_{\alpha}
-2\Delta^{R}_{-b/2}}\over{
|\xi-{\overline \xi'}|^{4\Delta^{N}_{\alpha}}}}{\cal F}_{\pm}(\eta).
\ee
Here, the conformal blocks are given by hypergeometric functions (which are known)
and:
\be\nn
\eta={(\xi-\xi')({\overline \xi}-{\overline \xi'})\over{(\xi-{\overline \xi'})
({\overline \xi}-\xi')}}.
\ee

In the cross channel, an equivalent expression for the two-point
function can be obtained as follows:
\be
G^{N}_{\alpha,-b/2}={|\xi'-\bar\xi'|^{2\Delta^N_\alpha-2\Delta^R_{-b/2}}
\over |\xi-\bar\xi'|^{4\Delta^N_\alpha}}
\left[B^{(N)}_{+}(\alpha){\widetilde{\cal F}_{+}}(\eta)
+B^{(N)}_{-}(\alpha){\widetilde{\cal F}_{-}}(\eta)\right]
\label{nscross}
\ee
where $\tilde{\cal F}(\eta)$ are again given by some known hypergeometric functions. The boundary structure constants $B^{(N)}_{\pm}$ can be determined
from the monodromy relations connecting ${\cal F}(\eta)$ and $\tilde{\cal F}(\eta)$.

The conformal block $\widetilde{\cal F_-}$ corresponds to the identity
boundary operator with dimension $0$ appearing in the boundary as
the bulk operator $R_{-b/2}$ approaches the boundary with $\eta\rightarrow 1$.
Another boundary operator $n_{-b}$ appearing as $R_{-b/2}$ approaches
the boundary generates the $\widetilde{\cal F_+}$ block.
As mentioned above, the geodesic distance to the boundary
on the pseudosphere is infinite.
Therefore, the two-point function in the LHS of (\ref{nsope}) can be
factorized into a product of two one-point functions and satisfies
\be\nn
B^{(N)}_{-}(\alpha)=U^N(\alpha)U^R(-b/2).
\ee
Combining all these and using (\ref{nscross}), we obtain the
following nonlinear functional equation in the $\eta\rightarrow 1$ limit:
\be
{\Gamma\left({1-b^2\over{2}}\right)U^{N}(\alpha)U^R\left(-{b\over{2}}\right)
\over{\Gamma(-b^2)\Gamma\left(\alpha b-{b^2\over{2}}+{1\over{2}}\right)}}
={U^{R}\left(\alpha-{b\over{2}}\right)\over{\Gamma\left(\alpha b-b^2\right)}}
+{\pi\mu b^2\ U^{R}\left(\alpha+{b\over{2}}\right)
\over{\gamma\left({1-b^2\over{2}}\right)\Gamma(\alpha b)
\left(\alpha b-{b^2\over{2}}-{1\over{2}}\right)}}.
\label{firsteq}
\ee

Analysis of the other two-point function:
\be\nn
G^{R}_{\alpha,-b/2}(\xi,\xi')=\langle R_{\alpha}(\xi)R_{-b/2}(\xi')\rangle
\label{rtwopt}
\ee
goes along the same line and leads to the second functional equation:
\be
{\Gamma\left({1-b^2\over{2}}\right)U^{R}(\alpha)U^R\left(-{b\over{2}}\right)
\over \Gamma(-b^2)\Gamma\left(\alpha b-{b^2\over{2}}\right)}=
{U^{N}\left(\alpha-{b\over{2}}\right)\over
\Gamma\left(\alpha b-b^2-{1\over{2}}\right)}
+{\pi\mu b^2\ U^{N}\left(\alpha+{b\over{2}}\right)\over{
\gamma\left({1-b^2\over{2}}\right)\Gamma\left(\alpha b+{1\over{2}}\right)}}.
\label{secondeq}
\ee

The SLFT satisfies the duality $b\to 1/b$.
This property requires considering another degenerate (R) operator
$R_{-1/2b}$ which generates two more functional equations
in addition to (\ref{firsteq}) and (\ref{secondeq}).
These additional equations can be obtained by just replacing the
coupling constant $b$ with $1/b$ and the paramter $\mu$ by
the ``dual'' $\widetilde\mu$ satisfying:
\be\label{dualmu}
\pi\widetilde\mu\gamma\left({Q\over 2b}\right)=
\left[\pi\mu\gamma\left({bQ\over 2}\right) \right]^{1/{b^2}}.
\ee
Therefore, the one-point functions $U^{N}(\alpha)$ and $U^{R}(\alpha)$
should satisfy four nonlinear functional equations.

We have found the solutions to these overdetermined nonlinear equations
as follows:
\bea
U^{N}_{mn}(\alpha)&=&{\sin\left({\pi Q\over{2b}}\right)
\sin\left({\pi bQ\over{2}}\right)
\sin\left[m\pi\left({Q\over{2b}}-{\alpha\over{b}}\right)\right]
\sin\left[n\pi\left({bQ\over{2}}-b\alpha\right)\right]
\over{
\sin\left({m\pi Q\over{2b}}\right)
\sin\left({n\pi bQ\over{2}}\right)
\sin\left[\pi\left({Q\over{2b}}-{\alpha\over{b}}\right)\right]
\sin\left[\pi\left({bQ\over{2}}-b\alpha\right)\right]}} U^{N}_{11}(\alpha)
\label{unmn}\\
\nn U^{R}_{mn}(\alpha)&=&{\sin\left({\pi Q\over{2b}}\right)
\sin\left({\pi bQ\over{2}}\right)
\sin\left[m\pi\left({Q\over{2b}}-{\alpha\over{b}}+{1\over{2}}\right)\right]
\sin\left[n\pi\left({bQ\over{2}}-b\alpha+{1\over{2}}\right)\right]
\over{ \sin\left({m\pi Q\over{2b}}\right)\sin\left({n\pi bQ\over{2}}\right)
\cos\left[\pi\left({Q\over{2b}}-{\alpha\over{b}}\right)\right]
\cos\left[\pi\left({bQ\over{2}}-b\alpha\right)\right] }} U^{R}_{11}(\alpha)
\quad
\eea
where the `basic' solutions are given by:
\bea
U^{N}_{11}(\alpha)&=&
\left[\pi\mu\gamma\left({bQ\over{2}}\right)\right]^{-\alpha/b}
{\Gamma\left({bQ\over{2}}\right)\Gamma\left({Q\over{2b}}\right){Q\over{2}}
\over{ \Gamma\left(-\alpha b+{bQ\over{2}}\right)
\Gamma\left(-{\alpha\over{b}}+{Q\over{2b}}\right)
\left({Q\over{2}}-\alpha\right) }}
\label{un11}\\
U^{R}_{11}(\alpha)&=&
\left[\pi\mu\gamma\left({bQ\over{2}}\right)\right]^{-\alpha/b}
{\Gamma\left({bQ\over{2}}\right)\Gamma\left({Q\over{2b}}\right){Q\over{2}}
\over{ \Gamma\left(-\alpha b+{bQ\over{2}}+{1\over{2}}\right)
\Gamma\left(-{\alpha\over{b}}+{Q\over{2b}}+{1\over{2}}\right)}}.
\label{ur11}
\eea

There are infinite number of possible solutions which are parametrized
by two integers $(m,n)$.
For these to be solutions, we find that
the two integers should satisfy $m-n=$ even.
The basic solutions, Eqs.(\ref{un11}) and (\ref{ur11}),
can be interpreted as the one-point functions of the bulk
operators $N_{\alpha}$ and $R_{\alpha}$ corresponding to the vacuum boundary conditions (BC),
the BC corresponding to the bulk vacuum operator $N_{0}$.
Then, the  general solutions (\ref{unmn})
can be identified with the one-point functions with the conformal
BC $(m,n)$ classified by Cardy \cite{cardy}.
Since $m-n=$ even, the one-point functions we obtained correspond to the
(NS)-type BCs only.
This seems consistent with the fact that only the (NS) boundary
operators arise when the (NS) or (R) bulk degenerate operators approach
the boundary corresponding to the vacuum BC.

We also note that the solutions(\ref{unmn}) satisfy the
so-called bulk ``reflection relations'':
\be\label{bulkrefrel}
U^{N}_{m,n}(\alpha)=D^{(N)}(\alpha)U^{N}_{m,n}(Q-\alpha),\quad
U^{R}_{m,n}(\alpha)=D^{(R)}(\alpha)U^{R}_{m,n}(Q-\alpha)
\ee
where $D^{(N)}(\alpha)$ and $D^{(R)}(\alpha)$ are the (NS) and
the (R) reflection amplitudes computed in the previous section (called $S^{(N,R)}$ there) and derived in \cite{my.no3}.

Now we turn to the computation of the bulk one-point function in the presence of a boundary.
We define the SLFT on half plane where superconformally
invariant boundary action is imposed by
choosing the following boundary action at $y=0$:
\begin{equation}\nn
{\cal L}_{B}={\mu_B\over{2}}
e^{b\phi/2}a(\psi-i\gamma{\overline\psi})(x)
\end{equation}
with $\gamma=\pm 1$ and the fermionic zero-mode $a$ satisfying \cite{gosh}:
\be
\nn \sigma^{(\pm)}=a\sigma^{(\mp)}\qquad{\rm and}\qquad
a^2=1.
\ee
This action includes additional boundary parameter $\mu_B$ which
generates continuous family of BCs.
The boundary equations of motion are given by:
\bea\nn
{1\over{2\pi}}\partial_{y}\phi&=&-{1\over{2}}b\mu_B a
(\psi-i\gamma{\overline\psi})e^{b\phi/2}\\
\nn {i\over{2\pi}}\psi&=&\mu_B e^{b\phi/2}\ a,\qquad
{i\over{2\pi}}{\overline \psi}=i\gamma\mu_B e^{b\phi/2}\ a.
\eea
Plugging these constraints back into the action, one can
simplify the boundary action:
\be\nn
{\cal L}_{B}=\mu_B e^{b\phi/2}a\psi.
\ee

One can see that physical quantities should contain only even powers of
$\mu_B$ because of the fermionic zero-mode.
While the bulk properties of the boundary SLFT should be identical,
we should define the boundary operators.
As in the bulk, there are two sectors, the (NS) and (R) boundary operators:
\be\nn
n_{\beta}=e^{\beta\phi/2}(x),\qquad
r_{\beta}=\sigma^{(\epsilon)} e^{\beta\phi/2}(x).\qquad
\ee
The procedure to derive the functional equations satisfied by
the bulk one-point functions are identical to that we used above.
Major difference arises when the bulk degenerate operator $R_{-b/2}$
approaches the boundary as $z\to{\overline z}$.
The LHS of Eq.(\ref{gna}) can be evaluated by the boundary OPE which
generates the boundary operator $n_0$ and $n_{-b}$.
We choose the identity operator $n_0$, or the boundary vacuum state,
since we are interested in the bulk one-point function.
The fusion of the degenerate field $R_{-b/2}$ can be computed by
a first order perturbation from the boundary action:
\bea\nn
{\cal R}^{(\epsilon)}(-b/2,Q)&=&-\mu_B
\int dx\langle R^{(\epsilon)}_{-b/2}\left({i\over{2}}\right) a\psi(x)
e^{b\phi_B /2}(x) e^{Q\phi_B /2}(\infty)\rangle\\
\nn &=&\mu_B\int dx |x-i/2|^{b^2-1}=
2\pi\mu_B{\Gamma(-b^2)\over{\Gamma\left({1-b^2\over{2}}\right)^2}}.
\eea

Again, the dependence on the superindex $\epsilon$ disappears so that
we can suppress it.
With the vacuum state on the boundary, the two-point function becomes
the bulk one-point function of the operator $N_{\alpha}$.
Equating this with the RHS of Eq.(\ref{gna}) gives the
functional equation:
\bea\nn
{2\pi\mu_{B}\over{\Gamma\left({1-b^2\over{2}}\right)}}U^{N}(\alpha)
&=&{\Gamma\left(\alpha b-{b^2\over{2}}+{1\over{2}}\right)\over{
\Gamma\left(\alpha b-b^2\right)}}U^{R}\left(\alpha-{b\over{2}}\right)+\\
&+&{\pi\mu b^2\Gamma\left(\alpha b-{b^2\over{2}}-{1\over{2}}\right)\over{
\gamma\left({1-b^2\over{2}}\right)\Gamma(\alpha b)}}
U^{R}\left(\alpha+{b\over{2}}\right).
\label{eqns}
\eea
Similar consideration for the $G^{R}_{\alpha,-b/2}$ leads to:
\bea\nn
{2\pi\mu_{B}\over{\Gamma\left({1-b^2\over{2}}\right)}}U^{R}(\alpha)
&=&{\Gamma\left(\alpha b-{b^2\over{2}}\right)\over{
\Gamma\left(\alpha b-b^2-{1\over{2}}\right)}}
U^{N}\left(\alpha-{b\over{2}}\right)+\\
&+& {\pi\mu b^2 \Gamma\left(\alpha b-{b^2\over{2}}\right)\over{
\gamma\left({1-b^2\over{2}}\right)\Gamma\left(\alpha b+{1\over{2}}\right)}}
U^{N}\left(\alpha+{b\over{2}}\right).
\label{eqr}
\eea
As before, one should consider the dual equations coming from the dual
degenerate operator $R_{-1/2b}$.

The solutions of Eqs.(\ref{eqns}) and (\ref{eqr}) can be found as:
\bea
\label{UN}
U^{N}(\alpha)&=&{\cal N}b\left[\pi\mu\gamma\left({bQ\over{2}}\right)
\right]^{{Q-2\alpha\over{2b}}}
\Gamma\left(\left(\alpha-{Q\over{2}}\right)b\right)
\Gamma\left(1+\left(\alpha-{Q\over{2}}\right){1\over{b}}\right)\\
\nn &\times&\cosh\left[\left(\alpha-{Q\over{2}}\right)\pi s\right]\\
\label{UR}
U^{R}(\alpha)&=&{\cal N}\left[\pi\mu\gamma\left({bQ\over{2}}\right)
\right]^{{Q-2\alpha\over{2b}}}
\Gamma\left(\left(\alpha-{b\over{2}}\right)b\right)
\Gamma\left(\left(\alpha-{1\over{2b}}\right){1\over{b}}\right)\\
\nn &\times&\cosh\left[\left(\alpha-{Q\over{2}}\right)\pi s\right],
\eea
where the normalization factor ${\cal N}$ is given by:
\be\nn
{\cal N}=\left[\pi\mu\gamma\left({bQ\over{2}}\right)
\right]^{-Q/2b}
\left[b\Gamma(-Qb/2)\Gamma(1-Q/2b)\cosh(Q\pi s/2)\right]^{-1}
\ee
so that $U^{N}(0)=1$.
Here, the boundary parameter $s$ is related to $\mu_B$ by:
\be\nn
{\mu_B^2\over{\mu b^2}}\sin\left({\pi bQ\over{2}}\right)=\cosh^2\left(
{\pi bs\over{2}}\right).
\ee

Notice that the solutions (\ref{UN}) and (\ref{UR}) are self-dual if the parameter $s$ is
invariant and $\mu\to{\widetilde\mu}$ as Eq.(\ref{dualmu}).
The continuous parameter $s$ coming from $\mu_B$ generates a continuous
family of conformally invariant BCs.
One can also check that these satisfy the bulk reflection relations
Eq.(\ref{bulkrefrel}).



\setcounter{equation}{0}
\section{$N=2$ superconformal theories}

This Section is devoted to the description of the $N=2$ superconformal theories. As in $N=1$, an effective approach to the $N=2$ minimal models is the Coulomb gas representation.
We  present a detailed discussion of the Coulomb gas construction for NS, R and T fields. The basic ingredient of the $N=2$ Coulomb gas representation is the system of two free scalar fields $\phi,\phi^\dagger$ and two free fermionic fields $\psi,\psi^\dagger$, with total central charge $c=3$. In the NS sector they can be combined into two free dimensionless chiral $N=2$ superfields $S^\pm$. The Rmond and twisted primary fields can be represented by vertex operators using the spin $\s_{1/8}^\pm$ and twisted $\s_{1/16}^T$
fields of the $c=3$ system. The non-trivial dynamics of the free field construction of the $N=2$ minimal models (reflected in the non-integer central charge $c_p$) is carried by two background charges $\b,\bar\b$ placed at infinity and by the corresponding screening operators.

Another approach to the $N=2$ minimal models, based on the $D_n$ parafermionic construction \cite{zfpf,zfpf1} of the $N=2$ superconformal algebra, has an advantage in comparison with the Coulomb gas method in the calculation of the 4-point functions and the structure constants of the 2D OPE algebras. The reason for that is in the relation of the $D_n$ parafermionic models with the $su(2)$ Wess-Zumino-Witten (WZW) models \cite{wzwit}. Then the problem of the computation of the 4-point functions of the NS and R fields reduces to the simple problem to express these functions in terms of the $su(2)$ WZW functions and the 4-point functions of the free field vertices $e^{i\a\phi(z)}$. The same is true also for the structure constants and the corresponding OPE algebras. The twisted fields in this language are realized in terms of the so called C-disorder fields of the parafermionic models.

An important point in our discussion of the $N=2$ minimal models is related to the origin of the $Z_{p+2}$ discrete symmetry \cite{gep5}. It turns out that in each $c_p$ model there exist a set of NS superfields $N^p_{-m}$ which together with the super-stress-energy tensor $\CW(z,\theta^+,\theta^-)$ close an OPE algebra of the $Z_{p+2}$ parafermionic type. We describe in details this $N=2$ super-parafermionic  models and their relation to the corresponding $N=2$ superconformal minimal models. It turns out that the super-parafermionic currents and the order-disorder fields are precisely the fields that play an important role in the construction of the vertices of the low-energy massless particles \cite{gep5}. So these, as well as our general results for the FR's, can be applied to the Gepner's tensor product construction. More precisely, we consider in this Section the three-generation $1^116^3$ Gepner model
of the heterotic string compactification \cite{gep5}. The main point of the Gepner's approach is the construction of the (moded out) tensor products of $N=2$ minimal models
($c=1+3\times 8/3=9$)
having the same discrete symmetries as the maximally symmetric  Calabi-Yau (C-Y) space. The generalized GSO projection relates specific combinations of the fields of these models with the free fields of the non-compactified degrees of freedom which reproduce the massless spectrum of the C-Y three-generation string model \cite{gep5}. An important step in the consistency checks of this model is the explicit construction of the low-energy effective cubic superpotential. In this Section we derive the exact Yukawa couplings for the $1^116^3$ model.

The rest of this Section is devoted to the discussion of the $N=2$ super Liouville field theory (SLFT). We show that it exhibits an interesting duality behaviour. Under the dual transformation $b\to 1/b$ the theory maps to a dual action which is another $N=2$ super CFT. The $N=2$ SLFT with a strong coupling can be described by the dual action perturbatively. We compute the reflection amplitudes (the 2-point functions) of the theory using functional relations derived from these actions. This procedure provides an exact relation between the two actions. An important generalization is the $N=2$ SLFT in the presence of a boundary. We use the modular  bootstrap method, which is a generalization of the Cardy formulation for the conformal BC's to the irrational CFT's. We compute the one-point function for general BC's parameterized by a continuous parameter (the so called FZZT branes). Using the one-point functions, we rederive the bulk reflection amplitudes and compare them with those obtained before. Furthermore, we provide a conformal bootstrap analysis based on the $N=2$ SLFT and its dual theory and confirm that the one-point functions obtained from the modular transformations are consistent with the bulk and boundary actions. As a byproduct, we obtain a relation between the continuous BC parameter and the boundary cosmological constants of the two dual theories. To find all the consistent conformal BC's of the $N=2$ SLFT is important since they describe the D-branes moving in the black hole background. In addition to the FZZT and the vacuum BC's there is an infinite number of discrete BC's
which are called ZZ branes in general. We find the ZZ brane solutions from the functional equations defined on the pseudosphere and discuss their implications for the $N=2$ SLFT.
We then perform modular bootstrap calculations for the degenerate fields of the theory which provide some consistency checks for the solutions.

At the end of this Section we find a set of higher equations of motion in $N=2$ SLFT. In doing that we used the one point function obtained above. As already mentioned this is an interesting question because of the fact that this theory has actually few properties in common with the $N=0,1$ SLFT's. One difference is the lack of the strong-week coupling duality mentioned above. Another important difference is in the spectrum of the degenerate representations \cite{bfk,vpz,myne6}. We will show that the $N=2$ SLFT still possesses higher equations of motion despite these differences.

The results of this Section have been published in \cite{my.no1}, \cite{myne6}-\cite{myne12}, (9.-16.).


\subsection{$N=2$ minimal models}

The set of generators of the $N=2$ extended superconformal algebra contains in addition to the stress-energy tensor $T(z)$ and the supercurrent $G^1(z)$ (generating $N=1$ SUSY) a $U(1)$ current $J(z)$ of conformal dimension 1 and its super-partner $G^2(z)$ of dimension $3/2$ (and also the corresponding right-moving modes).
It is convenient to use the $U(1)$ diagonal basis for the description of the NS and R sectors of the $N=2$ algebra: $G^\pm(z)={1\ov \sqrt{2}}(G^1\pm iG^2)$ so that:
\bea
\nn T(z_1)G^\pm(z_2)&=&{3\ov 2z_{12}^2}G^\pm(z_2)+{1\ov z_{12}}\p G^\pm(z_2)+\ldots\\
\nn J(z_1)G^\pm(z_2)&=&\pm {1\ov 2z_{12}}G^\pm(z_2)+\ldots\\
\nn G^+(z_1)G^-(z_2)&=&{2c\ov 3z_{12}^3}+{4\ov z_{12}^2}J(z_2)+{2\ov z_{12}}(T(z_2)+\p J(z_2))+\ldots\\
\nn G^+(z_1)G^+(z_2)&=&O(z_{12})=G^-(z_1)G^-(z_2).
\eea
Similarly to the case of $N=1$ SUSY one may choose periodic or antiperiodic boundary conditions for the supercurrents with corresponding Laurent expansions:
\be
\nn G^\pm(z_)=\sum_{n\in Z}z^{-n-2}G_{n+1/2}^\pm\qquad or\qquad G^\pm(z_)=\sum_{n\in Z}z^{-n-3/2}G_{n}^\pm.
\ee
The coefficients $G_r^\pm$ of these expansions close NS and R parts of the $N=2$ superalgebra respectively:
\bea\label{n2alg}
\left[L_m,G^{\pm}_{r}\right]&=&\left({m\over{2}}-r\right)G^{\pm}_{m+r},
\qquad \left[J_n,G^{\pm}_{r}\right]=\pm G^{\pm}_{n+r},\\
\nn \left\{G^{+}_{r},G^{-}_{s}\right\}&=&2L_{r+s}+(r-s)J_{r+s}
+{c\over{3}}\left(r^2-{1\over{4}}\right)\delta_{r+s},\quad
\left\{G^{\pm}_{r},G^{\pm}_{s}\right\}=0,\\
\nn \left[L_m,J_{n}\right]&=&-nJ_{m+n},\qquad
\left[J_m,J_{n}\right]={c\over{3}}m\delta_{m+n},
\eea
$r,s\in Z+1/2$ for the NS sector, $r,s\in Z$ for the R sector and:
\be
\nn T(z)=\sum_{n\in Z}z^{-n-2}L_n,\qquad J(z)=\sum_{n\in Z}z^{-n-1}J_n.
\ee

Due to the presence of the $U(1)$ current $J(z)$ we have one more possibility, that is to choose $Z_2$ twisted boundary conditions for this current $J(e^{2\pi i}z)=-J(z)$ and:
\be
\nn J(z)=\sum_{n\in Z}z^{-n-1/2}J_{n+1/2}.
\ee
Then, as a consequence, one of the currents $G^{1,2}(z)$ nust have periodic and the other one antiperiodic boundary conditions. The two choices are equivalent so we can take for example:
\be
\nn G^1(z)=\sum_{n\in Z}z^{-n-2}G_{n+1/2}^1,\qquad G^2(z)=\sum_{n\in Z}z^{-n-3/2}G_{n}^2
\ee
obeying the so called $N=2$ twisted superconformal algebra.

The Cartan subalgebra of the NS and R algebras is two-dimensional and consists of the generators $L_0$ and $J_0$. Therefore, the corresponding lowest weight representations (LWR's) for the fixed value of $c$ are labeled by two parameters: the conformal dimension $\D$ and the $U(1)$ charge $q$. The primary states $|\D,q>$ which generate these LWR's satisfy the conditions:
\bea
\nn L_0|\D,q>&=&\D |\D,q>,\qquad J_0|\D,q>=q |\D,q>,\\
\nn L_n|\D,q>&=&J_m|\D,q>=G_r^\pm|\D,q>=0,\quad n,m,r>0.
\eea

Let us consider the NS sector first. Introducing the $OSp(2|2)$ invariant vacuum of the the theory which belongs to the NS sector:
\be
\nn L_n|0>=J_{n+1}|0>=G_r^\pm|0>=0,\qquad n\ge -1 ,r\ge -1/2
\ee
we can realize the NS primary states $|\D,q>$ by acting on the vacuum with the NS primary superfields of dimension $\D$ and $U(1)$ charge $q$:
\be
\nn N(z,\theta^+,\theta^-)=\varphi(z)+\theta^+\psi^-(z)+\theta^-\psi^+(z)+\theta^+\theta^-\tilde\varphi(z).
\ee
The component fields have the following OPE's with the generators:
\bea
\nn G^\pm(z_1)\vf(z_2)&=&{1\ov z_{12}}\psi^\pm(z_2)+\ldots\\
\nn G^\pm(z_1)\psi^\mp(z_2)&=&{2(\D\pm q)\ov z_{12}^2}\vf(z_2)+{1\ov z_{12}}(\p\vf(z_2)\pm \tilde\vf(z_2))+\ldots\\
\nn G^\pm(z_1)\tilde\vf(z_2)&=&\mp{2(\D\pm q)\ov z_{12}^2}\psi^\pm(z_2)+\ldots.
\eea

The LWR in the NS sector of the $N=2$ superconformal algebra corresponding to the LW state $|\D,q>$ consists of all the linear combinations of the vectors of the form:
\be\nn
L_{\{-n_i\}}J_{\{-m_i\}}G^+_{\{-r_i\}}G^-_{\{-s_i\}}|\D,q>,\quad n_i,m_i,r_i,s_i\ge 0
\ee
where $\{n_i\}$ etc. is a multi-index and $k=\sum_in_i+\sum_im_i+\sum_ir_i+\sum_is_i$ and $m=\hf(\#G^+-\#G^-)$ are called level and relative charge of the state respectively. Because of the specific properties of the generators: $\{G^\pm_r,G^\pm_s\}=0$, it turns out that only the relative charges $0$ (at integer level $k$) and $\pm 1/2$ (at half-integer level)are allowed.

The two and three-point functions of the primary superfields $N(z,\theta^+,\theta^-)$ are determined (up to an arbitrary constant) by the conditions of $OSp(2|2)$ invariance i.e. the finite subalgebra of (\ref{n2alg}) spanned by $L_0,L_{\pm1},J_0,G^\pm_{\pm\hf}$. Using the corresponding superconformal Ward identities and the invariance of the vacuum, these conditions lead to a system of differential equations which can be solved (up to a constant). Then the corresponding two-point function for example has the form:
\be\label{tpf}
<N_1(z_1,\theta^+_1,\theta^-_1)N(z_2,\theta^+_2,\theta^-_2)>=C_2\tilde z_{12}^{-2\D_1}\left(1+2q_1{\theta^-_{12}\theta^+_{12}\ov \tilde z_{12}}\right)\d_{\D_1-\D_2}\d_{q_1+q_2}.
\ee

In the case of the 3-point functions one can see that there are three independent solutions corresponding to the following three possibilities for the total $U(1)$ charge: $q_1+q_2+q_3=0,\pm 1/2$ which are dictated by the $J_0$ invariance. Thus we have one even 3-point function:
\be\nn
<N_{\D_1}^{q_1}N_{\D_2}^{q_2}N_{\D_3}^{-q_1-q_2}>
\ee
and two odd ones:
\be\nn
<N_{\D_1}^{q_1}N_{\D_2}^{q_2}N_{\D_3}^{-q_1-q_2\pm 1/2}>
\ee
with three independent arbitrary structure constants $C_3, C_3^\pm$. The fact that we can have three different 3-point functions has an important consequence for the analysis of the fusion rules for the NS fields. As we shall show later it gives rise to three independent NS fusion rules- one even and two odd ones.

In this Section we will be interested in the so called degenerate unitary representations. Their main peculiarity is the existence of null-vectors at given levels and relative charges, i.e. states which are again primary. This property gives rise to a certain relation between the parameters characterizing these representations: the central charge $c$, the dimension $\D$ and the $U(1)$ charge q, the so called Kac formula, which completely classify them. For the NS sector of the $N=2$ superconformal algebra this formula reads \cite{bfk,vpz}:
\bea\label{ntrep}
c&=&3-{6\ov p+2}, \quad p=1,2,\ldots, \quad q_s={s\ov 2(p+2)},\\
\nn \D_{n1}^s&=&{(p+2-n)^2-s^2-1\ov 4(p+2)},\quad n=0,1,\ldots, p+1, \quad |s|\le p-n+1,\\
\nn\D_{n0}^s&=&{(n+|s|)^2-s^2-1\ov 4(p+2)}, \quad n+|s|\le p+1, \quad n=1,3,5,\ldots
\eea
Correspondingly, one can introduce two kinds of degenerate NS superfields: $N_{n1}^s$ which has a degeneracy at level $n$ with zero relative charge and $N_{n0}^s$ having level $n/2$ degeneracy with relative charge $\pm 1/2$.

The simplest null-vectors, at level $1/2$ and relative charge $\pm 1/2$, have the form:
\be\nn
G^\pm_{-\hf}|\D,q>=0
\ee
provided $\D=\pm q$. They generate the so called $N=2$ chiral (and anti-chiral)  superfields:
\bea\nn
N^+(z,\theta^+,\theta^-)&=&\varphi(z)+\theta^+\psi^-(z)-\theta^+\theta^-\p\vf(z),\quad D^+N^+=0,\\
\nn N^-(z,\theta^+,\theta^-)&=&\varphi(z)+\theta^-\psi^+(z)+\theta^+\theta^-\p\vf(z),\quad D^-N^-=0,
\eea
$D^+$ and $D^-$ are the $N=2$ supercovariant derivatives.

In the R sector we have in addition to $L_0$ and $J_0$ the zero modes $G_0^\pm$ of the supercurrents $G^\pm(z)$. As a consequence, for each $\D\ne c/24$ we have to consider two Ramond states $|\D,q>$ and $|\D,q+1/2>$. For the corresponding R fields $R_\D^q$ creating the R states from the NS vacuum: $|\D,q>_R=R_\D^q|0,0>$ we get the following WI:
\bea
\nn G^+(z_1)R_\D^q(z_2)&=&{\sqrt{2\D-{c\ov 12}}\ov z_{12}^{3/2}}(z_2)R_\D^{q+1/2}(z_2)+\ldots\\
\nn G^-(z_1)R_\D^{q+1/2}(z_2)&=&{\sqrt{2\D-{c\ov 12}}\ov z_{12}^{3/2}}(z_2)R_\D^{q}(z_2)+\ldots\\
\nn G^+(z_1)R_\D^{q+1/2}(z_2)&=&O(\sqrt{z_{12}})=G^-(z_1)R_\D^q(z_2)
\eea
The structure of the degenerate LWR representations in the R sector is very similar to that of the NS sector. The formula for the dimensions and charges of the degenerate fields is now:
\bea\label{rrep}
c&=&3-{6\ov p+2}, \quad p=1,2,\ldots, \quad q_r^s={s-r\ov 2(p+2)}+{r\ov 4},\quad r=\pm 1,\\
\nn \D_{n1}^{sr}&=&{(p+2-n)^2-(s-r)^2-1\ov 4(p+2)}+{1\ov 8},\quad n=1,\ldots, p+1, \quad |s|\le p-n+1,\\
\nn \D_{n0}^{sr}&=&{(n+|s-r|)^2-(s-r)^2-1\ov 4(p+2)}+{1\ov 8}, \quad n+|s-r|\le p+1, \quad n=0,2,4,\ldots
\eea
In the conformal family of the primary field $R_{n1}^{sr}(z)$ there is a degeneracy at level $n$ with relative charge $0$, while the null-vector for the field $R_{n0}^{sr}(z)$ is at level $n/2$ (which is integer in this case) with relative charge $\pm 1/2$.

In the twisted sector, because of the antiperiodic boundary conditions for the $U(1)$ current $J(z)$, it has no zero mode and hence the primary states are labeled by the value of the conformal dimension only: $L_0|\D>=\D|\D>$ and $|\D>$ is annihilated by all the positive modes. Similarly to the R sector of the $N=1$ SUSY $|\D>$ is doubly degenerated, i.e. to each primary state $|\D,+>$ there corresponds a state $|\D,->\sim G_0^2|\D,+>$ with the same dimension $\D$ which is again primary. Due to the properties of $G_0^2$ we have for the corresponding primary fields:
\be\nn
G^2(z_1)T^\pm_\D(z_2)=\sqrt{\D-{c\ov 24}}{1\ov z_{12}^{3/2}}T^\mp_\D(z_2)+\ldots.
\ee
It is clear that the twisted $N=2$ multiplet has more complicated structure. At level $1/2$ we have in general two independent descendants:
\bea\nn
J_{-1/2}|\D,\pm>&=&J_{-1/2}T^\pm_\D(0)|0>=t^\pm_{\D+1/2}(0)|0>,\\
\nn G^1_{-1/2}|\D,\pm>&=&T^\pm_{\D+1/2}(0)|0>
\eea
which are not primary states of the full $N=2$ algebra. Considering the $N=1$ subalgebra (which is generated by the supercurrent $G^1(z)$) we can combine the fields $T^\pm_\D(z)$ and $T^\pm_{\D+1/2}(z)$ in a $N=1$ superfield.

As in the NS and R sectors we are interested in the LWR's only. The series of dimensions of the degenerate primary fields is given by:
\be\nn
\D_n={\left({p+2\ov 2}-n\right)^2-1\ov 4(p+2)}+{1\ov 8},\qquad n=1,2,\ldots, p+2,
\ee
where the field $T_n(z)$ has degeneracy at level $n/2$ ($n\in Z_+$).

Let us now pass to the Coulomb gas representation of the $N=2$ supersymmetric models. It is slightly different from that we considered in the case of $N=1$ ones. It is based on the theory of two $N=2$ NS chiral superfields:
\bea\nn
S^+(z,\th^+,\th^-)&=&\phi^+(z)+\th^-\psi^+(z)+\th^-\th^+\p\phi^+(z)\\
\nn S^-(z,\th^+,\th^-)&=&\phi^-(z)+\th^+\psi^-(z)-\th^-\th^+\p\phi^-(z)
\eea
where $\phi^\pm$ and $\psi^\pm$ are free complex scalar and fermion fields respectively. The chirality is guaranteed by the covariant condition:
\be\nn
D^+S^-=0=D^-S^+.
\ee
The free action of the theory:
\be\nn
A(S^+,S^-)=\hf\int d^2z d^2\th d^2\bar\th S^+S^-
\ee
(here we denoted by $S^\pm$ the two-dimensional superfield) gives the propagators of the chiral fields. In terms of components we have:
\be\nn
<\psi^+(z_1)\psi^-(z_2)>={1\ov z_{12}},\qquad <\phi^+(z_1)\phi^-(z_2)>=-\ln{z_{12}}.
\ee

Analogously to the $N=1$ case we define the NS vertices as:
\be\label{nsv}
N_{\a,\bar\a}(z,\th^+,\th^-)=\exp{i(\a S^-(z,\th^+,\th^-)+\bar\a S^+(z,\th^+,\th^-))}.
\ee
They are labeled by two  real numbers $(\a,\bar\a)$ called charges. The corresponding improved action:
\be\nn
\CA(S^\pm,\tilde R)=A(S^+,S^-)+\int d^2z d^2\th d^2\bar\th (2\b \tilde R S^++2\bar\b \tilde R S^-)
\ee
where $\tilde R$ is the $N=2$ supersymmetric curvature (which could be placed at infinity), leads effectively to the extra vertex at infinity with charges $(-2\b,-2\bar\b)$. In what follows we made a special choice $\b=\bar\b$.

Similarly to the $N=1$ case, in order to get a consistent $N$-point function of the NS vertices, we must impose the neutrality condition:
\be\label{neut}
\sum_{i=1}^N\a_i=2\b=\sum_{i=1}^N\bar\a_i.
\ee
From the improved action one can extract the form of the generators of the $N=2$ superconformal symmetry:
\bea\label{gentwo}
J&=&\hf \psi^+\psi^--i\b\p\phi^-+i\b\p\phi^+,\\
\nn G^+=-\sqrt{2}\psi^+\p\phi^-&+&2\sqrt{2}i\b\p\psi^+,\qquad G^-=-\sqrt{2}\psi^-\p\phi^++2\sqrt{2}i\b\p\psi^-,\\
\nn T=-\p\phi^+\p\phi^-&+&\hf(\psi^+\p\psi^-+\psi^-\p\psi^+)+i\b\p^2\phi^++i\b\p^2\phi^-.
\eea
They close a $N=2$ superconformal algebra with a central charge:
\be\nn
c=3-24\b^2.
\ee
Thus, in order to describe the minimal models, the charge at infinity should be quantized:
\be\nn
\b^2={1\ov 4(p+2)},\qquad p=1,2,3,\ldots .
\ee

We first discuss in more details the Coulomb gas representation of the NS sector. From the explicit form of the NS vertex (\ref{nsv}) and the $N=2$ generators (\ref{gentwo}) one can extract the conformal dimension $\D$ and the $U(1)$ charge $q$ of the NS primary field:
\be\label{dq}
\D(\a,\bar\a)=\a\bar\a-\b(\a+\bar\a),\qquad q(\a,\bar\a)=\b(a-\bar\a).
\ee
We note some symmetries of these formulas which will be useful below. The change:
\be\nn
\a\rightarrow 2\b-\bar\a,\qquad \bar\a\rightarrow 2\b-\a
\ee
leads to the same dimension and charge $\D,q$ while the following two changes:
\be\nn
\a\rightarrow 2\b-\a,\qquad \bar\a\rightarrow 2\b-\bar\a
\ee
and:
\be\nn
\a\rightarrow \bar\a,\qquad \bar\a\rightarrow \a
\ee
give the same dimension $\D$ but but the opposite charge $-q$.

Let us introduce the following NS vertices:
\bea\label{scr}
V_{\a_+,\bar\a_+}&=&e^{i(\a_+S^-+\bar\a_+S^+)},\\
\nn V_{\a_-,0}&=&e^{i\a_-S^-},\qquad V_{0,\bar\a_-}=e^{i\bar\a_-S^+}
\eea
with charges $\a_\pm,\bar\a_\pm$ chosen in such a way that the corresponding integrals (called screening operators):
\bea\label{scrint}
Q_+&=&\oint_{C_+}dz d\th^+d\th^- V_{\a_+,\bar\a_+},\\
\nn Q_-&=&\oint_{C_-}dz d\th^+V_{\a_-,0},\qquad \bar Q_-=\oint_{\bar C_-}dz d\th^-V_{0,\bar\a_-}
\eea
are invariant under the $N=2$ superconformal transformations, i.e. have zero dimension and zero charge. This condition determines the charges $\a_\pm,\bar\a_\pm$ as a function of the charge at infinity:
\bea\label{chinf}
\a_+&=&2\b=\bar\a+,\qquad \D(\a_+)=0=q(\a_+),\\
\nn \a_-&=&-{1\ov 2\b}=\bar\a_-,\quad \D(\a_-)=\hf=\D(\bar\a_-),\quad q(\a_-)=-\hf==-q(\bar\a_-).
\eea

The analysis of the null-vectors in the Coulomb gas picture (constructed with the help of the screening operators) leads to the realization of the fields in the minimal models through the vertex operators (\ref{nsv}). It turns out that in this case the charges $\a,\bar\a$ are quantized. For example, the fields $N_{n1}^s$ (degenerated at level $n$ with relative charge zero) are represented by the vertex operator (\ref{nsv}) with charges:
\be\nn
\a_{n1}^s=\hf(1-n+s)\a_+-\hf\a_-,\qquad \bar\a_{n1}^s=\hf(1-n-s)\a_+-\hf\a_-.
\ee
Analogously, for the field $N_{n0}^s$ degenerated at level $n/2$ with relative charge $1/2$ we obtain:
\be\nn
\a_{n0}^s=\hf(1-n)\a_+,\quad \bar\a_{n1}^s=\hf(1-n-2s)\a_+,\quad s>0,
\ee
while for the one with relative charge $-1/2$ we have:
\be\nn
\a_{n0}^s=\hf(1-n+2s)\a_+,\quad \bar\a_{n1}^s=\hf(1-n)\a_+,\quad s<0.
\ee

In the Coulomb gas picture the correlation functions of the fields are obtained by inserting in the corresponding correlators of the vertices (\ref{nsv}) a proper number of screening operators (\ref{scrint}). They ensure that the neutrality condition (\ref{neut}) is satisfied. Also, they do not destroy the $N=2$ superconformal covariance of the functions since they are $N=2$ superconformally invariant. In the case of the 3-point functions this screening procedure allows one to recognize which of them do not vanish, or equivalently to obtain the corresponding FR's for the products of two arbitrary primary fields. As we discussed above, the general 3-point function of NS $N=2$ superfields contains one even and two odd ingredients (with charges $\pm 1/2$ respectively). This specific structure gives rise to three different FR's, one even and two odd, generated by the corresponding parts of this function. The even 3-point function is obtained by inserting an arbitrary number of even screening operators and an equal number of the two kinds of odd ones (in order to have an uncharged function). The odd 3-point function with charge $1/2$ ($-1/2$) can be obtain instead if we insert in the correlator  an arbitrary number of even screening operators $Q_+$ but one more of the screenings $Q_-$ ($\bar Q_-$) than $\bar Q_-$ ($Q_-$). Finally, combining  the even and odd FR's, we obtain the general ones, i.e. we obtain all the (families of) primary fields that appear in the OPE of two given superfields. We do not present here the explicit results for these FR's. This is because, as we will explain below, in the case of minimal $N=2$ models, there is a more convenient construction in terms of the so called parafermionic theories. The FR's have more compact form in that description so we leave the explicit expressions for the discussion of that construction. The same concerns also the case of the 4-point functions.

We turn now to the discussion of the Ramond sector. The generators of the $N=2$ superconformal  algebra are the same as in the NS sector (\ref{gentwo}). The only difference is that in this case $G^\pm(z)$ are antiperiodic fields and therefore we have to impose antiperiodic boundary conditions for the fields $\psi^\pm(z)$ too. Then the fields are no more combined in supermultiplets. Similarly to the $N=1$ case we define the R vertices as follows:
\be\label{rver}
R_{\a,\bar\a}^r(z)=\s^r e^{i(\a\phi^++\bar\a\phi^-)}(z),\qquad r=\pm
\ee
where the fields $\s^\pm(z)$ of dimension $\D=1/8$ and charge $q=\pm 1/4$ correspond to the lowest energy states in the R sector of the algebra of the complex fermion field $\psi^\pm(z)$. For the dimension and charge of the field $R_{\a,\bar\a}^r$ we obtain:
\be\nn
\D(\a,\bar\a,r)=\a\bar\a-\b(\a+\bar\a)+{1\ov 8},\qquad q(\a,\bar\a,r)=\b(a-\bar\a)+{r\ov 4}.
\ee
The symmetries of these formulas are similar to those of the NS sector:
\be\nn
\a\rightarrow 2\b-\bar\a,\quad \bar\a\rightarrow 2\b-\a,\quad r\rightarrow r
\ee
leads to $\D,q\rightarrow \D,q$, and:
\be\nn
\a\rightarrow 2\b-\a,\quad \bar\a\rightarrow 2\b-\bar\a,\quad r\rightarrow -r
\ee
or:
\be\nn
\a\rightarrow \bar\a,\quad \bar\a\rightarrow \a,\quad r\rightarrow -r
\ee
give the same dimension $\D$ but the opposite charge $-q$.

As in the $N=1$ case, the screening operators have the same form as in the NS sector. The only difference is that $\psi^\pm(z)$ are now antiperiodic. So the null-vector construction goes in the same way. As a result we obtain that the vertices representing the degenerate fields in the Ramond sector have a similar form. More precisely, the charges of the vertices of the fields $R_{n1}^{sr}(z)$ are obtained form those of $N_{n1}^{s}(z)$ by simply replacing $s$ with $s-r$. In the same way we obtain the charges of $R_{n0}^{sr}(z)$ from those of $N_{n0}^{s}(z)$.

In order to obtain the FR's for the R fields we have to recognize which of the 3-point functions $<R_1R_2N_x>$ are different from zero. In the Coulomb gas picture this means to obtain all the ways of  screening such function. We have to consider separately the two cases: $<R^rR^{-r}>$ and $<R^rR^{+r}>$. In the first case the corresponding R fields are constructed with the help of the fields $\s^{\pm r}$ respectively. The 3-point function is then proportional to:
\be\nn
<\s^r\s^{-r}\psi^+\psi^-\ldots>.
\ee
It is different from zero only if the number of screenings $Q_-$ is equal to the number of $\bar Q_-$, the number of the even screenings $Q_+$ remains arbitrary. Therefore the screening procedure in this case is exactly the same as in the case of the even FR's  in the NS sector. In the second case the corresponding 3-point function is proportional to:
\be\nn
<\s^r\s^{r}\psi^{r'}\ldots>
\ee
where $\psi^{r'}$ is one of the fields $\psi^\pm$ which makes this function neutral. Here we have two possibilities. The first one is to put one more screening operator $Q_-$
or $\bar Q_-$ and hence to have the same procedure as in the case of odd FR's in the NS sector. The second possibility is to implement the fact that the second component of the NS field (built with the help of some $\psi^\pm$) can contribute. The corresponding 3-point function is screened then with equal number of $Q_-$ and $\bar Q_-$. Therefore this possibility corresponds to the even screening procedure in the NS sector. As before, we do not present here the explicit expressions for the Ramond FR's and the corresponding structure constants. We postpone this problem to the parafermionic description of the $N=2$ minimal models.

Finally, we describe the construction of the twisted vertices. As we explained above we redefine the supercuurents such that $G^2\sim G^+-G^-$ is now antiperiodic. Also, the $U(1)$ current $J$:
\be\nn
J(z)={1\ov 4}\psi_1\psi_2 +i\b \p\phi_2
\ee
is antiperiodic in this sector (here we introduced the field $\phi_2$ which is a linear combination of $\phi^\pm$). It follows that the scalar field $\phi_2$ should be antiperiodic too. The other scalar field $\phi_1$ is periodic. Following the general idea, we define the twisted field $T(z)$ in terms of the lowest dimensional field in the twisted sector of the fermion $\psi_1,\psi_2$ theory called $\s_0^\psi$, the lowest twisted field of the $U(1)$ current $\p\phi_2$ - $\s_0^\phi$ and the exponential of the free scalar field $\phi_1$:
\be\nn
T_\a(z)=\s_0^\psi(z)\s_0^\phi(z)e^{i\a\phi_1(z)}.
\ee
One can check that all the properties of the T-primary field are satisfied. In particular, its dimension is given by:
\be\nn
\D(\a)=\a^2-2\a\b+{1\ov 8}
\ee
As in the NS and R sectors, the analysis of the null-vectors gives the quantized charge of the degenerate fields:
\be\nn
\a_n=\hf(1-n)\a_+-{1\ov 4}\a_-
\ee
and the corresponding field $T_n$ has a degeneracy at level $n/2$. The fusion rules are obtained analogously to NS and R sectors. The difference is that in the case of the lowest twisted field $\s_0^\psi$ we have the following FR's:
\be\nn
\s_0^\psi\s_0^\psi\sim 1+\s^\pm+\psi.
\ee
This confirms the fact that in the product of two twisted fields both Ramond and NS fields occur.

As we already mentioned above, the $N=2$ superconformal minimal models admit a representation in terms of the  $D_{2p}$ parafermionic (PF) theories. It is based on the observation \cite{zfpf,zfpf1} of the fact that the generators of the $N=2$ supersymmetric theory could be expressed in terms of the PF currents and a free scalar field:
\bea\label{pfc}
T&=&T_p+T_\vf,\qquad J={i\ov 2}{p\ov \sqrt{2p(p+2)}}\p\vf,\\
\nn G^+&=&\sqrt{{2p\ov p+2}}\psi_1\exp{\left(i{p+2\ov \sqrt{2p(p+2)}}\vf\right)},\\
\nn G^-&=&\sqrt{{2p\ov p+2}}\psi_1^\dagger\exp{\left(-i{p+2\ov \sqrt{2p(p+2)}}\vf\right)}
\eea
where $\psi_1$ and $\psi_1^\dagger$ are parafermionic currents with $Z_p$ charges $1$ and $p-1$ respectively, and the reflection $C$ acts as: $\psi_1\rightarrow\psi_1^\dagger$.

The central charge of these theories takes the values: $c_p=2{p-1\ov p+2},\quad p=1,2\ldots$. In (\ref{pfc}) $\vf$ denotes a free scalar field with central charge $1$. Therefore the central charge of the algebra of the currents (\ref{pfc}) is:
\be\nn
c=1+c_p={3p\ov p+2},\qquad p=1,2,\ldots
\ee
which coincides with that of the $N=2$ minimal models.

The lowest dimensional fields $\phi_m^l$ of the parafermionic theory have dimensions:
\be\nn
d_m^l=d_l+{l^2-m^2\ov 4p},\qquad -l\le m\le l
\ee
where:
\be\nn
d_l={l(p-l)\ov 2p(p+2)},\qquad l=0,1,\ldots,p
\ee
is the dimension of the corresponding  order parameter fields $\s_l=\phi_l^l$. The dimension $d_m^l$ has the following symmetries:
\bea\label{simd}
l&\rightarrow& l,\qquad m\rightarrow -m,\\
\nn l&\rightarrow& p-l,\qquad m\rightarrow p\pm m
\eea
where we identify $m=m+2p$ since m is the $Z_{2p}$ charge of the field $\phi_m^l$.

The primary fields in the $N=2$ theories are constructed from the lowest fields of the PF theory and exponentials of the free scalar field $\vf$. For the NS sector we have:
\bea\label{pfns}
N_m^l(z)&=&\phi_m^l(z)\exp{\left(i{m\ov \sqrt{2p(p+2)}}\vf(z)\right)},\\
\nn l&=&0,1,\ldots,p\quad m=-l,-l+2,\ldots,l.
\eea
The $U(1)$ charge of this field is:
\be\nn
q_m^l={m\ov 2(p+2)}
\ee
and its dimension is simply the sum of the dimensions of the two ingredients:
\be\nn
\D_m^l=d_m^l+{m^2\ov 2p(p+2)}={l(l+2)\ov 4(p+2)}-{m^2\ov 4(p+2)}.
\ee
To make a connection with the Coulomb gas representation of the NS sector we note that the first series $N_{n1}^s$ is obtained from this construction as:
\be\nn
s=m,\qquad n=p-l+1
\ee
and for the second one $N_{n0}^s$:
\be\nn
s=m,\qquad n=l-|m|+1.
\ee
The product with the supercurrents is:
\be\nn
G^\pm(z_1)N_m^l(z_2)=\sqrt{{2p\ov p+2}}{1\ov z_{12}}\phi_{m\pm 2}^l(z_2)e^{i\left(m\pm(p+2)/\sqrt{2p(p+2)}\right)\vf(z_2)}+\ldots.
\ee
Note that, due to the symmetries (\ref{simd}), the second component of the field $N_m^l$ has the form:
\be\nn
(N_m^l)^{II\pm}\sim \phi_{m\pm (p+2)}^{p-l}e^{i\left(m\pm(p+2)/\sqrt{2p(p+2)}\right)\vf}
\ee
and therefore it looks just like $N_{m\pm(p+2)}^{p-l}$ but without satisfying the conditions (\ref{pfns}).

The primary fields of the Ramond sector  are represented in a similar way:
\bea\label{pfr}
R_{m,\a}^l(z)&=&\phi_m^l(z)\exp{\left(i{m-\a p/2\ov \sqrt{2p(p+2)}}\vf(z)\right)},\\
\nn l&=&0,1,\ldots,p\quad m=-l,-l+2,\ldots,l,\quad \a=\pm 1.
\eea
The $U(1)$ charge is given by:
\be\nn
q_{m,\a}^l={2m-\a p\ov 4(p+2)}
\ee
and the conformal dimensions are:
\be\nn
\D_{m,\a}^l={l(l+2)\ov 4(p+2)}-{(m+\a)^2\ov 4(p+2)}+{1\ov 8}.
\ee
The fields (\ref{pfr}) reproduce the right analytic behaviour of the supercurrennts. For example:
\be\nn
G^+(z_1)R_{m,\a}^l(z_2)=\sqrt{{2p\ov p+2}}{1\ov z^{1+\a/2}_{12}}\phi_{m+2}^l(z_2)e^{i\left(m+2+(2-\a)p/2/\sqrt{2p(p+2)}\right)\vf(z_2)}+\ldots
\ee
shows that the fields $R_{m,\a}^l$ produce the right branch cut singularities of the supercurrent $G^+(z)$. Similar result holds also for the OPE with with $G^-(z)$. The expressions for the $U(1)$ charge and the dimension of the R fields coincide with those of the $N=2$ minimal models with the following identification:
\be\nn
s=m,\quad n=p-l+1,\quad r=-\a
\ee
for the series $R_{n1}^{rs}$ and:
\be\nn
s=m,\quad n=l-|m|+1,\quad r=-\a
\ee
for the second one $R_{n0}^{rs}$.

In order to construct the twisted fields we have to consider the product of a primary field $\vf^{(s)}(z)$ from the so called C-disorder sector of the PF models and the twisted field $\s_0^\vf(z)$ representing the lowest weight state of the $Z_2$ twisted current $\p\vf(z)$:
\be\label{tcon}
T_s(z)=\vf^{(s)}(z)\s_0^\vf(z).
\ee
The fields $\vf^{(s)}$ are characterized by their OPE's with the PF currents:
\be\nn
\psi_1(z)\vf^{(s)}(0)=z^{-\D_1}\sum_{n\in Z}z^{-n/2}A_{n/2}^{(1)}\vf^{(s)}(0)
\ee
and have dimensions:
\be\nn
\D^{(s)}={p-2+(p-2s)^2\ov 16(p+2)},\quad s=0,1,\ldots |p/2|.
\ee
The $Z_2$ twisted field $\s_0^\vf(z)$ has a dimension $1/16$ and is defined by the OPE:
\be\nn
\p\vf(z)\s_0^\vf(0)={1\ov \sqrt{z}}\s_1^\vf(0)+\ldots .
\ee
As it can be easily seen, the construction (\ref{tcon}) leads to the well known dimension of the twisted fields of the discrete unitary series:
\be\nn
\D_s={(p-2s)^2-4\ov 16(p+2)}+{1\ov 8},\quad s=0,1,\ldots |p/2|
\ee
and indeed reproduces the correct branch cut of the $U(1)$ current:
\be\nn
J(z_1)T_s(z_2)={1\ov \sqrt{z_{12}}}\sqrt{{p\ov 2(p+2)}}t^s_{\D+1/2}(z_2)+\ldots
\ee
where $t^s_{\D+1/2}(z)=\vf^{(s)}(z)\s_1^\vf(z)$.

In order to derive the FR's of the NS and R fields we need the corresponding FR's of the PF fields. The latter can be obtained using the relation between PF fields and the primary fields of the ${su}(2)$ WZW theory \cite{zfpf}:
\be\label{pfwz}
\Phi^j_m(z)=\phi_{2m}^{2j}(z)e^{i{m\ov \sqrt{p}}\vf(z)}.
\ee
Then the FR's of the PF and consequently of the fields in the $N=2$ superconformal theory follow directly from the known FR's of the fields in the WZW theory.

Investigating the FR's in the NS sector  one must keep attention that they have more complicated structure due to the fact that there exist three different 3-point functions of the NS superfields - one even and two odd ones. The meaning of the odd FR's in terms of component fields  is that in the product of two first components of given superfields the second component of the RHS superfield appears. Taking all this into account we obtain the following FR's in the NS sector:
\bea\label{nsfr}
N^{l_1}_{m_1}N^{l_2}_{m_2}&=&\sum_{l=|l_1-l_2|}^L [\Psi^l_m],\\
\nn L&=&\min{(l_1+l_2,2p-l_1-l_2)}
\eea
where:
\bea\nn
\Psi^l_m&=&(N^l_{m_1+m_2})^{even},\quad |m_1+m_2|\le l,\\
\nn \Psi^l_m&=&(N^{p-l}_{m_1+m_2\pm (p+2)})^{odd},\quad |m_1+m_2|>l.
\eea

Repeating the same procedure for the R sector we obtain the corresponding FR's for the R fields:
\bea\label{rafr}
R^{l_1}_{m_1,\a}R^{l_2}_{m_2,-\a}&=&\sum_{l=|l_1-l_2|}^L [\Psi^l_{m_1+m_2}],\\
\nn R^{l_1}_{m_1,\a}R^{l_2}_{m_2,\a}&=&\sum_{l=|l_1-l_2|}^L [\Psi^{p-l}_{m_1+m_2-\a p}]
\eea
where:
\bea\nn
\Psi^l_m&=&(N^l_m),\quad |m|\le l,\\
\nn \Psi^l_m&=&(N^{p-l}_{m\pm (p+2)})^{II,\pm},\quad |m|>l.
\eea

In the twisted sector the situation is more complicated. The product of two twisted $U(1)$ fields reproduces the exponents of the corresponding scalar field with the allowed charges. The exact FR's in the C-disorder PF sector however are not known exactly. We will not need the explicit FR's in the twisted sector in what follows so we omit the details.

We now turn to the computation of the 4-point correlation functions. In view of the construction presented above the latter can be expressed in terms of of the 4-point functions of the corresponding $su(2)$ fields $\Phi^j_m$. The most general correlation function of these fields is calculated in \cite{zfsu}. It is proportional, up to a standard powers of $z_k$ (coming from the conformal invariance) and the isospin $j_k$ (due to the $su(2)$ invariance), to a function $V_{j_1j_2j_3j_4}(z,\bar z;x,\bar x)$ with $z={z_{12}z_{34}\ov z_{14}z_{32}}$ and similarly for $x$ (here $x$ is the isospin variable). This function has the form:
\bea\label{vfn}
V_{j_1j_2j_3j_4}&=&N(j_1\ldots j_4)|z|^{{4j_1j_2\ov p+2}}|1-z|^{{4j_1j_3\ov p+2}}\int\prod_{l=1}^{2j_1}dt_ld\bar t_l|t_l-z|^{-{2\b_1\ov p+2}}\times\\
\nn &\times &|t_l|^{-{2\b_2\ov p+2}}|1-t_l|^{-{2\b_3\ov p+2}}|x-t_l|^2\prod_{i<j}|t_i-t_j|^{4\ov p+2},
\eea
\bea\nn
\b_1=j_1+j_2+j_3&+&j_4+1,\quad \b_2=p+j_1+j_2-j_33-j_4+1,\\
\nn \b_3&=&p+j_1-j_2+j_33-j_4+1.
\eea
The constant $N(j_1\ldots j_4)$ is given by:
\bea\nn
N^2(j_1\ldots j_4)&=&\left({\G({1\ov p+2})\ov \G({p+1\ov p+2})}\right)^{4j_1+2}{\G(1-{2j+1\ov p+2})P^2(j_1+j_2+j_3+j_4+1)\ov \G({2j_1+1\ov p+2})P^2(2j_1)}\times\\
\nn &\times&\prod_{n=2}^4 {\G(1-{2j_n+1\ov p+2})\ov \G({2j_n+1\ov p+2})}{P^2(-j_1+j_2+j_3+j_4-2j_n)\ov P^2(2j_n)},
\eea
where:
\be\nn
P(j)=\prod_{n=1}^j{\G({n\ov p+2})\ov \G(1-{n\ov p+2})}.
\ee
In two cases: $j_1=1/2$ and $j_4={p-1\ov 2}$ the integral in (\ref{vfn}) can be expressed in terms of hypergeometric functions. Using the PF construction of the NS and R fields we can write down their most general  4-point function.

From the 4-point functions we can also extract the structure constants of the OPE algebra. They appear in the explicit form of the FR's. In the NS case we have (we introduce here also the $\bar z$ dependence):
\be\nn
N_{m_1\bar m_1}^{l_1}(z_1,\bar z_1)N_{m_2\bar m_2}^{l_2}(z_2,\bar z_2)=\sum_l\sum_{m,\bar m=-l}^lC \begin{pmatrix}
 l & m & \bar m \\
  l_1 & m_1 &  \bar m_1  \\
   l_2 &m_2 &  \bar m_2
 \end{pmatrix}
|z_{12}|^{2(\D_l-\D_1-\D_2)}N_{m\bar m}^l(z_2,\bar z_2)+\ldots.
\ee
The meaning of the field $N_{m\bar m}^l$ in the RHS is clear from the FR's (\ref{nsfr}). The constant $C$ in the above expression exactly coincides with the corresponding structure constant and is given by:
\be\label{scons}
C\begin{pmatrix}
 l_1 & m_1 &  \bar m_1  \\
   l_2 &m_2 &  \bar m_2 \\
    l_3 & m_3 & \bar m_3
 \end{pmatrix}
=\begin{bmatrix}
 {l_1\ov2} & {l_2\ov2} & {l_3\ov2} \\
  {m_1\ov2} & {m_2\ov2} & {m_3\ov2}
 \end{bmatrix}
\begin{bmatrix}
 {l_1\ov2} & {l_2\ov2} & {l_3\ov2} \\
  {\bar m_1\ov2} & {\bar m_2\ov2} & {\bar m_3\ov2}
 \end{bmatrix}
\r({l_1\ov 2},{l_2\ov 2},{l_3\ov 2})
\ee
where the first two coefficients are the $3j$-Wigner symbols and:
\bea\nn
&&{\r^2\ov (l_1+1)(l_2+1)(l_3+1)}=\\
&=&{\G({p+3\ov p+2})\ov \G({p+1\ov p+2})}\prod_{k=1}^3{\G(1-{l_k+1\ov p+2})\ov \G(1+{l_k+1\ov p+2})}{\tilde P}^2({l_1+l_2+l_3\ov 2}+1)\prod_{k=1}^3{{\tilde P}^2({l_1+l_2+l_3\ov 2}-l_k)\ov {\tilde P}^2(l_k)}
\eea
where:
\be\nn
{\tilde P}(l)=\prod_{k=1}^l\G(1+{k\ov p+2}) \G^{-1}(1-{k\ov p+2}).
\ee
Exactly the same procedure goes also for the R sector. The structure constants are defined here by:
\bea\nn
R_{m_1\bar m_1,\a}^{l_1}(1)R_{m_2\bar m_2,-\a}^{l_2}(2)&=&\sum_l\sum_{m,\bar m=-l}^lC\begin{pmatrix}
 l & m & \bar m \\
  l_1 & m_1 &  \bar m_1  \\
   l_2 &m_2 &  \bar m_2
 \end{pmatrix}
|z_{12}|^{2(\D_l-\D_1-\D_2)}N_{m\bar m}^l(2)+\ldots\\
\nn R_{m_1\bar m_1,\a}^{l_1}(1)R_{m_2\bar m_2,\a}^{l_2}(2)&=&\sum_l\sum_{m,\bar m=-l}^lC\begin{pmatrix}
 l & m & \bar m \\
  l_1 & m_1 &  \bar m_1  \\
   l_2 &m_2 &  \bar m_2
 \end{pmatrix}
|z_{12}|^{2(\D_l-\D_1-\D_2)}N_{m-\a p,\bar m-\a p}^{p-l}(2)+\ldots
\eea
and coincide with (\ref{scons}) (which NS field appear above is dictated again by the FR's (\ref{rafr})).

We are going now to discuss in more details the discrete symmetries of the $N=2$ minimal models which we will need in the next Section. The original $Z_p$ symmetry, due to the PF construction, is lost because of the presence of the bigger $U(1)$ symmetry. It is known however \cite{gep5} that these models possess a bigger $Z_{p+2}$ discrete symmetry. In our considerations this fact can be explained as follows. Consider the special superfields $N^p_{-m}$ (i.e. $l=p$). They form a closed OPE algebra as it can be seen from the FR's obtained above:
\be\nn
N^p_{-m_1}N^p_{-m_2}=[N^p_{p+2-(m_1+m_2)}]^{odd}.
\ee
Denoting all the fields of this type as $N^p_{-(p+2)+2k},k=1,2,\ldots,p+1$ we can consider these FR's as a multiplication law of the discrete group $Z_{p+2}$ for representations with charges $k_1$ and $k_2$. Therefore these superfields (of fractional dimension $\D_k=k(p+2-k)/(p+2)-1/2$) generate a PF type symmetry of the $N=2$ minimal models.

To make this statement more clear let us consider the $N=2$ PF theory with a $Z_{p+2}$ discrete symmetry. It contains, in addition to the $N=2$ super stress-energy tensor $W(z,\theta^+,\theta^-)$ (which includes $T,J,G^{\pm}(z)$), the PF supercurrents $\Psi_k(z,\theta^+,\theta^-)$ carrying $Z_{p+2}$ charge $k=1,2,\ldots,p+1$ with dimension $\D_k$ and $U(1)$ charge $q_k$. According to the $Z_{p+2}$ symmetry they should close the following algebra:
\bea\nn
\Psi_{k_1}(z_1,\theta^+_1,\theta^-_1)\Psi_{k_2}(z_2,\theta^+_2,\theta^-_2)&=&C_{k_1,k_2}\tilde z^{\D_{k_1+k_2}-\D_{k_1}-\D_{k_2}-1/2}\CD_1\Psi_{k_1+k_2}(z_2,\theta^+_2,\theta^-_2)+\ldots,\\
\nn \Psi_{k}(z_1,\theta^+_1,\theta^-_1){\Psi}^\dagger_{k}(z_2,\theta^+_2,\theta^-_2)&=&\tilde z^{-2\D_k}+\tilde z^{-2\D_k+1}\CD_2W(z_2,\theta^+_2,\theta^-_2)+\ldots
\eea
where $\CD_1$ and $\CD_2$ are some $N=2$ super-covariant derivatives (which can be explicitly constructed) and ${\Psi}^\dagger_k$ is the conjugate of ${\Psi_k}$ with $Z_{p+2}$ charge $p+2-k$, dimension $\D_k$ and $U(1)$ charge $-q_k$.

These superconformal OPE's are consistent with the $Z_{p+2}$ symmetry if the dimensions $\D_k$ obey certain monodromy condition. The elementary solution of the corresponding equation is:
\be\nn
\D_k=k(p+2-k)/(p+2)-1/2.
\ee
In plus, we add the requirement that the $Z_{p+2}$ symmetry is consistent with the bigger $U(1)$ symmetry coming from the $N=2$ superconformal algebra. This imposes a condition on the possible $U(1)$ charges of the parafermions. Its simplest solution is:
\be\nn
q_k={k\ov p+2}-\hf.
\ee
We note that the dimensions and charges of the super-parafermions given above coincide with  those of the fields $N^p_{-(p+2)+2k}$. This supports the suspect that these fields indeed generate a $Z_{p+2}$ PF symmetry in the $p$-th minimal $N=2$ superconformal model. To complete this discussion we need also to compute the central charge of the $N=2$ PF theories. It turns out that it is exactly equal to:
\be\nn
c={3p\ov p+2},
\ee
i.e. the central charge of these super-parafermionic models coincides with the central charge of the $p$-th $N=2$ minimal model. This completes the proof of the equivalence between $Z_{p+2}$ $N=2$ super-parafermionic theories and $p$-th $N=2$ minimal models.

At the end, we would like also to describe the spectrum of the super PF theory. It is clear that it should have NS and R order parameters corresponding to the different choices of boundary conditions for the supercurrents $G^\pm(z)$. The OPE of these order parameters with the PF currents is defined by their monodromy properties and is well known. We compare the latter
with the OPE's in the $N=2$ minimal models:
\bea\nn
N^p_{-p}(z)N^k_{k}(0)&=&z^{-k/p+2}(N^k_{k+2})^{II}+\ldots\\
\nn N^p_{-p}(z)N^k_{-k}(0)&=&z^{k/p+2}(N^k_{k-2})^{II}+\ldots
\eea
where, according to our identification, the field $N^p_{-p}(z)$ corresponds to the first component of the super-parafermionic current $\Psi_1(z)$. The above OPE's suggest that the chiral superfields $N^k_{k}$ and $N^k_{-k}$ are the NS order parameters with dimensions and charges:
\bea\nn
d_k&=&{k\ov p+2}=q_k,\quad for\quad N^k_k=\s_k^{NS},\\
\nn d_k&=&{k\ov p+2}=-q_k,\quad for\quad N^k_{-k}={\s_k^{NS}}^\dagger.
\eea
Analogous calculation in the R sector leads to:
\bea\nn
N^p_{-p}(z)R^{k-1}_{k-1,1}(0)&=&z^{-k/p+2+1/2}(R^{k-1}_{k+1,1})^{II}+\ldots\\
\nn N^p_{p}(z)R^{k-1}_{k-1,1}(0)&=&z^{k/p+2-1/2}(R^{k-1}_{k-3,1})^{II}+\ldots.
\eea
Therefore, the fields that represent the order parameters of $Z_{p+2}$ charge $k$ are given in the minimal models by $R^{k-1}_{k-1,1},k=1,2,\ldots,p+1$. These are in fact all the Ramond fields with lowest dimension and $U(1)$ charge:
\be\nn
d_k={c\ov 24},\qquad q_k={k\ov 2(p+2)}-{1\ov 4}.
\ee
We thus obtained that the spectrum of the $N=2$ PF theories is given by the above expressions in the NS and R sectors. The other primary fields in the minimal models correspond to the descendants of the above primary fields of the PF theories with respect to the PF currents. Finally, we note that the fields in the twisted sector correspond to the C-disorder sector of the $N=2$ super-PF theories.

At the end of this Section we would like to discuss the renormalization group properties of the $N=2$ minimal models. In other words we would like to describe the RG flow of these models perturbed by the least relevant field. In the case of  $N=2$ minimal models the latter is constructed from the chiral and antichiral fields $N^p_{\pm p}$ of dimension $\D=1/2-1/(p+2)$ and $U(1)$ charge $q=\pm\D$. The suitable perturbation term, neutral and of dimension close to one, is therefore constructed out of the second components of such chiral fields. Explicitly we consider:
\be\nn
\CL=\CL_0+\int d^2z\Phi(z)
\ee
where $\CL_0$ represents the minimal model itself and the field $\Phi(z)$ is a combination of the second components:
\be\nn
\Phi=(N^p_p)^{II}+(N^p_{-p})^{II}\equiv \phi_++\phi_-.
\ee
It is neutral and has a dimension $\D=1-1/(p+2)=1-\e$. Similarly to what we did for the $N=1$ superconformal theories, we consider the case $p\rightarrow\infty$ and assume $\e=1/(p+2)$ to be a small parameter. Also, according to our parafermionic construction, we can express the perturbing field in terms of the PF currents and exponents of the scalar field as follows:
\bea\nn
(N^p_p)^{II}&=&\sqrt{{2p\ov p+2}}{\psi_1}^\dagger e^{-i{2\ov \sqrt{2p(p+2)}}}\equiv \phi_+,\\
\nn (N^p_{-p})^{II}&=&\sqrt{{2p\ov p+2}}{\psi_1} e^{i{2\ov \sqrt{2p(p+2)}}}\equiv \phi_-.
\eea
Our purpose now is to compute the beta-function of this theory and to check for an eventual fixed point. For that we need to compute the two-point function of the perturbing field up to a second order. The expansion was already written in (\ref{secex}). As in the case of $N=1$ theory we need the 3- and 4-point functions of the perturbing field. We note that, due to the FR's computed above, the 3-point function of the field $\Phi(z)$, and therefore the first term in (\ref{secex}), is identically zero. So we are left with the computation of the second order term only. This computation goes along the same lines as in the $N=1$ case. We need to compute the 4-point function of $\Phi(z)$ up to zeroth order in $\e$ and to integrate it in the safe region $\O_{l,l_0}$ far from the singularities. Near the singular points $0$, $1$ and $\infty$ we use the OPE's that we computed above.

The 4-point function of the perturbing field $\Phi(z)$ is expressed through the corresponding functions of the parafermionic fields which are known \cite{zfpf} and the trivial power-like contribution of the exponents. The final result is (up to zeroth order in $\e$):
\be\nn
<\Phi(x)\Phi(0)\Phi(1)\Phi(\infty)>=C|1+{1\ov x^2}+{1\ov (1-x)^2}|^2
\ee
where $C$ is some structure constant. We will not need its explicit expression here. The integration of this function over the safe region gives:
\be\nn
{2\pi^2\ov\e}\left({31\ov 16}+{1\ov l^2}+{1\ov 4l_0^2}\right).
\ee
From this we have to subtract the contribution of the lens-like region:
\be\nn
{\pi^2\ov\e}\left({31\ov 16}-{1\ov l^2}+{1\ov 2l_0^2}\right).
\ee
At the end, we add the result of the integration near the singular points:
\be\nn
2\left(-{\pi^2\ov l^2\e}\right)+{2\pi^2\ov \e}\left(-{1\ov 2l^2}+{1\ov 2l_0^2}\right)
\ee
corresponding to the integrals around $0$ (and $1$) and $\infty$ respectively. Summing all the contributions we get finally as a result:
\be\nn
{\pi^2\ov \e l_0^2}.
\ee
Two comments are in order. First, this result contains only the cut-off parameter and could be cancelled by adding an appropriate counterterm in the action. Second, the finite contribution is identically zero. This means that there is no contribution to the beta-function neither in the first nor in the second order. One can speculate that this is the case also in higher orders. This result leads us to the conclusion that there do not exits a nontrivial fixed point of the beta-function close to the UV one. If such a fixed point exists it should be due to some non-perturbative effects.


\subsection{Yukawa couplings for the three-generation string model}

Let us describe briefly the derivation of the Yukawa couplings for the special C-Y manifold constructed as a hypersurface in $CP^2\times CP^3$ \cite{cycstr}. It is defined by the zeroes of the polynomials:
\bea\nn
P_1&=&z_0^3+z_1^3+z_2^3+z_3^3=0,\\
\nn P_2&=&z_1x_1^3+z_2x_2^3+z_3x_3^3=0,\\
\nn z_i&\in & CP^3,\quad x_i\in CP^2.
\eea
It represents a C-Y manifold of complex dimension $3$ and Euler characteristic $\chi=-54$. We are particularly interested in the discrete symmetries of this manifold given by:

(a) $S_3$ group of permutations of the indexes $i=1,2,3: z_i\rightarrow z_{p(i)},\quad x_i\rightarrow x_{p(i)}$;

(b) $Z_3\times Z^3_9$ spanned by the transformations:
\bea\nn
z_0&\rightarrow&\exp{(2\pi i r_0/3)}z_0,\quad z_i\rightarrow\exp{(2\pi i r_i/3)}z_i,\\
\nn x_i&\rightarrow&\exp{(-2\pi r_i/9)}x_i.
\eea
The irrelevance of the overall phase in the discrete transformations of the polynomials determines the full group of the global discrete symmetries of this C-Y space to be\\
$G=S_3\times Z_3\times Z_9^3/Z_9$. We denote the charges of the different objects under $Z_3\times Z_9^3$ as a vector $m=(m_0,m_1,m_2,m_3)$.

It is known that the number of generations in the case of interest can be found enumerating the independent deformations of the complex structure of the corresponding C-Y manifold. They are given by all the possible (homogeneous) deformations of the defining polynomials. It turns out that they fit into $9$ "families". Taking also into account their possible\\
$Z_3\times Z_9^3$ charges we have $35$ independent polynomials and hence this C-Y sigma model has $35$ generations. From the topological properties it follows that there are also $8$ antigenerations. The Yukawa couplings of the massless matter fields are given by some topological formula. The invariance condition with respect to the discrete $G$-transformations determines them (up to an appropriate normalization of the matter superfields) to be:
\be\label{qt}
\l_{ijk}=\d_{mod (3)}(2+m_i^0+m_j^0+m_k^0)\prod_{r=1}^3\d_{mod (9)}(2+m_i^r+m_j^r+m_k^r)
\ee
where $\{m_i^r\}$ are the charges of the $i$-th polynomial. According to our normalization convention all the Yukawa couplings are either 1 or 0.

We now pass to the explicit evaluation of the Yukawa couplings in the framework of the $N=2$ superconformal model $1^1 16^3$ describing the corresponding C-Y manifold \cite{gep5}. The compactified part ($c=9$) of this model contains the tensor product (moded out by $G$-projection) of the $N=2$ superconformal minimal models with $p=1$ ($c=1$) and three copies of $p=16$ ($c=8/3$) \cite{gep5}. Each of these models possesses $Z_{p+2}$ discrete symmetry originated from its superparafermionic structure that we described in the previous Section. Then the total discrete symmetry group of the model $1^1 16^3$ is $Z_3\times Z_{18}^3$. The comactified part of the corresponding massless string vertices is constructed in terms of the superparafermionic order parameters (we list them again here for convenience):
\be\nn
NS: \s_k^N=N_k^k,\quad \D_k={k\ov 2(p+2)},\quad q_k={k\ov 2(p+2)}
\ee
with $Z_{p+2}$ charge $k$, and:
\be\nn
R: \s_{k+1}^R=R_k^k,\quad \D_k={c\ov 24},\quad q_k={k+1\ov 2(p+2)}-{1\ov 4}
\ee
with $Z_{p+2}$ charge $k+1$. Due to the $G$-projection only fields with even $Z_{18}$ charge appear in the massless spectrum of the $1^1 16^3$ model and therefore the effective discrete group becomes $Z_3\times Z_9^3$. The composite model obeys a new symmetry $S_3$ of permutation of the $p=16$ models. Finally, the condition for an integer $U(1)$ charge of the composite fields implies that the element $g_0=(1,1,1,1)\in Z_3\times Z_{18}^3$ acts trivially. Therefore the full group of discrete symmetries of the $1^1 16^3$ model exactly coincides with the one of the C-Y model considered above, i.e. $S_3\times Z_3\times Z_9^3/Z_9$.

According to the Gepner's construction the massless matter fields appear as $N=1$ space-time superfields in the ${\bf 27}$ (for the generations) or in $\bar{{\bf 27}}$ (for the antigenerations) of $E_6$. In fact the spinor ${\bf 16}_(q)$ vector ${\bf 10}_(q)$ and scalar ${\bf 1}_(q)$ representations of $SO(10)\times U(1)$ combine into the ${\bf 27}$ and
$\bar{{\bf 27}}$ of $E_6$:
\bea\nn
{\bf 27}&=&{\bf 1}_{(-1)}\oplus {\bf 16}_{(-1/4)}\oplus {\bf 10}_{(1/2)},\\
\nn \bar{{\bf 27}}&=&{\bf 1}_{(1)}\oplus {{\bar{\bf 16}}}_{(1/4)}\oplus {\bf 10}_{(-1/2)}.
\eea
Let us discuss first the generations. The vertex operators for the space-time spinor component of the matter superfield $S_a$ in ${\bf 1}_{(-1)}$ of $SO(10)\times U(1)$ is given by:
\be\label{verto}
\CV_1(z,\bar z)=\left( R_k^k\prod_{i=1}^3 R_{l_i}^{l_i}\right)(z)\left( N^m_{-m}\prod_{i=1}^3 N_{-n_i}^{n_i}\right)(\bar z)
\ee
satisfying the following conditions:

(1) $6k+\sum l_i=18$, i.e. $U(1)$ charge to be $q=-1/4$,

$\qquad 6m+\sum n_i=36$, i.e. $U(1)$ charge to be $q=-1$,

(2) $(m+k)mod (3)=(n_i+l_i)mod (9), i=1,2,3$,

(3) the left $(k,l_i)-$right $(m,n_i)$ 2D constructions for each individual model are restricted by the exceptional modular invariant at level $k=16$ of the underlying $su(2)$ Kac-Moody algebra \cite{gep5}.

The corresponding spinor components of $S_a$ in ${\bf 16}_{(-1/4)}$ and ${\bf 10}_{(1/2)}$ have a similar form:
\bea\label{othv}
\CV_{16}(z,\bar z)&=&(R_1^1(R_{16}^{16})^3)(\bar z)\CV_1(z,\bar z),\\
\nn \CV_{10}(z,\bar z)&=&(R_1^1(R_{16}^{16})^3)(\bar z)\CV_{16}(z,\bar z).
\eea
The vertex operators for the scalar components of $S_a$ can be obtained by acting on the spinor ones with the space-time SUSY charge $Q$ using the 2D OPE's.

In order to compare the geometrical description of the model with the algebraic $(1^1 16^3)$ one we have to make a correspondence between the vertices (\ref{verto}) and (\ref{othv})
and the polynomials comparing their $Z_3\times Z_9^3$ charges. As argued in \cite{gep5} the charges $\{Q_k\}$ to be compared are those of the scalars $\CV_{10}^{scalar}$ normalized as follows:
\be\label{nor}
m_0=Q_0 (mod (3)),\quad m_i=2Q_i (mod (9)),\quad i=1,2,3.
\ee
In the table below the nine "families" are represented by the spinor vertices $\CV_1(z,\bar z)$ and the relevant charges of $\CV_{10}^{scalar}(z,\bar z)$ and we used the notation
$R(0\quad 6\quad 6\quad 6)N_\pm(1\quad 10\quad 10\quad 10)$ for $(R_0^0(R_{6}^{6})^3)(z)(N_{\pm 1}^1(N_{\pm 10}^{10})^3)(\bar z)$, etc.

\begin{table}[h]
\begin{center}
\begin{tabular}{cccc}\hline
"Family"&$\CV_1(z,\bar z)$&$Z_3\times Z_9^3$ charges of $\CV_{10}^{scalar}$\\
\hline
1&$R(1\quad 12\quad 0\quad 0)N_-(0\quad 4\quad 16\quad 16)$&(1\quad 6\quad 0\quad 0)\\
2&$R(1\quad 8\quad 4\quad 0)N_-(0\quad 8\quad 12\quad 16)$&(1\quad -2\quad -1\quad 0)\\
3&$R(1\quad 6\quad 6\quad 0)N_-(0\quad 10\quad 10\quad 16)$&(1\quad 3\quad 3\quad 0)\\
4&$R(1\quad 4\quad 4\quad 4)N_-(0\quad 12\quad 12\quad 12)$&(1\quad -1\quad -1\quad -1)\\
5&$R(0\quad 12\quad 6\quad 0)N_-(1\quad 4\quad 10\quad 16)$&(0\quad 3\quad 6\quad 0)\\
6&$R(0\quad 10\quad 8\quad 0)N_-(1\quad 6\quad 8\quad 16)$&(0\quad 2\quad -2\quad 0)\\
7&$R(0\quad 10\quad 4\quad 4)N_-(1\quad 6\quad 12\quad 12)$&(0\quad 2\quad -1\quad -1)\\
8&$R(0\quad 8\quad 6\quad 4)N_-(1\quad 8\quad 10\quad 12)$&(0\quad 3\quad -2\quad -1)\\
9&$R(0\quad 6\quad 6\quad 6)N_-(1\quad 10\quad 10\quad 10)$&(0\quad 3\quad 3\quad 3)\\
\hline
\end{tabular}
\end{center}
\end{table}

The condition of $SO(10)\times U(1)$ invariance for the cubic superpotential restricts its form as follows \cite{gsw}:
\bea\nn
W&=& \sum_{ijk}( \l^{(1)}_{ijk} S^i_{{\bf 16}(-1/4)}S^j_{{\bf 16}(-1/4)}S^k_{{\bf 10}(1/2)}+\\
\nn &+& \l^{(2)}_{ijk} S^i_{{\bf 10}(1/2)}S^j_{{\bf 10}(1/2)}S^k_{{\bf 1}(-1)})
\eea
($i,j,k$ are the family indexes of the chiral superfields $S^i_a$). The explicit construction of the vertices (\ref{verto}) and (\ref{othv}) and the specific properties of the fields $R_1^1$ and $R^{16}_{16}$ discussed in the previous section lead to the following important equality between the Yukawa couplings:
\be\nn
\l^{(1)}_{ijk}=\l^{(2)}_{ijk}(\equiv \l_{ijk}=<ijk>).
\ee
The latter can be expressed as products of the 2D OPE structure constants of the $N=2$ fields from the compactified part of the vertices $\CV_{16}(z,\bar z)$ (see (\ref{othv}) and the above table):
\bea\label{sco}
\l_{ijk}&=&N_{ijk}\d(Q^0_i+Q^0_j+Q^0_k-1)\prod_{r=1}^3\d(Q^r_i+Q^r_j+Q^r_k-16),\\
\nn N_{ijk}&=&\prod_{l=1}^4<\s^i_{n_1l}(\infty)\s^j_{n_2l}(1){\s^{k\dagger}_{n_1+n_2,l}}(0)>.
\eea
The $\d$-function part in (\ref{sco}) is a direct consequence of the $N=2$ fusion rules derived in the previous Section. The three-point functions:
\be\nn
<\s^i_{n_1l}(\infty)\s^j_{n_2l}(1){\s^{k\dagger}_{n_1+n_2,l}}(0)>
\ee
represent the structure constants of the underlying parafermionic model. Putting together all these ingredients we get the following values for the non-vanishing Yukawa couplings:
\bea\label{yk}
<994>&=&k_1^3,\qquad <973>=k_1^2=<883>,\\
\nn <882>&=&k_1k_2,\qquad <884>=k_1k_2^2,\\
\nn <872>&=&k_2^2,\qquad <862>=k_2,\\
\nn <861>&=&<751>=<663>=<652>=1,\\
\nn <852>&=&<753>=<554>=k_1
\eea
where:
\bea\label{ykex}
k_1^2&=&{\G({1\ov 18})\G({13\ov 18})\G({11\ov 18})^2\ov \G({17\ov 18})\G({5\ov 18})\G({7\ov 18})^2},\\
\nn k_2^2&=&{\G({1\ov 18})\G({13\ov 18})^2\ov \G({17\ov 18})\G({5\ov 18})^2}.
\eea
Let us compare now these algebraic results with the quasi-topological ones (\ref{qt}). Taking into account the normalization condition (\ref{nor}) it is easy to see that the Yukawa couplings  of the C-Y model (\ref{qt}) and of the $1^1 16^3$ Gepner model (\ref{sco}) are equal up to the constants $N_{ijk}$. We can absorb these constants in the normalization of the corresponding polynomials. However, we have 9 families and 14 non-zero couplings and therefore the normalization should satisfy a non-trivial consistency condition. It turns out that in our case this condition is satisfied and the proper normalization can be chosen in the form:
\bea\label{norpo}
9^G&=&k_1^{4/3}k_2^{-2/3}9^{C-Y},\qquad 8^G=k_1^{1/3}k_2^{1/3}8^{C-Y},\\
\nn 7^G&=&k_1^{-2/3}k_2^{4/3}7^{C-Y},\qquad 6^G=k_1^{-2/3}k_2^{1/3}6^{C-Y},\\
\nn 5^G&=&k_1^{1/3}k_2^{-2/3}5^{C-Y},\qquad 4^G=k_1^{1/3}k_2^{4/3}4^{C-Y},\\
\nn 3^G&=&k_1^{4/3}k_2^{-2/3}3^{C-Y},\qquad 2^G=k_1^{1/3}k_2^{1/3}2^{C-Y},\\
\nn 1^G&=&k_1^{1/3}k_2^{-2/3}1^{C-Y}.
\eea
Then all the couplings (\ref{qt}) and (\ref{sco}) exactly coincide.

In the case of antigenerations the vertex representing the spinor component in ${\bf 1}_{(1)}$ of $SO(10)\times U(1)$ can be taken in the form:
\be\label{verta}
\bar\CV_1(z,\bar z)=\left( R_k^k\prod_{i=1}^3 R_{l_i}^{l_i}\right)(z)\left( N^m_{m}\prod_{i=1}^3 N_{n_i}^{n_i}\right)(\bar z).
\ee
The conditions $(1)$ and $(3)$ below (\ref{verto}) remain unchanged but in the second we replace $k$ to $-k$ and $l_i$ to $-l_i$. The scalar component of the superfield is again obtained by acting with the space-time supercharge. The spinor vertices $\bar\CV_{\bar {16}}$ and $\bar\CV_{10}$ are realized in terms of (\ref{verta}) as follows:
\bea\label{othva}
\bar\CV_{\bar {16}}(z,\bar z)&=&(R_0^0(R_0^0)^3)(\bar z)\bar\CV_1(z,\bar z),\\
\nn \bar\CV_{10}(z,\bar z)&=&(R_0^0(R_0^0)^3)(\bar z)\bar\CV_{\bar {16}}(z,\bar z).
\eea
All this leads to the construction of 4 families representing 8 antigenerations listed in the table below:

\begin{table}[h]
\begin{center}
\begin{tabular}{cccc}\hline
"Family"&$\bar\CV_1(z,\bar z)$&$\bar\CV_{10}^{scalar}$charges&Number of vertices\\
\hline
1&$R(1\quad 8\quad 2\quad 2)N_+(1\quad 14\quad 8\quad 8)$&(2\quad 6\quad 3\quad 3)&3\\
2&$R(0\quad 4\quad 4\quad 4)N_+(0\quad 12\quad 12\quad 12)$&(0\quad 0\quad 0\quad 0)&1\\
3&$R(0\quad 8\quad 8\quad 2)N_+(0\quad 14\quad 14\quad 8)$&(1\quad 6\quad 6\quad 3)&3\\
4&$R(0\quad 6\quad 6\quad 6)N_+(1\quad 10\quad 10\quad 10)$&(0\quad 0\quad 0\quad 0)&1\\
\hline
\end{tabular}
\end{center}
\end{table}

Applying the $N=2$ fusion rules  we get only two allowed Yukawa couplings for the above families of antigenerations: $<\bar 4\bar 4\bar 2>$ and $<\bar 4\bar 3\bar 1>$. It is straightforward to compute the first one using the same procedure as for the generations since it is connected to the correlation function of scalar ($\D=\bar\D$) $su(2)$ fields. There is an obstruction in the evaluation of $<\bar 4\bar 3\bar 1>$ since it involves left-right asymmetric fields $\Phi_{7,4}(z,\bar z)$ ($l=7,\bar l=4$) and $\Phi_{4,1}(z,\bar z)$
of spins $s_{7,4}=\D-\bar\D=2$ and $s_{4,1}=1$. However, using the correlation functions constructed in \cite{vpet}, one can find the values of the structure constants and the corresponding Yukawa couplings:
\be\label{yka}
<\bar 4\bar 4\bar 2>=k_1^3,\qquad <\bar 4\bar 3\bar 1>=k_3^3
\ee
where:
\be\label{ykaex}
k_3^2={\G({1\ov 18})\G({11\ov 18})\G({5\ov 6})\ov \G({17\ov 18})\G({7\ov 18})\G({1\ov 6})}.
\ee

Up to now we have calculated and compared the Yukawa couplings for the 27-generation C-Y model \cite{cycstr} and the corresponding tensor product Gepner model \cite{gep5}. Our final goal is to compute the Yukawa couplings for the three-generation $1^116^3$ model. As it is observed in \cite{cycstr,gep5}, the global automorphism group $G$ contains a subgroup $H=Z_3\times Z_3$ generated by the elements $h$ and $g$:
\bea\nn
h&:& z_i\rightarrow z_{i+1},\quad x_i\rightarrow x_{i+1},\\
g&=& (0,3,6,0)\in Z_3\times Z_9.
\eea
The three generation model is obtained by factoring out the 27-generation model by this $H=Z_3\times Z_3$ subgroup.

Let us consider first the action of the element $g$. By simply projecting the spectrum onto $g$-invariant states we find that 17 generations survive and all the antigenerations as well. More detailed analysis shows that the subgroup spanned by $g$ does not act freely on the C-Y manifold. Therefore we have to complete the spectrum by the corresponding twisted states. It turns out that the only family of (six) twisted generations that appears is represented by the following spinor vertices:
\be\nn
\CV^t_1(z,\bar z)=R(0\quad 8\quad 8\quad 2)(z)N_-(1\quad 8\quad 14\quad 8)(\bar z).
\ee
The vertices of the twisted family of 6 antigenerations are given by:
\be\nn
\bar\CV^{t_a}_1(z,\bar z)=R(0\quad 12\quad 0\quad 6)(z)N_+(1\quad 4\quad 16\quad 10)(\bar z).
\ee
Then the total number of generations becomes $23=17+6$, the antigenerations are $14=8+6$ and therefore at this stage the model contains $9=23-14$ net generations.

The next step is to divide by the subgroup generated by $h$. Since $h$ acts freely we have only to project the space-time onto $h$-invariant states. The result is as follows:

\noindent Generations ($\CV_1$ vertices):
\bea\label{leps}
L_1&=&R(1\quad 12\quad 0\quad 0)N_-(0\quad 4\quad 16\quad 16)+c.p., \\
\nn L_2&=&R(1\quad 6\quad 6\quad 0)N_-(0\quad 10\quad 10\quad 16)+c.p.,\\
\nn L_3&=&R(1\quad 4\quad 4\quad 4)N_-(0\quad 12\quad 12\quad 12),\\
\nn L_4&=&R(0\quad 12\quad 6\quad 0)N_-(1\quad 4\quad 10\quad 16)+c.p.,\\
\nn L_5&=&R(0\quad 6\quad 12\quad 0)N_-(1\quad 10\quad 4\quad 16)+c.p,\\
\nn L_6&=&R(0\quad 10\quad 4\quad 4)N_-(1\quad 6\quad 12\quad 12)+c.p.,\\
\nn L_7&=&R(0\quad 6\quad 6\quad 6)N_-(1\quad 10\quad 10\quad 10),\\
\nn L^t_8&=&R(0\quad 8\quad 8\quad 2)N_-(1\quad 8\quad 14\quad 8)+c.p.,\\
\nn L^t_9&=&R(0\quad 8\quad 8\quad 2)N_-(1\quad 14\quad 8\quad 8)+c.p.
\eea

\noindent Antigenerations ($\bar\CV_1$ vertices):
\bea\label{lepas}
L_1&=&R(1\quad 8\quad 2\quad 2)N_+(1\quad 14\quad 8\quad 8)+c.p.,\\
\nn L_2&=&R(1\quad 4\quad 4\quad 4)N_+(0\quad 12\quad 12\quad 12),\\
\nn L_3&=&R(0\quad 8\quad 8\quad 2)N_+(0\quad 14\quad 14\quad 8)+c.p.,\\
\nn L_4&=&R(0\quad 6\quad 6\quad 6)N_+(1\quad 10\quad 10\quad 10),\\
\nn L_5^t&=&R(0\quad 12\quad 0\quad 6)N_+(1\quad 4\quad 16\quad 10)+c.p.,\\
\nn L_6^t&=&R(0\quad 12\quad 6\quad 0)N_+(1\quad 4\quad 10\quad 16)+c.p..
\eea
It is obvious that the vertices (\ref{leps}) and (\ref{lepas}) represent the massless spectrum  of the three-generation model ($\# L-\# \bar L=3$).

The vertices (\ref{leps}) and (\ref{lepas}) correspond to the trivial embedding of the subgroup $H=Z_3\times Z_3$ in $E_6$ which leaves it unbroken, i.e. the entire $27$ ($\bar{27}$) contribute to the massless spectrum. As it is explained in \cite{gsw}, in the case of a non-trivial embedding (Wilson lines) only part of $27$ ($\bar{27}$) survives the compactification. In this case $E_6$ is broken to $SU_c(3)\times SU_L(3)\times SU_R(3)$. The 27 of $E_6$ decomposes as $(1,3,\bar 3)\oplus(3,\bar 3,1)\oplus (\bar 3,1,3)$. The superfields belonging to the color singlet $(1,3,\bar 3)$ contain leptons and Higgses while the color triplets $(3,\bar 3,1)$ and $(\bar 3,1,3)$ - the quarks and antiquarks. Therefore the vertices $L_i$ and $\bar L_i$ given by (\ref{leps}) and (\ref{lepas}) represent the leptons and Higgses from generations and antigenerations. The corresponding quark and antiquark vertices are given by:
\bea\label{qrk}
Q_1&=&R(1\quad 4\quad 8\quad 0)N_-(0\quad 12\quad 8\quad 16)+c.p.,\\
\nn \bar Q_1&=&R(1\quad 8\quad 4\quad 0)N_-(0\quad 8\quad 12\quad 16)+c.p.,\\
\nn Q_2&=&R(0\quad 10\quad 8\quad 0)N_-(1\quad 6\quad 8\quad 16)+c.p.,\\
\nn \bar Q_2&=&R(0\quad 8\quad 10\quad 0)N_-(1\quad 8\quad 6\quad 16)+c.p.,\\
\nn Q_3&=&R(0\quad 8\quad 6\quad 4)N_-(1\quad 8\quad 10\quad 12)+c.p.,\\
\nn \bar Q_3&=&R(0\quad 6\quad 8\quad 4)N_-(1\quad 10\quad 8\quad 12)+c.p..
\eea
The explicit construction of the vertices (\ref{leps}), (\ref{lepas}) and (\ref{qrk}) together with the exact values (\ref{yk}), (\ref{yka}) of the Yukawa couplings lead to the following general expression for the low-energy cubic superpotential:
\bea\label{lesp}
W&=&\sum_{ijk}[\l_{ijk}L_i L_j L_k+\bar\l_{ijk}\bar L_i \bar L_j \bar L_k+\\
\nn &+&(\mu_{ijk}Q_i Q_j Q_k+h.c.)+\rho_{ijk}Q_i \bar Q_j L_k],
\eea
where the new couplings $\l, \bar\l, \mu, \rho$ are linear combinations of the old ones (\ref{yk}), (\ref{yka}):

(1) Leptons and Higgses

(a) Generations:
\bea\nn
\l_{773}=k_1^3,\quad \l_{641}=1=\l_{651},\\
\nn \l_{762}=k_1^2,\quad \l_{652}=\l_{642}={1\ov 3}\l_{543}=k_1.
\eea

(b) Antigenerations:
\be\nn
\bar\l_{652}=3k_1,\quad \bar\l_{442}=k_1^3,\quad \bar\l_{431}=k_3^3.
\ee

(2) Quarks
\be\nn
\mu_{133}=k_1 k_2=\mu_{\bar 1\bar 3\bar 3},\quad \mu_{123}=k_2=\mu_{\bar 1\bar 2\bar 3}.
\ee

(3) Quarks-antiquarks-Higgses
\bea\nn
\rho_{3\bar 3 2}&=&k_1^2,\quad \rho_{3\bar 3 3}=3k_1 k_2^2,\\
\nn \rho_{3\bar 1 5}&=&k_1,\quad \rho_{3\bar 1 6}=k_2^2=\rho{1\bar 3 6},\\
\nn \rho_{2\bar 3 1}&=&\rho_{1\bar 2 5}=\rho_{\bar 2 3 1}=\rho_{2\bar 2 2}=\rho_{1\bar 2 4}=1.
\eea
In conclusion, we mention some specific features of the effective low-energy superpotential (\ref{lesp}) for the three-generation Gepner model with Planck scale group
$SU_c(3)\times SU_L(3)\times SU_R(3)$:

(a) the absence of quark antigenerations,

(b) small number of quark selfcouplings,

(c) the absence of Yukawa interactions for the twisted generations of leptons-Higgses.


\subsection{Duality in $N=2$ super-Liouville theory}

The action of the $N=2$ SLFT at the flat background is given by:
\beq
{\cal A}_{\rm I}(b)=\int d^2 z \left[\int d^4\theta SS^{\dagger}
+\mu\int d^2\theta e^{bS}+c.c.\right]
\label{actionone}
\eeq
where $S$ is a chiral superfield satisfying:
\beq\nn
D_{-}S={\overline D}_{-}S=0,\qquad D_{+}S^{\dagger}=
{\overline D}_{+}S^{\dagger}=0.
\eeq
As in the LFT and the $N=1$ SLFTs, one should introduce a background charge
$1/b$ so that the second term in Eq.(\ref{actionone}) becomes a screening
operator of the conformal field theory (CFT).
However, a fundamental difference arises since
the background charge is unrenormalized due to the $N=2$ supersymmetry.
For the LFT and the $N=1$ SLFTs, the background charge is
renormalized to $Q=1/b+b$ and the theories are invariant under the
dual transformation $b\to 1/b$.
This self-duality plays an essential role to determine
various exact correlation functions of those Liouville theories.
Unrenormalized, the $N=2$ SLFT is not self-dual.

This theory is a CFT with a central charge:
\beq\label{cc}
c=3+6/b^2.
\eeq
The primary operators of the $N=2$ SLFT are classified into
Neveu-Schwarz (NS) and Ramond (R) sectors and can be written
in terms of the (first) component fields as follows:
\begin{equation}
N_{\alpha{\overline\alpha}}=
e^{\alpha\varphi^{\dagger}+{\overline\alpha}\varphi},\qquad
R^{\pm}_{\alpha{\overline\alpha}}=
\sigma^{\pm}e^{\alpha\varphi^{\dagger}+{\overline\alpha}\varphi},
\label{primary}
\end{equation}
where $\sigma^{\pm}$ are the spin operators.
The conformal dimensions of these fields are given by:
\begin{equation}
\Delta^{N}_{\alpha{\overline\alpha}}=-\alpha{\overline\alpha}
+{1\over{2b}}(\alpha+{\overline\alpha}),\qquad
\Delta^{R}_{\alpha{\overline\alpha}}=\Delta^{N}_{\alpha{\overline\alpha}}+
{1\over{8}}
\label{deltal}
\end{equation}
and the $U(1)$ charges are:
\begin{equation}
Q^{N}_{\alpha{\overline\alpha}}=-{1\over{2b}}(\alpha-{\overline\alpha}),
\qquad
Q^{R\pm}_{\alpha{\overline\alpha}}=Q^{N}_{\alpha{\overline\alpha}}
\pm{1\over{4}}.
\label{u1charge}
\end{equation}
From these expressions one can notice that:
\beq
\alpha\to 1/b-{\overline\alpha},\qquad
{\overline\alpha}\to 1/b-\alpha
\label{reflection}
\eeq
do not change the conformal dimension and $U(1)$ charge.
From the CFT point of view, this means that
$N_{1/b-{\overline\alpha},1/b-\alpha}$ should be identified
with $N_{\alpha{\overline\alpha}}$ and similarly for the (R) operators,
up to normalization factors.
The reflection amplitudes are determined by these normalization factors.

Without the self-duality, it is possible that there exists a `dual' action
to (\ref{actionone}) whose perturbative (weak coupling) behaviour describes
the $N=2$ SLFT in the strong coupling region.
This action should be another CFT.
Our proposal for the dual action is as follows:
\begin{equation}
{\cal A}_{\rm II}(b)=\int d^2 z \int d^4\theta \left[SS^{\dagger}
+{\tilde\mu}e^{b(S+S^{\dagger})}\right]
\label{actiontwo}
\end{equation}
with the background charge $b$.
The $N=2$ supersymmetry is preserved because $S+S^{\dagger}$ is a
$N=2$ scalar superfield.
One can see that this action is conformal invariant because the
interaction term is a screening operator.
Our conjecture is that the two actions,
${\cal A}_{\rm I}(b)$ and ${\cal A}_{\rm II}(1/b)$ are equivalent.
To justify this conjecture, we will compute the reflection amplitudes
based on these actions and will compare them with some independent results.

As mentioned above, the reflection amplitudes of the
Liouville-type CFT are defined by linear
transformations between different exponential fields,
corresponding to the same primary field of the chiral algebra.
For simplicity, we will restrict ourselves to the case
$\alpha={\overline\alpha}$ in (\ref{primary})
where the $U(1)$ charge of the (NS) operators becomes $0$.
We will refer to this case as the `neutral' sector.
(From now on, we will suppress the second indices ${\overline\alpha}$.)
The physical states in this sector are given by:
\beq\nn
\alpha={1\over{2b}}+iP
\eeq
where $P$ is a real parameter.
This parameter is transformed by $P\to -P$ under
the reflection relation (\ref{reflection})
and can be thought of as a `momentum' which is reflected off from
a potential wall.

The two-point functions of the same operators can be expressed as:
\begin{eqnarray}\nn
\langle N_{\alpha}(z,{\overline z})N_{\alpha}(0,0)\rangle
&=&{D^{N}(\alpha)\over{|z|^{4\Delta^N_{\alpha}}}}\\
\nn \langle R^{+}_{\alpha}(z,{\overline z}) R^{-}_{\alpha}(0,0)\rangle
&=&{D^{R}(\alpha)\over{|z|^{4\Delta^R_{\alpha}}}}
\end{eqnarray}
where $\Delta^N_{\alpha},\Delta^R_{\alpha}$ are given by Eq.(\ref{deltal}).
The normalization factors $D^{N}(\alpha), D^{R}(\alpha)$ define the reflection amplitudes and should satisfy:
\beq\nn
D^{N}(\alpha)D^{N}(1/b-\alpha)=1,\qquad
D^{R}(\alpha)D^{R}(1/b-\alpha)=1.
\eeq
To find these amplitudes explicitly, we consider the operator
product expansions (OPE's) with degenerate operators.

The NS and R degenerate operators in the neutral sector are
$N_{\alpha_{nm}}$ and $R^{\pm}_{\alpha_{nm}}$ with integers $n,m$ and:
\begin{equation}\nn
\alpha_{nm}={1-n\over{2b}}-{mb\over{2}},\qquad n,m\ge 0.
\end{equation}
The OPE of a NS field with a degenerate operator $N_{-b/2}$ is
simply given by:
\begin{equation}
N_{\alpha}N_{-b/2}=N_{\alpha-b/2}+ C_{-}^{N}(\alpha)N_{\alpha+b/2}.
\label{opei}
\end{equation}
Here the structure constant can be obtained from the screening integral
as follows:
\begin{equation}\nn
C_{-}^{N}(\alpha)= \kappa_1
\g(1-\alpha b)\g(1/2-\alpha b-b^2/2)
\g(-1/2+\alpha b)\g(\alpha b+b^2/2)\,
\end{equation}
where:
\bea\nn
\kappa_1 ={\mu^2 b^4\pi^2\over{2}}
\ga(-b^2-1)  \ga\left(1+{b^2\over{2}}\right)
\ga\left({b^2\over{2}}+{3\over{2}}\right)
\eea
with $\ga(x)=\Ga(x)/\Ga(1-x)$ as usual.

To use this OPE, we consider a three-point function
$\langle N_{\alpha+b/2}N_{\alpha}N_{-b/2}\rangle$
and take the OPE of $N_{-b/2}$ with either $N_{\alpha+b/2}$ or
$N_{\alpha}$ using (\ref{opei}).
This leads to a functional equation:
\beq
C_{-}^{N}(\alpha)D^{N}(\alpha+b/2)=D^{N}(\alpha).
\label{relationi}
\eeq
This functional equation determines the NS reflection amplitude
in the form:
\begin{equation}
D^{N}(\alpha)=(\frac{\kappa_1}{b^4})^{-2\al/b}
\ga(2\al/b-1/b^2)
{\ga(b\al+1/2)\over{\ga(b\al)}}f(\al),
\label{eqone}
\end{equation}
with an arbitrary function $f(\al)$ satisfying $f(\al)=f(\al+b)$.
To fix this unknown function, we need an additional functional equation.
It is natural that this relation is provided by
the dual action ${\cal A}_{\rm II}(1/b)$.

For this purpose we consider OPE's with another degenerate operator, namely:
\begin{eqnarray}
N_{\alpha}R^{+}_{-1/2b}&=&R^{+}_{\alpha-1/2b}+
{\tilde C}_{-}^{N}(\alpha)R^{+}_{\alpha+1/2b}
\label{opeii}\\
R^{-}_{\alpha}R^{+}_{-1/2b}&=&N_{\alpha-1/2b}+
{\tilde C}_{-}^{R}(\alpha)N_{\alpha+1/2b}.
\label{opeiii}
\end{eqnarray}
The structure constants can be computed by the screening integrals
using the dual action ${\cal A}_{\rm II}(1/b)$.
The result is:
\begin{eqnarray}\nn
{\tilde C}_{-}^{N}(\alpha)&=&
\kappa_{2}(b){\gamma(2\al/b-1/b^2)\over{\gamma(2\al/b)}},\\
\nn {\tilde C}_{-}^{R}(\alpha)&=&
\kappa_{2}(b){\gamma(2\al/b-1/b^2+1)\over{\gamma(2\al/b+1)}}\,
\end{eqnarray}
where:
\beq\nn
\kappa_{2}(b)= \tilde \mu \pi \ga\left(\frac1{b^2}+1\right)\,.\quad
\eeq
These results are consistent with the $N=2$ minimal model results presented in Section 3.1.

Now we consider the three-point functions
$\langle R^{-}_{\alpha+1/2b}N_{\alpha}R^{+}_{-1/2b}\rangle$
and
$\langle N_{\alpha+1/2b}R^{-}_{\alpha}R^{+}_{-1/2b}\rangle$.
Taking the OPE of $R^{+}_{-1/2b}$ with one of the other two operators in the
correlation functions and using the OPE relations (\ref{opeii})
and (\ref{opeiii}),
we obtain an independent set of functional relations as follows:
\begin{eqnarray}
\label{relationiii}
{\tilde C}_{-}^{N}(\al)D^{R}(\al+1/2b)&=&D^{N}(\alpha),\\
\nn {\tilde C}_{-}^{R}(\al)D^{N}(\al+1/2b)&=&D^{R}(\alpha).
\end{eqnarray}
Solving for $D^{N}(\alpha)$, we find that the most
general solution of Eqs.(\ref{relationiii}) is:
\begin{equation}
D^{N}(\alpha)=\kappa_2^{-2\al b}{\Ga^2(\al b+1/2)\over{\Ga^2(\al b)}}
\ga(2\al/b-1/b^2)g(\al)
\label{eqtwo}
\end{equation}
where $g(\al)$ is another arbitrary function satisfying $g(\al)=g(\al+1/b)$.
Combining Eqs.(\ref{eqone}) and (\ref{eqtwo}),
and requiring the normalization $D^N (\alpha = \frac1{2b} ) =1 $,
we can determine the NS reflection amplitude completely as follows:
\begin{equation}
D^{N}(\alpha)=-\frac 2 {b^2}\, \kappa_2^{-2\al b+1}\,
\ga\left({2\al\over{b}}-{1\over{b^2}}\right)
\ga\left(\al b+{1\over{2}}\right)\ga(1-\al b)\,
\label{eqthree}
\end{equation}
where the two parameters in the actions, $\mu$ and $\tilde \mu$,
are related by:
\beq
\left({\kappa_1\over{b^4}}\right)^{1/b}=\kappa_2^b.
\eeq
The R reflection amplitude can be obtained from (\ref{relationiii}):
\begin{equation}
D^{R}(\alpha)=-\frac {b^2} 2\kappa_2^{-2\al b+1}\,
\ga\left({2\al\over{b}}-{1\over{b^2}}+1\right)
\ga\left(-\al b+{1\over{2}}\right) \ga(\al b).
\label{eqfour}
\end{equation}

To justify the reflection amplitudes derived above and
based on the conjectured action ${\cal A}_{\rm II}$, we can provide
several consistency checks.
It has been noticed that an integrable model with
two parameters proposed in \cite{bfat}
can have $N=2$ supersymmetry if one of the parameters take a special value.
This means that one can compute the reflection amplitudes of the
$N=2$ SLFT independently as a special case of those in \cite{bfat}.
Indeed, we have confirmed that the two results agree exactly.

Furthermore, one can check the reflection amplitude for specific values
of $\alpha$ directly from the action.
For example when $\alpha = \frac 1{2b} - \frac b2$ using the action ${\cal A}_{\rm I}(b)$. If instead $\al\to 0$, one can compute the two-point function
directly from the action ${\cal A}_{\rm II}(1/b)$. Both results agree with (\ref{eqthree}).


\subsection{One-point functions of $N=2$ super-Liouville theory with boundary}

The action of the $N=2$ SLFT with boundary in terms of component fields is given by:
\beaq
S&=&\int d^2z\Bigg[\frac{1}{2\pi}\left(\partial\phi^-\bar{\partial}\phi^+
+\partial\phi^+\bar{\partial}\phi^-
+\psi^-\bar{\partial}\psi^++\psi^+\bar{\partial}\psi^-
+\bar{\psi}^-\partial\bar{\psi}^++\bar{\psi}^+\partial\bar{\psi}^-\right)
\nonumber\\
&&+i\mu b^2\psi^-\bar{\psi}^-e^{b\phi^+}
+i\mu b^2\psi^+\bar{\psi}^+e^{b\phi^-}
+\pi\mu^2 b^2e^{b(\phi^++\phi^-)}\Bigg]+S_B,
\label{N2L}
\eeaq
where the boundary action is derived in \cite{ay}:
\beaq
S_B&=&\int_{-\infty}^{\infty}dx\Bigg[-\frac{i}{4\pi}
(\bar{\psi}^+\psi^-+\bar{\psi}^-\psi^+) +\frac{1}{2}a^-\partial_x a^+
\nonumber\\
&-&\frac{1}{2}e^{b\phi^+/2}\left(\mu_B a^+
+\frac{\mu b^2}{4\mu_B}a^-\right)(\psi^-+\bar{\psi}^-)
-\frac{1}{2}e^{b\phi^-/2}\left(\mu_B a^-
+\frac{\mu b^2}{4\mu_B} a^+\right)(\psi^++\bar{\psi}^+)\nonumber\\
&-&\frac{2}{b^2}\left(\mu_B^2+\frac{\mu^2b^4}{16\mu_B^2}\right)
e^{b(\phi^++\phi^-)/2}\Bigg].\label{baction}
\eeaq
Note that we slightly changed the notations with respect to the previous Section.

The stress tensor $T$, the supercurrent $G^{\pm}$ and
the $U(1)$ current $J$ are given by:
\beaq
&&T=-\partial\phi^-\partial\phi^+
-\frac{1}{2}(\psi^-\partial\psi^++\psi^+\partial\psi^-)
+{1\over{2b}}(\partial^2\phi^++\partial^2\phi^-),
\label{N2T}\\
&&G^{\pm}=\sqrt{2}i(\psi^{\pm}\partial\phi^{\pm}-{1\over{b}}
\partial\psi^{\pm}),\qquad
J=-\psi^-\psi^{+}+{1\over{b}}(\partial\phi^+-\partial\phi^-).
\label{N2J}
\eeaq
Using the mode expansions for the currents and their operator product
expansion, one obtains the $N=2$ super-Virasoro algebra with central charge $c=3+6/b^2$.

The primary fields from the NS and R sectors are written in the new notations as follows:
\begin{equation}
N_{\alpha{\overline\alpha}}=
e^{\alpha\phi^{+}+{\overline\alpha}\phi^{-}},\qquad
R^{\pm}_{\alpha{\overline\alpha}}=
\sigma^{\pm}e^{\alpha\phi^{+}+{\overline\alpha}\phi^{-}},
\label{primarys}
\end{equation}
The conformal dimensions are given by (\ref{deltal}). According to the normalization in this Section the $U(1)$ charges of NS and R fields are rescaled:
\begin{equation}
\omega={1\over{b}}(\alpha-{\overline\alpha}),
\qquad
\omega^{\pm}=\omega \pm{1\over{2}}.
\label{ucharge}
\end{equation}
It is more convenient to use a `momentum' defined by:
\beq
\alpha+{\overline\alpha}={1\over{b}}+2iP, \label{momentum}
\eeq
and the $U(1)$ charge $\omega$ instead of $\alpha,{\overline\alpha}$.
In terms of these, the conformal dimensions are given by:
\begin{equation}
\Delta^{NS}={1\over{4b^2}}+P^2+{b^2\omega^2\over{4}}.
\label{deltanew}
\end{equation}
From now on, we will denote a NS primary state by
$\vert[P,\omega]\rangle$ and an R state by
$\vert[P,\omega,\epsilon]\rangle$ with $\epsilon=\pm 1$.

In this Section we compute exact one-point functions of the NS and R
bulk operators $N_{\alpha{\overline\alpha}}$ and
$R^{\epsilon}_{\alpha{\overline\alpha}}$ of the $N=2$
SLFT with boundary.
The one-point functions are defined by:
\beq
\langle N_{\alpha{\overline\alpha}}(\xi,\bar\xi)\rangle
={U^{NS}(\alpha,{\overline\alpha})\over
{|\xi-\bar\xi|^{2\Delta^{NS}_{\alpha{\overline\alpha}}}}},\quad
{\rm and}\quad
\langle R^{\epsilon}_{\alpha{\overline\alpha}}(\xi,\bar\xi)\rangle
={U^{R}(\alpha,{\overline\alpha})\over
{|\xi-\bar\xi|^{2\Delta^R_{\alpha{\overline\alpha}}}}},
\eeq
with the conformal dimensions given in(\ref{deltal}).
We will simply refer to the coefficients $U^{NS}(\alpha,{\overline\alpha})$
and $U^R(\alpha,{\overline\alpha})$ as the one-point functions.

According to Cardy's formalism, one can associate
a conformal BC with each primary state \cite{cardy}.
For the $N=2$ SLFT, there will be an infinite number of conformal BCs.
These BCs can be constructed by the fusion process and related to
the one-point functions.
Let us begin with the `vacuum' BC which corresponds to the identity
operator.
First we introduce an amplitude as an inner product (or overlap) between
the Isibashi state of a primary state and the conformal boundary state:\footnote{
We denote a conformal BC in `bold face' like ${\mathbf 0}$ and
a conformal boundary state like $\vert({\mathbf 0})\rangle$.}
\beq\nn
\Psi_{\mathbf 0}^{NS}(P,\omega)=
\langle ({\mathbf 0})|[P,\omega]\rangle\rangle.
\eeq
Following Cardy and using the modular properties of the $N=2$ characters we find that the
amplitude satisfies the following relation:
\beq
\Psi_{\mathbf 0}^{NS}(P,\omega){\Psi_{\mathbf 0}^{NS}}^{\dagger}(P,\omega)
={\mathbf S}_{NS}(P,\omega)
\label{nsvac}
\eeq
where:
\be\nn
{\mathbf S}_{NS}(P,\omega)=
{\sinh(2\pi bP)\sinh\left({2\pi P\over{b}}\right)\over{
2b^{-1}\cosh\left(\pi bP+{i\pi b^2\omega\over{2}}\right)
\cosh\left(\pi bP-{i\pi b^2\omega\over{2}}\right)}}.
\ee
Since ${\Psi_{\mathbf 0}^{NS}}^{\dagger}(P,\omega)
=\Psi_{\mathbf 0}^{NS}(-P,\omega)$, one can solve this up to some
unknown constant as follows:
\beq
\Psi^{NS}_{\mathbf 0}(P,\omega)=\sqrt{{b^3\over{2}}}
\left(X_{NS}\right)^{{iP\over{b}}}
{\Gamma\left({1\over{2}}-ibP+{b^2\omega\over{2}}\right)
\Gamma\left({1\over{2}}-ibP-{b^2\omega\over{2}}\right)\over{
\Gamma\left(-{2iP\over{b}}\right)\Gamma\left(1-2ibP\right)}}.
\eeq
The unknown constant $X_{NS}$ does not depend on $P,\omega$
and can not be determined by the modular transformation alone.
We will derive this constant later in this Section by comparing with
the bulk reflection amplitudes.

Similarly, for the R sector we define the R amplitude by:
\beq\nn
\Psi_{\mathbf 0}^{R}(P,\omega)=
\langle ({\mathbf 0})|[P,\omega,\epsilon]\rangle\rangle
\eeq
which satisfies:
\beq
\Psi_{\mathbf 0}^{R}(P,\omega){\Psi_{\mathbf 0}^{R}}^{\dagger}(P,\omega)
={\mathbf S}_{R}(P,\omega)
\label{rvac}
\eeq
with:
\be\nn
{\mathbf S}_{R}(P,\omega)=
{\sinh(2\pi bP)\sinh\left({2\pi P\over{b}}\right)\over{
2b^{-1}\sinh\left(\pi bP+{i\pi b^2\omega\over{2}}\right)
\sinh\left(\pi bP-{i\pi b^2\omega\over{2}}\right)}}.
\ee
The solution is up to an unknown constant:
\beq\nn
\Psi^{R}_{\mathbf 0}(P,\omega)=-i\sqrt{{b^3\over{2}}}
\left(X_{R}\right)^{{iP\over{b}}}
{\Gamma\left(-ibP+{b^2\omega\over{2}}\right)
\Gamma\left(1-ibP-{b^2\omega\over{2}}\right)\over{
\Gamma\left(-{2iP\over{b}}\right)\Gamma\left(1-2ibP\right)}}.
\eeq
Again, the unknown constant $X_{R}$ will be fixed later.

Now we consider a continuous BC associated with a primary field.
This field should be NS and its $U(1)$ charge should be zero
because only the boundary neutral operators should appear.
So, we consider the character of a (NS) primary state
$\vert s\rangle\equiv\vert[s,0]\rangle$ and its modular transformation.
The parameter $s$ depends on the boundary parameter $\mu_B$ in (\ref{baction}).
We define an inner product between the conformal
boundary state and an Ishibashi state:
\beq\nn
\Psi_{\mathbf s}^{NS}(P,\omega)
=\langle({\mathbf s})|[P,\omega]\rangle\rangle.
\eeq
Following the previous analysis of the modular transformation one can find that:
\beq\nn
\Psi^{NS}_{\mathbf s}(P,\omega){\Psi_{\mathbf 0}^{NS}}^{\dagger}(P,\omega)
=b\cos(4\pi s P).
\eeq
Now acting by $\Psi_{\mathbf 0}^{NS}(P,\omega)$ on this and using
(\ref{nsvac}), we obtain
\beaq
\Psi^{NS}_{\mathbf s}(P,\omega)&=&b\Psi_{\mathbf 0}^{NS}(P,\omega)
{\cos(4\pi s P)\over{{\mathbf S}_{NS}(P,\omega)}}\nonumber\\
&=&\sqrt{2b^3}\left(X_{NS}\right)^{{iP\over{b}}}
{\Gamma\left(1+{2iP\over{b}}\right)\Gamma\left(2ibP\right)
\cos(4\pi s P)\over{\Gamma\left({1\over{2}}+ibP+{b^2\omega\over{2}}\right)
\Gamma\left({1\over{2}}+ibP-{b^2\omega\over{2}}\right)}}.
\label{ampns}
\eeaq
One can follow the same steps for the R sector which leads to:
\beq\nn
\Psi^{R}_{\mathbf s}(P,\omega){\Psi_{\mathbf 0}^{R}}^{\dagger}(P,\omega)
=b\cos(4\pi s P),
\eeq
where
\beq\nn
\Psi_{\mathbf s}^{R}(P,\omega)=\langle({\mathbf s})|[P,\omega,\epsilon]
\rangle\rangle.
\eeq
Using (\ref{rvac}) on this, we can obtain:
\beaq
\Psi^{R}_{\mathbf s}(P,\omega)&=&b\Psi_{\mathbf 0}^{R}(P,\omega)
{\cos(4\pi s P)\over{{\mathbf S}_{R}(P,\omega)}}\nonumber\\
&=&-i\sqrt{2b^3}\left(X_{R}\right)^{{iP\over{b}}}
{\Gamma\left(1+{2iP\over{b}}\right)\Gamma\left(2ibP\right)
\cos(4\pi s P)\over{\Gamma\left(1+ibP-{b^2\omega\over{2}}\right)
\Gamma\left(ibP+{b^2\omega\over{2}}\right)}}.
\label{ampr}
\eeaq

The amplitudes (\ref{ampns}) and (\ref{ampr}) we have obtained are
the one-point functions of the two sectors up to some normalization
constants.
To fix these constants, we recall the relation proved in \cite{carlew}:
\beq
U_{\mathbf k}(\phi)={\langle({\mathbf k})|\phi\rangle\rangle\over{
\langle({\mathbf k})|0\rangle\rangle}}
\label{carlew}
\eeq
where ${\mathbf k}$ is a conformal BC, $\phi$ a primary field,
and $|\phi\rangle\rangle$, its Isibashi state.
For the $N=2$ SLFT, this relation means:
\beq\nn
U^{NS}_{\mathbf s}(P,\omega)
={\Psi_{\mathbf s}^{NS}(P,\omega)\over{
\Psi_{\mathbf s}^{NS}(-i/2b,0)}},\qquad
U^{R}_{\mathbf s}(P,\omega)
={\Psi_{\mathbf s}^{R}(P,\omega)\over{
\Psi_{\mathbf s}^{NS}(-i/2b,0)}}.
\eeq
From (\ref{ampns}) and (\ref{ampr}) we can obtain
the one-point functions as follows:
\beaq
U^{NS}_{\mathbf s}(P,\omega)
&=&{\cal N}\left(X_{NS}\right)^{{iP\over{b}}}
{\Gamma\left(1+{2iP\over{b}}\right)\Gamma\left(2ibP\right)
\cos(4\pi s P)\over{\Gamma\left({1\over{2}}+ibP+{b^2\omega\over{2}}\right)
\Gamma\left({1\over{2}}+ibP-{b^2\omega\over{2}}\right)}},
\label{oneptns}\\
U^{R}_{\mathbf s}(P,\omega)&=&
{\cal N}\left(X_{R}\right)^{{iP\over{b}}}
{\Gamma\left(1+{2iP\over{b}}\right)\Gamma\left(2ibP\right)
\cos(4\pi s P)\over{\Gamma\left(1+ibP-{b^2\omega\over{2}}\right)
\Gamma\left(ibP+{b^2\omega\over{2}}\right)}}, \label{oneptr}
\eeaq
where the normalization coefficient ${\cal N}$ can be
fixed by:
\beq\nn
U^{NS}_{\mathbf s}(-i/2b,0)=1\to
{\cal N}=\left[
\left(X_{NS}\right)^{1/2b^2}\Gamma(1+b^{-2})
\cosh\left({2\pi s\over{b}}\right)\right]^{-1}.
\eeq

We will use now the reflection amplitudes found in the previous Section in order to fix the normalization of the one-point functions.
Remind that the reflection amplitudes are defined by
two-point functions of the same operators:
\beq\nn
\langle N_{\alpha{\overline\alpha}}(z,{\overline z})
N_{\alpha{\overline\alpha}}(0,0)\rangle
={D^{NS}(\alpha,{\overline\alpha})\over{
|z|^{4\Delta^{NS}_{\alpha{\overline\alpha}}}}},\qquad
\langle R^{+}_{\alpha{\overline\alpha}}(z,{\overline z})
R^{-}_{\alpha{\overline\alpha}}(0,0)\rangle
={D^{R}(\alpha,{\overline\alpha})\over{
|z|^{4\Delta^R_{\alpha{\overline\alpha}}}}}
\eeq
Reflection properties imply that, in general:
\beq
\langle N_{\alpha{\overline\alpha}}(z,{\overline z})\ldots\rangle
=D^{NS}(\alpha,{\overline\alpha})
\langle N_{{1\over{b}}-{\overline\alpha},{1\over{b}}-\alpha}
(z,{\overline z})\ldots\rangle
\label{refcorr}
\eeq
and similarly for the R sector.
Here the part $\ldots$ can be any products of the primary fields.
The reflection relations among the correlation functions can be used
for the simplest case, namely, the one-point functions.
In this case, the relation becomes:
\beaq\nn
\langle N_{\alpha{\overline\alpha}}(z,{\overline z})\rangle
&=&D^{NS}(\alpha,{\overline\alpha})
\langle N_{{1\over{b}}-{\overline\alpha},{1\over{b}}-\alpha}
(z,{\overline z})\rangle,\\
\nn \langle R_{\alpha{\overline\alpha}}(z,{\overline z})\rangle
&=&D^{R}(\alpha,{\overline\alpha})
\langle R_{{1\over{b}}-{\overline\alpha},{1\over{b}}-\alpha}
(z,{\overline z})\rangle.
\eeaq
These lead to the following equations:
\beq
{U^{NS}_{\mathbf s}(P,\omega)\over{
U^{NS}_{\mathbf s}(-P,\omega)}}=D^{NS}(P,\omega),\qquad
{U^{R}_{\mathbf s}(P,\omega)\over{
U^{R}_{\mathbf s}(-P,\omega)}}=D^{R}(P,\omega).
\label{therelation}
\eeq
For the neutral sector $\omega=0$, the reflection amplitudes
has been derived in the previous Section:
\beaq
D^{NS}(P,0)&=&-\kappa^{-2iP/b}
{\Ga\left(1+{2iP\over{b}}\right)\over{\Ga\left(1-{2iP\over{b}}\right)}}
{\Ga\left(1+iPb\right)\over{\Ga\left(1-iPb\right)}}
{\Ga\left({1\over{2}}-iPb\right)\over{\Ga\left({1\over{2}}+iPb\right)}},
\label{eqfive}\\
D^{R}(P,0)&=&\kappa^{-2iP/b}
{\Ga\left(1+{2iP\over{b}}\right)\over{\Ga\left(1-{2iP\over{b}}\right)}}
{\Ga\left(1-iPb\right)\over{\Ga\left(1+iPb\right)}}
{\Ga\left({1\over{2}}+iPb\right)\over{\Ga\left({1\over{2}}-iPb\right)}}.
\label{eqsix}
\eeaq
where:
\beq\nn
\kappa ={\mu^2 \pi^2\over{2}}
\ga(-b^2-1)  \ga\left(1+{b^2\over{2}}\right)
\ga\left({b^2\over{2}}+{3\over{2}}\right),
\eeq
with $\gamma(x)=\Gamma(x)/\Gamma(1-x)$ and the bulk
cosmological constant $\mu$ defined in (\ref{N2L}).

Inserting $\omega=0$ and using (\ref{oneptns}) and (\ref{oneptr}),
the reflection amplitudes in Eq.(\ref{therelation})
are indeed in exact agreement with (\ref{eqfive}) and (\ref{eqsix})
if and only if we identify the constants:
\beq\nn
X_{NS}=X_{R}=\left[2^{2b^2}\kappa\right]^{-1}.
\eeq
This provides a nontrivial check and completes
our derivation for the one-point functions.
Furthermore, we can use (\ref{therelation}) to compute
the reflection amplitudes for $\omega\neq 0$ case:
\beq
D^{NS}(P,\omega)=(2^{2b^2}\kappa)^{-2iP/b}
{\Ga\left(1+{2iP\over{b}}\right)\over{\Ga\left(1-{2iP\over{b}}\right)}}
{\Gamma\left(2ibP\right)\over{\Gamma\left(-2ibP\right)}}
{\Gamma\left({1\over{2}}-ibP+{b^2\omega\over{2}}\right)\over{
\Gamma\left({1\over{2}}+ibP+{b^2\omega\over{2}}\right)}}
{\Gamma\left({1\over{2}}-ibP-{b^2\omega\over{2}}\right)\over{
\Gamma\left({1\over{2}}+ibP-{b^2\omega\over{2}}\right)}}
\label{genrefns}
\eeq
and:
\beq
D^{R}(P,\omega)=(2^{2b^2}\kappa)^{-2iP/b}
{\Ga\left(1+{2iP\over{b}}\right)\over{\Ga\left(1-{2iP\over{b}}\right)}}
{\Gamma\left(2ibP\right)\over{\Gamma\left(-2ibP\right)}}
{\Gamma\left(1-ibP-{b^2\omega\over{2}}\right)\over{
\Gamma\left(1+ibP-{b^2\omega\over{2}}\right)}}
{\Gamma\left(-ibP+{b^2\omega\over{2}}\right)\over{
\Gamma\left(ibP+{b^2\omega\over{2}}\right)}}
\label{genrefr}.
\eeq
These results can be compared with those from the
two-parameter family models \cite{bfat}
and we checked that the two independent results match exactly.

To complete our derivation of the one-point functions, we should
relate the boundary parameter $s$ to the boundary cosmological
constant $\mu_B$ in (\ref{baction}). For this, we consider one-point function of a neutral NS field
$N_{\alpha\alpha}$:
\beq\nn
{\rm residue}\ {U^{NS}(\alpha)\over{{\cal N}}}
\bigg\vert_{\alpha=(b^{-1}-nb)/2}=
\langle e^{\alpha(\phi^{+}+\phi^{-})}\rangle=
\sum_{p,q}{1\over{p!q!}}\langle e^{\alpha(\phi^{+}+\phi^{-})}
V^p B^q\rangle_{0},
\eeq
where $V,B$ are the interaction terms in the bulk and boundary actions.
If we choose $n=1$ ($\alpha=1/2b-b/2$), all terms vanish except
$p=0,q=2$ which can be easily computed:
\be\nn
\big\langle e^{\alpha(\phi^{+}+\phi^{-})}\left(i/2\right)
B^2\big\rangle_0=8\pi{\overline\mu_B}^2\Gamma(-b^2)\gamma\left({1+b^2\over{2}}\right)
\sin\left(\pi{1+b^2\over{2}}\right)
\ee
with:
\beq
{\overline\mu_B}^2=\mu_B^2+\frac{\mu^2b^4}{16\mu_B^2}.
\label{bcosmo}
\eeq
The residue of (\ref{oneptns}) at $\alpha={\overline\alpha}=1/2b-b/2$
becomes:
\beq
{b\over{2}}(2^{2b^2}\kappa)^{1/2}{\Gamma(-b^2)\over{\Gamma
\left({1-b^2\over{2}}\right)^2}}\cosh(2\pi sb).
\eeq
Comparing these two, we find:
\beq
{\overline\mu_B}^2 ={\mu b\over{32\pi}}\cosh(2\pi sb).
\label{boundparam}
\eeq

We want now to compare these results with those obtained by the so called conformal bootstrap approach. It consists in deriving functional equations for the one-point functions in a way similar to what we discussed for $N=1$ case in Section 2. Namely, consider two-point functions of the neutral operators:
\beq\nn
G^{NS}_{\alpha}(\xi,\xi')=
\langle R^+_{-{1\over{2b}}}(\xi)N_\alpha(\xi')\rangle,\qquad
G^{R}_{\alpha}(\xi,\xi')=
\langle R^+_{-{1\over{2b}}}(\xi)R^-_\alpha(\xi')\rangle
\eeq
where
$R^+_{-1/2b}$ is a degenerate R operator, whose OPE's are given by:
\beaq\nn
R^+_{-{1\over{2b}}}N_\alpha &=&\left[R^+_{\alpha-{1\over 2b}}\right]
+ C^{NS}(\alpha) \left[R^+_{\alpha+{1\over 2b}}\right],\\
\nn R^+_{-{1\over{2b}}}R^-_\alpha &=&\left[N_{\alpha-{1\over 2b}}\right]
+ C^{R}(\alpha) \left[N_{\alpha +{1\over 2b}}\right].
\eeaq
Here the bracket [\ldots] represents the conformal family of a given primary
field and the structure constants have been computed in the previous Section
based on the dual $N=2$ SLFT:
\beaq\nn
C^{NS}(\alpha) &=&{\tilde\mu}\pi\gamma\left(1+b^{-2}\right)
{\Gamma\left({2\alpha\over{b}}-{1\over{b^2}}\right)
\Gamma\left(1-{2\alpha\over{b}}\right)\over{
\Gamma\left(1-{2\alpha\over{b}}+{1\over{b^2}}\right)
\Gamma\left({2\alpha\over{b}}\right)}},\\
\nn C^{R}(\alpha) &=&{\tilde\mu}\pi\gamma\left(1+b^{-2}\right)
{\Gamma\left(1+{2\alpha\over{b}}-{1\over{b^2}}\right)
\Gamma\left(-{2\alpha\over{b}}\right)\over{
\Gamma\left(-{2\alpha\over{b}}+{1\over{b^2}}\right)
\Gamma\left(1+{2\alpha\over{b}}\right)}},
\eeaq
where ${\tilde\mu}$, the cosmological constant of the dual theory,
has been related there to that of the $N=2$ SLFT.

The two-point functions can be written as:
\beaq\nn
G^{NS}_{\alpha}(\xi,\xi')
&=&U^{R}\left(\alpha-{b\over{2}}\right)
{\cal G}^{NS}_{+}(\xi,\xi')+C^{NS}(\alpha)U^{R}
\left(\alpha+{b\over{2}}\right) {\cal G}^{NS}_{-}(\xi,\xi')\\
\nn G^{R}_{\alpha}(\xi,\xi')
&=&U^{NS}\left(\alpha-{b\over{2}}\right)
{\cal G}^{R}_{+}(\xi,\xi')+C^{R}(\alpha)U^{NS}
\left(\alpha+{b\over{2}}\right) {\cal G}^{R}_{-}(\xi,\xi')
\eeaq
where ${\cal G}_{\pm}(\xi,\xi')$'s are expressed in terms of the special conformal blocks:
\beq
{\cal G}^{NS}_{\pm}(\xi,\xi')={|\xi'-{\overline \xi'}|^{2\Delta^{NS}_{\alpha}
-2\Delta^{R}_{-b/2}}\over{|\xi-{\overline \xi'}|^{4\Delta^{NS}_{\alpha}}}}
{\cal F}^{NS}_{\pm}(\eta),\quad
{\cal G}^{R}_{\pm}(\xi,\xi')={|\xi'-{\overline \xi'}|^{2\Delta^{R}_{\alpha}
-2\Delta^{NS}_{-b/2}}\over{|\xi-{\overline \xi'}|^{4\Delta^{R}_{\alpha}}}}
{\cal F}^{R}_{\pm}(\eta)
\nonumber
\eeq
with
\beq
\eta={(\xi-\xi')({\overline \xi}-{\overline \xi'})\over{(\xi-{\overline \xi'})
({\overline \xi}-\xi')}}.
\nonumber
\eeq
These conformal blocks are expressed in terms of some known hypergeometric functions.

On the other hand, one can compute the two-point functions as
both $R^{+}_{-1/2b}$ and $N_{\alpha}$ or $R^{-}_{\alpha}$
approach the boundary.
The fusion of the degenerate operator with the boundary is described
by a special bulk-boundary structure constant which could be computed
as a boundary screening integral with one insertion of the boundary
interaction of the dual $N=2$ theory if it were known.
Since we can not fix it, we denote the unknown constant just as
${\cal R}(-1/2b)$.
Then, we can obtain the system of functional relations as follows:
\beaq\nn
{\cal R}\left(-{1\over{2b}}\right)
U^{NS}(\alpha)&=&{\Gamma(1-{1\over{b^2}}+{2\alpha\over b})
\Gamma(-{2\over b^2})\over \Gamma(1-{2\over b^2}+{2\alpha\over b})
\Gamma(1-{1\over b^2})} U^R\left(\alpha-{1\over 2b}\right)\nonumber\\
\nn &+&C^{NS}(\alpha){\Gamma(1+{1\over b^2}-{2\alpha\over b})\Gamma(-{2\over b^2})
\over \Gamma(1-{2\alpha\over b})\Gamma(1-{1\over b^2})}
U^R\left(\alpha+{1\over 2b}\right)\\
\nn {\cal R}\left(-{1\over{2b}}\right)
U^R(\alpha) &=&{\Gamma({2\alpha\over b}-{1\over b^2})
\Gamma(-{2\over b^2})\over \Gamma ({2\alpha\over b}-{2\over b^2})
\Gamma (1-{1\over b^2})}U^{NS}\left(\alpha-{1\over 2b}\right)\nonumber\\
\nn &+&C^{R}(\alpha){\Gamma({1\over b^2}-{2\alpha\over b})\Gamma(-{2\over b^2})
\over \Gamma ({-2\alpha\over b})\Gamma(1-{1\over b^2})}
U^{NS}\left(\alpha+{1\over 2b}\right).
\eeaq
Although we do not know the bulk-boundary structure constant,
we can eliminate it by taking the ratio of the above equations and
find one relation which is completely fixed.
It can be shown that the one-point functions (\ref{oneptns})
and (\ref{oneptr}) indeed satisfy this relation.
This means not only that the one-point functions obtained from the modular
bootstrap procedures are consistent with the $N=2$ SLFT actions,
but also that the $N=2$ theory proposed in the previous Section is indeed
dual to the $N=2$ SLFT.
Furthermore, we can find the bulk-boundary structure constant as follows:
\beq\nn
{{\cal R}\left(-{1\over{2b}}\right)\Gamma\left(1-{1\over{b^2}}\right)
\over{\Gamma\left(-{2\over{b^2}}\right)
\sqrt{{\tilde\mu}\pi\gamma\left(1+{1\over{b^2}}\right)}}}
=\cosh\left({2\pi s\over{b}}\right).
\eeq
Along with (\ref{boundparam}),
this equation relates the boundary cosmological constant of
the $N=2$ SLFT with that of the dual $N=2$ theory.

\subsection{ZZ-branes of $N=2$ super-Lioville theory}

Remind that the primary fields of NS and R sectors are expressed in terms of vertex operators:
\begin{equation}
N_{\alpha{\overline\alpha}}=
e^{\alpha\phi^{+}+{\overline\alpha}\phi^{-}},\qquad
R^{(\pm)}_{\alpha{\overline\alpha}}=
\sigma^{\pm}e^{\alpha\phi^{+}+{\overline\alpha}\phi^{-}}
\label{primaryy}
\end{equation}
where $\sigma^{\pm}$ are the spin operators. The conformal dimensions and the $U(1)$ charges of the primary fields
$N_{\alpha{\overline\alpha}}$ and $R^{(\pm)}_{\alpha{\overline\alpha}}$ were obtained above as (\ref{deltal}) and (\ref{ucharge}) respectively.

Among the primary fields there is a series of degenerate fields of the $N=2$ SLFT.
In this Section we divide these fields into three classes.
Class-I degenerate fields are given by:
\begin{eqnarray}\nn
N_{m,n}^{\omega}&=&N_{\alpha_{m,n}^{\omega},{\overline\alpha}_{m,n}^{\omega}},
\qquad
R_{m,n}^{(\ep)\omega}=R^{(\ep)}_{\alpha_{m,n}^{\omega},
{\overline\alpha}_{m,n}^{\omega}},\\
\alpha_{m,n}^{\omega}&=&{1-m+\omega b^2\over{2b}}-{nb\over{2}},\qquad
{\overline\alpha}_{m,n}^{\omega}={1-m-\omega b^2\over{2b}}-{nb\over{2}},
\ \ m,n\in {\mathbf Z}_{+}.
\label{degtypeI}
\end{eqnarray}
$N_{m,n}^{\omega}$ and $R_{m,n}^{(\ep)\omega}$ are degenerate at level $mn$
where the corresponding null states turn out to be:
\beq
N_{m,-n}^{\omega},\qquad{\rm and}\qquad R_{m,-n}^{(\ep)\omega}.
\label{nullone}
\eeq
As an example, consider the most simple case $N_{1,1}^{\omega}$ with
the conformal dimension\\ $b^2(\omega^2-1)/4-1/2$ and $U(1)$ charge $\omega$.
After simple calculation, one can check that:
\beq\nn
\left[{b^2\over{2}}(1-\omega^2)J_{-1}+G^{+}_{-1/2}G^{-}_{-1/2}
-(1-\omega)L_{-1}\right]\vert N_{1,1}^{\omega}\rangle
\eeq
is annihilated by all the positive modes of the $N=2$ super CFT.
Since this state has the $U(1)$ charge $\omega$ and dimension $+1$ more than
that of $N_{1,1}^{\omega}$, it corresponds to
$\vert N_{1,-1}^{\omega}\rangle$ up to a normalization constant.
One can continue this analysis to higher values of $m,n>1$ to confirm the
statement of Eq.(\ref{nullone}).
Notice that the null state structure changes dramatically for $\omega=\pm n$
case.
The field $N_{m,n}^{\pm n}$ has a null state $N_{m,-n}^{\pm n}$ at level $mn$.
This $N_{m,-n}^{\pm n}$ field is in fact a class-II degenerate field
which we will explain next and has infinite number of null states.
Therefore, we exclude the case of $\omega=\pm n$ from class-I fields.

The second class of degenerate fields is denoted by
$N_{m}^{\omega}$ and $R_{m}^{(\ep)\omega}$ and comes in two subclasses,
namely, class-IIA and class-IIB.
These are given by:
\beaq
{\rm Class-IIA}&:&\qquad N_{m}^{\omega}=N_{\alpha_{m}^{\omega},
{\overline\alpha}_{m}^{0}}\qquad
R_{m}^{(+)\omega}=R^{(+)}_{\alpha_{m}^{\omega},{\overline\alpha}_{m}^{0}},
\quad \om>0
\label{degtypeIIa}\\
{\rm Class-IIB}&:&\qquad {\tilde N}_{m}^{\omega}=
N_{\alpha_{m}^{0},{\overline\alpha}_{m}^{\omega}}\qquad
R_{m}^{(-)\omega}=R^{(-)}_{\alpha_{m}^{0},{\overline\alpha}_{m}^{\omega}},
\quad \om<0.
\label{degtypeIIb}
\eeaq
Here we have defined:
\beq\nn
\alpha_{m}^{\omega}\equiv{1-m+2\omega b^2\over{2b}},\qquad
{\overline\alpha}_{m}^{\omega}\equiv{1-m-2\omega b^2\over{2b}}
\eeq
with $m$ a positive odd integer for the NS sector and
even for the R sector.

These fields have null states at level $m/2$ which can be
expressed again by Eq.(\ref{degtypeIIa}) with
$\omega$ shifted by $+1$ for class-IIA
and by Eq.(\ref{degtypeIIb}) with $\omega$ shifted by $-1$ for class-IIB.
For $m=1$, these fields become either chiral or
anti-chiral field which are annihilated by $G^{\pm}_{-1/2}$,
respectively.
For $m=3$, one can construct a linear combination of descendants:
\beq
\left[\left(\omega-{2\over{b^2}}+1\right)G^{+}_{-3/2}
-G^{+}_{-1/2}L_{-1}+G^{+}_{-1/2}J_{-1}\right]\vert
N_{3}^{\omega}\rangle
\label{nullex}
\eeq
which satisfies the null state condition.
Since this state has $U(1)$ charge $\omega+1$
and dimension $3/2$ higher than that of $N_{3}^{\omega}$, it is
straightforward to identify it as $N_{3}^{\omega+1}$ up to a
normalization constant.
However, it is not the end of the story in this case.
The $N_{3}^{\omega+1}$ field is again degenerate at level $3/2$
because a linear combination of its descendants, exactly
Eq.(\ref{nullex}) with $\omega$ shifted by $+1$, satisfies the
null state condition.
This generates $N_{3}^{\omega+2}$ and it continues infinitely.
This infinite null state structure holds for any odd integer $m$.

This can be illustrated by semi-infinite sequence:
\beaq\nn
{\rm Class-IIA}&:& N_{m}^{\omega}\to N_{m}^{\omega+1}\to
N_{m}^{\omega+2}\to\ldots\\
\nn {\rm Class-IIB}&:&{\tilde N}_m^{\omega}\to {\tilde N}_m^{\omega-1}\to
{\tilde N}_m^{\omega-2}\to\ldots.
\eeaq
This works similarly for the R sector.
For example, the null state of the $m=2$ R field is given by:
\beq\nn
G^{\pm}_{-1}\vert R_{2}^{(\pm)\omega}\rangle.
\eeq

We need to deal with class-II neutral ($\omega=0$) NS fields separately.
For example, consider the $N_{3}^{0}$ which has two null states:
\beq
\left[\left(1-{2\over{b^2}}\right)G^{\pm}_{-3/2}
-G^{\pm}_{-1/2}L_{-1}+G^{\pm}_{-1/2}J_{-1}\right]\vert N_{3}^{0}\rangle,
\label{nullexi}
\eeq
which should be identified with $N_{3}^{1}$ and ${\tilde N}_{3}^{-1}$,
respectively.
We will call these neutral NS degenerate fields as class-III and denote them by:
\beq
{\rm Class-III}:\qquad N_{m}=N_{\alpha_{m}^{0}{\overline\alpha}_{m}^{0}}.
\label{degtypeIII}
\eeq
The null state structure of the class-III fields has an infinite sequence
in both directions:
\beq
{\rm Class-III}:\qquad \ldots\leftarrow{\tilde N}_m^{-2}\leftarrow
{\tilde N}_m^{-1}\leftarrow N_{m}\to N_{m}^{1}\to N_{m}^{2}\to\ldots.
\label{neutralseq}
\eeq
The identity operator is the most simple class-III field with $m=1$.

The degenerate fields are playing an essential role in
both conformal and modular bootstraps.
As we will see shortly, some simple degenerate fields satisfy
relatively simple operator product expansions and make the conformal
bootstrap viable.

In this Section we are interested in the $N=2$ SLFT on Lobachevskiy plane
or pseudosphere which is the geometry of the infinite constant negative
curvature surface.
As in the $N=1$ case, using conformal bootstrap we derive and solve
nonlinear functional equations which can provide discrete BCs.

The classical equations of motion for the $N=2$ SLFT can be derived from
the (bulk part of) Lagrangian (\ref{N2L}):
\beaq\nn
\partial\bar\partial\phi^{\pm}&=& \pi\mu b^3\left[\pi\mu
e^{b(\phi^{+}+\phi^{-})}+i\psi^{\pm}{\bar\psi}^{\pm}e^{b\phi^{\mp}}\right]\\
\nn \partial{\bar\psi}^{\pm}&=& i\pi \mu b^2 e^{b\phi^{\pm}}\psi^{\mp},\qquad
\bar\partial\psi^{\pm}= -i\pi \mu b^2 e^{b\phi^{\pm}}{\bar\psi}^{\mp}.
\eeaq
Assuming that the fermionic fields vanish in the classical limit, we can
solve the bosonic fields classically:
\beq\nn
e^{\varphi(z)}={4R^2\over{(1-|z|^2)^2}},
\eeq
where $\varphi=b(\phi^{+}+\phi^{-})$ and $R^{-2}=4\pi^2\mu^2 b^4$.
The parameter $R$ is interpreted as the radius of the pseudosphere
in which the points at the circle $|z|=1$ are infinitely
far away from any internal point.
This circle can be interpreted as the ``boundary'' of the pseudosphere.
This boundary has a different class of BC's.
For the $N=2$ SLFT, we will call the discrete BCs as ZZ-branes following \cite{boot2}
and show that these correspond to the degenerate fields of the $N=2$ SLFT.

Let us consider a two-point function of
a neutral degenerate field $N_{-b/2}$ and a general neutral field $N_{\alpha}$:
\footnote{
We will suppress one of the indices of the fields since ${\overline\alpha}=\alpha$.}
\beq
G_{\al}(\xi,\xi')=\langle N_{-b/2}(\xi)N_\val(\xi')\rangle.
\label{twoptps}
\eeq
Using the OPE of the two fields, we can express this two-point function as:
\beq\nn
G_{\al}(\xi,\xi')=U^{NS}\left(\alpha-{b\over{2}}\right)
{\cal G}^{NS}_{1}(\xi,\xi')+C_{--}(\alpha)U^{NS}
\left(\alpha+{b\over{2}}\right) {\cal G}^{NS}_{3}(\xi,\xi')
\eeq
where the structure constant $C_{--}$ is known and given by gamma-functions. The ${\cal G}^{NS}_{i}(\xi,\xi')$'s are expressed
in terms of the special conformal blocks:
\beq
{\cal G}^{NS}_{i}(\xi,\xi')={|\xi'-{\bar \xi'}|^{2\Delta^{NS}_{\alpha}
-2\Delta^{NS}_{-b/2}}\over{|\xi-{\bar \xi'}|^{4\Delta^{NS}_{\alpha}}}}
{\cal F}^{NS}_{i}(\eta),\quad i=1,2,3
\nonumber
\eeq
These conformal blocks can be determined by two-fold Dotsenko-Fateev integrals \cite{df}, the index $i$ denotes the three independent integration contours between
the branching points $0,\eta,1,\infty$. The conformal blocks ${\cal F}^{NS}_i(\eta)$ are regular at $\eta=0$.
Since we are interested in the limit $\eta\rightarrow 1$,
we need to introduce another blocks ${\tilde{\cal F}}^{NS}_{i}(\eta)$ which are well defined in that limit. The monodromy relations between the conformal blocks are given by \cite{df}:
\beq\nn
{\cal F}^{NS}_i(\eta)=\sum_{j=1}^{3}\val_{ij}{\tilde {\cal F}}^{NS}_j(\eta),
\eeq
where again $\val_{ij}$ are known and expressed in terms of gamma-functions.

On the pseudosphere geometry, as the two fields approach the boundary $\eta\rightarrow 1$,
the distance between the two points become infinite due to the singular metric.
This means that the two-point function is factorized into a product of two
one-point functions.
For example, the two-point function in (\ref{twoptps}) becomes:
\beq\nn
G_{\al}(\xi,\xi')={|\xi'-\bar\xi'|^{2\Delta^{NS}_\val-2\Delta^{NS}_{-b/2}}\over
|\xi-\bar\xi'|^{4\Delta^{NS}_\val}}U^{NS}(-b/2)U^{NS}(\val)
{\tilde{\cal F}}^{NS}_3 (\eta).
\eeq
Comparing these two results, we can obtain the following
nonlinear functional equation for $U(\val)$:
\beaq
{\cal C}_1U^{NS}(-b/2)U^{NS}(\val)&=&{\Ga(\val b-{b^2\over 2})
\Ga(\val b+{1\over 2})\over \Ga(\val b)\Ga(\val b-{b^2\over 2}-{1\over 2})}
U^{NS}\left(\val-{b\over 2}\right)\nonumber\\
&+&2^{-2-2b^2}\pi^2 b^4\mu^2{\Ga(\val b-{1\over 2})\Ga(\val b+{b^2\over 2})
\over \Ga(\val b)\Ga(\val b+{b^2\over 2}+{1\over 2})}
U^{NS}\left(\val +{b\over 2}\right)
\label{psi}
\eeaq
with:
\beq
{\cal C}_1={\sqrt{\pi}\Ga(-{b^2\over 2})\over{
\Ga(-1)\Ga(-{b^2\over 2}-{1\over 2})}}.
\label{c1}
\eeq

Similarly, using the OPE's of the degenerate field $N_{-{1\ov b}}$ with arbitrary primary fields $N_{\a,\bar\a}$ and $R_{\a,\bar\a}$ and the dual $N=2$ action defined in Section (3.3), we can derive functional equations:
\beaq
{\cal C}_2{\tilde U}^{NS}(-1/b,0)U^{NS}(\val,\bar\val)&=&
{\Ga({\val+\bar\val\over b}-{1\over b^2}+1)\over{
(1-\val b)\Ga({\val+\bar\val\over b}-{2\over b^2})}}
U^{NS}\left(\val-{1\over b},\bar\val\right)\nonumber\\
&-&{\tilde\mu}'
{(\bar\alpha b)\Ga({\val+\bar\val\over b}-{1\over b^2})\over{
\Ga(1+{\val+\bar\val\over b})}}
U^{NS}\left(\val,\bar\val+{1\over b}\right),
\label{psiv}\\
{\cal C}_2{\tilde U}^{NS}(-1/b,0)U^{R}(\val,\bar\val)&=&
{\Ga({\val+\bar\val\over b}-{1\over b^2}+1)\over{
({3\over{2}}-\val b)\Ga({\val+\bar\val\over b}-{2\over b^2})}}
U^{R}\left(\val-{1\over b},\bar\val\right)\nonumber\\
&-&{\tilde\mu}'
{(\bar\val b+{1\over{2}})\Ga({\val+\bar\val\over b}-{1\over b^2})
\over{\Ga(1+{\val+\bar\val\over b})}}
U^{R}\left(\val,\bar\val+{1\over b}\right)
\label{psv}
\eeaq
with ${\cal C}_2=\Ga(1-{1\over{b^2}})/\Ga(-{2\over{b^2}})$ and ${\tilde\mu}'$ is given by ${\tilde\mu}$, the cosmological constant of
the dual $N=2$ theory:
\beq\nn
{\tilde\mu}'=4\pi{\tilde\mu}\gamma(1+b^{-2})b^{-4}.
\eeq
Here we have denoted one-point functions of the class-II degenerate field in
terms of ${\tilde U}^{NS}$ since they are in principle different from the
one-point functions of general fields. Notice that ${\cal C}_1$ contains $\Gamma(-1)$ in Eq.(\ref{c1}) which arises
from singular monodromy transformation.
We can remove this singular factor by redefining $U^{NS(R)}/\Gamma(-1)\to U^{NS(R)}$.
Notice that this redefinition does not change Eqs.(\ref{psiv}) and (\ref{psv})
since they are linear in $U^{NS(R)}$ if assuming that ${\tilde U}^{NS}$ is regular.

The solutions to these equations can be expressed in terms of two integers
$m,n\ge 1$ as follows:
\beaq
U^{NS}_{mn}(\val,\bar\al)&=&{\cal N}_{mn}(\pi\mu)^{-{\al+{\bar\al}\over{b}}}
{\Ga({\al+{\bar\al}\over{b}}-{1\over b^2}+1)\Ga(b(\al+{\bar\al})-1)\over
\Ga(\al b)\Ga({\bar\al} b)}\nonumber\\
&\times&\sin\left[{\pi m\over b}\left(\al+{\bar\al}-{1\over b}\right)\right]
\sin\left[\pi nb\left(\al+{\bar\al}-{1\over b}\right)\right]
\label{ZZNS}
\\
U^{R}_{mn}(\val,\bar\al)&=&{\cal N}_{mn}(\pi\mu)^{-{\al+{\bar\al}\over{b}}}
{\Ga({\al+{\bar\al}\over{b}}-{1\over b^2}+1)\Ga(b(\al+{\bar\al})-1)\over
\Ga(\al b-1/2)\Ga({\bar\al} b+1/2)}\nonumber\\
&\times&\sin\left[{\pi m\over b}\left(\al+{\bar\al}-{1\over b}\right)\right]
\sin\left[\pi nb\left(\al+{\bar\al}-{1\over b}\right)\right],
\label{ZZR}
\eeaq
with the normalization factors given by:
\beq\nn
{\cal N}_{mn}=(-1)^n {4b^2\over{\Ga(-1/b^2)}}
{\cot(\pi nb^2)\over{\sin(\pi m/b^2)}}.
\eeq
This class of solutions will be associated with conformal BCs corresponding
to the class-I neutral degenerate fields.
It turns out that the conformal bootstrap equations do not allow
discrete BCs corresponding to non-neutral degenerate fields.
One possible explanation is that non-neutral BCs will introduce a boundary
field which will not produce the identity operator when fused with bulk
degenerate fields as they approach the boundary.

It is interesting to notice that the following one-point functions:
\beaq
U^{NS}_m(\alpha,\bar{\alpha})&=&{\cal N}_m(\pi\mu)^{-\frac{\alpha+
\bar{\alpha}}{b}}\frac{\Gamma(1-\alpha b)\Gamma(1-\bar{\alpha}b)}
{\Gamma(-\frac{\alpha+\bar{\alpha}}{b}+\frac{1}{b^2})\Gamma(2-b(\alpha
+\bar{\alpha}))}\nonumber\\
&\times&\frac{\sin\left[\frac{\pi m}{b}(\alpha+\bar{\alpha}-\frac{1}{b})\right]}
{\sin\left[\frac{\pi}{b}(\alpha+\bar{\alpha}-\frac{1}{b})\right]}
\label{UNSm}\\
U^R_m(\alpha,\bar{\alpha})&=&{\cal N}_m(\pi\mu)^{-\frac{\alpha+\bar{\alpha}}{b}}
\frac{\Gamma(\frac{3}{2}-\alpha b)\Gamma(\frac{1}{2}-\bar{\alpha}b)}
{\Gamma(-\frac{\alpha+\bar{\alpha}}{b}+\frac{1}{b^2})\Gamma(2-b(\alpha
+\bar{\alpha}))}\nonumber\\
&\times&\frac{\sin\left[\frac{\pi m}{b}(\alpha+\bar{\alpha}-\frac{1}{b})\right]}
{\sin\left[\frac{\pi}{b}(\alpha+\bar{\alpha}-\frac{1}{b})\right]}
\label{URm}\\
{\cal N}_m&=&\frac{\pi}{\Gamma(-\frac{1}{b^2}+1)}\frac{1}
{\sin(\frac{\pi m}{b^2})}
\eeaq
satisfy Eqs.(\ref{psiv}) and (\ref{psv}).
Although they do not satisfy Eq.(\ref{psi}), hence not complete solutions,
this class of solutions turns out to be consistent with modular bootstrap equations and
we will show that they correspond to the class-III BCs.

We want now to derive the modular bootstrap equations based on
the modular properties of degenerate characters.
We will derive the boundary amplitudes and show that they are consistent with the one-point
functions derived before. Among the conformal BC's of the $N=2$ SLFT, we concentrate on those associated
with the degenerate fields.
Following the modular bootstrap formulation, we can compute a boundary amplitude
as the inner product between
the Ishibashi state of a primary state and the conformal boundary state.

Let us start with the class-I BC's. The boundary amplitudes are defined by:
\beq\nn
\Psi_{mn\omega}^{NS}(P,\omega)=\langle m,n,\omega\vert N_{[P,\omega]}\rangle\rangle.
\eeq
From the modular transformation properties of the class-I character we can obtain:
\beq
\Psi^{NS}_{mn\omega}(P,\omega')\Psi^{NS\dag}_{\bf 0}(P,\omega')=
2b\sinh(2\pi mP/b)\sinh(2\pi nbP)e^{-\pi ib^2\omega\omega'}.
\label{NSde1Psi}
\eeq
Since the vacuum boundary amplitude $\Psi^{NS}_{\bf 0}(P,\omega)$ was obtained in the previous Section, we get from (\ref{NSde1Psi}) that:
\beaq
\Psi^{NS}_{mn\omega}(P,\omega')&=&
\sqrt{{8\over{b}}}
\left(\pi\mu\right)^{-{2iP\over{b}}}
{\Gamma\left({2iP\over{b}}\right)\Gamma\left(1+2ibP\right)
\over{\Gamma\left({1\over{2}}+ibP+{b^2\omega\over{2}}\right)
\Gamma\left({1\over{2}}+ibP-{b^2\omega\over{2}}\right)}}\nonumber\\
&\times&\sinh(2\pi mP/b)\sinh(2\pi nbP)
e^{-\pi ib^2\omega\omega'}.
\label{NSIres}
\eeaq
This solution coincides with (\ref{ZZNS}),
the ZZ-brane solution with BC $(m,n,\omega=0)$.
This provides an important consistency check between the conformal
and modular bootstraps.

We pass now to the class-II BC's. Denoting the class-II boundary state as $\vert m,\omega\rangle$,
we can define the following boundary amplitude:
\beq\nn
\Psi_{m\omega}^{NS}(P,\omega')=
\langle m,\omega\vert N_{[P,\omega']}\rangle\rangle.
\eeq
Comparing this with the modular transformation of the class-II characters we obtain:
\beq
\Psi^{NS}_{m\omega}(P,\omega')\Psi^{NS\dag}_{\bf 0}(P,\omega')=
S_{m\omega}(P,\omega'),
\label{NSde2Psi}
\eeq
where $S_{m\omega}(P,\omega')$ is the modular $S$-matrix component.
From this, one can solve for $\Psi^{NS}_{m\omega}(P,\omega')$.
Instead of presenting details for this case, we will analyze a
more interesting case, namely the neutral ($\omega=0$) class-III BCs.

For a class-III (neutral) boundary state $\vert m\rangle$,
we can define:
\beq\nn
\Psi_{m}^{NS}(P,\omega)=\langle m\vert N_{[P,\omega]}\rangle\rangle.
\eeq
In the same way as before it follows that:
\beq
\Psi^{NS}_m(P,\omega)={\Psi_{\mathbf 0}^{NS}}(P,\omega)
\frac{\sinh(2\pi mP/b)}{\sinh(2\pi P/b)}.
\label{NSde2cPsi}
\eeq
The solution (\ref{NSde2cPsi}) coincides with the one-point function (\ref{UNSm}).

One can perform similar analysis for the R sector. For the class-I amplitudes for example one obtains:
\beq\nn
\Psi^{R}_{mn\omega}(P,\omega')\Psi^{R\dag}_{\bf 0}(P,\omega')
=2b\sinh(2\pi mP/b)\sinh(2\pi nbP)e^{-\pi ib^2\omega\omega'}
\eeq
from which we can find:
\beaq\nn
\Psi^{R}_{mn\omega}(P,\omega')&=&
-i\sqrt{{8\over{b}}}
\left(\pi\mu\right)^{-{2iP\over{b}}}
{\Gamma\left({2iP\over{b}}\right)\Gamma\left(1+2ibP\right)
\over{\Gamma\left(ibP+{b^2\omega\over{2}}\right)
\Gamma\left(1+ibP-{b^2\omega\over{2}}\right)}}\nonumber\\
\nn &\times&\sinh(2\pi mP/b)\sinh(2\pi nbP)
e^{-\pi ib^2\omega\omega'}.
\eeaq

It is straightforward to continue this analysis for the class-II and class-III
BCs and their mixed BCs for the R sector.

\subsection{Higher equations of motion in $N=2$ super-Liouville field theory}

Let us remind that the $N=2$ SLFT is based on the
Lagrangian:
\bea\nn
\CL &=& {1\ov
{2\pi}}\left(\p\phi^-\bp
\phi^+ +\p\phi^+\bp\phi^- +\psi^-\bp\psi^+
+\psi^+\bp\psi^- +\bps^-\p\bps^+ +\bps^+\p\bps^- \right)+\\
\nn &+&i\mu b^2\psi^-\bps^-e^{b\phi^+}
+i\mu b^2\psi^+\bps^+e^{b\phi^-}+\pi\mu^2 b^2 e^{b\phi^+ +b\phi^-}
\eea
where $(\phi^\pm, \psi^\mp)$ are the components of a chiral
$N=2$ supermultiplet, $b$ is the coupling constant and $\mu$ is the
cosmological constant. It is invariant under the $N=2$
superconformal algebra (\ref{n2alg}) with central charge $c=3+{6\ov b^2}$. In (\ref{n2alg}) the left handed generators appear, there are in addition the
right handed ones $\bar L_n$, $\bar J_n$, $\bar G^{\pm}_r$ closing
the same algebra. The basic objects of interest here are the primary fields in the NS sector defined by the vertices $N_{\a,\bar\a}=e^{\a\phi^++\bar\a\phi^-}$,
the corresponding states being annihilated by the positive modes. There are in addition also Ramond primary fields $R_{\a,\bar\a}$ but we will not
be concerned with them in this Section. Let us remind also that the conformal dimension and
the $U(1)$ charge of the primary fields are:
\be\label{delta}
\Delta_{\a,\bar\a}=-\a\bar\a+{1\ov {2b}}(\a+\bar\a),\hskip1cm
\o={1\ov b}(\a-\bar\a) .
\ee
As we explained in the previous Section, among the primary fields there is a series of degenerate
fields of the $N=2$ SLFT. They are characterized by the fact that at
certain level of the corresponding conformal family a new primary
field (i.e. annihilated by all positive modes) appears. Such fields
were divided there in three classes.

Class I degenerate fields $N_{m,n}^\o=N_{\a_{m,n}^\o,\bar\a_{m,n}^\o}$ are given by (\ref{degtypeI}) where $m,n$ are positive
integers. $N_{m,n}^\o$ is degenerate at level $mn$ and relative
$U(1)$ charge zero. The irreducibility of the corresponding
representations is assured by imposing the null-vector condition
$D_{m,n}^\o N_{m,n}^\o=0$, $\bar D_{m,n}^\o N_{m,n}^\o=0$, where
$D_{m,n}^\o$ is a polynomial of the generators in (\ref{n2alg}) of degree
$mn$ and has $U(1)$ charge zero. It is normalized by choosing the
coefficient in front of $(L_{-1})^{mn}$ to be $1$.
Let us give some
examples of the corresponding null-operators:
\bea\nn
D^\o_{1,1} &=&L_{-1}-\half b^2(1+\o)J_{-1}+{1\ov \o-1}\gph\gmh, \\
\nn D_{1,2}^\o &=&L^2_{-1}+b^2L_{-2}-b^2(1+\o)L_{-1}J_{-1}+
{b^2\ov{2}}\left(1+\o-b^2(2+\o)\right)J_{-2}+\\
\nn &+&{b^4\ov 4}\o(\o+2)J^2_{-1}+{2\ov
{\o-2}}L_{-1}\gph\gmh-{b^2\o\ov {\o-2}}J_{-1}\gph\gmh-\\
\nn &-&{b^2\ov{2}}
\gph\gmi_{-{3\ov 2}}+{b^2\ov{2}}{{\o+2}\ov {\o-2}}\gpl_{-{3\ov
2}}\gmh, \\
\nn D_{2,1}^\o &=&L^2_{-1}+{1\ov
b^2}L_{-2}-b^2(1+\o)L_{-1}J_{-1}+\half \left(b^2(1+\o)-\o-2\right)J_{-2} +\\
\nn &+&{1\ov 4}\left(b^4(\o+1)^2-1\right)J^2_{-1}+{2b^4\o\ov
b^4(\o-1)^2-1}L_{-1}\gph\gmh-\\
\nn &-&{{b^2+b^6(\o^2-1)}\ov
b^4(\o-1)^2-1}J_{-1}\gph\gmh -{{b^4(\o+1)+b^2-2}\ov
2+2b^2(\o-1)}\gph\gmi_{-{3\ov 2}}+\\
\label{nullmn} &+&{{2-b^2+b^4(\o-1)\left(1+b^2(\o+1)\right)}\ov 2(b^4(\o-1)^2-1)}\gpl_{-{3\ov
2}}\gmh .
\eea

The second class of degenerate fields is
denoted by $N_m^\o$ and comes in two subclasses IIA and IIB introduced above in (\ref{degtypeIIa}) and (\ref{degtypeIIb}) respectively. Here $m$ is an odd positive
integer number and the level of degeneracy of $N_m^\o$ is ${m\ov
2}$, relative charge $\pm 1$. In this case the operator $D_m^\o$ is
a polynomial of ``degree'' $m/2$, the coefficient in front of
$L_{-1}^{{{m-1}\ov 2}}G^\pm_{-\half}$ is chosen to be $1$. Analogously
to the class I we have to impose $D_m^\o N_m^\o=\bar D_m^\o
N_m^\o=0$. Here are the first examples for class IIA fields:
\bea\nn
D_1^\o &=&G^+_{-\half},\\
\nn D_3^\o
&=&L_{-1}G^+_{-\half}-J_{-1}G^+_{-\half}+\left({2\ov {b^2}}-\o\right)G^+_{-{3\ov
2}},\\
\nn D_5^\o &=&L_{-1}^2G^+_{-\half}+ \left({4\ov
{b^2}}-\o-1\right)L_{-2}G^+_{-\half}-3L_{-1}J_{-1}\gpl_{-\half}+2J_{-1}^2\gph+\\
\nn &+&\left({5\ov 2}-{6\ov {b^2}}+{3\ov 2}\o\right)J_{-2}\gpl_{-\half}+\left(1+{6\ov
{b^2}}-2\o\right)L_{-1}\gpl_{-{3\ov 2}}+4\left(\o-{3\ov
{b^2}}\right)J_{-1}\gpl_{-{3\ov 2}}-\\
\label{nullm}&-&{\half}\gpl_{-{3\ov
2}}\gph\gmi_{-\half} + \left({24\ov {b^4}}-{14\o\ov
{b^2}}+2\o^2-1\right)\gpl_{-{5\ov2}}.
\eea
The null-operators for class IIB
fields are obtained from (\ref{nullm}) by changing $G^\pm\to G^\mp$ and
$\o\to-\o$.

We remind that a special case of Class IIA (B) fields are the chiral
(antichiral) fields with $m=1$. The Class II fields having $U(1)$
charge zero are classified in a separate Class III fields. The simplest
$m=1$ field here represents the identity operator.

Let us now
consider, for a further use, the norms of the states created by
applying the null-operators on primary states $|\a\rangle$. As explained
above, such sates should vanish at $\a=\a_M^\o$. Taking the first
terms in the corresponding Taylor expansion, we define:
\bea\nn
 r_M^\o &=&\partial_\alpha\langle\a,\bar\a|D^{\o\dagger}_M
 D_M^\o|\a,\bar\a\rangle|_{\a=\a_M^\o,\bar\a=\bar\a_M^\o},\\
\label{defr}\bar r_M^\o &=&\partial_{\bar\alpha}\langle\a,\bar\a|D^{\o\dagger}_M
D_M^\o|\a,\bar\a\rangle|_{\a=\a_M^\o,\bar\a=\bar\a_M^\o}
\eea
for both
classes of representations, $M=m$ or ($m,n$), where $D_M^\o$ is the
corresponding null-operator and $D^{\o\dagger}_M$ is defined as
usual through $L^\dagger_n=L_{-n}$, $J^\dagger_n=J_{-n}$,
$(G^{\pm}_r)^\dagger=G^\mp_{-r}$.

One can compute ``by hand'' the
first few $r$'s. With the use of the explicit form of the null-operators
(\ref{nullmn}) we find for the class I fields:
\bea\nn
r_{1,1}^\o &=&{1\over b}{(1+b^2)(1+\o)\over (-1+\o)},\\
\nn r_{1,2}^\o &=&{-2\over b}{(1-b^2)(1+b^2)(1+2b^2)(2+\o)\over
(-2+\o)},\\
\nn r_{1,3}^\o &=&{12\over
b}{(1-2b^2)(1-b^2)(1+b^2)(1+2b^2)(1+3b^2)(3+\o)\over (-3+\o)},\\
\nn r_{2,1}^\o &=&{2\over
b^5}{(1-b^2)(1+b^2)(2+b^2)(-1+b^2+b^2\o)(1+b^2+b^2\o)\over
(-1-b^2+b^2\o)(1-b^2+b^2\o)}
\eea
and $\bar r_{m,n}^\o=r_{m,n}^\o$  for all the examples above. Based
on these expression we propose for the general form of $r_{m,n}^\o$:
\be\label{rmno}
r_{m,n}^\o=\bar
r_{m,n}^\o=\prod_{l=1-m}^m\prod_{k=1-n}^n\left({l\over b}+kb\right)
\prod_{l=1-m,\ {\rm mod}\ 2}^{m-1}\left({{l-(n+\o)b^2}\over
l+(n-\o)b^2}\right).
\ee
Similarly, from (\ref{nullm}) we have for the class IIA:
\bea\nn
\bar r_1^\omega &=&2\left({1\over b}-\omega b\right),\\
\nn \bar r_3^\o
&=&{2\over b^5}(2-b^2\o)(3-b^2\o)(2-b^2-b^2\o),\\
\nn \bar r_5^\o
&=&{8\over
b^9}(3-b^2\o)(4-b^2\o)(5-b^2\o)(3-b^2-b^2\o)(4-b^2-b^2\o),\\
\nn r_m^\o &=&0,\quad m=1,3,5,7.
\eea
These expressions can be fitted in a general form of $r_m^\o$ and
$\bar r_m^\o$:
\bea\nn
r_m^\o &=& 0,\\
\label{rmo} \bar r_m^\o &=&2\G^2\left({{m+1}\over 2}\right) b^{1-m}
\prod_{l={{m+1}\over 2}}^{m}\left({l\over b}-b\o\right) \prod_{l={{m+1}\over
2}}^{m-1} \left({l\over b}-b(\o+1)\right).
\eea
For the class IIB fields one
obtains $\bar r^\o_m=0$ and $r_m^\o$ is as $\bar r_m^\o$ in (\ref{rmo})
with the change $\o\to -\o$.

Let us now
introduce the so called logarithmic fields. They are defined as:
$$
N'_{\a,\bar\a}=\p_\a N_{\a,\bar\a},\hskip1cm \bar
N'_{\a,\bar\a}=\p_{\bar\a}N_{\a,\bar\a}.
$$
One can introduce also the logarithmic primary fields corresponding
to degenerate fields by:
\be\label{log}
{N'}_M^{\o}
=N'_{\a,\bar\a} |_{\a=\a_M^\o,\bar\a=\bar\a_M^\o},\hskip1cm \bar {N'}_M^{\o}
=\bar N'_{\a,\bar\a}|_{\a=\a_M^\o,\bar\a=\bar\a_M^\o}
\ee
where $M$
is ($m,n$) for class I and $M$ is $m$ for class II fields
respectively. The basic statement about the fields (\ref{log}) is that:
\be\label{logpr}
\tilde N_M^\o=\bar D_M^\o D_M^\o {N'}^{\o}_M ,\hskip.5cm
\tilde{\bar N}_M^\o=\bar
 D_M^\o D_M^\o \bar {N'}^{\o}_M
\ee
with $D_M^\o$, $\bar D_M^\o$ as in (\ref{nullmn}) , (\ref{nullm}) are again
primary. The proof of this statement goes along the same lines as
for $N=0,1$ SLFT \cite{eqm,n1eqm}.

Comparing the dimension and $U(1)$ charge for class I
fields: $\tilde\Delta_{m,n}= \Delta_{m,n}+mn,\ \tilde\o=\o$ we
conclude that the fields (\ref{logpr}) are proportional to $N_{m,-n}^\o$.
Thus, we arrive at the higher equations of motion (HEM) for the
class I fields:
\be\label{hemmn}
\bar D_{m,n}^\o D_{m,n}^\o
{N'}^{\o}_{m,n}=B_{m,n}^\o N_{m,-n}^\o,\qquad \bar D_{m,n}^\o D_{m,n}^\o
\bar {N'}^{\o}_{m,n}=\bar B_{m,n}^\o N_{m,-n}^\o .
\ee
For class IIA (B) the dimension of the resulting primaries
in (\ref{logpr}) is $\tilde\Delta_m^\o=\Delta_m^\o+{m\ov 2}$, the $U(1)$
charges are $\tilde\o=\o+1$ ($\tilde\o=\o-1$) respectively, and the
HEMs in this case are:
\be\label{hemm}
\bar D_{m}^\o D_{m}^\o
{N'}^{\o}_{m} =B_{m}^\o N_{m}^{\o\pm 1},\qquad \bar D_{m}^\o D_{m}^\o
\bar {N'}^{\o}_{m} =\bar B_{m}^\o N_{m}^{\o\pm 1}.
\ee
Computation of
$B_{m,n}^\o$ ($\bar B_{m,n}^\o$) and $B_m^\o$ ($\bar B_m^\o$) is the
final goal of this Section. HEMs (\ref{hemmn}) and (\ref{hemm}) are to be
understood in an operator sense, i.e. they should hold for any
correlation function. Here we will insert them into the simplest
one-point function on the so called Poincar${\rm\acute{e}}$ disk (or Lobachevski plain) \cite{boot2}.
In this case we have:
\be\nn
\langle B_1|\bar D_M^\o D_M^\o N^{'\o}_M\rangle =\langle B_1|\tilde
N_M^\o\rangle,\qquad \langle B_1|\bar D_M^\o D_M^\o \bar N^{'\o}_M\rangle =\langle
B_1|\tilde{\bar N}_M^\o\rangle .
\ee
The boundary state $\langle B_1|$ corresponds to the identity boundary
conditions on the Poincar${\rm\acute{e}}$ disc. It enjoys $N=2$
superconformal invariance:
\be\nn
\la B_1|\bar G^{\pm}_r =-i\la B_1|G^\mp_{-r}= -i\la
B_1|(G^{\pm}_{r})^\dagger,\quad \la B_1|\bar L_n,=\la
B_1|(L_{n})^\dagger,\quad
\la B_1|\bar J_n =\la
B_1|(J_{n})^\dagger.
\ee
(so called A-type boundary conditions, see e.g. \cite{ay}).

With the definition of $r$'s in
(\ref{defr}) the HEMs (\ref{hemmn}) and (\ref{hemm}) take the form:
\be\label{rbumn}
r_{m,n}^\o U_1(m,n;\o) =B_{m,n}^\o U_1(m,-n;\o),\quad \bar
r_{m,n}^\o U_1(m,n;\o) =\bar B_{m,n}^\o U_1(m,-n;\o)
\ee
for class
I, and:
\be\label{rbum}
r_m^\o U_1(m,\o) =iB_m^\o U_1(m,\o\pm
1),\quad \bar r_m^\o U_1(m,\o) =i\bar B_m^\o U_1(m,\o\pm 1)
\ee
for
class II. Here $U_1$ is the one-point function for ``identity
boundary conditions'' of the corresponding field. In (\ref{rbum}) the factor $i$'s
appear because the class II null-operators are fermionic, and $+$
($-$) refers to class IIA (IIB).

The one-point function on the Poincar${\rm\acute{e}}$ disk
for identity (or vacuum) boundary conditions in $N=2$ SLFT was obtained in
the previous Section. Let us remind its general form:
\be\nn
U_1(\a,\bar\a)= \G (b^{-2})(\pi\mu)^{-{1\ov b}(\a+\bar\a)}
{\G(1-\a b)\G(1-\bar\a b)\over\G(-{{\a+\bar\a}\over b}+{1\over
b^2})\G(2-b(\a+\bar\a))}.
\ee
With the specific values (\ref{degtypeI}) the ratio of one-point functions of
class I fields then is:
\bea\nn
{U_1(m,n;\o)\over U_1(m,-n;\o)}&=& (\pi\mu)^{2n}{\g(1+m-nb^2)\over
\prod_{k=-n}^{n-1}({m\over b^2}+k) \prod_{l=-m}^{m}(l+nb^2)}
{\g({{1-m}\over 2}+(n-\o){b^2\over 2})\over \g({{1-m}\over
2}-(n+\o){b^2\over 2})}\times\\
\nn &\times& \prod_{l=1-m,\ {\rm mod}\ 2}^{m-1}\left({{l+(n-\o)b^2}\over
l-(n+\o)b^2}\right)
\eea
and for the HEM coefficient we obtain:
\bea\nn
&B_{m,n}^\o&=\bar B_{m,n}^\o=r_{m,n}^\o {U_1(m,n;\o)\over U_1(m,-n;\o)}=\\
\label{bmno}&=&(\pi\mu)^{2n}b^{1+2n-2m}\g(m-nb^2) { \g({{1-m}\over
2}+(n-\o){b^2\over 2})\over \g({{1-m}\over 2}-(n+\o){b^2\over 2})}
\prod_{l=1-m}^{m-1}\prod_{k=1-n}^{n-1}\left({l\over b}+k b\right)
\eea
where we impose that $(k,l)= (0,0)$ is excluded in the product.

Analogously for class IIA fields:
\be\nn
{U_1(m,\o)\over U_1(m,\o+1)}=\pi\mu b\ { \prod_{l={{m+1}\over
2}}^{m-1}({l\over b}-b(\o+1))\over \prod_{l={{m+1}\over
2}}^m({l\over b}-b\o)}
\ee
and:
\bea\nn
B_m^\o &=&0,\\
\label{bmo}\bar B_m^\o &=&-i\bar
r_m^\o{U_1(m,\o)\over U_1(m,\o+1)}=-2\pi i\mu
b^{2-m}\G^2\left({{m+1}\over 2}\right)\prod_{l={{m+1}\over 2}}^{m-1}\left({l\over
b}-b(\o+1)\right)^2.
\eea
For class IIB $B$ and $\bar B$ are exchanged and $\o$ is replaced by
$-\o$. Equalities (\ref{bmno}) and (\ref{bmo}) are the main results of this Section.

We want now to check these results in the classical limit. The latter is defined as $b\to 0$: $b\phi\to\vp$, $\b\psi\to\psi$,
 $\pi\mu b^2\to M$ the Lagrangian $\CL\to {1\ov {2\pi
b^2}}\CL$. The corresponding equations of motion are given by:
\be\label{em}
\bp\psi^{\pm} =-iM\bps^{\mp} e^{\vp^{\pm}},\quad
\p\bps^{\pm}=iM\psi^{\mp}e^{\vp^{\pm}},\quad \p\bp\vp^{\pm}
=iM\psi^{\pm}\bps^{\pm}e^{\vp\mp} +M^2e^{\vp^++\vp^-}.
\ee
The holomophic currents:
\bea\nn
T &=&-\p\vp^-\p\vp^+-{1\over
2}(\psi^-\p\psi^++\psi^+\p\psi^-)+{1\over 2}(\p^2\vp^++\p^2\vp^-),\\
\label{curr}S^{\pm} &=&-i\sqrt{2} (\psi^{\pm}\p\vp^{\pm}-\p\psi^{\pm}),\qquad
J=\p\vp^+-\p\vp^--\psi^-\psi^+,
\eea
are conserved by $\bp
T=\bp S^{\pm}=\bp J=0$ on the equations of motion and similarly for
the antiholomorphic ones. One has to introduce also the generators
of $N=2$ supersymmetry $G^{\pm}$ and $\bar G^{\pm}$:
\bea\nn
G^{\pm}\vp^{\mp} &=&i\sqrt{2}\psi^{\pm}, \hskip1cm
G^{\pm}\vp^{\pm}=0,\\
\label{suseq}\bar G^{\pm}\vp^{\mp}
&=&i\sqrt{2}\bar\psi^{\pm}, \hskip1cm \bar G^{\pm}\vp^{\pm}=0
\eea
obeying the algebra:
\bea\nn
\{G^+,G^-\} &=&2\p,  \hskip
1cm \{G^{\pm},G^{\pm}\}=\{\bar G^{\pm},\bar G^{\pm}\}=0,\\
\label{alg}\{\bar
G^+,\bar G^-\} &=&2\bp, \hskip1cm \{G,\bar G\}=0.
\eea
For the class
IIA fields only the chiral fields, $N_1^\o=e^{\o b\phi^+}$,
has a classical limit. Their HEMs take the form:
\be\nn
\bar G_{-\half}^+ G_{-\half}^+\phi^+N_1^\o =0,\qquad
\bar G_{-\half}^+ G_{-\half}^+\phi^-N_1^\o =B_1^\o N_1^{\o+1},
\ee
where $B_1^{\o}=-2\pi i\mu b$ can be read from (\ref{bmo}). In the
classical limit along with the analogous HEMs for class IIB
anti-chiral fields with $\o=0$, these become:
\be\nn
\bar G^{\pm}G^{\pm}\vp^{\mp}=-2iMe^{\vp^{\pm}}.
\ee
Together with (\ref{suseq}) and the algebra (\ref{alg}) these relations encode the
equations of motion (\ref{em}).

From the class I fields only the series $N_{1,n}^\o$ has a
classical limit, the simplest ``classical null-operators'' being:
\bea\nn
D_{1,1}^{\o(cl)} &=&\p-\half (\o+1)J+{1\ov {\o-1}}G^+G^-,\\
\nn D_{1,2}^{\o(cl)} &=&\p^2-(\o+1)J\p-\half(\o +2)\p J+{1\ov
4}\o(\o+2)J^2+{2\ov {\o-2}}G^+G^-\p-{\o\ov {\o-2}}JG^+G^- - \\
\nn &-&\half S^-G^++\half {{\o +2}\ov {\o-2}}S^+G^- .
\eea
It is easy to check, using the algebra (\ref{alg}) and the explicit form
of the currents (\ref{curr}) , that the classical expressions of the
corresponding null-vector conditions are:
\be\nn
D_{1,1}^{\o(cl)} e^{(\half (\o-1)\vp^+ -\half (\o
+1)\vp^-)}=0,\qquad D_{1,2}^{\o(cl)} e^{(\half (\o-2)\vp^+ -\half (\o
+2)\vp^-)}=0.
\ee
The same is of course  true also for $\bar D_{1,1}^{\o(cl)}$, $\bar
D_{1,2}^{\o(cl)}$. Then, with the help of (\ref{suseq}) and the equations of
motion (\ref{em}), we find that the classical HEMs then take the form:
\bea\nn
\bar D_{1,1}^{\o(cl)} D_{1,1}^{\o(cl)} \vp^\pm e^{(\half
(\o-1)\vp^+-\half (\o+1)\vp^-)} &=&{{\o+1}\ov {\o-1}}M^2 e^{(\half
(\o+1)\vp^+ -\half (\o-1)\vp^-)},\\
\bar D_{1,2}^{\o(cl)}
D_{1,2}^{\o(cl)}\vp^\pm e^{(\half (\o-2)\vp^+-\half (\o+2)\vp^-)}
&=&-2{{\o+2}\ov {\o-2}}M^4 e^{(\half (\o+2)\vp^+ -\half (\o-2)\vp^-)}.
\eea
This is in a perfect agreement with (\ref{hemmn}) if we take into account
that the classical limit, $b\to 0$, of $B_{1,n}^\o=\bar B_{1,n}^\o$
from (\ref{bmno}) is:
\be\nn
B_{1,n}^\o\to (-1)^{n+1}{{\o+n}\ov {\o-n}}n!(n-1)!\ b^{-1}(\pi\mu
b^2)^{2n}.
\ee

\setcounter{equation}{0}
\section{General $\hs$ coset models}

We start this Section with the description of the "fine structure" of the $\hat{su}(2)$ coset family of minimal models $M(k,l)={\hat{su}(2)_k\times \hat{su}(2)_l/ \hat{su}(2)_{k+l}}, k,l=1,2,\ldots$ \cite{gko}. The main statement in what follows is that a general $l$'th family of minimal models $M(k,l), l>1,k=1,2,\ldots$ can be realized as a projected tensor product of consequent Virasoro minimal models $M(k,1)\equiv M(k)$. We show that all the data for the general $M(k,l)$ model - the primary fields, the conformal blocks and the 4-point functions, the structure constants, the fusion algebra etc. - can be expressed explicitly in terms of the corresponding data from the Virasoro minimal models only. In this sense, all the minimal models $M(k,l)$ with $l>1$ are reducible. More precisely, we propose a construction  of $M(k,l)$ in terms of a recursive projected product of lower level $l$ models. By iterating this recursive construction we arrive at the projected product of Virasoro minimal models. The crucial role is played by the projection ${\bf P}$, i.e. the restriction of the products of Virasoro primary fields of the form $\phi_{rp_1}^k\phi_{p_1p_2}^{k+1}\ldots\phi_{p_{l-1}s}^{k+l-1}$ only. In particular, in computing the four-point functions only the products of conformal blocks corresponding to such products of fields are allowed. Still, we show that this is enough to construct monodromy invariant correlation functions. In this way we obtain the corresponding structure constants as products of the structure constants of the Virasoro models. One could wonder how general this procedure of reducing and solving a general coset model in terms of lower level coset models only is. Based on our experience with a variety of other coset constructions, our conjecture for an arbitrary (symmetric) coset series of models  $G(k,l)={\hat g_k\times \hat g_l/ \hat g_{k+l}}$ ($\hat g_k$ denotes level $k$ of the affine algebra $\hat g$) is that $G(k,l)$ is reducible to the products of the first level models only.

We are next interested in the calculation of the matrix of anomalous dimensions and the corresponding mixing of certain fields in the case of general $\hat{su}(2)$ coset models perturbed by the least relevant field in the second order of the perturbation theory. In this Section we extend the results of \cite{pogsc2} and \cite{myne5} (presented in Section 2) to these models denoted as $M(k,l)$.
The first order corrections are obtained in \cite{myne33}. It is shown that there exists an infrared (IR)fixed point of the renormalization group flow which coincides with the model $M(k-l,l)$. As it was demonstrated in Section 2 the calculation up to second order is difficult. The problem is that one needs the corresponding 4-point functions which are not known exactly.
Basic ingredients for the computation of the 4-point functions are the conformal blocks. They are quite complicated objects and a close form is not known. In this Section we use the strategy explained above. Namely, we use the fact that the structure constants and the conformal blocks for the general $\hat{su}(2)$ coset models $M(k,l)$ at some level $l$ can be obtained recursively from those of those of the lower levels or finally from the Virasoro minimal models by certain projected tensor product. We use this construction here to define the perturbing field and the other fields in consideration. It turns out that we are able to compute the necessary structure constants and conformal blocks up to the desired order. There is an alternative approach to the calculation of the mixing matrix in the perturbed CFT models, the so called RG domain wall \cite{gai,ibr}.It was shown in \cite{pogsc2,ppog1} for the Virasoro case and in \cite{ppog2} for the supersymmetric extension that there is an agreement between the results obtained by such construction and the perturbative calculations up to the second order. Moreover, as it was shown in Section 2, this mixing matrix do not depend on $\epsilon$ and is exactly the same in both theories. We show here that this is the case also for the
general $\hat{su}(2)$ coset models perturbed by the least relevant field.

The results of this Section have been published in \cite{myne33}, \cite{myne13}-\cite{myne151}, (17.-21.).

\subsection{Fusion of conformal models}

We start with the description of the $\hs$ coset family of minimal models:
\be\label{coset}
M(k,l)={\hat{su}(2)_k\times \hat{su}(2)_l\over \hat{su}(2)_{k+l}}
\ee
where $k,l=1,2,\ldots$ \cite{gko}. They are conformal field theories with a central charge given by:
\be\label{cenc}
c={3kl(k+l+4)\over (k+2)(l+2)(k+l+2)}={3l\over l+2}\(1-{2(l+2)\over (k+2)(k+l+2)}\).
\ee
The main statement in this Section is that a general $l$-th family of minimal models $M(k,l)$ with $l>1, k=1,2,\ldots$ can be realized as a projected tensor product of consequent Virasoro minimal models $M(k,1)\equiv M(k)$. We will show that all the data for a general $M(k,l)$ model - the primary fields, the conformal blocks and the four-point functions, the structure constants, the fusion algebra etc. - can be expressed explicitly in terms of the corresponding data from the Virasoro minimal models only. In this sense all the minimal models $M(k,l)$ are reducible.

More precisely, for general $l$ we state that:
\be\label{state}
M(l-1,1)\times M(k,l)={\bf P}(M(k,1)\times M(k+1,l-1)).
\ee
We introduced here explicitly the projection ${\bf P}$. In terms of primary fields it projects from the space of all product of fields to the subspace where only a product of fields with the same internal indexes are allowed.
By iterating eq. (\ref{state}) we arrive at:
\bea\label{totpr}
M(k,l)&\times&{\bf P}\left(M(1,1)\times M(2,1)\times\ldots\times M(l-1,1)\right)=\\
\nn &=&{\bf P}\left(M(k,1)\times M(k+1,1)\times\ldots\times M(k+l-1,1)\right).
\eea
Eq. (\ref{totpr}) means that any model $M(k,l), l>1$ can be constructed and explicitly solved in terms of Virasoro models only. Note that we have imposed the projection ${\bf P}$ on the LHS of (\ref{totpr}) too. Our statement is that Eqs. (\ref{state}) and (\ref{totpr}) imply that for any field from $M(k,l)$ one can find fields from $M(l-1,1)$, $M(k)$ and $M(k+1,l-1)$ such that the (projected) products of the fields have the same correlation functions. Furthermore, where there is no projection ${\bf P}$ (like between $M(k,l)$ and
$M(k,l)$), the monodromy invariant 2D correlation function of the product of the fields factorizes into the product of the correlation functions. A crucial role is played by the projection ${\bf P}$. In particular, in computing the 4-point functions only the products of of conformal blocks corresponding to the projected product of fields are allowed. We will show that this is enough to construct monodromy invariant correlation functions. In this way we will obtain the the corresponding structure constants as products of the structure constants of the Virasoro models.

One could wonder what is the origin of the reducibility of the models $M(k,l), l>1$. A formal answer is that it follows from the obvious coset identities:
\bea\nn
{\hat{su}(2)_1\times \hat{su}(2)_{l-1}\ov  \hat{su}(2)_{l}}&\times& {\hat{su}(2)_l\times \hat{su}(2)_{k}\ov  \hat{su}(2)_{k+l}}\\
\nn ={\hat{su}(2)_1\times \hat{su}(2)_{k}\ov  \hat{su}(2)_{k+1}}&\times& {\hat{su}(2)_{l-1}\times \hat{su}(2)_{k+1}\ov  \hat{su}(2)_{k+l}}.
\eea
We will give a precise meaning to this statement below.

Consider first the problem of the realization of the $M(k,l)$ chiral algebra and its field representations in the space ${\bf P}(M(k)\times M(l-1,k+1))$. We begin with the simplest case $l=2$ (i.e. the superconformal models)). The natural candidates for for the generators of the $N=1$ superconformal algebra in ${\bf P}(M(k)\times M(k+1))$ are the fields $\psi_p=\phi^k_{1p}\phi^{k+1}_{p1}, p=2,3\ldots$ with dimensions $\D_p=\hf(p-1)^2$, their derivatives, and the stress-energy tensors $T^{(k)}, T^{(k+1)}$ of $M(k)$ and $M(k+1)$. Define the following field combinations:
\bea\label{isi}
\psi&=&\phi^k_{12}\phi^{k+1}_{21},\\
\nn T^I&=&{k+2\ov 4(k+5)}T^{(k)}+{k+4\ov 4(k+1)}T^{(k+1)}+\hf \sqrt{{3k(k+6)\ov 4(k+1)(k+5)}}\phi^k_{13}\phi^{k+1}_{31},
\eea
\bea\label{sus}
&G&=i\sqrt{{1\ov (k+2)(k+4)}}\(k\phi^k_{12}\p\phi^{k+1}_{21}-(k+6)\p\phi^k_{12}\phi^{k+1}_{21}\),\\
\nn T^{SUSY}&=&{3(k+6)\ov 4(k+5)}T^{(k)}+{3k\ov 4(k+1)}T^{(k+1)}-\hf \sqrt{{3k(k+6)\ov 4(k+1)(k+5)}}\phi^k_{13}\phi^{k+1}_{31}.
\eea
We want to shaw that:

\no (i) $T^I$ and $\psi$ generate the usual Ising model algebra with central charge $c=\hf$;

\no (ii)  $ T^{SUSY}$ and $G$ are the generators of the $N=1$ superconformal algebra with\\ $c=3/2-12/(k+2)(k+4)$;

\no (iii) the $(T^I,\psi)$ and $(T^{SUSY},G)$ algebras are in direct product.

Let us start with (i). Using the OPE's of $M(k)$ and $M(k+1)$ models we have to prove that $\psi$ and $T^I$ given by (\ref{isi}) satisfy the well known OPE's:
\bea\nn
T^I(z_1)T^I(z_2)&=&{1\ov 4z_{12}^4}+{2\ov z_{12}^2}T^I(z_2)+{1\ov z_{12}}\p T^I(z_2)+\ldots,\\
\nn T^I(z_1)\psi(z_2)&=&{1\ov 2z_{12}^2}\psi(z_2)+{1\ov z_{12}}\p \psi(z_2)+\ldots,\\
\nn \psi(z_1)\psi(z_2)&=&{1\ov z_{12}}+2z_{12}T^I(z_2)+\ldots.
\eea
To do this we have to implement the projection ${\bf P}$ in the OPE's and in the construction of the conformal blocks of the primary fields in terms of $M(k)\times M(k+1)$ blocks. We address here the specific problem of constructing the 4-point functions and the OPE's of the currents using the conformal blocks of the ingredients $\phi^k_{1p}$ and $\phi^{k+1}_{p1}, p=1,2,3\ldots$. According to the construction (\ref{isi}), the 4-point function of $\psi(z)$ can be written as a sum of products of the conformal blocks $I_i^k$ of $\phi^k_{12}$ and $I_j^{k+1}$ of $\phi^{k+1}_{21}$:
\be\nn
F_\psi(z)\equiv <\psi(0)\psi(z)\psi(1)\psi(\infty)>=(z(1-z))^{-1}\sum_{i,j=1}^2 Y_{ij}I_i^k(z)I_j^{k+1}(z).
\ee
The condition for the monodromy invariance of $F_\psi(z)$ at $z=0$ implies that $Y_{12}=0=Y_{21}$ and we obtain:
\bea\nn
F_\psi(z)&=&(z(1-z))^{-1}\bigg( F(-{k\ov k+3},{1\ov k+3},{2\ov k+3};z)F(-{k+6\ov k+3},-{1\ov k+3},-{2\ov k+3};z)+\\
\label{fpsi} &+&Y_{22}z^2F({k+2\ov k+3},{1\ov k+3},{2k+4\ov k+3};z)F({k+4\ov k+3},-{1\ov k+3},{2k+8\ov k+3};z)\bigg),
\eea
where $Y_{11}=1$ is a normalization condition. Considering the small distance behaviour $z\to 0$ of Eq. (\ref{fpsi}) we conclude that the first term gives rise to $\phi_{11}^k\phi_{11}^{k+1}(0)$ in the OPE $\psi(z)\psi(0)$ and the second one to $\phi_{13}^k\phi_{31}^{k+1}(0)$, i.e. the terms $\phi_{11}^k\phi_{31}^{k+1}(0)$ and $\phi_{13}^k\phi_{11}^{k+1}(0)$ are projected out. We therefore see that in this case applying the projection ${\bf P}$ is the same as requiring monodromy invariance around $z=0$ for the 4-point functions.

The monodromy invariance around $z=1$ fixes:
\be\nn
Y_{22}=C_{(12)(12)(13)}^kC_{(21)(21)(31)}^{k+1}={3k(k+6)\ov 4(k+1)(k+5)}
\ee
where $C_{(12)(12)(13)}^k$ and $C_{(21)(21)(31)}^{k+1}$are the well known Virasoro structure constants.
One can then show, using some nontrivial identities between the hypergeometric functions, that the correlation function (\ref{fpsi}) coincides with the 4-point function of the free Majorana field $\psi(z)$ given by:
\be\nn
 <\psi(0)\psi(z)\psi(1)\psi(\infty)>=(z(1-z))^{-1}(1-z+z^2).
\ee
From here we can obtain the following OPE:
\bea\label{opsi}
\psi(z)\psi(0))={1\ov z}&+&2z\bigg({\D_{12}^k\ov c(k)}T^k(0)+{\D_{21}^{k+1}\ov c(k+1)}T^{k+1}(0)+\\
\nn &+&\hf\sqrt{Y_{22}}\phi^k_{13}\phi^{k+1}_{31}(0)\bigg)+\ldots .
\eea
We see that the structure constant in front of the $\phi^k_{13}\phi^{k+1}_{31}$ term is $\sqrt{C_{(12)(12)(13)}^kC_{(21)(21)(31)}^{k+1}}$, a square root of what one would naively expect. To understand this, remember that the OPE's should always be thought of as operations performed within well-defined correlation functions. Since the currents are distinguished from the usual scalar fields by having well defined 1D (dependent only on $z$, i.e. with only left-moving fields) correlation functions, their 1D OPE's are well-defined. In the present context, the currents are realized as sums of products of ordinary conformal fields whose only well-defined correlation functions (and therefore OPE's and structure constants) are two-dimensional. Still, it can be proved that the particular combinations used to construct the currents can have well-defined 1D correlation functions.
The monodromy invariance around $z=1$ of the 1D 4-point functions of the currents results in the structure constants of 1D OPE being constructed from the square roots of the standard 2D structure constants. Heuristically, one could think of the square root appearing since only the left moving fields contribute to the OPE.

Let us return to the proof of the statement (i) and consider the OPE $T^I\psi$. Keeping in mind the above discussion and using the Virasoro Ward identities we find that:
\bea\nn
T^I(z)\psi(0)&=&\bigg({(\D_{12}^k)^2\ov c(k)}+{(\D_{21}^{k+1})^2\ov c(k+1)}+{3k(k+6)\ov 8(k+1)(k+5)}\bigg){1\ov z^2}\psi(0)+\ldots\\
\nn&\equiv& {1\ov 2z^2}\psi(0)+\ldots .
\eea

In proving (ii)and (iii) we follow the same procedure, i.e. we start with the constructions (\ref{isi}) and (\ref{sus}) and perform the OPE's. In these OPE's we keep only the terms consistent with the projection ${\bf P}$ and use the square roots of the 2D structure constants in 1D OPE's. Applying this to the product of two supercurrents we obtain the well-known OPE:
\bea\nn
G(z)G(0))={k(k+6)\ov (k+2)(k+4)}{1\ov z^3}&+&{2\ov z}\bigg({3(k+6)\ov 4(k+5)}T^k(0)+{3k\ov 4(k+1)}T^{k+1}(0)-\\
\nn &-&\hf\sqrt{Y_{22}}\phi^k_{13}\phi^{k+1}_{31}(0)\bigg)+\ldots
\eea
implying $c(2,k)=3/2-12/(k+2)(k+4)$. Analogous calculations for $\psi(z)G(0))$ and $T^I(z)T^{SUSY}(0)$ show that no singular terms appear in these OPE's, i.e. the Ising model algebra $(T^I,\psi)$ and the $N=1$ superconformal algebra are in fact in direct product.

It remains to consider the supercurrent Ward identities and the properties of the primary fields. In terms of the latter ${\bf P}$ projects from the space of all products of fields $\{\phi_{rq}^k\phi_{ps}^{k+1}\}=M(k)\times M(k+1)$ to the subspace:
\be\nn
{\bf P}\left(M(k)\times M(k+1)\right)=\{\phi_{rp}^k\phi_{ps}^{k+1}\},\quad p=1,\ldots,k+2,
\ee
which is isomorphic to the representation space $M(1)\times M(2,k)$. This isomorphism is based on the following simple relations between the dimensions of the primary fields from two consequent Virasoro minimal models, $N=1$ superconformal models and the Ising model $M(1)$:
\bea\label{dno}
\D_{rp}(1,k)&+&\D_{ps}(1,k+1)-\D_{rs}^{NS}(2,k)=\hf (p-\hf(r+s))^2,\quad r-s\in 2Z,\\
\nn \D_{rp}(1,k)&+&\D_{ps}(1,k+1)-\D_{rs}^{R}(2,k)=\hf (p-\hf(r+s))^2-{1\ov 16},\quad r-s\in 2Z+1.
\eea
This leads us to suggest the following construction:
\bea\nn
N_{rs}&=&\phi_{r,\hf(r+s)}^k\phi_{\hf(r+s),s}^{k+1},\\
\label{cono}\s R^i_{rs}&=&\phi_{r,\hf(r+s\mp 1)}^k\phi_{\hf(r+s\mp 1),s}^{k+1},\quad i=1,2.
\eea
Here $\s$ is the Ising field with dimension $\D=1/16$, $N_{rs}$ and $R^i_{rs}$ are the NS and R fields of the $N=1$ superconformal minimal model. The rest of the products $\phi_{rp}^k\phi_{ps}^{k+1}$ for $p\ne (r+s)/2$ or $(r+s\mp 1)/2$ correspond to the descendants of the primary fields (\ref{cono}).

Let us consider now the transformation properties of the fields (\ref{cono}). Starting with the NS sector, we have to  find a realization of the second component $N_{rs}^{II}$ with dimension $\D_{rs}(2,k)+1/2$, consistent with the constructions for $G$ and $N_{rs}$, i.e. satisfying the OPE's:
\bea\nn
G(z)N_{rs}(0)&=&{1\ov z}N_{rs}^{II}(0)+\ldots,\quad r-s\in 2Z,\\
\nn G(z)N_{rs}^{II}(0)&=&{2\D_{rs}(2,k)\ov z^2}N_{rs}(0)+{1\ov z}\p N_{rs}(0)+\ldots.
\eea
The result is:
\bea\nn
N_{rs}^{II}(k)&=&a_-(k)\phi_{r,\hf(r+s)-1}^k\phi_{\hf(r+s)-1,s}^{k+1}+a_+(k)\phi_{r,\hf(r+s)+1}^k\phi_{\hf(r+s)+1,s}^{k+1},\\
\nn a_\mp(k)&=&{1\ov\sqrt{(k+2)(k+4)}}\bigg(k(\D_{\hf(r+s\mp 1),s}^{k+1}-\D_{21}^{k+1}-\D_{\hf(r+s),s}^{k+1})-\\
\nn &-&(k+6)(\D_{r,\hf(r+s)\mp 1}^k-\D_{12}^k-\D_{r,\hf(r+s)})\bigg)\times\\
\nn \qquad\quad &\times& \(C^k_{(12)(r,\hf(r+s))(r,\hf(r+s)\mp 1)}C^{k+1}_{(21)(\hf(r+s),s)(\hf(r+s)\mp 1,s)}\)^{1/2}.
\eea
For example, for $r=1, s=3$ we obtain the field driving the RG flow $M(2,k)\to M(2,k-2)$ described in Section 2:
\be\label{nofl}
N_{13}^{II}=\sqrt{{k\ov(k+4)}}\(\sqrt{{k\ov 2(k+1)}}\phi_{11}^k\phi_{13}^{k+1}+\sqrt{{k+2\ov 2(k+1)}}\phi_{13}^k\phi_{33}^{k+1}\).
\ee
Note that there is one more field with the same dimension:
\be\nn
\psi N_{13}={1\ov 2(k+1)}\((k+2)\phi_{11}^k\phi_{13}^{k+1}+k\phi_{13}^k\phi_{33}^{k+1}\),
\ee
but only $N_{13}^{II}$ defined as (\ref{nofl}) has all the properties of the second component of $N_{13}$.

To conclude the discussion of the supercurrent Ward identities we turn to the R sector. Using (\ref{cono}) one can show that:
\be\nn
G(z)R^{1(2)}_{rs}(0)=\sqrt{\D_{rs}-{c\ov 24}}{1\ov z^{3/2}}R^{2(1)}_{rs}(0)+\ldots.
\ee

What we have done so far is still not enough to prove that $N_{rs}$ and $R_{rs}$ constructed above obey all the required null-vector properties. We have to show that their fusion rules, structure constants and 4-point functions coincide with the ones for the $N=1$ minimal models. We will address these questions below.

In extending the discussion of the current algebra and the Ward identities to the higher level coset models one encounters some difficulties. This motivates the change of the strategy we will use for $l>2$ which will entail abandoning the study of the current algebra and focusing on the direct construction of monodromy invariants. One difficulty comes from the fact that even the dimensions of all the currents are not known - for $l\ge 5$ there seem to exist additional currents over the stress-tensor $T$ and the well known current $A(z)$ of dimension $\D_A=(l+4)/(l+2)$. Another difficulty is that the dimension of $A(z)$ stops being a multiple of $1/2$ for $l>2$. As a consequence its own algebra is not well understood.

Therefore, for the study of the higher level models we adopt a different strategy. We will start by constructing all the primary fields for any $l$ in terms of the projected products of the Virasoro fields since their conformal blocks, structure constants, etc., are fully understood and explicitly calculated. Then we proceed with the calculation of the corresponding conformal blocks and their monodromy-invariant combinations for higher levels. That will allow us to obtain their fusion rules and the structure constants. In the cases where there are previous results to compare with, e.g. $l=2,4$, our results will be confirmed.

We will limit ourselves to the fields which are primary with respect to the stress-tensor and the eventual additional currents present in the higher level models. The descendants will be considered later on. The primary fields of the model $k$ at level $l$ are $\phi_{mn}(l,k)$ with conformal dimensions given by \cite{kmq,rava}:
\bea\label{diml}
\D_{m,n}(l,k) &=&{((k+2+l)m-(k+2) n)^2-l^2\over 4l(k++2)(k+2+l)}+{s(l-s)\over 2l(l+2)},\\
\nn s &=&|m-n|( mod (l)),\hskip1cm 0\le s\le l,\\
\nn &1&\le m\le k+1, \hskip1cm 1\le n\le k+l+1.
\eea
If $n-m\in lZ$ the expression for $\D_{mn}$ simplifies since $s(l-s)=0$. For $l=2$ such fields belong to the NS sector, for general $l$ we will call such sector the "vacuum sector". Since it is significantly simpler we present the construction first for these fields.

It is easy to check that:
\be\nn
\D_{mn}(l,k)=\D_{mx}(1,k)+\D_{xn}(l-1,k+1),
\ee
if $x=(1/l)(n+(l-1)m)$. This identity leads us to write:
\be\label{vafi}
\phi_{mn}(l,k)=\phi_{mx}(1,k)\phi_{xn}(l-1,k+1).
\ee
Two remarks are in order: first, note that $M(l-1,1)$ from the LHS of eq. (\ref{state}) contributes the identity field to the LHS of (\ref{vafi}); second, as will become clear later on, the products $\phi_{mt}(1,k)\phi_{tn}(l-1,k+1)$ with $t\ne x$ represent (part of) descendants of $\phi_{mn}(l,k)$. Since:
\be\nn
n-{n+(l-1)m\ov l}\in (l-1)Z
\ee
we can immediately iterate (\ref{vafi}) and finally obtain $\phi_{mn}(l,k)$ written in terms of the Virasoro fields:
\bea\label{fivi}
\phi_{mn}(l,k)&=&\prod_{i=0}^{l-1}\phi_{k_ik_{i+1}}(1,k+i),\\
\nn k_i&=&{in+(l-i)m\ov l},\quad n-m\in lZ.
\eea

Furthermore, starting from eq. (\ref{fivi}) one can reach any other projected product:
\be\label{fiit}
\phi_{m\hat k_1}(1,k)\phi_{\hat k_1\hat k_2}(1,k+1)\ldots \phi_{\hat k_{l-1}n}(1,k+l-1)
\ee
by changing $k_1$ into $\hat k_1$, $k_2$ into $\hat k_2$, etc. Similarly to eq. (\ref{dno}), the dimension of the field (\ref{fiit}) is higher than the one of (\ref{fivi}) by a multiple  of $1/2$. We interpret products as (\ref{fiit}) as (part of) descendants of $\phi_{mn}(l,k)$ with respect to $T, G(l=2)$, any other additional currents or a product of these currents.

To summarize, among all the products (\ref{fiit}) we search for the one with the lowest dimension to identify it with the primary field $\phi_{mn}(l,k)$. Minimizing the dimension is equivalent to minimizing $S=\sum_{i=0}^{l-1}(k_i-k_{i+1})^2$. If $m-n\in lZ$ there is a unique solution (with $k_0=m, k_l=n$) that gives $S=lK^2$ ($K=(1/l)(n-m)$), namely equidistant $k_i$'s , $k_i=m+iK$, as in (\ref{fivi}).

Turning to the nonvacuum sector (Ramond and analogous), namely fields $\phi_{mn}(l,k)$ with $m-n\notin lZ$, we will see that things stop being so simple. We begin by deriving the expression for a product of Virasoro fields having the required (minimal) dimension. Unfortunately, it will become obvious that for nonvacuum sectors that expression is not unique. We will study the field $\phi_{mn}(l,k)$ with $n-m=lZ\mp s$, $1\le s\le l-1$. (A part of the reason for the non-uniqueness should be already obvious, namely, if $n-m\in lZ\mp s$ then also $n-m\in lZ\pm (l-s)$.) Some straightforward algebra shows that:
\be\nn
\D_{s,s+1}(1,l-1)+\D_{mn}(l,k)=\D_{my}(1,k)+\D_{yn}(l-1,k+1),
\ee
where $y=(1/l)((l-1)m+n\mp (l-s))$. Therefore we write:
\be\label{vafiy}
\phi_{s,s+1}(1,l-1)\phi_{mn}(l,k)=\phi_{my}(1,k)\phi_{yn}(l-1,k+1).
\ee
This time $M(1,l-1)$ contributes a nontrivial field $\phi_{s,s+1}(1,l-1)$. Analogously to the discussion for the vacuum sector, $n-m\in lZ\mp s$ implies:
\be\nn
n-{n+(l-1)m\mp (l-s)\ov l}\in (l-1)Z\mp (s-1).
\ee
The field from $M(k+1,l-1)$ is again from nonvacuum sector, even though one step closer to the vacuum sector. We again iterate the process and after $s$ steps arrive at the general formula for the nonvacuum sector  fields:
\bea\nn
&&\phi_{12}(1,l-s)\phi_{23}(1,l-s+1)\ldots \phi_{s,s+1}(1,l-1)\phi_{mn}(l,k)=\\
\nn &&=\phi_{1,s+1}(s,l-s)\phi_{mn}(l,k)=\prod_{i=0}^{l-1}\phi_{k_ik_{i+1}}(1,k+i),\\
\label{fivin} &&n-m\in lZ\mp s,\qquad 1\le s\le l-1
\eea
where $k_i={in+(l-i)m+d_i^s\ov l}$ and $d_i^s=\mp i(l-s)$ if $i\le s$, $d_i^s=\mp s(l-i)$ if $i>s$.

After some simple combinatorics one can notice that there are $\begin{pmatrix}l\\s \end{pmatrix}$
different products of the Virasoro fields that have the same dimension as those of $\phi_{1,s+1}(s,l-s)\phi_{mn}(l,k)$. To understand the origin of this degeneracy, note that $\phi_{1,s+1}(s,l-s)$ in (\ref{fivin}) represents actually (according to (\ref{totpr})) the product:
\be\nn
\phi_{11}(1,1)\ldots\phi_{11}(1,l-s-1)\phi_{12}(1,l-s)\phi_{23}(1,l-s+1)\ldots \phi_{s,s+1}(1,l-1).
\ee
There are exactly $\begin{pmatrix}l-1\\s \end{pmatrix}$ such projected products of the fields from:
\be\nn
{\bf P}\left(M(1,1)\times(M(1,1)\times \ldots\times M(l-1,1)\right)
\ee
that have the same dimension. Moreover, we could view the original  field $\phi_{mn}(l,k)$ as belonging to the $l-s$ sector. In that case the field $\phi_{1,l-s+1}(l-s,s)$ would appear in (\ref{fivin}) which in its turn would represent $\begin{pmatrix}l-1\\l-s \end{pmatrix}$ products like:
\be\nn
\phi_{11}(1,1)\ldots\phi_{11}(1,s-1)\phi_{12}(1,s)\phi_{23}(1,s+1)\ldots \phi_{l-s,l-s+1}(1,l-1).
\ee
Finally, since:
\be\nn
\begin{pmatrix}l\\s \end{pmatrix}=\begin{pmatrix}l-1\\s \end{pmatrix}+\begin{pmatrix}l-1\\l-s \end{pmatrix}
\ee
we conclude that all the degeneracy of the RHS of (\ref{fivin}) is accounted for by the degeneracy of the LHS.

We now turn to the explicit construction of the four-point correlation functions for arbitrary fields from a higher level model. We start with the simplest example of the function of the same fields:
\be\label{fpfs}
G(z,\bar z)=<\phi_{mn}(l,k)(0)\phi_{mn}(l,k)(z)\phi_{mn}(l,k)(1)\phi_{mn}(l,k)(\infty)>
\ee
where $n-m\in lZ$. As already mentioned there are two basic steps in the calculation of $G$. First, one obtains the conformal blocks, i.e. the linearly independent solutions of the differential equations obeyed by the correlation function, and second, one combines them in a monodromy invariant expression which is the correlation function. According to (\ref{fivi}) $\phi_{mn}(l,k)$ is a product of Virasoro fields  and therefore the conformal blocks for (\ref{fpfs}) will be products of the Virasoro conformal blocks. Of course, only certain products of conformal blocks will survive the projection ${\bf P}$.

It is known that the conformal blocks of the correlation function of the Virasoro fields:
\be\nn
G_V(z,\bar z)=<\phi_{rs}(1,k)(0)\phi_{rs}(1,k)(z)\phi_{rs}(1,k)(1)\phi_{rs}(1,k)(\infty)>
\ee
can be obtained with Coulomb gas technic as certain multi-contour integrals \cite{df}, denoted as $I_{ij}^k(a,a';z)$, $i=1,\ldots,r$, $j=1,\ldots,s$ where:
\bea\nn
a&=&2\a_-\a_{rs},\qquad a'=2\a_+\a_{rs},\\
\nn \a_{rs}&=&\hf((1-r)\a+(1-s)\a_-),\\
\nn \a_+^2&=&{k+3\ov k+2},\quad \a_-^2={k+2\ov k+3},\quad \a_+\a_-=-1.
\eea
In order to preserve the projection ${\bf P}$ in the intermediate channel we allow only products of conformal blocks of the form:
\be\label{cbpr}
I_{i_0i_1}^kI_{i_1i_2}^{k+1}\ldots I_{i_{l-1}i_l}^{k+l-1}.
\ee
Having obtained the conformal blocks, we want to construct their monodromy invariant combinations. We start with the simple example $l=2$, i.e. 4-point functions of of NS fields $\phi_{mn}(2,k)=\phi_{mx}(1,k)\phi_{xn}(1,k+1)$, $x=\hf(m+n)$. The task is to find the coefficients $X_{i_0i_1i_2,j_0j_1j_2}$ such that:
\be\nn
G(z,\bar z)=\sum_{\begin{matrix}i_0,j_0=1,...,m\\i_1,j_1=1,...,x\\i_2,j_2=1,...,n\end{matrix}}X_{i_0i_1i_2,j_0j_1j_2}I_{i_0i_1}^kI_{i_1i_2}^{k+1}(z)\overline{I_{j_0j_1}^kI_{j_1j_2}^{k+1}(z)}
\ee
is monodromy invariant. In other words, we want $G(z,\bar z)$ to be well-defined, that is, single valued in the complex plane. Since the conformal blocks only have poles at $z=0,1$ and $\infty$, $G(z,\bar z)$ will be single valued everywhere if it is invariant under analytic continuation in $z$  along a contours surrounding $z=0$ and $z=1$.

As usual, the calculation around $z=0$ is straightforward and leads to the following form of the 4-point function:
\be\label{zer}
G(z,\bar z)=\sum_{\begin{matrix}i_0=1,...,m\\i_1,j_1=1,...,x\\i_2=1,...,n\end{matrix}}X_{i_0i_1j_1i_2}I_{i_0i_1}^kI_{i_1i_2}^{k+1}(z)\overline{I_{i_0j_1}^kI_{j_1i_2}^{k+1}(z)}.
\ee
We turn to the analytic continuation around $z=1$. First, we remind that the Virasoro conformal blocks $I_{ij}^{(rs)}(a,a';z)$ could be rewritten as:
\be\nn
I_{ij}^{(rs)}(a,a';z)=\sum_{\begin{matrix}p=1,...,r\\q=1,...,s\end{matrix}}\a^{rs}_{ij,pq}(a,a')I_{pq}^{(rs)}(a,a';1-z)
\ee
where the $\a$-matrices are well known \cite{df}. To study (\ref{zer}) under $(1-z)\to (1-z)e^{2\pi i}$ we use these $\a$-matrices. The result is:
\be\nn
G(z,\bar z)=\sum X_{i_0i_1j_1i_2}\a_{i_0i_1,ef}^k\a_{i_1i_2,gh}^{k+1}\a_{i_0j_1,rs}^k\a_{j_1i_2,tu}^{k+1}I_{ef}^kI_{gh}^{k+1}(1-z)\overline{I_{rs}^kI_{tu}^{k+1}(1-z)}
\ee
(summation over the repeated indexes is assumed). There are two requirements that have to be satisfied. First, the $\a$-transformation should not take us outside of the subspace defined by ${\bf P}$. Second, $G((z,\bar z)$  should be of the same form with respect to $I(1-z)$ as eq. (\ref{zer}) is to $I(z)$ is order to insure invariance under the monodromy transformation around $z=1$. Using some special properties of the $\a$-matrices \cite{df}, we thus arrive at the final form of our monodromy invariant correlation function:
\be\label{ffg}
G(z,\bar z)=\sum_{i_0,i_1,j_1,i_2} X_{i_0i_1}^kX_{j_1i_2}^{k+1}I_{i_0i_1}^kI_{i_1i_2}^{k+1}(z)\overline{I_{i_0j_1}^kI_{j_1i_2}^{k+1}(z)}
\ee
where the coefficients $X^k_{ij}$ are those defining the monodromy invariant Virasoro 4-point function \cite{df}.
Since the $X$'s (up to certain normalization constants) define the structure constants, we can already see that the NS structure constants for $M(k,2)$ will be given by certain products of the Virasoro structure constants for $M(k,1)$ and $M(k+1,1)$.

Before turning to the structure constants we would like to generalize the simple example that led to (\ref{ffg}). Let us consider the correlation function (\ref{fpfs}). The relevant conformal blocks are of the form (\ref{cbpr}). An analysis similar to the one for $l=2$ shows that only the terms in the correlation function of the form:
\be\nn
I_{i_0i_1}^kI_{i_1i_2}^{k+1}\ldots I_{i_{l-1}i_l}^{k+l-1}(z)\overline{I_{i_0j_1}^kI_{j_1j_2}^{k+1}\ldots I_{j_{l-1}i_l}^{k+l-1}(z)}
\ee
are invariant under $z\to ze^{2\pi i}$. Furthermore, the properties of the $\a$-matrices that led us to (\ref{ffg}) do not depend on the level, indexes or the value of $k$. Thus, we conclude that the monodromy invariant correlation function will be again of the same form:
\be\label{finf}
G(z,\bar z)=\sum X_{i_0i_1}^kX_{j_1i_2}^{k+1}\ldots X_{j_{l-1}i_l}^{k+l-1}I_{i_0i_1}^k\ldots I_{i_{l-1}i_l}^{k+l-1}(z)\overline{I_{i_0j_1}^k\ldots I_{j_{l-1}i_l}^{k+l-1}(z)}
\ee
We can generalize further and discuss asymmetrical correlation functions:
\be\nn
G_a(z,\bar z)=<\prod_{a=1}^4\phi_{m_an_a}(l,k)(z_a,\bar z_a)>,\quad m_a-n_a\in lZ.
\ee
Now $I,\a$ and $X$ depend on three sets of parameters:
\be\nn
a_i=2\a_-\a_{m_in_i},\quad a'_i=2\a_+\a_{m_in_i},\quad i=1,2,3.
\ee
It is straightforward to go over the arguments and convince ourselves that there are no significant changes.

Turning to the nonvacuum sectors, we want to calculate the 4-point function of $\phi_{mn}(l,k)$ where $n-m\in lZ\mp s, 1\le s\le l-1$. From (\ref{fivin}) we know that the product\\ $\phi_{1,s+1}(s,l-s)\phi_{mn}(l,k)$ can be expressed as various products of the Virasoro fields. The construction of the 4-point functions of these products of Virasoro fields proceeds as above and we conclude that the 4-point function of $\phi_{1,s+1}(s,l-s)\phi_{mn}(l,k)$ has the form (\ref{finf}). Furthermore, since there is no projection between $\phi_{1,s+1}(s,l-s)$ and $\phi_{mn}(l,k)$, the 4-point function of the product factorizes into the product of the 4-point functions of the corresponding fields.

Now we want to use the construction of the monodromy invariant 4-point functions performed above for the study of the fusion algebras and the structure constants for the higher level models. We will limit our considerations here to the vacuum sector fields only.

Let us start with the $N=1$ supersymmetric theory, i.e. take $l=2$. The NS fields (vacuum sector) are constructed as:
\be\label{noco}
\phi_{mn}(2,k)=\phi_{m,\hf(m+n)}^k\phi_{\hf(m+n),n}^{k+1},\qquad n-m\in 2Z.
\ee
All the other combinations $\phi_{mp}^k\phi_{pn}^{k+1}$ belong to the descendants of $\phi_{mn}(2,k)$ with the dimension $\D_{mn}(2,k)+\hf(p-\hf(m+n))^2$. Since we have seen that the conformal blocks used in our constructions are projected products of $(k,1)$ and $(k+1,1)$ conformal blocks, the fusion algebra will follow the same recipe.

For the study of the fusion rules it suffices to consider only the diagonal terms ($i_1=j_1$) in (\ref{ffg}). The other terms correspond to descendants with respect to the stress tensor or $G(z)G(\bar z)$ which will be already accounted for in the fusion rules. Considering the diagonal terms only is equivalent to using the Virasoro fusion rules \cite{bpz}:
\be\nn
\phi_{m_1n_1}^k\phi_{m_2n_2}^{k}=\sum_{r=|m_1-m_2|+1}^{\min(m_1+m_2-1,2(k+2)-m_1-m_2-1)}
\sum_{s=|n_1-n_2|+1}^{\min(n_1+n_2-1,2(k+3)-n_1-n_2-1)}\phi_{rs}^k,
\ee
($r$ and $s$ advance in steps of 2), for each of the fields in (\ref{noco}) and then imposing the projection by identifying the middle indices. The result is:
\bea\nn
&&\phi_{m_1n_1}(2,k)\phi_{m_2n_2}(2,k)=\\
&&\nn =\sum_{p=|m_1-m_2|+1}^{\min(m_1+m_2-1,2(k+2)-m_1-m_2-1)}
\sum_{q=|\hf(m_1+n_1)-\hf(m_2+n_2)|+1}^{\min(\hf(m_1+n_1+m_2+n_2)-1,2(k+3)-\hf(m_1+n_1+m_2+n_2)-1)}\times\\
&&\nn \times \sum_{r=|n_1-n_2|+1}^{\min(n_1+n_2-1,2(k+4)-n_1-n_2-1)}\phi_{pq}^k\phi_{qr}^{k+1}.
\eea
The remaining problem is to identify the products $\phi_{pq}^k\phi_{qr}^{k+1}$ with the super-Virasoro fields. First we note that:
\bea\nn
r-p&=&|n_1-n_2|-|m_1-m_2|(mod (2))=\\
\nn &=&n_1-m_1-(n_2-m_2))(mod (2)) \in 2Z.
\eea
Therefore $\phi_{pq}^k\phi_{qr}^{k+1}$ stands for a NS field $\phi_{pr}(2,k)$ or its descendant. It is a simple exercise to fix the range of $q$ and identify its minimal values. This will distinguish between the primary field or the second component of a latter. The final conclusion is that the following NS fusion rules hold:
\bea\nn
&&\phi_{m_1n_1}(2,k)\phi_{m_2n_2}(2,k)=\\
&&\nn =\sum_{r=|m_1-m_2|+1}^{\min(m_1+m_2-1,2(k+2)-m_1-m_2-1)}
\sum_{s=|n_1-n_2|+1}^{\min(n_1+n_2-1,2(k+4)-n_1-n_2-1)}\phi_{rs}^{(II)}(2,k)
\eea
where $n_i-m_i\in 2Z$ and $\phi_{rs}^{(II)}=N_{rs}$ if $r+s-|m_1+n_1-m_2-n_2|\in 4Z+2$, $\phi_{rs}^{(II)}=N_{rs}^{II}$ if $r+s-|m_1+n_1-m_2-n_2|\in 4Z$ in agreement with our results in Section 2.

Let us sketch briefly the calculations for the next level $l=3$.  We study the fusions of two vacuum sector fields like:
\be\label{ntco}
\phi_{mn}(3,k)=\phi_{m,{1\ov 3}(n+2m)}^k\phi_{{1\ov 3}(n+2m),{1\ov 3}(2n+m)}^{k+1}\phi_{{1\ov 3}(2n+m),n}^{k+2},\qquad n-m\in 3Z.
\ee
Detailed analysis, similar to that of the previous $l=2$ case, leads to a new situation for $l=3$, namely, it happens that in the product of two vacuum sector fields a nonvacuum sector field, or a descendant of such field, appears. Omitting the details we present the final result:
\bea\nn
&&\phi_{m_1n_1}(3,k)\phi_{m_2n_2}(3,k)=\\
&&\nn =\sum_{r=|m_1-m_2|+1}^{\min(m_1+m_2-1,2(k+2)-m_1-m_2-1)}
\sum_{s=|n_1-n_2|+1}^{\min(n_1+n_2-1,2(k+5)-n_1-n_2-1)}\phi_{rs}^{(d)}(3,k)
\eea
where $r$ and $s$ advance in steps of 2 and $\phi_{rs}^{(d)}(3,k)$ is the primary field $\phi_{rs}(3,k)$ if $k-l\in 3Z$ and its descendant with respect to the current $A(z)$ (of dimension $7/5$ here) otherwise.

One proceeds in the same way for the next levels $l=4,5,..$. We will not present the explicit calculations for them here. Finally, we arrive at the following general vacuum sector fusion rules:
\bea\label{frfi}
&&\phi_{m_1n_1}(l,k)\phi_{m_2n_2}(l,k)=\\
&&\nn =\sum_{r=|m_1-m_2|+1}^{\min(m_1+m_2-1,2(k+2)-m_1-m_2-1)}
\sum_{s=|n_1-n_2|+1}^{\min(n_1+n_2-1,2(k+l+2)-n_1-n_2-1)}\phi_{rs}^{(d)}(l,k)
\eea
where $m_i-n_i\in lZ$, $r,s$ advance in steps of 2, and $\phi_{rs}^{(d)}(l,k)$ is the primary field $\phi_{rs}(l,k)$ if $r-s=lK, K\in Z$ and $K-1/l\(|n_1+(l-1)m_1-(n_2+(l-1)m_2)|-|m_1-m_2|\)\in 2Z$, and its descendant with respect to $A_{(l+4)/(l+2)}$, one of the additional currents appearing for $l\ge 5$, or some product of those currents otherwise.

All the results obtained above demonstrate that, as long as we stay within the part of the fusion rules that maps the primary vacuum sector fields into the primary vacuum sector fields, we have full control, for any $l$. We will use that now to obtain explicit expressions for the structure constants connecting three vacuum sector primary fields for any $l$.

The structure constants appear as a limit of the modromy invariant 4-point functions. The conformal blocks whose limit one is taking are nothing but the products of the Virasoro conformal blocks. Since for the primary fields from the vacuum sector the mapping from the $l$-th level fields into the products of the Virasoro fields remains strictly one-to-one and does not involve any nontrivial fields from $M(1,l-1)$ and/or additional currents, it is obvious that the $l$-th level vacuum sector structure constants are given by the products of the Virasoro structure constants. Explicitly, for the fields $\phi_{m_an_a}(l,k), a=1,2,3$ from (\ref{fivi}) where:
\be\nn
(n_3-m_3)-\(|n_1+(l-1)m_1-(n_2+(l-1)m_2)|-|m_1-m_2|\)\in 2lZ
\ee
the structure constants are given by:
\be\label{strc}
C_{(m_1n_1)(m_2n_2)(m_3n_3)}=\prod_{i=0}^{l-1}C_{(k_i^1k_{i+1}^1)(k_i^2k_{i+1}^2)(k_i^3k_{i+1}^3)}(1,k+i).
\ee

\subsection{Second order RG flow}

In this Section we want to discuss the renormalization group properties of the $\hs$ coset models $M(k,l)$ defined by (\ref{coset}) (we assume here that $k$ and $l$ are integers and $k>l$). We remind that it is written in terms of $\hat{su}(2)_k$ WZNW models with current $J^a$, $k$ is the level. The latter are CFT's with a stress tensor expressed through the currents by the Sugawara construction:
\be\label{suga}
T_k(z)={1\ov k+2}\((J^0)^2+\hf J^+J^-+\hf J^-J^+\).
\ee
The central charge of the corresponding Virasoro algebra is
$c_k={3k\ov k+2}$.
The energy momentum tensor of the coset (\ref{coset}) is then given by: $T=T_k+T_l-T_{k+l}$ in obvious notations.
It defines a Virasoro algebra with central charge that can be read from this construction and is given by (\ref{cenc}).
The dimensions of the primary fields $\phi_{m,n}(l,p)$ of the "minimal models" (rational CFT) were written in (\ref{diml}) ($m,n$ are integers). We want to slightly change the notations in what follows, introducing ${\bf p=k+2}$. The dimensions then become:
\bea\label{dimp}
\D_{m,n}(l,p) &=&{((p+l)m-p n)^2-l^2\over 4lp(p+l)}+{s(l-s)\over 2l(l+2)},\\
\nn s &=&|m-n|( mod (l)),\hskip1cm 0\le s\le l,\\
\nn &1&\le m\le p-1, \hskip1cm 1\le n\le p+l-1.
\eea
As we mentioned in the previous Section, it is known \cite{kmq,rava,argy} that the theory $M(k,l)$ possesses a symmetry generated by a "parafermionic current" $A(z)$ of dimension $\D_A={l+4\over l+2}$.
We shall present an explicit construction of this current below. Here we just mention that under this symmetry
the primary fields (\ref{dimp}) are divided in sectors labeled by the integer $s$.
The branching of the current $A(z)$ on the field (or state) of sector $s$ can be written symbolically as \cite{kak,wyl}:
\be\label{branch}
A_{-m-{(s+2)\over (l+2)}}|s>=|s+2>,\hskip.5cm A_{-m}|s>=|s>,\hskip.5cm A_{-m-{(l+2-s)\over (l+2)}}|s>=|s-2>.
\ee
In this Section we prefer to use the description of the theory $M(k,l)$ presented above, namely we will define the fields, correlation functions, structure constants etc. using the construction (\ref{state}) and the specific projection ${\bf P}$.

Let us now define the model. We consider the CFT $M(k,l)$ perturbed by the least
relevant field.
Our goal here
is to find the $\b$-function and investigate its eventual fixed point up to second order in the perturbation theory.
In addition, we want to describe also the mixing of certain fields under the RG flow.

Let us briefly sketch the constructions. The perturbed theory is
described by the Lagrangian:
$$
\CL(x)=\CL
_0(x)+\l \tilde\phi_{1,3}(x)
$$
where $\CL_0(x)$ describes the theory $M(k,l)$ itself. We
identify the field $\tilde\phi_{1,3}$ with the first descendent of
the corresponding primary field (\ref{dimp}) with respect to the current
$A(z)$. In fact, in view of (\ref{dimp}) $\phi_{1,3}$ belongs to the sector
$|2>$ and has a descendent belonging to sector $|0>$ due to the last
of (\ref{branch}). The dimension of this first descendent is therefore
(for $s=2$):
\be\label{delt}
\D=\D_{1,3}+{l\over l+2}=1-{2\over p+l}=1-\e.
\ee
In this and in the next Section we consider the case $p\rightarrow\infty$ and
assume that $\e={2\over p+l}\ll 1$ is a small parameter.

Following our constructions of the previous Section we find it more convenient here to define the field $\tilde\phi_{1,3}$
alternatively in terms of lower level fields:
\be\label{field}
\tilde\phi_{1,3}(l,p)=a(l,p)\phi_{1,1}(1,p)\tilde\phi_{1,3}(l-1,p+1)+b(l,p)\phi_{1,3}(1,p)\phi_{3,3}(l-1,p+1).
\ee
Here the field $\phi_{3,3}(l,p)$ is just a primary field constructed as:
\be\label{fitri}
\phi_{3,3}(l,p)=\phi_{3,3}(1,p)\phi_{3,3}(l-1,p+1)
\ee
with dimension from (\ref{dimp}). It is straightforward to check that the field (\ref{field}) has a correct dimension (\ref{delt}).
The coefficients $a(l,p)$ and $b(l,p)$ as well as the structure constants of the fields involved in the constructions (\ref{field}) and (\ref{fitri}) can be found by demanding the closure of the fusion rules:
\bea\label{frn}
\tilde\phi_{1,3}(l,p)\tilde\phi_{1,3}(l,p) &=&1+\CC_{(13)(13)}^{(13)}\tilde\phi_{1,3}(l,p)+\CC_{(13)(13)}^{(15)}(l,p)\tilde\phi_{1,5}(l,p),\\
\nn \phi_{3,3}(l,p)\phi_{3,3}(l,p) &=&1+\CC_{(33)(33)}^{(13)}(l,p)\tilde\phi_{1,3}(l,p)+\CC_{(33)(33)}^{(33)}(l,p)\phi_{3,3}(l,p)+\\
\nn &+&\CC_{(33)(33)}^{(15)}(l,p)\tilde\phi_{1,5}(l,p).
\eea
We found that:
$$
a=\sqrt{{(l-1)(p-2)\ov l(p-1)}},\qquad b=\sqrt{{p-l-2\ov l(p-1)}},
$$
the structure constants are just a special case of those listed in Appendix A.

We introduced explicitly here the descendent field:
\be\label{five}
\tilde\phi_{1,5}(l,p)=x'(l,p)\phi_{1,1}(1,p)\tilde\phi_{1,5}(l-1,p+1)+y'(l,p)\phi_{1,3}(1,p)\tphi_{3,5}(l-1,p+1).
\ee
of dimension $\tilde \D_{1,5}=2-{6\ov p+l}$.
The coefficients and the structure constants involving this field are found from the closure of (\ref{frn}):
\bea\label{confive}
x' &=&\sqrt{{(l-2)(p-3)\ov l(p-1)}}, \qquad y'=\sqrt{{2(p+l-3)\ov l(p-1)}},\\
\nn \CC_{(33)(33)}^{(15)}(l,p) &=&-\sqrt{{2l(l-1)\ov (p-2)(p-3)(p+l-3)(p+l-4)}} \tilde\CG_3(p+l-1),\\
\nn \CC_{(13)(13)}^{(15)}(l,p) &=&(p+l-2)\sqrt{{2(l-1)(p-3)\ov l(p+l-3)(p+l-4)(p-2)}} \tilde\CG_3(p+l-1)
\eea
where the function $\tilde\CG_n(p+l-1)$ is defined in Appendix A.

The mixing of the fields along the RG flow is connected to the two-point function. Up to the second order of the perturbation theory it is given by (\ref{secex}):
\bea\nn
<\phi_1(x)\phi_2(0)>&=&<\phi_1(x)\phi_2(0)>_0-\l\int <\phi_1(x)\phi_2(0)\tilde\phi(y)>_0 d^2y+\\
\nn &+&{\l^2\ov 2}\int <\phi_1(x)\phi_2(0)\tilde\phi(x_1)\tilde\phi(x_2)>_0 d^2x_1 d^2x_2 +\ldots
\eea
where $\phi_1$, $\phi_2$ can be arbitrary fields of dimensions $\D_1$, $\D_2$.

As it was was explained in Section 2.2 one can use the transformation properties of the fields to bring the double integral to the semi-factorized form (\ref{bint}) where $I(x)$ (\ref{ix}) is expressed through the hypergeometric functions which are fully under control. Also, in regularizing the integral we follow the procedure described in Section 2.2 \cite{pogsc2}. For convenience we will just change the notations for the additional parameter and the ultraviolet cut-off introduced there and call them $r$ and $r_0$ respectively in what follows (this is since we want to preserve the notation $l$ for the level of the coset model). Thus, for example, the safe region, far from singularities will be denoted as $\Omega_{r,r_0}$.

Let us consider the correlation function that enters the
integral (\ref{bint}). The basic ingredients for the computation of the
four-point correlation functions are the conformal blocks. These are
quite complicated objects in general and closed formulae were not
known. Recently, it was argued that they coincide (up to factors)
with the instanton partition function of certain $N=2$ YM theories.
Here we adopt another strategy,
namely, we find the expressions for
the conformal blocks up to a sufficiently high level in order to have a guess for
the limit $\e\rightarrow 0$.

Let us remind that, according to the construction (\ref{state}) presented in the previous Section, any field $\phi_{m,n}(l,p)$ (or its descendent) can be expressed recursively as a product of lower level fields. Therefore the corresponding conformal blocks will be a product of lower level conformal blocks. Due to the RHS of (\ref{state}) only certain products of conformal blocks will survive the projection ${\bf P}$. We would like to be more explicit here, so let us define the conformal block at level $l$ by:
$$
F_l(r,s)=<\phi_{i_1,j_1}(x)\phi_{i_2,j_2}(0)|_{r,s}\phi_{i_3,j_3}(1)\phi_{i_4,j_4}(\infty)>_l
$$
where in the notation  we omitted the "external" fields and $r,s$ stands for the internal channel field $\phi_{r,s}$. The latter could be a primary field from (\ref{dimp})
or a descendent like those defined in (\ref{field}) and (\ref{five}) (which could be identified with some descendent with respect to the current $A$). Which internal field can appear in the conformal block is defined by the fusion rules. The latter can be obtain recursively as it was explained in the previous Section.

The conformal block is a chiral object, i.e. it depends only on the chiral coordinate $x$. It can be expanded as:
\be\label{exp}
F(x)=x^{\D_{rs}-\D_{i_1j_1}-\D_{i_2j_2}}\sum_{N=0}^\infty x^N F_N
\ee
where $N$ is called level (not to be confused with the level $l$ of $M(k,l)$) and we omitted the indexes.

In order to preserve the projection ${\bf P}$ in the intermediate channel, we allow only products of conformal blocks of the form:
\bea\label{cbprod}
&<&\phi_{i_1,j_1}(x)\phi_{i_2,j_2}(0)|_{r,t}\phi_{i_3,j_3}(1)\phi_{i_4,j_4}(\infty)>_1\times\\
\nn &\times&<\phi_{k_1,l_1}(x)\phi_{k_2,l_2}(0)|_{t,s}\phi_{k_3,l_3}(1)\phi_{k_4,l_4}(\infty)>_{l-1}\times\\
\nn &\times&\sqrt{\CC_{(i_1j_1)(i_2j_2)}^{rt}\CC_{(i_3j_3)(i_4j_4)}^{rt}\CC_{(k_1l_1)(k_2l_2)}^{ts}\CC_{(k_3l_3)(k_4l_4)}^{ts}}.
\eea
Namely, only products of conformal blocks that involve the same internal indexes are allowed. Note that we included explicitly the corresponding structure constants. This is needed because they give different relative contribution on the subsequent levels in the expansion (\ref{exp}). The overall constant will define the actual structure constant. Also, as it was discussed above, we take square roots of the structure constants because our considerations are chiral, i.e. depend only on the chiral coordinate $x$. Then, the true structure constant will be a square of the resulting one in (\ref{cbprod}).

Actually, we consider below descendent fields which are some linear combinations like (\ref{field}). Therefore we will have a linear combinations of products (\ref{cbprod}). We give more details of the explicit construction of the conformal blocks in consideration in Appendixes B and C.

The conformal blocks are in general quite complicated objects.
Fortunately, in view of the renormalization scheme and the
regularization of the integrals, we need to compute them here only
up to the zero-th order in $\e$. This simplifies significantly the
problem.

Once the conformal blocks are known, the correlation function of spinless fields for our $M(k,l)$ models is written as:
$$
\sum_{r,s} C_{rs}|F(r,s)|^2
$$
where the range of $(r,s)$ depends on the fusion rules and $C_{rs}$ is the
corresponding structure constant (we omitted the external indexes). The structure constants for the fields of interest are listed in Appendix A.

Our strategy here is to compute the conformal blocks recursively up to sufficiently high level. In addition we impose the condition of the crossing symmetry of the corresponding correlation function and the correct behaviour near the singular points 1 and $\infty$.

We turn now to the computation of the $\beta$-function and the fixed point. For the computation of the $\b$-function up to the second order, we need
the four-point function of the perturbing field.
As explained in Appendix B  there are three ``channels" (or intermediate fields) in this
conformal block corresponding to the identity $\phi_{1,1}$,
$\tilde\phi_{1,5}$ and to $\tilde\phi$ itself. The explicit expression for
the correlation function is (B.8):
\bea\nn
&<&\tilde\phi(x)\tilde\phi(0)\tilde\phi(1)\tilde\phi(\infty)>=\\
\nn &=& \left|{(1 - 2 x + ({5\ov 3}+{4\ov 3l}) x^2 - ({2\ov 3}+{4\ov 3l}) x^3 + {1\ov 3} x^4)\ov x^2 (1 - x)^2}\right|^2+{16\ov 3l^2}\left| {(1 - {3\ov 2} x + {(l+1)\ov 2} x^2 - {l\ov 4} x^3)\ov x (1 - x)^2}\right|^2+\\
\nn &+& {5\ov 9}\left({2(l-1)\ov l}\right)^2\left| {(1 - x +{l\ov 2(l-1)} x^2)\ov (1 - x)^2}\right|^2.
\eea
In Appendix B we checked explicitly the crossing symmetry and the $x\rightarrow 1$ limit of this function.
In order to compute the $\beta$-function and the fixed point to the second order
we just have to integrate the above function.

The integration over the safe region $\Omega_{r,r_0}$ goes in exactly the same way as for the $N=1$ ($l=2$) case so we omit here the details. To compute the integrals near the singular points $0,1$ and $\infty$ we use again the OPE:
$$
\tilde\phi(x)\tilde\phi(0)=(x\bar x)^{-2\Delta}(1+\ldots)
+ C_{(1,3)(1,3)}^{(1,3)}(x\bar x)^{-\Delta}(\tilde\phi(0)+\ldots)
$$
which follows from the definition (\ref{frn}). The channel $\tilde\phi_{1,5}$ gives after integration a term proportional
to $r/r_0$ which is negligible. The structure constant is a particular case of those presented in Appendix A.
Its value is:
$$
 C_{(1,3)(1,3)}^{(1,3)}={4\ov l\sqrt 3} - 2 \sqrt 3 \e
$$
to the first order in $\e$.

Putting altogether, we obtain the finite part of the integral:
$$
{80\pi^2\ov 3l^2 \e^2}-{88\pi^2\ov l\e}.
$$
We notice that, although the single integrals give different results, the final answer matches perfectly the known $l=1$ \cite{pogsc2} and $l=2$ (Section 2.2) cases.

Taking into account also the first order term (whose calculation is straightforward and proportional to
the above structure constant), we get the final
result (up to the second order) for the two-point function of the
perturbing field:
\bea\label{twopt}
G(x,\l)&=&<\tphi (x)\tphi(0)>=\\
\nn &=&(x\bar x)^{-2+2\e}\left[1-\l {4\pi\ov \sqrt 3}\({2\ov l\e}-3\)(x\bar x)^\e+{\l^2\ov 2}\({80\pi^2\ov 3l^2 \e^2}-{88\pi^2\ov l\e}\)(x\bar x)^{2\e}
+\ldots\right].
\eea

The expression for the $\b$-function in this renormalization scheme was already given in Section 2.2 and reads:
$$
\beta(g)=\e\l{\p g \ov\p\l}=\e\l\sqrt{ G(1,\l)}
$$
where $G(1,\l)$ is given by (\ref{twopt}) with $x=1$. One can invert this and compute the bare coupling constant and the $\beta$-function in terms of the renormalized coupling constant $g$:
\bea\label{barel}
\l&=&g+g^2{\pi\ov \sqrt 3}\left({2\ov l\e}-3\right)+g^3{\pi^2\ov 3}\left({4\ov l^2\e^2}-{10\ov l\e}\right)+{\cal O}(g^4),\\
\nn \beta(g)&=&\e g-g^2{\pi\ov\sqrt 3}({2\ov l}-3\e)-{4\pi^2\ov 3l}g^3+{\cal O}(g^4).
\eea
In this calculations, we keep only the relevant terms by assuming the
coupling constant $\l$ (and $g$) to be order of ${\cal O}(\e)$.

A non-trivial IR fixed point occurs at the zero of the $\beta$-function:
\be\label{fxl}
g^*={l\sqrt{3}\ov 2\pi}\e(1+{l\ov 2}\e).
\ee
It corresponds to the IR CFT  $M(k-l,l)$ as can be seen from the central charge difference:
$$
c^*-c=-{4(l+2)\ov l}\pi^2\int_0^{g^*}\beta(g)d g=-l(1+{l\ov 2})\e^3-{3l^2\ov 4}(l+2)\e^4+{\cal O}(\e^5).
$$
The anomalous dimension of the perturbing field becomes:
$$
\D^*=1-\p_g\beta(g)|_{g^*}=1+\e+l\e^2+{\cal O}(\e^3)
$$
which matches with that of the field $\phi_{3,1}(l,p-l)$ of $M(k-l,l)$ (a particular case of the fields defined immediately below).

Let us now define recursively, in analogy with the fields $\tilde\phi_{1,3}(l,p)$ and $\phi_{3,3}(l,p)$, the following descendent fields:
\bea\label{defn}
\tilde\phi_{n,n+2}(l,p)&=&x(l,p)\phi_{n,n}(1,p)\tilde\phi_{n,n+2}(l-1,p+1)+\\
\nn &+&y(l,p)\phi_{n,n+2}(1,p)\phi_{n+2,n+2}(l-1,p+1),\\
\nn \tilde\phi_{n,n-2}(l,p)&=&\tilde x(l,p)\phi_{n,n}(1,p)\tilde\phi_{n,n-2}(l-1,p+1)+\\
\nn&+&\tilde y(l,p)\phi_{n,n-2}(1,p)\phi_{n-2,n-2}(l-1,p+1)
\eea
and the primary field
\be\label{defnn}
\phi_{n,n}(l,p)=\phi_{n,n}(1,p)\phi_{n,n}(l-1,p+1).
\ee
The dimensions of these fields are:
\bea\label{dimen}
\tilde\D_{n,n\pm 2} &=&1+{n^2-1\ov 4p}-{(2\pm n)^2-1\ov 4(p+l)}=1-{1\pm n\ov 2}\e+O(\e^2),\\
\nn \D_{n,n} &=&{n^2-1\ov 4p}-{n^2-1\ov 4(p+l)}={(n^2-1)l\ov 16}\e^2+O(\e^3).
\eea
They are analogs of the (descendants of the) NS fields of the $N=1$ super conformal theory($l=2$) and the fields from $S$ or $D$-sectors of $4/3$-parafermionic theory ($l=4$).

Two remarks are in order. First, similarly to $\tilde\phi_{1,3}(l,p)$ and $\phi_{3,3}(l,p)$ the fields defined above belong to the zero charge, or "vacuum sector" of the current $A(z)$. The arguments for that go along the same lines. Second, the fields (\ref{defn}) and the derivative of (\ref{defnn}) have dimensions close to one and therefore can mix. To ensure this we ask that their fusion rules with the perturbing field are closed. This requirement defines the coefficients in (\ref{defn}) and the corresponding structure constants.
So we impose the conditions:
\bea\label{frnl}
\tilde\phi_{1,3}(l,p)\tilde\phi_{n,n+2}(l,p) &=&\CC_{(13)(nn+2)}^{(nn)}(l,p)\phi_{n,n}(l,p)+\CC_{(13)(nn+2)}^{(nn+2)}(l,p)\tilde\phi_{n,n+2}(l,p),\\
\nn \phi_{3,3}(l,p)\phi_{n,n}(l,p) &=&\CC_{(33)(nn)}^{(nn+2)}(l,p)\tilde\phi_{n,n+2}(l,p)+\CC_{(33)(nn)}^{(nn)}(l,p)\phi_{n,n}(l,p).
\eea
Using the constructions (\ref{field}), (\ref{fitri}) and (\ref{defn}), (\ref{defnn}), we obtain functional equations for the coefficients and the structure constants:
\bea\label{eone}
a &x& \CC_{(13)(nn+2)}^{(nn+2)}(l-1,p+1)+b x  \CC_{(13)(nn)}^{(nn)}(1,p)\CC_{(33)(nn+2)}^{(nn+2)}(l-1,p+1)+\\
\nn &+&b y  \CC_{(13)(nn+2)}^{(nn)}(1,p)\CC_{(33)(n+2n+2)}^{(nn+2)}(l-1,p+1)=x \CC_{(13)(nn+2)}^{(nn+2)}(l,p),
\eea
\bea\label{etwo}
a &y& \CC_{(13)(n+2n+2)}^{(n+2n+2)}(l-1,p+1)+b x  \CC_{(13)(nn)}^{(nn+2)}(1,p)\CC_{(33)(nn+2)}^{(n+2n+2)}(l-1,p+1)+\\
\nn &+&b y  \CC_{(13)(nn+2)}^{(nn+2)}(1,p)\CC_{(33)(n+2n+2)}^{(n+2n+2)}(l-1,p+1)=y \CC_{(13)(nn+2)}^{(nn+2)}(l,p),
\eea
\bea\label{etri}
a &x& \CC_{(13)(nn+2)}^{(nn)}(l-1,p+1)+b x  \CC_{(13)(nn)}^{(nn)}(1,p)\CC_{(33)(nn+2)}^{(nn)}(l-1,p+1)+\\
\nn &+&b y  \CC_{(13)(nn+2)}^{(nn)}(1,p)\CC_{(33)(n+2n+2)}^{(nn)}(l-1,p+1)= \CC_{(13)(nn+2)}^{(nn)}(l,p)
\eea
from the first of (\ref{frnl}) and
\bea\label{efor}
& \CC_{(33)(nn)}^{(nn)}(1,p)\CC_{(33)(nn)}^{(nn+2)}(l-1,p+1)=x \CC_{(33)(nn)}^{(nn+2)}(l,p),\\
\nn & \CC_{(33)(nn)}^{(nn+2)}(1,p)\CC_{(33)(nn)}^{(n+2n+2)}(l-1,p+1)=y \CC_{(33)(nn)}^{(nn+2)}(l,p),\\
\nn & \CC_{(33)(nn)}^{(nn)}(1,p)\CC_{(33)(nn)}^{(nn)}(l-1,p+1)= \CC_{(33)(nn)}^{(nn)}(l,p)
\eea
from the second one. In all these equations $x$, $y$, $a$ and $b$ are at values $(l,p)$. Note that $x^2+y^2=1$ (as well as $a^2+b^2=1$)  by normalization.

In order to solve these functional equations we use the fact that we know the value of the structure constants $\CC(1,p)$, i.e. the Virasoro ones. Also, by construction, the fields $\phi_{3,3}(l,p)$ and $\phi_{n,n}(l,p)$ are primary. Therefore their structure constants are just a product of lower level ones, as can be seen from the last of the equations (\ref{efor}). Finally, one can use the knowledge of the solutions for $l=1,2,4$ \cite{df,pogzam,pog43}. With all this, we can make a guess and check it directly. We will present the result for the structure constants in Appendix A.

Our goal in this Section is the computation of the matrix of anomalous dimensions and the corresponding mixing matrix of the fields (\ref{defn}) and (\ref{defnn}) up to the second order of the perturbation theory. For that purpose we compute their two-point functions up to second order and the corresponding integrals.

\noindent\rb\hskip.3cm {\bf Function} $<\tphi_{n,n+2}(1)\tphi_{n,n+2}(0)>$

The corresponding function in the second order of the perturbation theory can be found in Appendix C (C.1).
After transformation $x\rightarrow 1/x$ it becomes:
\bea\nn
&<&\tilde\phi(x)\tilde\phi_{n,n+2}(0)\tilde\phi_{n,n+2}(1)\tilde\phi(\infty)>=
\left|{(l - (2l+4) x + (5l+4) x^2 - 6l x^3 +3l x^4)\ov 3l x^2 (1 - x)^2}\right|^2+\\
\nn &+&{8(n+3)\ov 3l^2(n+1)}\left| {(l -2(l+1) x +6 x^2 -4x^3)\ov 4 x^2 (1 - x)^2}\right|^2+\\
\nn &+& \left({2(l-1)\ov l}\right)^2{(n+3)(n+4)\ov 18 n(n+1)}\left| {(l +2(1-l) x +2(l-1) x^2)\ov 2(l-1)x^2 (1 - x)^2}\right|^2.
\eea
The integration of this function is very similar to that we did in the case of the computation of the $\b$-function. It goes along the same lines of the $l=1$  and $l=2$ cases so we do not present here the detailed calculation. The only difference is in the structure constants needed in the OPE's around $0$, $1$ and $\infty$. They are given in Appendix A:
\bea\label{str}
(C_{(13)(nn+2)}^{(nn+2)})^2&=&{4(n+3)^2\ov 3l^2 (n+1)^2}-{4 (n+2) (n+3)^2 \e\ov 3 l(n+1)^2}+O(\e^2),\\
\nn (C_{(13)(nn+2)}^{(nn)})^2&=&{n+2\ov 3n}+O(\e^2).
\eea
The final result of the integration is:
$$
{8\pi^2 (20 + 143 n + 121 n^2 + 33 n^3 + 3 n^4)\ov
 3l^2 n ( n+1) (n+3)^2 \e^2}-
{4\pi^2(n+5) (8 + 151 n + 143 n^2 + 45 n^3 + 5 n^4) \ov 3l n (n+1) (n+3)^2 \e}.
$$
This is in perfect agreement with $l=1$ and $l=2$ cases.

\noindent\rb\hskip.3cm {\bf Function} $<\tphi_{n,n+2}(1)\tphi_{n,n-2}(0)>$

The relevant four-point function in this case in the zeroth order of $\e$ is given by (C.3). Transforming $x\rightarrow
{1\ov x}$, one obtains:
$$
<\tphi(x)\tphi_{n,n+2}(1)\tphi_{n,n-2}(0)\tphi(\infty)>={1\ov 3} \sqrt{{(n^2-4)\ov n^2}}\left|{ 1\ov l x^2(1 - x)^2 }(l -2(l-1) x +2(l-1) x^2)\right|^2.
$$

Again, the integration over the safe region and lens-like region is very similar to $l=1$ and $l=2$ cases. The same is true also for the singular points where we have to take the structure constant:
$$
C_{(13)(nn-2)}^{(nn)}={n-2\ov 3n}+O(\e^2).
$$
Collecting all the integrals leads to the final result:
$$
{320 (1 - l \e) \pi^2\ov 3 l^2\e^2 n (n^2-9) \sqrt{n^2-4}}
$$
which again matches with Virasoro and superconformal cases.

\noindent\rb\hskip.3cm {\bf Function} $<\phi_{n,n}(1)\tphi_{n,n+2}(0)>$

The four point function differs only in the structure constant (C.4):
$$
<\tphi(x)\phi_{n,n}(1)\tphi_{n,n+2}(0)\tphi(\infty)>={4\ov 3l}\sqrt{{n+2\ov n}}|x|^{-2}.
$$

Therefore the calculations are exactly the same. Also, the necessary structure constants for the calculation around singular points were already presented above. This leads to a final result:
$$
{4 (n-1)\pi^2\ov 3 l(n+3) (n+5)}\sqrt{{n+2\ov n}}\left[-22 -6 n + \e(-2 (n+5) (3 n+11)+l(46+n(n+15)))\right].
$$

\noindent\rb\hskip.3cm {\bf Function} $<\phi_{n,n}(1)\phi_{n,n}(0)>$

Finally, we need the function $<\tphi(x)\phi_{n,n}(1)\phi_{n,n}(0)\tphi(\infty)>$.
As it is shown in Appendix C this function happens to coincide exactly with the one found in \cite{pogsc2} and in Section 2.2 and is given explicitly by (C.5).
Therefore almost all integrals are the same. The only exception  is
the integral around $\infty$ due to the different structure constants:
$$
C_{(13)(nn)}^{(nn)}C_{(13)(13)}^{(13)}={(n^2-1)  \e^2\ov 6  }(1-(l-2)\e).
$$
With this, the result is:
$$
{(n^2-1)\pi^2 \ov 12}(2 +(8-3l) \e).
$$

Since the dimension of the field $\phi_{n,n}$ is close to zero, it doesn't mix with other fields.
Therefore, we need to compute only its anomalous dimension.
Taking into account also the first order
contribution, the final result for the two-point function is:
\bea\nn
G_n(x,\l)=<\phi_{n,n}(x)\phi_{n,n}(0)>&=&(x\bar
x)^{-2\D_{n,n}}\left[1 - \l \left({\sqrt 3 l\pi\ov 24} (n^2-1) \e(2+(l+4)\e) \right)(x\bar x)^\e\right.\\
\nn &+&\left.{\l^2\ov 2} \left({\pi^2\ov 12} (2 +(8-3l)  \e) (n^2-1)\right)(x\bar x)^{2\e}+...\right].
\eea

Computation of the anomalous dimension goes in exactly the same way as for the perturbing field:
\bea\nn
\D_{n,n}^g &=&\D_{n,n}-{\e\l\ov 2}\p_\l G_n(1,\l)=\\
\nn &=&\D_{n,n}+{\sqrt 3\pi g l\ov 48} \e^2 (2 + (l+4) \e) (n^2-1)+{\pi^2 g^2\ov 24} \e^2(l-4) (n^2-1)
\eea
where we again kept the appropriate terms of order $\e\sim g$. Then, at the fixed point (\ref{fxl}), this becomes:
$$
\D_{n,n}^{g^*}=
{(n^2-1)l(4 \e^2 + 6l \e^3 + 7l^2 \e^4+...)\ov 64}
$$
which coincides up to the desired order with the dimension of the field $\phi_{n,n}(l,p-l)$
of the model $M(k-l,l)$.

We turn now to the computation of the matrix of anomalous dimensions. We use the same renormalization scheme that was presented in Section 2.2 for the case of $N=1$ supersymmetric models ($l=2$). Let us remind that the matrix of anomalous dimensions is defined as:
\be\label{anol}
\G=B\hat\D B^{-1}-\e\l B\p_\l B^{-1}
\ee
where $\hat\D=diag(\D_1,\D_2)$ is a
diagonal matrix of the bare dimensions.
The matrix $B$, defined as the multiplicative renormalization $\phi^g_\a=B_{\a\b}(\l)\phi_\b$, is
computed from the matrix of the bare two-point functions (see Section 2.2).

We computed above some of the entries of the $3\times 3$ matrix
of two-point functions in the second order. This matrix is obviously
symmetric. It turns out also that the remaining functions
$<\tphi_{n,n-2}(1)\tphi_{n,n-2}(0)>$ and
$<\phi_{n,n}(1)\tphi_{n,n-2}(0)>$ can be obtained from the computed
ones $<\tphi_{n,n+2}(1)\tphi_{n,n+2}(0)>$ and
$<\phi_{n,n}(1)\tphi_{n,n+2}(0)>$ by just taking $n\rightarrow -n$.

As we did for $l=2$ case, let us combine the fields in consideration in a vector with components:
$$
\phi_1=\tphi_{n,n+2},\quad
\phi_2=(2\D_{n,n}(2\D_{n,n}+1))^{-1}\p\bar\p \phi_{n,n},\quad
\phi_3=\tphi_{n,n-2}.
$$
The field $\phi_2$ is normalized so that its bare two-point function is $1$. It is straightforward to
modify the functions involving $\phi_2$ taking into account the derivatives and the normalization.

The matrix of the two-point functions up to the second order in the perturbation expansion was written in (\ref{tpfo})
where the first order term is proportional to the structure constant (\ref{fir}).

Collecting all the dimensions and structure
constants, we get:
\bea\nn
C^{(1)}_{1,1}&=&-{2 (n+3) (-2 +l\e(n+2) ) \pi\ov \sqrt 3 l\e (n+1)},\quad
C^{(1)}_{1,2}={8 (-2 +l \e) \sqrt{{n+2\ov n}} \pi\ov \sqrt 3 l \e (n+1) (n+3)},\quad
C^{(1)}_{1,3}=0,\\
\nn C^{(1)}_{2,2}&=&{16  \pi\ov \sqrt 3  l(n^2-1) \e} - {4  (n^2+1) \pi\ov
 \sqrt 3 (n^2-1) },\quad
C^{(1)}_{2,3}={8 (-2 + l\e) \sqrt{{n-2\ov n}} \pi\ov \sqrt 3 l\e (n-3) (n-1)},\\
\nn C^{(1)}_{3,3}&=&{-2 (n-3) (-2 +  l\e(2-n)) \pi\ov \sqrt 3 l\e (n-1)}
\eea
for the first order, and:
\bea\nn
C^{(2)}_{1,1}&=&{8 (20 + 143 n + 121 n^2 + 33 n^3 + 3 n^4) \pi^2\ov
 3l^2 n (n+1) (n+3)^2 \e^2}-\\
 \nn &-&{4 (n+5) (8 + 151 n + 143 n^2 + 45 n^3 + 5 n^4) \pi^2\ov
 3l n (n+1) (n+3)^2 \e},\\
\nn C^{(2)}_{1,2}&=&-{64 \sqrt{{n+2\ov n}} (3 n+11) \pi^2\ov
  3l^2 (n+1) (n+3) (n+5) \e^2} +{ 32 \sqrt{{n+2\ov n}} (57 + 18 n + n^2) \pi^2\ov
 3l (n+1) (n+3)(n+5) \e },\\
\nn C^{(2)}_{1,3}&=&{320 (1 - l \e) \pi^2\ov 3l^2 \e^2 n (n^2-9) \sqrt{n^2-4}},\\
\nn C^{(2)}_{2,2}&=&{128 \pi^2\ov 3l^2 (n^2-1) \e^2} - {16 (n^2+19) \pi^2\ov
 3l (n^2-1) \e},\\
\nn C^{(2)}_{2,3}&=&-{64 \sqrt{{n-2\ov n}} (3 n-11) \pi^2\ov
  3l^2 n-1) (n-3) (n-5)  \e^2} -  {32 \sqrt{{n-2\ov n}} (57 - 18 n + n^2) \pi^2\ov
 3l (n-1) (n-3) (n-5)  \e},\\
\nn C^{(2)}_{3,3}&=& -{8 (-20 + 143 n - 121 n^2 + 33 n^3 - 3 n^4) \pi^2\ov
  3l^2 n (n-1) (n-3)^2 \e^2} +\\
\nn  &+& {4 (n-5) (8 - 151 n + 143 n^2 - 45 n^3 + 5 n^4) \pi^2\ov
 3l n (n-1) (n-3)^2 \e}
\eea
for the second one.

Now we can apply the renormalization procedure of Section 2.2 and obtain
the matrix of anomalous dimensions (\ref{anol}). The bare coupling constant $\l$ is
expressed through $g$ by (\ref{barel}) and the bare dimensions, up to order
$\e^2$. The computation goes analogously to $l=2$ case so we omit the details here.

Evaluating this matrix at the fixed point (\ref{fxl}), we get:
\bea\nn
\G_{1,1}^{g^*}&=&1 + {(20 - 4 n^2) \e\ov 8 (n+1)} + {l(39 - n - 7 n^2 + n^3) \e^2\ov
 16 (n+1)},\\
\nn \G_{1,2}^{g^*}&=&\G_{2,1}^{g^*}={(n-1) \sqrt{{n+2\ov n}} \e(1+l\e)\ov n+1},\\
\nn \G_{1,3}^{g^*}&=&\G_{3,1}^{g^*}=0,\\
\nn \G_{2,2}^{g^*}&=&1 + {4 \e\ov n^2-1} + {l(65 - 2 n^2 + n^4) \e^2\ov 16 (n^2-1)},\\
\nn \G_{2,3}^{g^*}&=&\G_{3,2}^{g^*}={\sqrt{{n-2\ov n}} (n+1) \e(1+l\e)\ov n-1},\\
\nn \G_{3,3}^{g^*}&=&1 + {(n^2-5) \e\ov 2 (n-1)} + {l(-39 - n + 7 n^2 + n^3) \e^2\ov
 16 (n-1)}
\eea
whose eigenvalues are (up to order $\e^2$):
\bea\nn
\D_1^{g^*}&=&1 +  {1 + n\ov 2} \e + {l(7 +8 n + n^2)\ov 16} \e^2,\\
\nn \D_2^{g^*}&=&1 + {l(n^2-1)\ov 16} \e^2,\\
\nn \D_3^{g^*}&=&1 + {1-n\ov 2} \e +  {l(7 - 8 n + n^2)\ov 16} \e^2.
\eea

This result coincides with the dimensions $\tilde\D_{n+2,n}(l,p-l)$, $\D_{n,n}(l,p-l)+1$ and\\ $\tilde\D_{n-2,n}(l,p-l)$ of the model $M(k-l,l)$ up to this order.
The corresponding normalized eigenvectors should be identified with the fields of $M(k-l,l)$:
\bea\label{mixl}
\tphi_{n+2,n}(l,p-l)&=&{2 \ov n (n+1)}\phi_1^{g^*} + {2
\sqrt{{n+2\ov n}}\ov n+1}\phi_2^{g^*} + {\sqrt{n^2-4}\ov
n}\phi_3^{g^*},\\
\nn \phi_2(l,p-l)&=&-{2 \sqrt{{n+2\ov n}}\ov n +
1}\phi_1^{g^*} -{n^2-5\ov n^2+1}\phi_2^{g^*} +{2\sqrt{{n-2\ov n}}\ov n-1}\phi_3^{g^*},\\
\nn \tphi_{n-2,n}(l,p-l)&=&{\sqrt{n^2-4}\ov n}\phi_1^{g^*}  - { 2
\sqrt{{n-2\ov n}}\ov n-1}\phi_2^{g^*} +{ 2\ov n(n-1)}\phi_3^{g^*}.
\eea
We used as before the notation $\tphi$ for the descendent field defined as in (\ref{defn}) and:
$$
\phi_2(l,p-l)={1\ov 2\D_{n,n}^{p-l}(2\D_{n,n}^{p-l}+1)}\p\bar\p
\phi_{n,n}(l,p-l)
$$
is the normalized derivative of the corresponding primary field. We notice that these eigenvectors are finite
as $\e\rightarrow 0$ with exactly the same entries as in $l=1$ and  $l=2$ minimal models. We will show in the next Section that they are also in agreement with those computed using the domain wall construction. This is our main result in this Section.

\subsection{RG domain wall}

In the previous Section we proved that the coset CFT $M(k,l)$ perturbed by the field $\tilde\phi_{1,3}$ has a nontrivial fixed point corresponding to $M(k-l,l)$ up to the second order of the perturbation theory. We also found the mixing coefficients for certain fields between the UV $\CT_{UV}=M(k,l)$ and the IR $\CT_{IR}=M(k-l,l)$ theories.

Few years ago Gaiotto constructed a nontrivial conformal interface (RG domain wall) encoding the UV-IR map resulting through the RG flow described above \cite{gai} . Let us briefly recall the construction. Gaiotto considered a theory consisting of a IR $M(k-l,l)$ theory in the upper half plain and a UV $M(k,l)$ in the lower one. The conformal interface between the two CFT models is equivalent to some conformal boundary for the direct product of the theories $\CT_{UV}\times \CT_{IR}$:
$$
{\hat{su}(2)_k\times \hat{su}(2)_l\ov  \hat{su}(2)_{k+l}}\times {\hat{su}(2)_{k-l}\times \hat{su}(2)_l\ov  \hat{su}(2)_{k}}\sim {\hat{su}(2)_{k-l}\times \hat{su}(2)_l\times \hat{su}(2)_l\ov  \hat{su}(2)_{k+l}}.
$$
Note that two factors of $\hat{su}(2)_l$ appear at the RHS and therefore the theory possesses a natural $Z_2$ symmetry. In \cite{gai} it was shown that the desired boundary of the theory:
$$
\CT_B= {\hat{su}(2)_{k-l}\times \hat{su}(2)_l\times \hat{su}(2)_l\ov  \hat{su}(2)_{k+l}}
$$
acts as a $Z_2$ twisting mirror. Explicitly, this RG boundary is given by:
\be\label{rgb}
|\tilde B>=\sum_{s,t}\sqrt{S^{(k-l)}_{1,t}S^{(k+l)}_{1,s}}\sum_d|t,d,d,s;\CB,Z_2\gg
\ee
where the indices $t,d,s$ of the Ishibashi states refer to the representations of $\hat{su}(2)_{k-l}$, $\hat{su}(2)_{l}$,$\hat{su}(2)_{k+l}$ respectively and $S^{(k)}_{n,m}$ are the modular matrices of the $\hat{su}(2)_k$ WZNW model:
$$
S^{(k)}_{n,m}=\sqrt{{2\ov k+2}}\sin {\pi nm\ov k+2}.
$$
In this construction, the coefficients (\ref{mixl}) of the UV-IR map are expressed in terms of the one point functions of the theory $\CT_{UV}\times \CT_{IR}$ in the presence of the RG boundary. So we need the explicit expression of the states corresponding to the fields $\phi^{IR}\phi^{UV}$ in terms of the states of the coset theory $\CT_B$.

Basic ingredient of the latter is the $\hat{su}(2)_k$ WZNW with a current $J$. As we mentioned above it is a CFT with central charge $c_k={3k\ov k+2}$. The primary fields $\phi_{j,m}$ and the corresponding states $|j,m>$ are labeled by the (half)integer spin $j$ and its projection $m=-j,-j+1,...,j$. Their conformal dimensions are given by:
\be\label{dimj}
\D_j={j(j+1)\ov k+2}.
\ee
The representations are defined by the action of the currents on these states:
\bea\label{act}
J_0^{\pm}|j,m>&=&\sqrt{j(j+1)-m(m\pm 1)}|j,m\pm 1>,\\
\nn J_0^{0}|j,m>&=&m|j,m>.
\eea

Following \cite{ppog2} let us denote by $K(z)$ and $\tilde K(z)$ the WZNW currents of $\hat{su}(2)_l$ entering the cosets of the IR and UV theories respectively. We reserve the notion $J(z)$ for the current of $\hat{su}(2)_{k-l}$ entering the IR coset. The corresponding energy momentum tensors can be expressed in terms of these currents using (\ref{suga}). For example we can write symbolically the IR stress tensor as:
\be\label{tir}
T_{ir}={1\ov k-l+2}J^2+{1\ov l+2}K^2-{1\ov k+2}(J+K)^2
\ee
and similarly for the UV one. Finally, we impose the condition that the state of the coset $\CT_B$ be a highest weight state of the diagonal current $J+K+\tilde K$.

Now we are in a position to compare the mixing coefficients in (\ref{mixl}) with the corresponding one-point functions of the domain wall construction. Actually, we found it easier to compute the one-point functions of the other components of the corresponding multiplets. Namely, we shall consider the mixing of the "first components" given by the primary fields $\phi_{n,n\pm 2}$ and the first descendent of $\phi_{n,n}$ with respect to the current $A(z)$. Indeed, since $\phi_{n,n}$ belongs to the "vacuum sector" the current $A(z)$ is not branched around it and the dimension of the descendent $\tilde\phi_{n,n}=A_{-{2\ov l+2}}\phi_{n,n}$ is:
$$
\tilde\D_{n,n}={2\ov l+2}+{n^2-1\ov 4p}-{n^2-1\ov 4(p+l)}.
$$
So all these fields have dimension close to ${2\ov l+2}$ in the limit $p\rightarrow\infty$. Suppose they mix in the same way like it was in the case $l=2$ for example \cite{ppog2}.
We want to compare the corresponding one point functions  with the coefficients in (\ref{mixl}).

We shall need therefore the explicit construction for the current $A(z)$. It goes in a way very similar to that of \cite{ppog2} (see also \cite{argy}). Consider for example the IR model. As in \cite{ppog2} we take:
\be\label{cur}
A(z)=C_a J^a(z)\phi_{1,-a}(z)+D_a K^a_{-1}\phi_{1,-a}(z)
\ee
where $\phi_{1,m}(z)$ is a spin 1 field of the level $l$ WZNW theory with a current $K(z)$ and there is a summation over the index $a=\pm 1,0$. Indeed, the dimension of this current is:
$$
\D_A=1+{2\ov l+2}={l+4\ov l+2}.
$$
The coefficients $C_a$, $D_a$ are fixed by the requirement that the respective state be the highest weight state of the diagonal current algebra $J+K$. We get:
\bea\label{cocur}
D_+&=&{\k\over\sqrt{2}},\qquad D_0=\k,\qquad D_-=-{\k\over\sqrt{2}},\\
\nn C_+&=&-\k {l+4\ov (k-l)\sqrt{2}},\quad C_0=-\k {l+4\ov (k-l)},\quad C_-=\k {l+4\ov (k-l)\sqrt{2}}
\eea
where $\k$ is a normalization constant. Since below we shall normalize the corresponding states we don't need it explicitly here.
It is straightforward to make a similar construction for the UV coset with obvious change of currents and levels.

Now we can pass to the computation of the one-point functions of the fields $\phi^{ir}\phi^{uv}$ and compare them with the corresponding coefficients in (\ref{mixl}).

Let us first start though with the field $\phi_{n,n}^{uv}$ itself. As we showed above it flows to the field $\phi_{n,n}^{ir}$ in the infrared. So we need to find the state in $\CT_B$ corresponding to $\phi_{n,n}^{ir}\phi_{n,n}^{uv}$. For this we need to match their conformal dimensions and to ensure that the state is a highest weight state of the diagonal current $J+K+\tilde K$. The dimension of the primary field $\phi_{n,n}$ can be read from (\ref{dimen}). For the product of the IR and UV fields we have:
\be\label{nnd}
\D_{n,n}^{ir}+\D_{n,n}^{uv}={n^2-1\ov 4(k-l+2)}-{n^2-1\ov 4(k+l+2)}.
\ee
It is easy to identify the corresponding state with:
$$
|{n-1\ov 2},{n-1\ov 2}>|0,0>|0,0>
$$
where the three states correspond to $\hat{su}(2)$ of levels $k-l$ (with current $J$), IR level $l$ (with current $K$) and UV level $l$ (with current $\tilde K$) respectively. Indeed, this state is obviously a spin ${n-1\ov 2}$  highest weight state of $J+K+\tilde K$ and its dimension:
$$
\D_{{n-1\ov 2}}^J+\D_0^K+\D_0^{\tilde K}-\D_{{n-1\ov 2}}^{J+K+\tilde K}
$$
coincides with (\ref{nnd}). It is obvious that this state is invariant under the $Z_2$ action, i.e. the exchange of the second and third factors. So the overlap of this state with its $Z_2$ image is just equal to $1$ and therefore:
\be\label{nnf}
<\phi_{n,n}^{ir}\phi_{n,n}^{uv}|RG>={\sqrt{S_{1,n}^{(k-l)}S_{1,n}^{(k+l)}}\ov S_{1,n}^{(k)}}=1+{3l^2\ov 4k^2}+O({1\ov k^3}).
\ee
This confirms that up to the leading order in $k\rightarrow\infty$ the field  $\phi_{n,n}^{uv}$ flows to $\phi_{n,n}^{ir}$.

Let us now find, for example, the state corresponding to $\phi_{n+2,n}^{ir}\phi_{n,n+2}^{uv}$. The dimensions can be found from (\ref{dimen}) and we have:
$$
\D_{n+2,n}^{ir}+\D_{n,n+2}^{uv}={4\ov l+2}+{(n+1)(n+3)\ov 4(k-l+2)}-{(n+1)(n+3)\ov 4(k+l+2)}.
$$
The corresponding state should have the form:
\be\label{pp}
\sum_{\a,\b=\pm 1,0}C_{\a\b}|{n+1\ov 2},{n+1\ov 2}-\a-\b>|1,\a>|1,\b>.
\ee
The coefficients $C_{\a\b}$ are obtained by imposing the condition that (\ref{pp}) has a correct IR dimension and is a highest weight state of $J+K+\tilde K$. We obtain:
$$
C_{++}=-{1\ov \sqrt{n}}C_{0+},\qquad C_{-+}=-{\sqrt{{n+1\ov 2}}}C_{0+}
$$
and all the other coefficients vanish. The overall normalization fixes:
$$
C_{0+}^2={2n\ov (n+1)(n+2)}.
$$
Taking the overlap of the state (\ref{pp}) with its $Z_2$ image we find:
\be\label{ppf}
<\phi_{n+2,n}^{ir}\phi_{n,n+2}^{uv}|RG>={2\ov (n+1)(n+2)}{\sqrt{S_{1,n+2}^{(k-l)}S_{1,n+2}^{(k+l)}}\ov S_{1,n}^{(k)}}={2\ov n(n+1)}+O({1\ov k^2}).
\ee

The other calculations go in the same way, we present here just the result:
\bea\label{others}
<\phi_{n-2,n}^{ir}\phi_{n,n-2}^{uv}|RG>&=&{2\ov (n-1)(n-2)}{\sqrt{S_{1,n-2}^{(k-l)}S_{1,n-2}^{(k+l)}}\ov S_{1,n}^{(k)}}={2\ov n(n-1)}+O({1\ov k^2}),\\
\nn <\phi_{n+2,n}^{ir}\phi_{n,n-2}^{uv}|RG>&=&{\sqrt{S_{1,n+2}^{(k-l)}S_{1,n-2}^{(k+l)}}\ov S_{1,n}^{(k)}}={\sqrt{n^2-4}\ov n}+O({1\ov k^2}),\\
\nn <\phi_{n-2,n}^{ir}\phi_{n,n+2}^{uv}|RG>&=&{\sqrt{S_{1,n-2}^{(k-l)}S_{1,n+2}^{(k+l)}}\ov S_{1,n}^{(k)}}={\sqrt{n^2-4}\ov n}+O({1\ov k^2}).
\eea

Consider in more details the functions involving the descendent field $\tilde\phi_{n,n}$. Let us first consider $\phi_{n,n}^{ir}\phi_{n,n+2}^{uv}$:
\be\label{desd}
\D_{n,n}^{ir}+\D_{n,n+2}^{uv}={2\ov l+2}+{n^2-1\ov 4(k-l+2)}-{(n+1)(n+3)\ov 4(k+l+2)}.
\ee
The corresponding state is:
\be\label{stat}
|{n-1\ov 2},{n-1\ov 2}>|0,0>|1,1>
\ee
(because the spin $1$ term in (\ref{desd}) refers to UV level $l$ current $\tilde K$). Using the explicit expression of the current it is easy to find that for the descendent we have:
\bea\nn
A_{-{2\ov l+2}}|{n-1\ov 2},{n-1\ov 2}>|0,0>|1,1>&=&C_a J_0^a|{n-1\ov 2},{n-1\ov 2}>|1,-a>|1,1>+\\
\nn &+&D_a K_0^a|{n-1\ov 2},{n-1\ov 2}>|1,-a>|1,1>
\eea
where the coefficients are given by (\ref{cocur}). This gives:
\bea\nn
A_{-{2\ov l+2}}|{n-1\ov 2},{n-1\ov 2}>|0,0>|1,1>&=&\k {(l+4)\ov (k-l)}\sqrt{{n-1\ov 2}}|{n-1\ov 2},{n-3\ov 2}>|1,1>|1,1>-\\
\nn &-&\k {(l+4)\ov (k-l)}{n-1\ov 2}|{n-1\ov 2},{n-1\ov 2}>|1,0>|1,1>.
\eea
The normalization condition is:
\be\label{norm}
{\k^2(l+4)^2\ov (k-l)^2}{n^2-1\ov 4}=1.
\ee
Thus, for the one-point function we get:
\be\label{nnex}
<\tilde\phi_{n,n}^{ir}\phi_{n,n+2}^{uv}|RG>={2\ov n+1}{\sqrt{S_{1,n}^{(k-l)}S_{1,n+2}^{(k+l)}}\ov S_{1,n}^{(k)}}={2\ov n+1}\sqrt{{n+2\ov n}}+O({1\ov k^2}).
\ee

The other calculations are similar and finally we get:
\bea\label{nnrest}
<\tilde\phi_{n,n}^{ir}\phi_{n,n-2}^{uv}|RG>&=&-{2\ov n-1}{\sqrt{S_{1,n}^{(k-l)}S_{1,n-2}^{(k+l)}}\ov S_{1,n}^{(k)}}=-{2\ov n-1}\sqrt{{n-2\ov n}}+O({1\ov k^2}),\\
\nn <\phi_{n-2,n}^{ir}\tilde\phi_{n,n}^{uv}|RG>&=&-{2\ov n-1}{\sqrt{S_{1,n-2}^{(k-l)}S_{1,n}^{(k+l)}}\ov S_{1,n}^{(k)}}=-{2\ov n-1}\sqrt{{n-2\ov n}}+O({1\ov k^2}),\\
\nn <\phi_{n+2,n}^{ir}\tilde\phi_{n,n}^{uv}|RG>&=&{2\ov n+1}{\sqrt{S_{1,n+2}^{(k-l)}S_{1,n}^{(k+l)}}\ov S_{1,n}^{(k)}}={2\ov n+1}\sqrt{{n+2\ov n}}+O({1\ov k^2}),\\
\nn <\tilde\phi_{n,n}^{ir}\tilde\phi_{n,n}^{uv}|RG>&=&{n^2-5\ov n^2-1}{\sqrt{S_{1,n}^{(k-l)}S_{1,n}^{(k+l)}}\ov S_{1,n}^{(k)}}={n^2-5\ov n^2-1}+O({1\ov k^2}).
\eea
We see that all these results (\ref{pp}),(\ref{others}),(\ref{nnex}) and (\ref{nnrest}) are in a perfect agreement with the leading order calculations (\ref{mixl}) presented in the previous section.

\setcounter{equation}{0}
\section{Perturbation of 2D CFT and hidden symmetries of the related 2D integrable field theories}

In this Section we present a general framework to investigate the symmetries and related charges in 2D integrable field theories.
We start with the construction of some non-trivial conserved quantities in simple models like the massive free Majorana fermions and their $O(N)$ generalization.
A more powerful strategy put forward in \cite{blz} consists in trying to understand the link between the particle description and the field theory one by implementing a quantum inverse scattering method (QISM) for CFT. Indeed, off-criticality the Virasoro symmetry is lost, and the QISM results in the most hopeful method applicable to compute physical quantities of the theory. The proposal in \cite{blz} is then first to map the CFT data into a QISM structure at criticality and later study how to leave the critical point by suitable perturbation of this structure.

As mentioned in the Introduction, of special interest is the calculation of the VEV's of the descendants fields in the integrable theories providing the next to leading order in the UV limit of the two-point function. In this Section we calculate the second and third order descendant fields in the Bullough-Dodd (BD) model. It should be stressed that differently from the SG model the third level VEV is non-zero due to the existence of a local conserved current of spin 3 in the BD model. This model has attracted big interest, in particular in connection with perturbed minimal models: $c<1$ minimal CFT perturbed by the operators $\Phi_{12},\Phi_{21}$ or $\Phi_{15}$ can be obtained by a quantum group restriction of the imaginary BD model \cite{smir,kmmus} with special value of the coupling. We use this property to deduce the VEV's $<L_{-2}\bar L_{-2}\Phi_{lk}>$ and $<L_{-3}\bar L_{-3}\Phi_{lk}>$ in the mentioned
perturbed minimal models.

We further present another idea for the investigation of symmetries and corresponding charges in the 2D integrable theories. It is based on a generalization of the so called dressing symmetry transformations \cite{djk,semt,baber}. In fact, our basic objects will be the transfer matrix $T(x,\lambda)$ which generates the dressing and the resolvent $Z(x,\lambda)$, the dressed generator of the underlying symmetry. Although it is clear from the construction that our method is applicable to any generalized KdV hierarchy \cite{dsoc}, we will be concerned with the semiclassical limit \cite{ger,kuper,blz} of minimal CFT's \cite{bpz}, namely the $A_1^{(1)}$-and $A_2^{(2)}$-KdV systems \cite{dsoc}.
We present an alternative approach to the description of the spectrum of local fields in the classical limit of certain 2d integrable theories. It is possible to generalize the aformentioned dressing transformations. The idea is that we may dress not only the generators of the underlying Kac-Moody algebra but also differential operators in the spectral parameter $\lambda^m\p_\lambda^n$ forming a $w_\infty$ algebra. The corresponding vector fields close a $w_\infty$ algebra as well with a Virasoro subalgebra (for $n=1$) made up of quasi-local and non-local transformations. Finally, we present a construction of a Virasoro symmetry directly in the sine-Gordon theory. Although we are of course interested in the quantum theory, we restrict ourselves to the classical picture. Also, we are mainly concerned here with the construction of this symmetry in the case of the $N$-soliton solutions.
One of the reasons for this is that the symmetry in this case is much simpler realized - in particular it becomes local contrary to the field theory realization.

The results of this Section have been published in \cite{myne16}-\cite{myne24}, (22.-30.).

\subsection{Off critical current algebras}

The conformal models belong to the big family of the relativistic
integrable models. The key point in the construction of the spectrum and correlation functions
of the fields in this class of integrable models (IM's) is the appearance of the infinite dimensional
symmetry provided by the Virasoro algebra.
One could wonder whether analogous {\it infinite
symmetries algebra} approach works in the case of the {\it nonconformal
IM's}, say - sin-Gordon, massive fermions, affine Toda models etc. As it is
known, the integrability of all these models is based on the existence of
an infinite set of conserved charges ( CC ):
\bea\label{pss}
P_s &=&\oint T_{2s}dz-\oint \Theta_{2s-2}d\bar z,\qquad
\bar P_s=\oint\bar T_{2s}d\bar z-\oint\Theta_{2s-2}dz\\
\nn T^{\mu_1\ldots\mu_{2s}} &=&(T_{2s}, \bar T_{2s}, \Theta_{2s-2}),\qquad
\bar\partial T_{2s}=\partial \Theta_{2s-2}\qquad
\partial\bar T_{2s}=\bar\partial\Theta_{2s-2}
\eea
they have. However, the algebra of the $P_s$ ($\bar P_s$) is {\it abelian}: $[P_s,P_{s'}]=0=[P_s,\bar P_{s'}]$
and this is an obstacle in using these symmetries for the
calculation of the exact correlation functions of the model. Therefore, the
question one has to answer first is {\it whether} $P_s$ exhaust all the
conserved charges of these models,{\it i.e.}

\no 1) are there more (nontrivial) conservation laws?

\no 2) if so, is it the algebra of the new conserved charges {\it nonabelian} ?

In this Section we give an explicit construction of noncommuting CC's describing such infinite symmetries of the IM's. Our starting
point is the fact that almost all relativistic (~nonconformal ) IM's can
be represented as an appropriate perturbation of certain conformal models
\cite{zpert}, {\it i.e.}:
\be\label{pert}
S_{IM}=S_{conf}+g\int\Phi_\Delta(z,\bar z)d^2z .
\ee
This suggests that the desired new charges (if they exist) should be
realized as specific combinations of the higher momenta of the conserved
tensors ( $T_{2s}$, $\bar T_{2s}$, $\Theta_{2s-2}$ ):
\bea\label{mom}
{\cal F}^{(n)}_{2s} &=&\sum_{k=1}^s\bigg\{ \alpha _k(g)z^{2k-1+n}
\bar z^{\gamma (k,n)}T_{2k}(z,\bar z) + \bar\alpha _k(g)\bar z^{2k-1+n}
z^{\bar\gamma (k,n)}\bar T_{2k}(z,\bar z)\\
\nn &+&\beta _k(g)z^{\delta (k,n)}\bar z^{\bar\delta (k,n)}
\Theta _{2k-2}(z,\bar z)\bigg\}
\eea
such that:
$$
\bar\partial{\cal F}^{(n)}_{2s}=\partial{\cal G}^{(n)}_{2s-2}
$$
where
$
\alpha_k(g)=\alpha_k g^{{s-k\over 1-\Delta}},\quad
\beta_k(g)=\beta_k g^{{s-k\over 1-\Delta}}.
$

The {\it crucial observation} that simplifies the construction of the new
conservation laws ${\cal F}_{2s}^{(n)}$ is the following {\it criterion
of existing of such quantities}: {\it If the conservation laws of the\\ spin
- 2s} ( $s>1$ ) {\it tensors} $T_{2s}$ {\it are in the form}:
\bea\label{claws}
\bar\partial T_{2s} &=&\partial^{2s-1}\Theta+g^p\sum_{l=1}^{s-1}
A_l\partial^{2(s-l)-1}T_{2l}\\
\nn \partial\bar T_{2s} &=&\bar\partial^{2s-1}\Theta+g^p\sum_{l=1}^{s-1}
A_l\bar\partial^{2(s-l)-1}\bar T_{2l}\\
\nn \bar\partial T &=&\partial\Theta,\qquad\qquad\partial T=\bar
\partial \Theta,\qquad\qquad p={1\over 1-\Delta},
\eea
{\it then there exist} $4s-3$ {\it new conserved currents} ${\cal F}_{2s}^{
(n)}(n=1,2,\ldots,4s-3)$ {\it for each fixed} $s=1,2,\ldots$ . In
words, the existence of new conserved charges:
\be\label{lns}
L_{-n}^{(2s)}=\int{\cal F}_{2s}^{(n)}dz-
\int{\cal G}_{2s-2}^{(n)}d\bar z\ \quad
\bar L_{-n}^{(2s)} =\int\bar{\cal F}_{2s}^{(n)}-
\int\bar{\cal G}_{2s-2}^{(n)}dz
\ee
{\it is hidden in the specific form of the traces} $\Theta_{2s-2}$ {\it of
the traditional conserved currents} $T_{2s}$:
$$
\Theta_{2s-2}=\partial^{2s-2}\Theta+g^p\sum_{l=1}^{s-1}A_l
\partial^{2(s-l)-2}T_{2l} .
$$

Turning back to our problem of {\it constructing noncommuting conserved
charges} for the IM's given by (\ref{pert}) we have to check whether exist models
which satisfy our criterion, {\it i.e.} their standard $T_{2s}$ -
conservation laws to be in the form (\ref{claws}).

The simplest case is the set of models obtained by $\Phi_{\Delta_{1,3}}$ perturbations of the conformal minimal models
($c_p=1-{6\over (p+1)(p+2)}$, $\Delta_{1,3}(p)={p\over p+2}$)
(see \cite{zpert}).
They have all the $T_{2s}$, $\bar T_{2s}$, $s=1,2,\ldots$ conserved. The
first model ($p=2$) of this set is the thermal perturbation of the Ising
model which in the continuum limit coincides with the theory of free massive
Majorana fermion ( $\psi$, $\bar\psi$ ):
\be\label{eqm}
\bder \psi=m\bar\psi\qquad\qquad \der\bpsi=-m\psi
\ee
\be\label{tpsi}
T={1\over 2}\psi\der\psi ,\qquad \bar T={1\over 2}\bar\psi\bder\bar\psi ,
\qquad \Theta=m\bar\psi\psi .
\ee

To find the explicit form of $\Theta_{2s-2}$ in this case it is better to use
the equation of motion (\ref{eqm}) instead of the conformal perturbative
technics. The corresponding conservation tensors of spin
$2s$ can be taken in the form
$
T_{2s}=\psi\partial^{2s-1}\psi ,\quad s=2,3,\ldots .
$

Simple computations based on the eq. (\ref{eqm}) leads to the following desired
form of the\\ $T_{2s}$ - conservation laws:
\bea\label{isclaw}
\bder T_4 &=&\der^3\Theta +2m^2\der T\\
\nn \bder T_6 &=&\der^5\Theta+m^2 (\der T_4+4\der^3 T)\\
\nn \bder T_8 &=&2\der^7\Theta+m^2(\der T_6+3\der^3T_4+2\der^5T)
\eea
etc.
The conclusion is that this model satisfies our criterion
and therefore it
has $4s-3$ new conservation laws for each $s=1,2,\ldots$.
The corresponding conserved charges $L_{-n}^{(2s)}$, $\bar L_{-n}^{(2s)}$,
$0\le n\le 2s-1$, can be derived as integrals of certain momenta of the conserved currents.
One could try to construct them order by order but this turns to be inconvenient for deriving (or guessing) the general form of $\lch {-n}{2s}$ and
for computing their algebra as well. For these purposes it is better to have
$\lch {-n}{2s}$'s as differential operators acting on $\psi$ and
$\bar\psi$. One can do this in few steps. We first exclude the time derivatives $\der_t\psi$ and then take
$t=0$ ( $z=t+x$, $\bar z=t-x$, $\der=\der_x$ ) in the first few examples. The next step is to derive the momentum space form of $\lch {-n}{2s}$
by substituting the standard creation and annihilation operators $a^{\pm}(p)$ decomposition of $\psi$ and $\bar\psi$. This allows us to make {\it a conjecture}
about the general form of all the $\lch {-k}{2s}$ ( $0\le k\le 2s-1$ ):
\be\label{gen}
[\lch {-k}{2s},\psi]=
{-i\over 2}\left[(\bar z\bder -z\der)_{2s-1-k}+
(\bar z\bder -z\der -2s+1)_{2s-1-k}\right]\der^k\psi ,
\ee
where $(A)_p=A(A+1)\ldots (A+p-1)$. In order to prove our conjecture we have to be able to
derive from (\ref{gen}) the integral form of $\lch {-k}{2s}$ and to show that the integrands are
conserved quantities. It exists, however, an indirect way to prove that
(\ref{gen}) are conserved charges, namely we can prove that they are generators of symmetries of the action (\ref{pert}). Let us first check whether
the simplest nontrivial charge $\lch {-2}4$ leaves invariant the action:
\be\label{actis}
S=\int\left( -{1\over 2}\psi\bder\psi+{1\over 2}\bar\psi\der\bar\psi+
m\bar\psi\psi\right)d^2z\equiv\int{\cal L}d^2z .
\ee
By using (\ref{gen}) and:
$$
[\lch {-2}4 ,\bar\psi]=(\bar z\bder -z\der -{1\over2})\der^2\bar\psi
$$
one can verify easily that:
$$
[\lch {-2}4 ,{\cal L}]=\der A+\bder B
$$
for some $A$ and $B$. Therefore $\lch {-2}{4}$ is a generator of a specific new symmetry of
(\ref{actis}) . The same is true for $\blch {-2}{4}$. Together with the Lorentz
rotation generator:
$$
L_0=\int(zT+\bar z\Theta)dz-\int(\bar z\bar T+z\Theta)d\bar z
$$
they close an $SL(2,R)$ - algebra. One can repeat this
calculation with $\lch {-1}{4}$, $\blch {-1}{4}$, $\lch 04$,
$\lch {-2}{6}$ etc. and the result is always that these
charges commute with the action (\ref{actis}). As it becomes clear from
this discussion, the proof that $\lch {-k}{2s}$ given by (\ref{gen}) are
conserved charges is equivalent to the statement that $[\lch {-k}{2s},S]=0$.
To prove it we have to make one more conjecture, namely:
\bea\label{conjec}
[\lch {-k}{2s},\bar\psi]
&=&{-i\over 2}\left[(\bar z\bder -z\der +1)_{2s-1-k}
+(\bar z\bder -z\der -2s+2)_{2s-1-k}\right]\der^k\bar\psi\\
\nn 0 &\le& k\le 2s-1.
\eea
The remaining part of the proof is a straightforward but tedious higher
derivative calculus.

To make complete our study of the conserved charges of the off - critical
Ising model we have to find the general form of the ``conjugated charges''
$\blch {-k}{2s}$. By arguments similar to the ones presented above we
arrive to the following result:
\be\label{blss}
\blch {-k}{2s}={1\over 2}\left[(\bar z\bder -z\der
+\bar\alpha-2s+k+2)_{2s-1-k}+(\bar z\bder-z\der +\bar\alpha+k+1)_{2s-1-k}
\right]\bder^k,
\ee
where
$\bar\alpha =-1\quad$ for $\quad \psi$ and
$\bar\alpha =0 \quad$ for $\quad \bar\psi$.
Our claim is that (\ref{gen}), (\ref{conjec}) and (\ref{blss}) {\it do exhaust all the local
symmetries} (i.e. local conserved charges) of the action (\ref{actis}).

Studying the conformal
limit of $\lch {-k}{2s}$ we have realized that the ``conformal''
$W_\infty$ algebra has a specific subalgebra ${\cal P}W_\infty(V)$
spanned by:
\bea\nn
{\cal L}_{-k}^{(2s)} &=&\fr 12\left[(\tilde L_0-\fr 12)_{2s-1-k}+
(\tilde L_0+2s-\fr 32)_{2s-1-k}\right]L_{-1}^k\\
\nn 0 &\le& k\le 2s-1,\qquad \tilde L_0=z\der+\fr 12.
\eea

Having at hand the explicit form (\ref{gen}), (\ref{conjec}) and (\ref{blss})
of $\lch {-k}{2s}$ and $\blch {-k}{2s}$ we are prepared to
compute their algebra. As we have already mentioned $\lch {-2}4$,
$\lch {-1}4$ and $L_0$ close an $SL(2,R)$ algebra. Two more $SL(2,R)$
algebras are spanned by $\bar L_{-1}$, $\lch {-1}4$, $L_0$ and $L_{-1}$,
$\blch {-1}4$, $L_0$. Passing to the general case let us first try to find the
structure of the ``left'' algebra, i.e.:
\be\label{lef}
\left[\lch {-k_1}{2s_1},\lch {-k_2}{2s_2}\right]=
\sum_{r=1}^{s_1+s_2-1}g^{s_1s_2}_{2r}(k_1,k_2)L_{-k_1-k_2}^{2(s_1+s_2-r)}.
\ee
The simplest way to prove (\ref{lef}) and to compute the structure constants
$g^{s_1s_2}_{2r}(k_1,k_2)$ is based on the following ``conformal''
decomposition of the generators $\lch {-k}{2s}$ in terms of the conformal
generators ${\cal L}_{-k}^{(2s)}$:
\bea\label{decomp}
\lkts  &=&\sum_{l=0}^{2s-1-k}\begin{pmatrix}
 2s-1-k \\ l
 \end{pmatrix}
(\bz\bder^2+\alpha\bder)^l{\cal L}^{(2s)}_{-k-l}(-m^2)^{-l} \\
\nn \alpha &=&0\quad for\quad\psi\quad and \quad\alpha=1\quad for\quad
\bar\psi.
\eea
The fact that the operators
$
S_l=(\bz\bder^2+\alpha\bder)^l=(\bz\bder +\alpha)_l\bder^l
$
are commuting, i.e.\\ $[S_{l_1},S_{l_2}]=0$ reduces the computation of the
structure constants $g^{s_1s_2}_{2r}(k_1,k_2)$ to the conformal ones
$C^{s_1s_2}_{2r}(k_1,k_2)$ ($k_i\le 2s_i-1$):
\be
\left[{\cal L}^{(2s_1)}_{-k_1},{\cal L}^{(2s_2)}_{-k_2}\right]=
\sum_{r=1}^{s_1+s_2-1}C_{2r}^{s_1s_2}(-k_1,-k_2){\cal
L}_{-k_1-k_2}^{2(s_1+s_2-r)} .
\ee\label{cstr}
Actually, it turns out that:
\bea\nn
g^{s_1s_2}_{2r}(k_1,k_2) &=&C^{s_1s_2}_{2r}(-k_1,-k_2)\\
\nn 0\le &k_i&\le 2s_i-1
\eea
The identical statement holds for the algebra of $\blch {-k}{2s}$'s
as well. The
conclusion is that the algebra we are looking for has as subalgebras two
incomplete
($0\le k\le 2s-1$) $W_\infty$ algebras which do not commute between themselves.

The general structure of the remaining ``left - right'' commutators:
\be\label{lrcomm}
\left[ \lkts ,\blch {-l}{2p} \right]=\sum_{r=0}^{s+p-k-2}\bar g_{2r}^{sp}
(k,l)(m^2)^k\bar L_{k-l}^{2(s+p-k-1-r)}
\ee
(if $k<l$) is a consequence of the explicit form (\ref{gen}) and (\ref{blss}) of the
generators. In order to calculate $\bar g_{2r}^{sp}(k,l)$ we first commute
$L_{-1}^k$ and $\bar L_{-1}^l$ to the right
and then expand the both sides of
(\ref{lrcomm}) in powers of $X=L_0+\fr 12$. In doing this we have to know the
coefficients $B_m^{N,a}$ in the power expansion of $(X+a)_{N+1}$:
$$
\left(X+a\right)_{N+1}=\sum_{m=0}^{N+1}B_m^{N,a}X^m .
$$
A simple combinatorial analysis \cite{myne16} leads to the following form of
$B_m^{N,a}$:
$$
B_m^{N,a}=\fr 1{3.2^{m+1}}(N+2-m)_mA_m^{N,a}(N+2a) .
$$
The general solution for $A_m^{N,a}$ is a specific linear combination of the Bernuli polynomials of degree $m$.

The LHS of (\ref{lrcomm}) contains 8 terms of the form:
$$
(X+a)_{N+1}(X+a)_{M+1}=\sum_{k+0}^{N+M+2}Y_R^{(N+M+2-k)}(aN|bM)X^k
$$
where:
$$
Y_L^{(m)}(aN|bM)=\sum B_k^{a,N}B_{m-k}^{b,M} .
$$
Denote by $Y_L^{(m)}$ the sum of the contributions of all the 8 terms in
the LHS. One can perform the same calculations for the RHS. Equating the left and right hand sides one can derive the following recursive relations for the
structure constants.:
$$
Y_L^{(2n-1)}=2\sum_{r=0}^{n-1}\bar g_{2r}^{sp}(k|l)Y_R^{(2n-2-2r)}.
$$

Inspired by the observation \cite{ith} that the off-critical
$XY$- model has as dynamical symmetries two different Virasoro algebras we now look for Virasoro algebras generated by specific combinations of
$\lkts$ and $\blkts$. One can easily verify
that $\cL_n$ given by:
\bea\nn
\cL_n &=&[\lo -\fr 12]_{n+1}\der^n ,\qquad n\ge -1\\
\nn [A]_k &=&A(A-1)\ldots(A-k+1) ,\qquad [A]_0=1
\eea
(note that $\der^{-1}\psi =-1/m^2\bder \psi$) close a (incomplete) Virasoro algebra. One more
incomplete Virasoro algebra $\bar V$ is generated by:
\be\nn
\bar\cL_n=[\lo +\fr 12]_{n+1}\bder^n ,\qquad n\ge -1 .
\ee
The third Virasoro algebra $V_c$ spanned by:
\bea\nn
\fr {(-m^2)^{1-s}}2 \lch {-2s+2}{2s} &=&\fr {(-m^2)^{1-s}}2
(\lo -\fr {2s-1}2)\der^{2s-2}
\equiv\fr
{(-m^2)^{1-s}}2(L_0-s+1)L_{-1}^{2s-2}\\
\nn \fr {(-m^2)^{1-s}}2 \blch {-2s+2}{2s} &=&\fr {(-m^2)^{1-s}}2
(\lo +\fr {2s-3}2 )\bder^{2s-2}
\equiv
\fr {(-m^2)^{1-s}}2 (L_0+s-1)\bar L_{-1}^{2s-2}
\eea
$ s=1,2,\ldots$,
plays in our opinion the major role for the exact integrability. Using once more the formal identity $\bar L_{-
1}=-m^2(L_{-1})^{-1}$ we can rewrite the $V_c$ - generators in an
unique formula:
$$
V_n=\fr 12(-m^2)^n(L_0-n)L_{-1}^{2n} ,\qquad -\infty\le n\le\infty .
$$
As in the case of the massive Dirac fermion \cite{ith} one is expecting that
$V_c$ has nonzero central charge. One could check this by calculating the
commutator $\left[ \lch {-4}6 ,\blch {-4}6\right]$. The result is:
$$
\left[ \lch {-4}6 ,\blch {-4}6\right]=-8m^8L_0+m^8 .
$$
In terms of the Virasoro algebra it means that the central charge is:
$$
c=\fr 12 ,
$$
i.e. the massive Majorana fermion has the same central charge as the
massless one. This is in agreement with the result $c=
1$ for the Dirac fermions.

An important consequence of the fact that $\lch {-k}{2s}$ and $\blkts$ are
generators of the symmetries of the action (\ref{actis}) is the following infinite
set of Ward identities for the $n$-point functions of $\psi(z,\bz)$
and $\bar\psi(z,\bz)$:
\bea\label{grf}
 \left\langle 0\left | \lkts
\Pi_{i=1}^M\psi(z_i,\bz_i)\Pi_{j=1}^N\bar\psi(z_j,\bz_j)\right |
0\right\rangle &=&0,\\
\nn \left\langle 0\left |
\Pi_{i=1}^M\psi(z_i,\bz_i)\Pi_{j=1}^N\bar\psi(z_j,\bz_j)\blkts\right |
0\right\rangle &=&0.
\eea
The condition for the invariance of the vacuum:
$
\lkts |0\rangle =0=\langle 0|\blkts
$
together with eqs. (\ref{gen}),
(\ref{conjec}) and (\ref{blss}) lead to the following system of
differential equations for\\ $G_{MN}(z_l,\bz_l)= \left\langle 0\left |
\Pi_{i=1}^M\psi(z_i,\bz_i)\Pi_{j=1}^N\bar\psi(z_j,\bz_j)\right |
0\right\rangle$:
\bea\label{gmn}
&&\bigg\{ \sum _{i=1}^M\left[ (\loi )_{2s-1-k} +(\loi
-2s+1)_{2s-1-k}\right] \der _i^k+\\
\nn &&+ \sum _{j=1}^N\left[ (\loi +1)_{2s-1-k} +(\loi
-2s+2)_{2s-1-k}\right] \der _j^k\bigg\}  G_{MN}(z_l,\bz _l)=0.
\eea
A similar set of equations can be obtained
from the condition of $\blkts$ - symmetry of
$G_{MN}$.
Restricting ourselves to the case of 2 - point functions ($M+N=2$) we are
going to demonstrate that the Poincare invariance ($L_{-1}$, $L_0$,
$\bar L_{-1}$) and the new $SL(2,R)$ symmetries ($\lch {-2}4$,
$\blch {-2}4$, $L_0$) are sufficient to fix uniquely $G_{20}$, $G_{02}$ and
$G_{11}$ - functions. The relativistic invariance requires:
$$
G_{20} =m\sqrt {\fr {\bz}z }g_{20}(y) ,\quad
G_{02} =m\sqrt {\fr z{\bz} }g_{02}(y) ,\quad
G_{11} =img_{11}(y),\quad y=m\sqrt {-4z\bz} .
$$
The condition of $\lch {-2}4$ - invariance of $G_{20}$ leads to the
following third order differential equation:
\be\label{dif}
y^3g_{20}^{'''}+2y^2g_{20}^{''}-y(y^2+1)g_{20}^{'}-(y^2+1)g_{20}=0 .
\ee
It happens that one can solve (\ref{dif}) in terms of $K_1(y)$ - Bessel function.
This reflects the fact that (\ref{dif}) can be obtained as a consequence of the
$K_1$ - Bessel equation:
$$
y^2g''_{20}+yg'_{20}-(y^2+1)g_{20}=0
$$
and a specific third order equation:
\be\label{bess}
y^3g'''_{20}-y(y^2+3)g'_{20}+(y^2+3)g_{20}=0 .
\ee
The eq.(\ref{bess}) follows from the standard recursive relations for
$K_{\pm 1}$, $K_0$ and $K_2$ - Bessel functions. The $\lch {-2}4$ - Ward
identity imposes the eq. (\ref{bess}) only. Repeating the same analysis for
$G_{02}$ and $G_{11}$ we find that
$
g_{02}(y)=K_1(y)
$
and that $g_{11}$ satisfy the $K_0$ - Bessel equation:
$$
yg''_{11}+g'_{11}-yg_{11}=0 ,
$$
i.e. $g_{11}=K_0(y)$.

To make
complete our discussion of the off-critical Ising model we have to
mention that as in the conformal case the WI's (\ref{grf}), (\ref{gmn}) are fixing
uniquely the 2- and 3-point functions only. The calculation of, say, the 4-
point function (using only the symmetries of the model) requires more
information about the representations of the algebra (\ref{lef}), (\ref{lrcomm})
we are using.
One could expect that the null-vector conditions
for the off-critical
Virasoro algebra spanned by $\lch {-2s+2}{2s}$, $L_0$ and
$\blch {-2s+2}{2s}$ will be sufficient to fix uniquely the corresponding 4-
point functions ($M+N=4$).

As a generalization, let us consider a $k=1$ $O(n)$-WZW models which represent free fermions \cite{witt}. Their massive
perturbation is described by the action:
\be\label{1.1}
S=\int {1\over 2}\left( i\overline \psi^i \not \! \partial \psi^i + m\overline
\psi ^i\psi^i\right) d^2z.
\ee
Our problem is to construct explicitly all the conserved charges of the models
given by (\ref{1.1}), i.e. - $O(n)$-Majorana massive fermions $\overline\psi^i(z,\overline z)\,(i=1, \cdots, n)$.
One could expect that the case of $n$ massive fermions
in the $O(n)$-vector representation is a straightforward generalization of the
results for one massive fermion. There exist however few important
differences. The first is that together with $T_{2s}=\delta_{ij}T^{ij}_{2s}$
and $J^{ij}_{2s-1}$ we have to consider all components of the symmetric
conserved tensor $T^{ij}_{2s}$. The second very important point is that the
algebra of the standard conserved charges:
\bea\label{3.1}
P^{ij}_s&=&\int T^{ij}_{2s}dz - \int \theta ^{ij}_{2s-2}d\overline
z\quad ,\quad \overline P^{ij}_s=\int \overline T^{ij}_{2s}d\overline z -
\int \theta ^{ij}_{2s-2}dz\\
\nn Q^{ij}_s&=&\int J^{ij}_{2s-1}dz - \int \widetilde\theta ^{ij}_{2s-3}d\overline z
\quad ,\quad \overline Q^{ij}_s=\int \overline J^{ij}_{2s-1}d\overline z -
\int \widetilde
\theta ^{ij}_{2s-3}dz
\eea
is {\bf nonabelian}. Its abelian subalgebra is spanned by
$P_s=\delta_{ij}P^{ij}_s$ and $\overline P_s=\delta_{ij} \overline P_s^{ij}$.
In order to find this algebra, it is better to realize $P^{ij}_s\, ,\,
Q^{ij}_s$
etc in terms of differential operators. Following the standard massive fermion
technology we start with the conserved tensors:
\bea\label{3.2}
T^{ij}_{2s}&=&\psi^i\partial ^{2s-1}\psi^j + \psi^j\partial
^{2s-1}\psi^i\quad ,\quad T^{ij}_2={1\over 2}(\psi^i\partial \psi^j +
\psi^j\partial \psi^i)\\
\nn J^{ij}_{2s-1}&=&{1\over 2}\left(\psi^i\partial ^{2s-2}\psi^j - \psi^j\partial
^{2s-2}\psi^i \right)
\eea
and similar expressions for $\overline T^{ij}_{2s}$ and $\overline
J^{ij}_{2s-1}$. Using the equations of motion:
$$
\overline \partial \psi^k=m\overline \psi^k \quad ,\quad \partial \overline
\psi^k=-m\psi^k
$$
one can show that (\ref{3.2}) are indeed conserved tensors.
Using as before the formal identity $\overline \partial =-m^2\partial ^{-1}$, one can
write $(m^2)^{-s}Q_s^{ij}\equiv \widetilde Q_s^{ij}$ and $(m^2)^{-s}\overline
Q^{ij}_s\penalty-200\equiv\widetilde Q_{-s}^{ij}$ as a unique object
$\widetilde Q_s^{ij}\,
(-\infty \le s \le \infty)$.
One can show that the latter {\it generates the $O(n)$-Kac-Moody algebra}.
The total algebra is a {\it subalgebra $GL(n,R)_{mod\, 2}$
of the $\widehat {GL}(n,R)$-Kac-Moody algebra } spanned
by $\widetilde P^{ij}_{2s-1} \equiv P^{ij}_s$
and $\widetilde Q^{ij}_{2s}=Q^{ij}_s$, i.e. the {\it closed subalgebra of
symmetric
generators $P^{ij}$ with odd indices and antisymmetric generators $Q^{ij}$ with
even indices}.

Following the previous discussion we are interested mainly in the possible Virasoro subalgebras of full algebra of symmetries.
How to construct the Virasoro charges for one massive
fermion we already know from the
off-critical Ising model case. In order to generalize it for the
$O(n)$-massive fermions we have to find specific combinations of the ``higher
momenta" of the $T^{ij}_{2s},
J^{ij}_{2s-1},\theta^{ij}$ and $\widetilde \theta^{ij}$ to be conserved.
It turns out that we can construct
$(4s-3){n(n+1)\over 2}$ new symmetric charges  $L^{ij(2s)}_{-n},
\overline L^{ij(2s)}_{-n}\, (0\le n\le 2s-1)$ and
$(4s-5){n(n-1)\over 2}$ antisymmetric ones  $Q^{ij(2s-1)}_{-k}, \overline
Q^{ij(2s-1)}_{-k}\, (0\le k\le 2s-2)$ for each
$s=2,3, \cdots$.
The simplest one is the generalization  of the Lorentz rotation
$L_0={1\over 2}\delta^{ij}L_0^{ij}$:
$$
L_0^{ij}=\int (zT^{ij}_2 + \overline z\theta^{ij})dz - \int (\overline z
\overline T^{ij}_2 + z\theta^{ij})d\overline z\quad .
$$
The next ones are straightforward $O(n)$-matrix generalizations of the
corresponding one fermion charges
$L^{(2s)}_{-2s+2}={1\over 2}\delta^{ij}
 L^{ij(2s)}_{-2s+2}$
and we can take them in the following differential form:
\bea\label{3.14}
\left[L^{ij(2s)}_{-2s+2},\psi^k(z,\overline z)\right]&=&
-i(\delta^{ik}\delta^{jl}
+ \delta^{il}\delta^{jk})\left( \overline z\overline \partial -z\partial
-{2s-1\over 2} \right)\partial ^{2s-2}\psi^l\\
\nn \left[\overline L^{ij(2s)}_{-2s+2},\psi^k(z,\overline z)\right]&=&
-i(\delta^{ik}\delta^{jl} +
\delta^{il}\delta^{jk})\left(\overline z\overline \partial -z\partial
+{2s-3\over 2} \right)\overline\partial^{2s-2} \psi^l
\eea
The proof that they are indeed the conserved charges we are looking for is
again based on the fact that they do generate new symmetries of the action
(\ref{1.1}), i.e.
$$
[L^{ij(2s)}_{-2s+2},S]=0=[\overline L^{ij(2s)}_{-2s+2},S].
$$

The question about the algebra of these new symmetries is now in order.  By
direct calculations, using (\ref{3.14}) one can see
that $L^{ij(2s)}_{-2s+2}$ and $ \overline L^{ij(2s)}_{-2s+2}$ {\it does not
close} an algebra. It is necessary to consider together with them the first
momenta
${Q}^{ij(2s-1)}_{-2s+3}$ of the current
$J^{ij}_{2s-1}$. Before doing this we should mention that the traces
$L(\bar L)^{(2s)}_{-2s+2}={1\over 2}\delta_{ij}L(\bar L)^{ij(2s)}_{-2s+2}$
{\it do close} an algebra which coincides with the off-critical Virasoro
algebra $V_c$  of the off-critical Ising model. One could
wonder what is then the algebra of $\widetilde Q^{ij}_s$ and  these Virasoro
generators:
$$
V_k={1\over 4}(-m^2)^k\delta_{ij}L^{ij(2k+2)}_{-2k}\quad ,\quad
V_{-k}={1\over 4}(-m^2)^k\delta_{ij}\overline L^{ij(2k+2)}_{-2k}.
$$
As one could expect, the result of simple computations is the larger current
algebra $V_c\subset\!\!\!\!\!\!\times \widehat O_n $:
\bea\label{3.15}
\quad\Bigl[V_{m_1},V_{m_2}\Bigr] &=& (m_1-m_2)V_{m_1+m_2}+{n\over
24}m_1(m_1^2-1)\delta_{m_1+m_2}\\
\nn \quad\Bigl[V_{m_1},\widetilde Q_{m_2}^{ij}\Bigr] &=&
-m_2\widetilde Q^{ij}_{m_1+m_2}\\
\nn \quad\Bigl[\widetilde Q_{m_1}^{ij},\widetilde Q_{m_2}^{kl}\Bigr] &=&\delta^{ik}
\widetilde
Q^{jl}_{m_1+m_2} + \delta^{jl}\widetilde Q^{ik}_{m_1+m_2} -
\delta^{il}\widetilde
Q^{jk}_{m_1+m_2} - \delta^{jk}\widetilde Q^{il}_{m_1+m_2} +\\
\nn &+& {n\over 2} m_1
\delta_{m_1+m_2}(\delta^{ik}\delta^{jl} - \delta^{il}\delta^{jk}) .
\eea
We have enlarged  in this way the known symmetries of the action (\ref{1.1}) to the
$V_c\subset\!\!\!\!\!\!\times  \widehat O(n)$-algebra.

Turning back to our problem of constructing the first momenta of the current
 $J^{ij}_{2s-1}$ we arrive at the following general differential form for
$Q(\bar Q)^{ij(2s-1)}_{-2s+3}$:
\bea\label{3.16}
\left[Q^{ij(2s-1)}_{-2s+3},\psi^k(z, \overline z)\right]&=&-i(\delta^{ik}\delta^
{jl} -\delta^{il}\delta^{jk})(\overline z\overline \partial - z \partial
-s+1)\partial ^{2s-3}\psi^l(z,\overline z)\\
\nn \left[\overline Q^{ij(2s-1)}_{-2s+3},\psi^k(z,\overline z)\right]&=&-i(\delta^
{ik}\delta^{jl} -\delta^{il}\delta^{jk})(\overline z\overline \partial -
z \partial + s-2)\overline \partial ^{2s-3}\psi^l(z,\overline z).
\eea

Considering $Q(\bar Q)^{ij(2s-1)}_{-2s+3}$
together with
$L(\bar L)^{ij(2s)}_{-2s+2}$,
$Q(\bar Q)^{ij}_{s}
\equiv Q(\bar Q)^{ij(2s-1)}_{-2s+2}$,
$P(\bar P)^{ij}_{s}\equiv L(\bar L)^{ij(2s)}_{-2s+1}$
we are expecting them to close an algebra. However this is not the case. One
can easily check using (\ref{3.16}) that the commutator
$[Q^{ij(2s_1-1)}_{-2s_1+3},Q^{kl(2s_2-1)}_{-2s_2+3}]$ contains higher momenta
of $J^{ij}_{2s-1}$ and $T^{ij}_{2s} $ as well.
Therefore {\it the algebra of the first
momenta} of $T^{ij}_{2s}$ and $J^{ij}_{2s-1}$ {\it is not closed}. Involving
the higher momenta of $T^{ij}_{2s}\, ,\, J^{ij}_{2s-1}$ we are constructing in
this way an algebra of the $W_\infty (\widehat G_n)$-type. We address here the question about its subalgebras. Up to
now we have constructed two such subalgebras: $\widehat{GL}(n,R)_{{\rm mod} 2}$
and ${\rm Vir} \subset\!\!\!\!\!\!\times\widehat O(n)$
of
eq. (\ref{3.15}). Deriving the missing commutator:
$$
\left[ V_{m_1},\widetilde P^{ij}_{m_2}\right] = -(m_2-1/2)\widetilde
P^{ij}_{m_1+m_2}
$$
we can unify them in an unique current algebra, namely: ${\rm Vir}
\subset\!\!\!\!\!\!\times \widehat {GL}(n,R)_{{\rm mod} 2}$. Are there more
subalgebras of this type? As in the cases of one and two
fermions one could expect to find two incomplete $(n\ge -1)$ Virasoro
subalgebras. In our case they are generated by a specific combination of
$\delta^{ij} L(\bar L)^{ij(2k)}_{-s+1}$ and $ \delta^{ij}
L(\bar L)^{ij}_{-1}\equiv P(\bar P)_1$:
$$
{\mathop{\cal L}^{(-\!\!-)}}_n =\left[ \overline z\overline
\partial -z\partial \pm 1/2\right] _{n+1}{\mathop{\partial}^{(-\!\!-)}} ^n
\quad .
$$
Do they have an $\widehat O(n)$-Kac-Moody counterpart? Actually, one can easily guess the general form of the $\widehat O(n)$ generators:
\be\label{3.21}
({\cal Q}^{ij}_s)_{kl}=({\cal Q}^{ij}_0)_{kl}
\left[ \overline z \overline \partial - z\partial -1\right] _n\partial^n \quad
,\quad n\ge 0\quad
\ee
which indeed close $\widehat O(n)$-Kac-Moody
algebra. Similarly, the conserved charges:
$$
(\overline {\cal Q}^{ij}_s)_{kl}=({\cal Q}^{ij}_0)_{kl}
\left[ \overline z \overline \partial - z\partial \right] _n\overline
\partial^n \quad ,\quad n\ge 0
$$
generate one more $\widehat O(n)$-current algebra. These two algebras however
do not mutually commute.

The algebras of symmetries of (\ref{1.1}) we have
described up to now are sufficient for the calculation of the correlation
functions. We shall mention here the
following simple and {\it remarkable fact}: the $Q^{ij(3)}_{-1}$ (or $\overline
Q^{ij(3)}_{-1}$) W.I.'s for the 2-point function:
$$
g^{lm}(z_1,z_2\vert\overline z_1,\overline
z_2)=\left\langle\psi^l(z_1,\overline z_1) \psi^m(z_2,\overline
z_2)\right\rangle
$$
coincide with the $K_1$-Bessel equation.

Taking into account the Poincar\'e invariance (i.e., $L_0$, $L(\bar L)_{-1}={1\over 2}\delta^{ij}L(\bar L)_{-1}^{ij}$)
 we get:
$$
g^{lm}(z_1,z_2\vert\overline z_1,\overline z_2)=\delta^{lm}\sqrt{\overline
z_{12}\over z_{12}} K(x)\quad ,\quad x=\sqrt{-4m^2z_{12}\overline z_{12}}.
$$
We next require the  $Q_{-1}^{ij(3)}$-Ward identity:
$$\left\langle Q_{-1}^{ij(3)}\psi^l(z_1,\overline z_1)\psi^m(z_2,\overline
z_2)\right\rangle =0 \quad .
$$
As a consequence of (\ref{3.16}) and $Q^{ij(3)}_{-1}$-invariance of the vacua we
obtain the following equation:
$$
\left(\overline z_{12}+{2\over m^2}\partial_{12}+{1\over
m^2}z_{12}\partial_{12}^2\right)\sqrt{ \overline
z_{12}\over z_{12}} K(x)=0
$$
which is equivalent to the  $K_1$-Bessel equation:
$$
x^2K^{\prime\prime}(x)+xK^\prime (x)-(x^2+1)K(x)=0.
$$

\subsection{Quantum inverse scattering description of conformal minimal models}

In this Section we present an alternative description of CFT which is a variation of the quantum inverse scattering method. The basic objects are the monodromy matrix and its trace
generating the integrals of motion and encoding the basic CFT data.
It is known \cite{ger,kuper} that the $A_1^{(1)}$ KdV system describes the classical limit ($c\to -\infty$) of the 2D CFT. There exists one more possible description of this theory.
It is based on the generalized KdV system attached to $A_2^{(2)}$ which also yields the classical limit of Virasoro as Poisson bracket structure.

Consider the
generalized KdV equations corresponding to the two vertices $c_0$ and $c_1$
of the Dynkin diagram of $A_2^{(2)}$ in the Drinfeld-Sokolov \cite{dsoc}
classification:
\bea
c_0 &:& \partial_t U = \de^5 U + 5 U \de^2 U +5 \de U \de^2 U + 5 U^2 \de U
\nonumber \\
c_1 &:& \partial_t U = \de^5 U + 10U\de^3 U + 25 \de U \de^2 U + 20 U^2 \de U
\label{dieci}
\eea
As the usual KdV, both equations (\ref{dieci}) are Hamiltonian.
Their second Hamiltonian structures are associated with the
Hamiltonians:
\eq\nn
H^{(0)} = 3(\de U)^2 -16 U^3 \virg H^{(1)}= 3 (\de U)^2 - U^3.
\en
Here and in the following, the superscript in parenthesis $^{(0)}$ and
$^{(1)}$ refer to the $c_0$ and $c_1$ cases respectively.
The crucial observation is that the Poisson bracket algebra of the fields
$U(u)$ corresponding to these two second hamiltonian structures
coincides with the classical ($c\to -\infty$) limit of the Virasoro algebra:
\eq
\{U(u),U(v)\}=2(U(u)+U(v))\delta'(u-v)+\delta'''(u-v).
\label{due-bis}
\en
The systems
(\ref{dieci}) describe isospectral deformations of {\em third}
order differential operators:
\eq
L^{(0)} = \de^3 + U \de + \de U - \lambda^3 \virg
L^{(1)} = \de^3 + U \de - \lambda^3.
\label{tredici}
\en
Eqs. (\ref{dieci}) can be obtained directly by
{\em reduction} of the Boussinesq equation, which describes the classical
limit of CFT having extended $W_3$-algebra symmetry \cite{kuper}. There are two
consistent reductions of Boussinesq equation: $W=\de U$ and $W=0$,
leading to the first and second equation of (\ref{dieci}) respectively.
However this
observation is valid only at the classical level, since $(A_2^{(2)})_q$,
which is relevant for the quantum case, is an essentially nonlinear deformation
of $A_2^{(2)}$, and not just a twist of $(A_2^{(1)})_q$.
Being integrable, the equations (\ref{dieci})
possess an infinite number of conserved IM $I_s^{(i)}$, $i=0,1$ having
spin $s=1,5 \bmod 6$ \cite{syff}.
One can compute them using the Lax pair
representations of (\ref{dieci}) and show that the Poisson
bracket algebra they close is abelian $\{I_k^{(i)},I_l^{(i)}\}=0$, $i=0,1$.
These IM should obviously be the classical limit
of the corresponding
quantum conserved charges of CFT, and indeed they happen to coincide with
the classical limit of the quantum IM written in \cite{kuper} for the
Boussinesq system, once the reductions $W=\de U$ (for $c_0$) and $W=0$
(for $c_1$) are enforced.

Let us consider the first order matrix realization of (\ref{tredici}):
\eq
\cL=\de - \phi'(u)h - (e_0+e_1)
\label{quattordici}
\en
where $\phi(u)$ is related to $U(u)$ by the Miura
transformation \cite{miura} $U(u)=-\phi'(u)^2 -\phi''(u)$.
Written in the canonical gradation of $A_2^{(2)}$ \cite{dsoc} eq.(\ref{quattordici})
defines the Lax representation for the generalized {\em modified} KdV
(mKdV) corresponding to the algebra $A_2^{(2)}$
and $h,e_0,e_1$ are the Cartan-Chevalley
generators of $A_2^{(2)}$ level 0 algebra:
\eq
e_0 = \left( \begin{array}{ccc} 0 & 0 & \lambda \\
                                0 & 0 & 0 \\
                                0 & 0 & 0 \end{array} \right) \virg
e_1 = \left( \begin{array}{ccc} 0 & 0 & 0 \\
                                \lambda & 0 & 0 \\
                                0 & \lambda & 0 \end{array} \right) \virg
h   = \left( \begin{array}{ccc} 1 & 0 & 0 \\
                                0 & 0 & 0 \\
                                0 & 0 &-1 \end{array} \right)
\label{trentuno}
\en
By choosing instead to
represent the $h,e_0,e_1$ matrices in the two possible standard
gradations ($c_0$ or $c_1$), one obtains that the first component of
eq.(\ref{quattordici}) satisfies the first and second of
(\ref{tredici}) respectively.

The expressions (\ref{tredici}) are obtained if one takes $h,e_0,e_1$ in the
fundamental representation. One can however give meaning to (\ref{quattordici})
for general representations of $A_2^{(2)}$.
The irreducible representations $\pi_s$ relevant here are labelled by
an integer $s\geq 0$. From the solution to the equation
$\cL\Psi(u)=0$, the monodromy matrix can be easily written:
\eq
\bM_s(\lambda) = \pi_s \left\{ e^{2\pi ikh} \cP \exp \lambda \int_0^{2\pi} du
(e^{-2\phi(u)}e_0 + e^{\phi(u)}e_1) \right\}.
\en
Its ``improved'' form
$\bL_s(\lambda) = \pi_s(e^{-i\pi k h}) \bM_s(\lambda)$
satisfies the Poisson bracket algebra:
\eq
\{\bL_s(\lambda) \stackrel{\otimes}{,} \bL_{s'}(\mu)\}=
[r_{s,s'}(\lambda\mu^{-1}),\bL_s(\lambda)\otimes \bL_{s'}(\mu)]
\label{diciannove}
\en
where $r_{s,s'}$ is the classical r-matrix associated with $A_2^{(2)}$ \cite{jim}.
It follows from (\ref{diciannove}) that the trace of the monodromy matrix
$\bT_s(\lambda)=\mbox{Tr} \bM_s(\lambda)$ closes an abelian Poisson
bracket algebra
\eq
\{\bT_s(\lambda),\bT_{s'}(\mu)\}=0
\en
One can check that this $\bT$-operator in the fundamental representation
($\bT_1$) is
indeed the generating function of the infinite number of classical IM
of the $A_2^{(2)}$ mKdV.

Let us turn now to the quantum case. The quantization procedure \cite{fatluk} consists
essentially in using the quantum deformations $(A_2^{(2)})_{q_{\pm}}$
instead of $A_2^{(2)}$, where:
\eq\nn
q_{\pm}=
e^{i\pi\beta_{\pm}^2} \virg
\beta_{\pm}=\sqrt{\frac{1-c}{24}} \pm \sqrt{\frac{25-c}{24}}
\virg \beta_+=\frac{1}{\beta_-}
\en
and a free scalar field:
\eq\label{cinque}
\phi(u)=Q + Pu + i\sum_{n \not= 0} \frac{a_n}{n} e^{inu}
\en
\eq\nn
[Q,P]=i{\beta_{\pm}^2\ov 2} \virg
[a_n,a_m]={\beta_{\pm}^2\ov 2}n\delta_{n,-m}.
\en
The Miura transformation translates, at the quantum level,
into the celebrated Feigin-Fuchs construction of the CFT through the
screened free boson (\ref{cinque}):
\eq\nn
-\beta_{\pm}^2 T(u) =
:\phi'(u)^2: + (1-\beta_{\pm}^2) \phi''(u) + \frac{\beta_{\pm}^2}{24}.
\en
Following \cite{fatluk,blz} we define
the quantum monodromy matrix and the $\bL$-operator as follows
\eq
\bL_s(\lambda)=\pi_s\left\{e^{i\pi Ph}{\cal P}\exp \lambda \int du
(:e^{-2\phi}: q^h e_0 + :e^{\phi}: q^{-h/2} e_1)\right\},
\label{ventuno}
\en
\eq\nn
\bM_s(\lambda)=\pi_s(e^{i\pi Ph})\bL_s(\lambda)
\en
where $\phi(u)$ is a free massless scalar field like (\ref{cinque}),
and $e_0,e_1,h$
are now Cartan-Chevalley generators of the affine quantum algebra
$(A_2^{(2)})_q$ for $q=e^{i\pi\beta^2}$:
$$
[e_i,f_j]=\delta_{ij}[h_j] \virg [h_i,e_j]=a_{ij}e_j \virg [h_i,f_j]=-a_{ij}f_j
\virg i,j=0,1
$$
\eq\nn
h=h_0=-2h_1 \virg a_{00}=a_{11}=2 \virg a_{01}=-4 \virg
a_{10}=-1
\en
where $[a]=\frac{q^a-q^{-a}}{q-q^{-1}}$.
We shall comment later on the relation between
$\beta$ and $c$.
Similarly to the classical case we can give meaning to
(\ref{ventuno}) in any
irreducible representation of $(A_2^{(2)})_q$.

We briefly describe here these representations. Denote the basic vector of the representation $\pi_s$ as
$|j,m\rangle$, $j=0,\frac{1}{2},1,...,\frac{s}{2}$, $m=-j,-j+1,...,j$.
We define the action of the generators of $(A_2^{(2)})_q$ on this basis by:
\bea
h|j,m\rangle &=& 2m |j,m\rangle \nonumber \\
e_0|j,m\rangle &=& \sqrt{[j-m][j+m+1]}|j,m+1\rangle \nonumber \\
f_0|j,m\rangle &=& \sqrt{[j+m][j-m+1]}|j,m-1\rangle \label{ventiquattro}\\
e_1|j,m\rangle &=& \sqrt{e(j)[j-m+1]}|j+\half,m-\half\rangle +
                   \sqrt{e(j-\half)[j+m]}|j-\half,m-\half
                   \rangle \nonumber \\
f_1|j,m\rangle &=& \sqrt{e(j)[j+m+1]}|j+\half,m+\half\rangle +
                   \sqrt{e(j-\half)[j-m]}|j-\half,m+\half
                   \rangle \nonumber
\eea
and:
\eq\nn
e(j)=\frac{[j+1][j+\frac{1}{2}]}{[\frac{1}{2}][2j+2][2j+1]}
\{[{s\ov 2}+1] + [{s\ov2}+\half] - [j+1] - [j+\half]\}
\en
is the solution of the recursive equation:
\eq
\frac{[2j]}{[j]}e(j-\half) - \frac{[2j+2]}{[j+1]}e(j) = 1 \virg
e({s\ov 2})=0.
\label{venticinque}
\en
One can verify by direct calculation that the definition (\ref{ventiquattro})
indeed ensures the closing of the $(A_2^{(2)})_q$ algebra provided
(\ref{venticinque}) is satisfied.

Let us now return to the operator (\ref{ventuno}). It can be shown that
$\bL_s(\lambda)$ so constructed satisfies the quantum Yang-Baxter equation:
\eq
\bR_{ss'}(\lambda\mu^{-1})(\bL_s(\lambda)\otimes\buno)
(\buno\otimes\bL_{s'}(\mu)) = (\buno\otimes\bL_{s'}(\mu))
(\bL_s(\lambda)\otimes\buno) \bR_{ss'}(\lambda\mu^{-1})
\label{ventisei}
\en
where now $\bR_{ss'}$ is the quantum
$\bR$-matrix associated with $(A_2^{(2)})_q$.

The definition (\ref{ventuno}) is understood
in terms of power series expansion in $\lambda$:
\eq
\bL_s(\lambda)=\pi_s\left\{e^{i\pi Ph} \sum_{k=0}^{\infty} \lambda^k
\int_{2\pi\geq u_1 \geq ... \geq u_k \geq 0} du_1 ... du_k K(u_1) ... K(u_k)
\right\}
\label{ventotto}
\en
where:
\eq\nn
K(u)=:e^{-2\phi(u)}:q^h e_0 + :e^{\phi(u)}: q^{-h/2} e_1
\en
Similarly to the case considered in \cite{blz}
an estimate of the singularity properties of the integrands shows
that the integrals in (\ref{ventotto})
should be convergent for $\beta^2<\half$
and need regularization for $\beta^2\geq\half$. The analytic properties of the
eigenvalues of $\bT_s$
are strongly influenced by this regularization.

A direct consequence of (\ref{ventisei}) is that
the trace of the quantum monodromy matrix:
\eq
\bT_s(\lambda) \equiv \mbox{\rm Tr} \bM_s(\lambda)
\label{trenta}
\en
defines a commuting operator $[\bT_s(\lambda),\bT_{s'}(\mu)]=0$ which
is the generator of quantum local and non-local IM.
In the case of the fundamental representation
$\pi_1$, one easily computes $\bT_1(\lambda)$ in terms of power
series expansion around $\lambda=0$:
\eq
\bT_1(\lambda) = 2 \cos 2\pi P + \sum_{n=1}^{\infty} \lambda^{3n} Q_n
\label{trentadue}
\en
where:
\bea
Q_n &=& q^{3n/2} \int_{2\pi\geq u_1 \geq ... \geq u_{3n} \geq 0}
du_1 ... du_{3n}~\times \label{trentatre}\\
&\times&\left\{e^{2i\pi P} :e^{-2\phi(u_1)}::e^{\phi(u_2)}::e^{\phi(u_3)}: ...
:e^{-2\phi(u_{3n-2})}::e^{\phi(u_{3n-1})}::e^{\phi(u_{3n})}:\right.
\nonumber\\
&+& \left. e^{-2i\pi P} :e^{\phi(u_1)}::e^{\phi(u_2)}::e^{-2\phi(u_3)}: ...
:e^{\phi(u_{3n-2})}::e^{\phi(u_{3n-1})}::e^{-2\phi(u_{3n})}: \right\}
\nonumber
\eea
are the non-local IM. As a consequence of (\ref{trentadue}),
$\bT_1(\lambda)$ is
an entire function of $\lambda^3$. One can show that it also exhibits an
essential singularity at infinity. The analysis of the corresponding
asymptotic expansion should involve a hard Bethe Ansatz calculation.
Our expectation, of course,
is that the coefficients in this expansion should be given
by the quantum version of the local IM \cite{kuper}.

In the general case, eq.(\ref{trenta}) can be computed using the so-called
$\bR$-fusion procedure \cite{kul}.
Here we give only the first non-trivial terms
in the $\lambda$-expansion:
\eq
\bT_s(\lambda)=\frac{\sin\frac{s+2}{2}x \sin\frac{s+1}{2}x}
{\sin x \sin\frac{x}{2}} + \lambda^3 A_s(x,a) Q_1 + O(\lambda^6)
\label{trentaquattro}
\en
where $x=2\pi P$, $a=\pi \beta^2$ and:
\bea
A_s(x,a)&=&\sum_{l=0}^s \frac{1}{8\sin x \sin a \sin\sfrac{a}{4}}
\left[\frac{\sin (x-a)(l+1)}{\sin(x-a)}-
  \frac{\sin(x+a)(l+1)}{\sin(x+a)}\right] \nonumber \\
&\times& \frac{\cos\sfrac{a}{2}\sin\sfrac{a}{2}(s+\sfrac{3}{2})-
      \cos\sfrac{a}{2}\sin\sfrac{a}{2}(l+1)}
     {\cos\sfrac{a}{2}(l+1)\cos\sfrac{a}{2}l}.
\label{trentacinque}
\eea
One can show, using the explicit form (\ref{trentaquattro}) that
$\bT_s(\lambda)$ satisfies
(at least to order $\lambda^3$) the fundamental
relation:
\eq
\bT_s(q^{1/6}\lambda) \bT_s(q^{-1/6}\lambda) =
\bT_{s+1}(\lambda) \bT_{s-1}(\lambda)+
\bT_s(q^{1/3\beta^2}\lambda).
\label{trentasei}
\en
The very nice result is that this equation coincides with the one
conjectured in a completely different fashion in \cite{kun}.
Conversely, it is
interesting to note that, assuming (\ref{trentasei}) as correct and
expanding in $\lambda$, one gets, for each order in the expansion,
new curious identities.

The possible choices of $\beta$ in the $(A_2^{(2)})_q$ case are
dictated by adapting the classical limit of the $A_1$ Feigin-Fuchs
construction to the two possible choices of Miura transformations,
labeled by $c_0$ and $c_1$. Moreover, the classical limit can be
realized in two ways, sending $\beta_+\to\infty$ or
$\beta_-\to\infty$. This gives 4 possibilities in total. One can see
that both operators $\int_0^{2\pi}du:e^{-2\phi}:$ and
$\int_0^{2\pi}du:e^{\phi}:$ commute with the IM \cite{kuper}. Following
the same reasoning as in the $A_1^{(1)}$ case, if one
chooses $:e^{-2\phi}:$ as screening operator, then $:e^{\phi}:$ is the
perturbing field. The two parametrizations with $\beta_{\pm}$
correspond to the two possible choices for the screening operator and
give $:e^{\phi}:=\phi_{1,2}$ and $:e^{\phi}:=\phi_{2,1}$ respectively.
However, one is also free to choose $:e^{\phi}:$ as screening
operator. This leads to the identification of $:e^{-2\phi}:$ with
$\phi_{1,5}$ for $\beta_-$ and $\phi_{5,1}$ for $\beta_+$.

In general (\ref{trentasei}) could be considered as a
recursive relation for $\bT_s(\lambda)$. For $q$ root of 1 however
the quantum group truncation operates and (\ref{trentasei})
becomes a closed system of functional equations. This important fact
allows one to do a crucial conjecture: the
solutions of (\ref{trentasei}) having the suitable asymptotic behavior
and analytic properties (see \cite{blz} for details) are the whole set of
eigenvalues $t_s(\lambda)$ of $\bT_s(\lambda)$ in the Hilbert space of the
model. Therefore the system (\ref{trentasei}), together with its analog presented in \cite{blz}, provide a
complete description of the (chiral) Hilbert space of $c<1$ RCFT's, i.e.
minimal models. These ideas were further developed in \cite{dfmr1}-\cite{dfmr4}.

\subsection{Expectation values of descendent fields in the Bullough-Dodd model and related perturbed conformal field theories}

The purpose of this Section is to calculate the VEV's of the simplest
non-trivial descendent fields in the Bullough-Dodd (BD) model which is
generally described by the following action in the Euclidean space :
\beqa
{\cal A}_{BD} = \int d^2x \big[\frac{1}{16\pi}(\partial_\nu\varphi)^2 +
\mu e^{b\varphi} + \mu' e^{-\frac{b}{2}\varphi}\big].\label{actionBD}
\eeqa
Here, the parameters $\mu$ and $\mu'$ are introduced, as the two operators do
not renormalize in the same way, on the contrary to any simply-laced affine
Toda field theory. This model has attracted over the years a certain
interest, in particular in connection with perturbed minimal models :
 $c<1$ minimal CFT perturbed  by the operators
$\Phi_{12}$, $\Phi_{21}$ or $\Phi_{15}$ can be obtained by a quantum group (QG) restriction
of imaginary Bullough-Dodd model \cite{smir,kmmus,flzz} with
special values of the coupling. We will use this property to
deduce the VEV's of the descendents in the
following perturbed minimal models :
\beqa
{\cal{A}} &=& {\cal{M}}_{p/p'} + {\lambda} \int d^2x
\Phi_{12}\ ,\label{action}\\
{\hat{\cal{A}}} &=& {\cal{M}}_{p/p'} +
{\hat\lambda} \int d^2x \Phi_{21}\ \label{actiontilde}\\
\mbox{or}\ \ \ \ \ \ {\tilde{\cal{A}}} &=& {\cal{M}}_{p/p'} +
{\tilde\lambda} \int d^2x \Phi_{15}\ ,\label{actionhat}
\eeqa
where  we denote respectively $\Phi_{12}$, $\Phi_{21}$ and $\Phi_{15}$
as specific primary operators of the unperturbed minimal model
${\cal{M}}_{p/p'}$ and the
parameters $\lambda$, $\hat\lambda$ and $\tilde\lambda$
\ characterize the strength of the perturbation.

Similarly to the ShG model \cite{frad}, the BD model can be regarded as a
relevant perturbation of a Gaussian CFT. We remind that in this free field theory, the field
is normalized such that:
\beqa\nn
<\varphi_(z,{\ovz})\varphi(0,0)>_{Gauss}=-2\log(z{\ovz}).
\eeqa
and we have the classical equation of motion :
\beqa
\partial {\overline \partial}\varphi = 0.\label{motion}
\eeqa
Instead of considering the action (\ref{actionBD}) we turn directly to the
case of an imaginary coupling constant which is the most interesting for
our purpose. The perturbation is then relevant if \
$0<\beta^2<1$ ($b=i\beta$). Although the model (\ref{actionBD})
for real coupling is very different from the one with imaginary coupling
in its physical content, there are good
reasons to believe that the expectation values obtained in the real
coupling case provide also the expectation values for the imaginary coupling.
The calculation of the VEVs in both cases ($b$ real or imaginary) within
 the standard perturbation theory agree through the identification
 $b=i\beta$ \cite{flzz}. With this substitution in (\ref{actionBD}),
the general short distance OPE for two arbitrary primary fields
$e^{i\alpha_1\varphi}(x)$ and $e^{i\alpha_2\varphi}(y)$ takes the form :
\beqa
e^{i\alpha_1\varphi}(x)e^{i\alpha_2\varphi}(y) &=&
\sum_{n=0}^{\infty}\big\{
C_{\alpha_1\alpha_2}^{n,0}(r)e^{i(\alpha+n\beta)\vph}(y)+ \ ... \big\}\nonumber\\
&+&\sum_{n=1}^{\infty}\big\{{C'}_{\alpha_1\alpha_2}^{\ n,0}(r)e^{i(\alpha-\frac{n\beta}{2})\vph}(y)+ \ ...\big\}\nonumber \\
&+&
\sum_{n=1}^{\infty}\big\{D_{\alpha_1\alpha_2}^{n,0}(r)e^{i(\alpha+(n-\frac{1}{2})\beta)\vph}(y)+
 \ ...\big\}\label{opep}
\eeqa
where $\alpha=\alpha_1+\alpha_2$, \ $r=|x-y|$ and  the dots in each term
stand for the contributions of the descendants of each field. The
different coefficients in eq. (\ref{opep}) are computable within the
conformal perturbation theory (CPT) \cite{alz,df}. We obtain :
\beqa
C^{n,0}_{\alpha_1\alpha_2}(r)&=&
 {\mu}^n r^{4\alpha_1\alpha_2+4n\beta(\alpha_1+\alpha_2)+2n(1-\beta^2)+2n^2\beta^2}
f^{n,0}_{\alpha_1\alpha_2}\big(\mu(\mu')^2
r^{6-3\beta^2}\big)\label{coef};\\
{C'}_{\alpha_1\alpha_2}^{\ n,0}(r)&=&
{\mu'}^n
r^{4\alpha_1\alpha_2-2n\beta(\alpha_1+\alpha_2)+2n(1-\frac{\beta^2}{4})+
\frac{n^2\beta^2}{2}}
{f'}^{n,0}_{\alpha_1\alpha_2}\big(\mu(\mu')^2 r^{6-3\beta^2}\big);\nonumber\\
D_{\alpha_1\alpha_2}^{\ n,0}(r)&=&
{\mu'}{\mu}^n
r^{4\alpha_1\alpha_2+4(n-\frac{1}{2})\beta(\alpha_1+\alpha_2)+2n(1-2\beta^2)+2+2n^2\beta^2}
{g}^{n,0}_{\alpha_1\alpha_2}\big(\mu(\mu')^2 r^{6-3\beta^2}\big)\nonumber
\eeqa
where any function $h\in\{f,{f'},g\}$ admits a power series expansion :
\beqa
h^{n,0}_{\alpha_1\alpha_2}(t)=\sum_{k=0}^{\infty}h^{n,0}_{k}(\alpha_1,\alpha_2)t^k.\label{serie}
\eeqa
Each coefficient in (\ref{coef}) is expressed in terms of Coulomb type
integrals. The corresponding leading terms are respectively given by :
\beqa
f^{n,0}_{0}(\alpha_1,\alpha_2)&=&j_n(\val_1\beta,\val_2\beta,\beta^2)\ \ \ \ \mbox{for}\ \ \ n\neq
0\ ;\label{func}\\
{f'}^{n,0}_{0}(\alpha_1,\alpha_2)&=&j_n(-\frac{\val_1\beta}{2},-\frac{\val_2\beta}{2},
\frac{\beta^2}{4})\ ;\nonumber\\
{g}^{n,0}_{0}(\alpha_1,\alpha_2)&=&
{\cal F}_{n,1}(\val_1\beta,\val_2\beta,\beta^2)\nonumber
\eeqa
where we introduced the Dotsenko-Fateev integrals $j_{n}(a,b,\rho)$ and ${\cal F}_{n,m}(a,b,\rho)$ \cite{df}.
The integrals $j_n(a,b,\rho)$ have been evaluated explicitly in
\cite{df} with the result :
\beqa
j_{n}(a,b,\rho)&=&\pi^n\big(\gamma(\rho)\big)^{-n}\prod_{k=1}^n \ \gamma(k\rho)
\times \label{jn}\\
&&\ \ \ \prod_{k=0}^{n-1}\gamma(1+2a+k\rho) \gamma(1+2b+k\rho) \gamma(-1-2a-2b-(n-1+k)\rho).\nonumber
\eeqa
As we already mentioned, the
next sub-leading terms in (\ref{opep}) involve the descendent
fields. There are four independent second-level descendent fields in BD :
\beqa
&&(\partial\vph)^2({\overline\partial}\vph)^2e^{i\alpha\vph}\ ;\ \ \ \ \
(\partial\vph)^2({\overline\partial}^2\vph)e^{i\alpha\vph}\ ;\label{4fields}\\
&&(\partial^2\vph)({\overline\partial}\vph)^2e^{i\alpha\vph}\ ;\ \ \ \ \
(\partial^2\vph)({\overline\partial}^2\vph)e^{i\alpha\vph}.\nonumber
\eeqa
Similarly to the SG (or ShG) case, using (\ref{motion}) it is easy to
show that linear combinations of these descendent fields can be written in
terms of total derivatives of local fields. As a result,
the VEVs of the composite fields (\ref{4fields}) can all be expressed in terms of
a single VEV, say :
\beqa
<(\partial\vph)^2({\overline\partial}\vph)^2e^{i\alpha\vph}>_{BD}.\label{twovev}
\eeqa

Let us make an important observation. The second sub-leading terms in the OPE
(\ref{opep}) appear to be the third order descendants of the primary fields.
Analogously to the previous discussion linear combinations of them can be
expressed in terms of total derivatives of some local fields. As before, all the
corresponding VEVs can be expressed through
$<(\partial\varphi)^3(\bar\partial\varphi)^3e^{i\alpha\varphi}>$. Unlike the SG
case, it is non-vanishing due to the absence of a conserved charge of spin 3 in
the BD model. We will consider in more details this VEV later in this Section.

One can now write the short-distance expansion for the
two-point function :
\beqa\nn
{\cal G}_{\val_1\val_2}(r)=<e^{i\val_1\vph}(x)e^{i\val_2\vph}(y)>_{BD}\ \ \ \ \mbox{with} \ \ r=|x-y|
\eeqa
by taking the expectation value of the r.h.s. of the OPE (\ref{opep}) in the BD
model with imaginary coupling. Due to the previous discussion,
the first non-vanishing contribution of the VEVs of lowest descendent
fields in the r.h.s. of the VEV of (\ref{opep}) correspond to the following terms :
\beqa\nn
&&C^{n,2}_{\val_1\val_2}(r)
<(\partial\vph)^2({\overline\partial}\vph)^2
e^{i(\alpha+n\beta)\vph}>_{BD}\ ; \\
&&{C'}^{\ n,2}_{\val_1\val_2}(r)
<(\partial\vph)^2({\overline\partial}\vph)^2
e^{i(\alpha-\frac{n\beta}{2})\vph}>_{BD}\ ; \nonumber\\
&&D^{n,2}_{\val_1\val_2}(r)
<(\partial\vph)^2({\overline\partial}\vph)^2
e^{i(\alpha+(n-\frac{1}{2})\beta)\vph}>_{BD}\ ,\nonumber
\eeqa
respectively. These coefficients also admit expansion similar to eqs.
 (\ref{coef}), (\ref{serie}) and (\ref{func}).
Finally, the short-distance ($r\rightarrow 0$)
expansion of the two-point correlation function in
the BD model with imaginary coupling writes :
\beqa
&&{\cal G}_{\val_1\val_2}(r)= {\cal G}_{\val_1+\val_2} r^{4\val_1\val_2}
\Big\{ 1 + {\cal F}_{1,2}(\val_1\beta,\val_2\beta,\beta^2)
\mu(\mu')^2r^{6-3\beta^2}  +\frac{(\val_1\val_2)^2}{4} {\cal H}(\val_1+\val_2)r^4
\nonumber \\
&& \ \ \ \ \ \ \qquad \ \  \ \qquad \qquad \ - \frac{\alpha_1^2\alpha_2^2(\alpha_1\!-\!\alpha_2)^2}{144}{\cal
K}(\alpha_1+\alpha_2)r^{6}
 + O\big(\mu^2(\mu')^4r^{12-6\beta^2}\big) \Big\}\nonumber\\
&& + \sum_{n=1}^{\infty} {\mu}^n
r^{4\alpha_1\alpha_2+4n\beta(\alpha_1+\alpha_2)+2n(1-\beta^2)+2n^2\beta^2}
j_n(\val_1\beta,\val_2\beta,\beta^2)\nonumber \\
&& \ \ \ \ \ \ \ \ \  \ \ \ \ \ \ \  \ \ \ \ \  \ \ \ \ \ \ \ \ \ \ \ \
\ \ \ \ \ \ \  \ \ \ \ \ \ \   \times \ \ {\cal G}_{\val_1+\val_2+n\beta}
\ \Big\{\ 1+\ O\big(\mu(\mu')^2r^{6-3\beta^2}\big)\Big\}\nonumber \\
&& + \sum_{n=1}^{\infty} {\mu'}^n
r^{4\alpha_1\alpha_2-2n\beta(\alpha_1+\alpha_2)+2n(1-\frac{\beta^2}{4})+\frac{n^2\beta^2}{2}}
j_n(-\frac{\val_1\beta}{2},-\frac{\val_2\beta}{2},\frac{\beta^2}{4})\label{twop} \\
&& \ \ \ \ \ \ \ \ \  \ \ \ \ \ \ \  \ \ \ \ \  \ \ \ \ \ \ \ \ \ \ \ \
\ \ \ \ \ \ \  \ \ \ \ \ \ \   \times \ \  {\cal G}_{\val_1+\val_2-\frac{n\beta}{2}}
\ \Big\{\ 1+\ O\big(\mu(\mu')^2r^{6-3\beta^2}\big) \Big\}\nonumber \\
&& + \sum_{n=1}^{\infty} {\mu}^n{\mu'}
r^{4\alpha_1\alpha_2+4(n-\frac{1}{2})\beta(\alpha_1+\alpha_2)+2n(1-2\beta^2)+2+2n^2\beta^2}
{\cal F}_{n,1}(\val_1\beta,\val_2\beta,\beta^2)\nonumber \\
&& \ \ \ \ \ \ \ \ \  \ \ \ \ \ \ \  \ \ \ \ \  \ \ \ \ \ \ \ \ \ \ \ \
\ \ \ \ \ \ \  \ \ \ \ \ \ \   \times \ \  {\cal G}_{\val_1+\val_2+(n-\frac{1}{2})\beta}
\ \Big\{\ 1+\ O\big(\mu(\mu')^2r^{6-3\beta^2}\big) \Big\}\nonumber
\eeqa
where we defined ${\cal H}(\alpha)$ and ${\cal K}(\alpha)$
by the ratios :
\beqa
&&{\cal H}(\alpha)=
\frac{<(\partial\vph)^2({\overline\partial}\vph)^2e^{i\alpha\vph}>_{BD}}
{<e^{i\alpha\vph}>_{BD}}\ ,\label{H}\\
&&{\cal K}(\alpha)=
\frac{<(\partial\vph)^3({\overline\partial}\vph)^3e^{i\alpha\vph}>_{BD}}
{<e^{i\alpha\vph}>_{BD}}\label{Ka}
\eeqa
and\ ${\cal G}_\val=<e^{i\alpha\vph}>_{BD}$ is the VEV of the exponential
field in the BD model. A closed analytic expression for this latter VEV
has been proposed in \cite{flzz}:
\beqa
<\!e^{i\alpha\vph}\!>_{BD}\!\!\!&=&\!\!\Big[\frac{\mu'}{\mu}\frac{2^{\frac{-\beta^2}{2}}
\Gamma(1+\beta^2)\Gamma(1-\frac{\beta^2}{4})}{\Gamma(1-\beta^2)
\Gamma(1+\frac{\beta^2}{4})}\Big]^{\frac{2\alpha}{3\beta}}
\Big[\frac{m\Gamma(1-\frac{\beta^2}{6-3\beta^2})\Gamma(\frac{2}{6-3\beta^2})}
{2^{\frac{2}{3}}\sqrt{3}\Gamma(\frac{1}{3})}\Big]^{-\alpha\beta+2\alpha^2}
\times\nonumber \\
\!\!\!&&\!\!\!\exp\Big[\int^{+\infty}_0 \frac{dt}{t}
\Big(\frac{\sinh((2-\beta^2)t)\Psi(t,\alpha)}{\sinh(3(2-\beta^2)t)\sinh(2t)\sinh(\beta^2t)}-2\alpha^2e^{-2t}\Big)\Big]
\label{VEVBD}
\eeqa
where:
\beqa
\Psi(t,\alpha)&=&-\sinh(2\alpha\beta t)\big(\sinh((4-\beta^2-2\alpha\beta)t)
-\sinh((2-2\beta^2+2\alpha\beta)t)+\nonumber\\
&&\sinh((2-\beta^2-2\alpha\beta)t)-\sinh((2-\beta^2+2\alpha\beta)t)-\sinh((2+\beta^2-2\alpha\beta)t)\big).\nonumber
\eeqa
Its integral representation is well defined if :
\beqa
-\frac{1}{2\beta}\ <\ {\mathfrak R}e(\alpha)\ <\ \frac{1}{\beta}\label{cond}
\eeqa
and obtained by analytic continuation outside this domain.

It is then straightforward to obtain the result associated with the action
(\ref{actionBD}) i.e. for {\it real} values of the coupling constant $b$ which
follows from the obvious substitutions :
\beqa
&&\beta\rightarrow -ib ;\ \ \ \val_1\rightarrow -ia_1\ ;\ \ \ \val_2\rightarrow
-ia_2\ ;\label{substit}\\
&&\mu\rightarrow -\mu\ ;\ \ \ \ \mu'\rightarrow -\mu' \ .\nonumber
\eeqa
In the (Gaussian) free field theory, the composite fields
(\ref{4fields}) are spinless with scale dimension :
\beqa
D\equiv \Delta+{\overline \Delta}=2\alpha^2+4.\label{dim}
\eeqa
For generic value of the coupling $\beta$ some divergences arise in
the VEVs of the fields
(\ref{4fields}) due to the perturbation in (\ref{actionBD}) with
imaginary coupling. They are generally cancelled if we add specific
counterterms which contain spinless local fields with cutt-off
dependent coefficients. For $0<\beta^2<1$ the perturbation becomes
relevant and a finite number of lower scale dimension couterterms are
then sufficient. However, this procedure is regularization scheme dependent,
i.e. one can always add finite counterterms. For generic values of
$\alpha$ this ambiguity in the definition of the renormalized expression
for the fields (\ref{4fields}) can be eliminated by fixing their scale
dimensions to be (\ref{dim}). In the BD model
with imaginary coupling, this situation arises if two fields, say
${\cal O}_{\val}$ and  ${\cal O}_{\val'}$, satisfy the resonance condition :
\beqa\nn
D_\val=D_{\val'} + 2n(1-\beta^2) + 2m(1-\frac{\beta^2}{4})\ \ \ \ \mbox{with}\
\ \ \ (n,m)\in {\mathbb N}
\eeqa
associated with the ambiguity :
\beqa\label{reson}
{\cal O}_{\val}\longrightarrow{\cal O}_{\val} + {\mu}^{n}{\mu'}^{m}{\cal O}_{\val'}.
\eeqa
In this specific case one says that the renormalized
field ${\cal O}_\val$ has an $(n|m)$-th {\it resonance} \cite{frad} with the field ${\cal O}_{\val'}$.
Due to the condition (\ref{cond}) and using (\ref{dim}) we find
immediately that a resonance can appear between the descendent
field $(\partial\varphi)^2(\bar\partial\varphi)^2e^{i\alpha\varphi}$ and the
following primary fields :
\beqa
(i) \ \ \ \ &&e^{i(\val+\beta)\vph}\ \ \ \ \mbox{i.e.}\ \  (n|m)=(1|0) \ \ \
\mbox{for}\ \ \ \alpha=\frac{1}{2\beta}\ ;\label{nm}\\
(ii) \ \ \ \ &&e^{i(\val+2\beta)\vph}\ \ \ \mbox{i.e.}\ \  (n|m)=(2|0) \ \ \
\mbox{for}\ \ \ \alpha=-\frac{\beta}{2}\ ;\nonumber\\
(iii) \ \ \ \ &&e^{i(\val-\beta)\vph}\ \ \ \ \mbox{i.e.}\ \  (n|m)=(0|2) \ \ \
\mbox{for}\ \ \ \ \alpha=\frac{\beta}{4}\ ;\nonumber\\
(iv) \ \ \ \ &&e^{i(\val+\frac{\beta}{2})\vph}\ \ \ \ \mbox{i.e.}\ \  (n|m)=(1|1) \ \ \
\mbox{for}\ \ \ \ \alpha=\beta\ .\nonumber
\eeqa
If we now look at the expression (\ref{twop}), we notice that the
contribution (\ref{H}), brought by the second level descendent field,
and that of any of the
exponential fields in  $(i)$, $(ii)$, $(iii)$ and $(iv)$, have
the same power behaviour in $r$
($r^{4\val_1\val_2 + 4}$) at short-distance for
the corresponding values of $\alpha$ in  (\ref{nm}). The integrals which
appear in these contributions and their corresponding poles are,
respectively :
\beqa
&&j_1(\val_1\beta,\val_2\beta,\beta^2)\ \ \ \ \ \ \ \ \ \ \ \ \mbox{with the pole}\ \ \
\alpha=\frac{1}{2\beta}\ ;\label{jpole}\\
&&j_2(\val_1\beta,\val_2\beta,\beta^2)\ \ \ \ \ \ \ \ \ \ \ \ \mbox{with
the pole}\ \ \
\alpha=-\frac{\beta}{2}\ ;\nonumber\\
&&j_2(-\frac{\val_1\beta}{2},-\frac{\val_2\beta}{2},\frac{\beta^2}{4})\ \ \ \ \
\mbox{with the pole}\ \ \
\alpha=\frac{\beta}{4}\ ;\nonumber\\
&&{\cal F}_{1,1}(\val_1\beta,\val_2\beta,\beta^2)\ \ \ \ \ \ \ \ \
 \mbox{with the pole}\ \ \
\alpha=\beta\ .\nonumber
\eeqa
By analogy with the SG (or ShG) model, one expects that the VEV
(\ref{H}) (and similarly for the real coupling case) exhibits, at least,
the same poles in order that the divergent contributions compensate each
other. This last requirement leads for instance to the relations:
\beqa
(i')\ \ \ && {\cal R}es_{\alpha=\frac{1}{2\beta}}{\cal H}(\val) =
8\pi\beta^3\mu\frac{{\cal G}_{\val+\beta}}{{\cal
G}_\val}|_{\val=\frac{1}{2\beta}}\ ;\label{residu}\\
(ii')\ \ \ && {\cal R}es_{\alpha=-\frac{\beta}{2}}{\cal H}(\val) =
-32\pi^2\beta^3\mu^2\frac{\gamma(2\beta^2)}{\gamma(\beta^2)}\gamma(-1-\beta^2)\frac{{\cal G}_{\val+2\beta}}{{\cal
G}_\val}|_{\val=-\frac{\beta}{2}}\ ;\nonumber\\
(iii')\ \ \ && {\cal R}es_{\alpha=\frac{\beta}{4}}{\cal H}(\val) =
4\pi^2\beta^3{\mu'}^2\frac{\gamma(\beta^2/2)}{\gamma(\beta^2/4)}\gamma(-1-\beta^2/4)\frac{{\cal G}_{\val-\beta}}{{\cal
G}_\val}|_{\val=\frac{\beta}{4}}\ ;\nonumber\\
(iv')\ \ \ && {\cal R}es_{\alpha=\beta}{\cal H}(\val) =
-\frac{4}{(\alpha_1\alpha_2)^2}\mu{\mu'}\frac{{\cal
G}_{\val+\frac{\beta}{2}}}{{\cal G}_\val}|_{\val=\beta}{\cal
R}es_{\alpha=\beta}\
{\cal F}_{1,1}(\alpha_1\beta,\alpha_2\beta,\beta^2)\ .\nonumber
\eeqa
These last conditions will be used to
fix the normalization of the VEV (\ref{H}). Let us now turn to the evaluation of
(\ref{H}) which plays an important role in the two-point function (\ref{twop}).

The BD model (\ref{actionBD}) can be regarded as two different
perturbations of the Liouville field theory \cite{flzz}. First, one can consider
the Liouville action :
\beqa
{\cal A}_{L}^{(1)} = \int d^2x \big[\frac{1}{16\pi}(\partial_\nu\varphi)^2 +
\mu e^{b\varphi}\big].\label{liouvo}
\eeqa
The perturbation is then identified with $e^{-\frac{b}{2}\vph}$.
Alternatively, we can take :
\beqa
{\cal A}_{L}^{(2)} = \int d^2x \big[\frac{1}{16\pi}(\partial_\nu\varphi)^2 +
\mu' e^{-\frac{b}{2}\varphi}\big] \label{liouvtwo}
\eeqa
as the initial action and consider $e^{b\varphi}$ as a perturbation.
Using the first picture, the holomorphic stress-energy tensor :
\beqa
T(z)=-\frac{1}{4}(\partial\vph)^2 +
\frac{Q}{2}\partial^2\vph\label{strt}
\eeqa
ensures the local conformal invariance of the Liouville field theory
(\ref{liouvo}) and similarly for the anti-holomorphic part. The
exponential fields $e^{a\vph}$ are spinless primary fields with
conformal dimension :
\beqa\nn
\Delta=a(Q-a).
\eeqa
The property of reflection relations which relates operators with the
same quantum numbers is a characteristic of the CFT. Using the conformal perturbation theory
(CPT) framework, one expects that similar relations are also satisfied in
the perturbed case (\ref{actionBD}). With the change $b\rightarrow -b/2$
in (\ref{strt}) and using the
second picture (\ref{liouvtwo}), one assumes that the VEV of the exponential field
$<e^{a\vph}>_{BD}$ satisfies simultaneously the following
two functional equations:
\beqa
&&<e^{a\vph}>_{BD}\ \ \ = \ \ R(a)<e^{(Q-a)\vph}>_{BD}\ ;\label{sys}\\
&&<e^{-a\vph}>_{BD}\ =\ \ R'(a)<e^{(-Q'+a)\vph}>_{BD}\nonumber
\eeqa
with:
\beqa
Q=\frac{1}{b}+b\ \ \ \ \ \mbox{and}\ \ \ \ \ Q'=\frac{2}{b}+\frac{b}{2}.
\eeqa
The functions $R(a)$, $R'(a)$ are the ``reflection amplitudes''.
An exact expression for $R(a)$ was presented in
\cite{boot1}. $R'(a)$ is obtained from $R(a)$ by the substitutions
$b\rightarrow\frac{b}{2}$ and $\mu\rightarrow\mu'$. Under certain
assumptions about the analytic properties of the VEV, the system (\ref{sys})
was solved and the VEV for these exponential
fields was derived in \cite{flzz}.

Let us denote the descendent fields :
\beqa
L_{[n]}{\overline L}_{[m]}
e^{a\vph}\equiv L_{-n_1}...L_{-n_1}{\overline L}_{-m_1}...{\overline L}_{-m_K} e^{a\vph}\label{Lnm}
\eeqa
where $[n]=[-n_1,...,-n_N]$ \ and \ $[m]=[-m_1,...,-m_K]$ are arbitrary strings
and $L_n$, ${\overline L}_n$ are the standard Virasoro generators.
The descendent fields (\ref{Lnm}) and the ones obtained after the
reflection $a\rightarrow Q-a$ possess the same quantum numbers.
Consequently, using the arguments of \cite{flzz,frad} based on the CPT
framework, one also expects that their VEVs in the perturbed theory
(\ref{actionBD}) satisfy the following ``reflection relation'' :
\beqa
<L_{[n]}{\overline L}_{[m]}e^{a\vph}>_{BD}=R(a)<L_{[n]}
{\overline L}_{[m]}e^{(Q-a)\vph}>_{BD}.\label{refLnm}
\eeqa
However, it is more convenient to use the basis :
\beqa
(\partial^{n_1}\vph)...(\partial^{n_N}\vph)
({\overline\partial}^{m_1}\vph)...({\overline\partial}^{m_K}\vph)e^{a\vph}.\label{basis}
\eeqa
Using (\ref{motion}) we get:
\beqa
<L_{-2}{\overline
L}_{-2}e^{a\vph}>_{BD}=\frac{1}{16}\big(1+2a(Q+2a)\big)^2
<(\partial\vph)^2({\overline\partial}\vph)^2e^{a\vph}>_{BD}\label{L2L2}
\eeqa
which leads to the following reflection relation :
\beqa\nn
&&\big(1+2a(Q+2a)\big)^2
<(\partial\vph)^2({\overline\partial}\vph)^2e^{a\vph}>_{BD}=\\
&&\ \ \ \ \ \ \ \ \ \ \ \ \ \ \ \ \ \ \ \ \ \ \  \ \ \big(1+2(Q-a)(3Q-2a)\big)^2
<(\partial\vph)^2({\overline\partial}\vph)^2e^{(Q-a)\vph}>_{BD}\nonumber
\eeqa
One can also consider the second picture (\ref{liouvtwo}) where the Liouville
theory has coupling $-\frac{b}{2}$ instead of $b$ and is perturbed by
 $e^{b\vph}$. If we define the analytic continuation of (\ref{H}) :
\beqa
H(a)=\frac{<(\partial\vph)^2({\overline\partial}\vph)^2e^{a\vph}>_{BD}}
{<e^{a\vph}>_{BD}},\label{Hreal}
\eeqa
then the two different pictures provide us the following two functional
relations :
\beqa
H(a)&=&\Big[\frac{(2b+3/b-2a)(3b+2/b-2a)}{(b+2a)(1/b+2a)}\Big]^2H(Q-a),\label{ref}\\
H(-a)&=&\Big[\frac{(b+6/b-2a)(3b/2+4/b-2a)}{(b/2+2a)(2/b+2a)}\Big]^2H(-Q'+a).\nonumber
\eeqa
Notice that these equations are invariant with respect to the symmetry
$b\rightarrow-\frac{2}{b}$ with $a\rightarrow -a$ in agreement with the
well-known self-duality of the BD-model.

As was shown above, the solution of these functional
equations should exhibits, at least, the poles (\ref{jpole}) through the
identification $b=i\beta$ and $a=i\alpha$. Since the solution
of (\ref{ref}) is defined up to a multiplication constant, we naturally
choose to fix it by imposing eqs. (\ref{residu}). We find that the
``minimal'' solution which follows from these constraints is :
\beqa
H(a)&=&-\Big[\frac{m\Gamma(\frac{b^2}{h})\Gamma(\frac{2}{h})}{\Gamma(\frac{1}{3}){\sqrt 3}\ 2^{2/3+3/2}(Q+Q')^2}
\Big]^4\times
\frac{\gamma^2(\frac{1}{3})}{\gamma(\frac{2b^2}{h})\gamma(\frac{4}{h})}\label{VEVHreal}\\ \nonumber\ \ \
&\times& \ \
\gamma\big(\frac{2ba+4}{h}\big)\gamma\big(\frac{-2ba-b^2}{h}\big)
\gamma\big(\frac{2ba+3+b^2}{h}\big)\gamma\big(\frac{-2ba-1}{h}\big)\\ \nonumber \ \ \
&\times& \ \
\gamma\big(\frac{-2ba+2b^2}{h}\big)\gamma\big(\frac{2ba-2}{h}\big)
\gamma\big(\frac{-2ba+2+3b^2/2}{h}\big)\gamma\big(\frac{2ba-b^2/2}{h}\big)
\eeqa
where $h=6+3b^2$ is the ``deformed'' Coxeter number \cite{corr}. Here we
have used the exact relation between the parameters $\mu$ and $\mu'$ in
the action (\ref{actionBD}) and the mass of the particle $m$ \cite{flzz}:
\beqa
m=\frac{2\sqrt 3 \Gamma(1/3)}{\Gamma(1+b^2/h)\Gamma(2/h)}
\big( -\mu\pi\gamma(1+b^2)\big)^{1/h} \big(
-2\mu'\pi\gamma(1+b^2/4)\big)^{2/h}.\label{massmu}
\eeqa
It is
easy to see (taking in account also (\ref{cond}))that for $b=i\beta$ and $a=i\val$, ${\cal H}(\alpha)$  possess
poles located at :
\beqa
\alpha_0\in\{-\frac{\beta}{2},\frac{1}{2\beta},\frac{\beta}{4},
\beta\}.
\eeqa
Accepting the conjecture (\ref{VEVHreal}) and using eq. (\ref{L2L2}) for $a=0$ one
can easily deduce for instance :
\beqa
<T{\overline T}>_{BD}\ \equiv\ <L_{-2}{\overline L}_{-2}{\mathbb I}>_{BD}\
=\ -\pi^2f^2_{BD}\label{TTbar}
\eeqa
where:
\beqa\nn
f_{BD}&=&\frac{m^2}{16{\sqrt 3}\sin(\frac{\pi b^2}{h})\sin(\frac{2\pi}{h})}
\eeqa
is the bulk free energy of the BD model \cite{flzz}.

Let us now turn to the computation of the expectation values of the descendent fields in $\Phi_{12}$,
$\Phi_{21}$ and $\Phi_{15}$ perturbed minimal models.

For imaginary value of the coupling $b=i\beta$, $\mu\rightarrow-\mu$ and
$\mu'\rightarrow -\mu'$ the action of the BD model (\ref{actionBD})
becomes complex.
Whereas it is not clear if it can be defined as a QFT, this model is
known to be integrable and its $S$-matrix was constructed in
\cite{smir}. It is known that this model possess a quantum group symmetry
 $U_q(A_2^{(2)})$ with deformation parameter
$q=e^{i\frac{\pi}{\beta^2}}$ \cite{smir}. An important role is played by
one of its subalgebras $U_q(sl_2)\subset U_q(A_2^{(2)})$. Following \cite{smir}, we can restrict the Hilbert space of states of the
complex BD model at special values of the coupling constant, more precisely
when $q$ is a root of unity, i.e. for :
\beqa\nn
\beta^2=\frac{p}{p'} \ \ \ \ \ \ \ \ \ \ \mbox{or} \ \ \ \ \ \ \ \ \ \beta^2=\frac{p'}{p} \
\ \ \ \ \ \
\mbox{with} \ \ \ 1<p<p'
\eeqa
relative prime integers, in which case the complex BD is identified with the
perturbed minimal models (\ref{action}) or (\ref{actiontilde}),
respectively. In the following, $\Phi_{lk}$ will denote a primary field
of the minimal model ${\cal M}_{p/p'}$.

It is then
straightforward to get the VEV in the model associated with the action
 (\ref{action}) :
\beqa
\frac{<0_s|L_{-2}{\overline L}_{-2}\Phi_{lk}|0_s>}{<0_s|\Phi_{lk}|0_s>}
&=& -\Big[\frac{{\sqrt
3}\pi(\xi+2)M\Gamma(1+\frac{2+2\xi}{3\xi+6})}
{\Gamma(\frac{1}{3})2^{2/3+1/2}\Gamma(\frac{\xi}{3\xi+6})} \Big]^4
\frac{\gamma^2(1/3)}{\gamma(-\frac{2\xi}{3\xi+6})
\gamma(\frac{4+4\xi}{3\xi+6})}\nonumber\\
&&\ \ \ \ \ \ \ \ \ \ \ \ \ \ \ \ \ \ \ \ \ \ \ \ \ \
\times\ \ \ {\cal W}_{12}((\xi+1)l-\xi k)\label{12}
\eeqa
where we denote:
\beqa
\xi=\frac{p}{p'-p}.\label{xi}
\eeqa
Here:
\beqa
{\cal W}_{12}(\eta)=\frac{1}{\xi^{2}(\xi+1)^{2}}
\times w(\eta;\ 5+4\xi,\ 4+2\xi,\ -1-2\xi,\ 1+\xi/2;\ 3\xi+6)\nonumber
\eeqa
and we introduce the useful notation :
\beqa
w(\eta;a_1,a_2,a_3,a_4;g)=\prod_{i=1}^{4}
\gamma\big(\frac{a_i+\eta}{g}\big)\gamma\big(\frac{a_i-\eta}{g}\big).\nonumber
\eeqa
We also use
the particle-breather identification :
\beqa
m=2M\sin\big(\frac{\pi\xi}{3\xi+6}\big).\label{mass}
\eeqa
Here $|0_s>$ is one of the degenerate ground
states of the QFT (\ref{action}). Taking $\Phi_{lk}$ in (\ref{12}) to be the
identity operator, it is easy to get :
\beqa\nn
<T{\overline T}>=-\frac{\pi^2M^4}{48}\frac{\sin^2(\frac{\pi\xi}{3\xi+6})}
{\sin^2(\frac{\pi(2\xi+2)}{3\xi+6})}\ .
\eeqa

In the second restriction $\beta^2=p'/p$, which leads to the
action (\ref{actiontilde}). Along the same line as for the $\Phi_{12}$ perturbation we obtain the
following expression for the VEV in
the model associated  with this action :
\beqa
\frac{<0_s|L_{-2}{\overline L}_{-2}\Phi_{lk}|0_s>}{<0_s|\Phi_{lk}|0_s>}
&=& -\Big[\frac{{\sqrt
3}\pi(1-\xi)M\Gamma(1-\frac{2\xi}{3-3\xi})}
{\Gamma(\frac{1}{3})2^{2/3+1/2}\Gamma(-\frac{\xi+1}{3-3\xi})} \Big]^4
\frac{\gamma^2(1/3)}{\gamma(\frac{2\xi+2}{3-3\xi})
\gamma(\frac{-4\xi}{3-3\xi})}\nonumber\\
&&\ \ \ \ \ \ \ \ \ \ \ \ \ \ \ \ \ \ \ \ \ \ \ \ \ \
 \times\ \ \ {\cal W}_{21}((\xi+1)l-\xi k)\nonumber
\eeqa
with:
\beqa
{\cal W}_{21}(\eta)=\frac{1}{\xi^{2}(\xi+1)^{2}}
\times w(\eta;\ 1-4\xi,\ 2-2\xi,\ 1+2\xi,\ 1/2-\xi/2;\ 3-3\xi)\nonumber
\eeqa
where $|0_s>$ is one of the degenerate ground
states of the QFT (\ref{actiontilde}).

Another subalgebra of $U_q(A^{(2)}_2)$ is the subalgebra
$U_{q^4}(sl_2)$.
One can again restrict the phase space of the complex BD with respect to this
subalgebra for a special value of the coupling :
\beqa\nn
\beta^2=\frac{4p}{p'} \ \ \ \ \ \  \ \ \ \mbox{with} \ \ \ \ \ \ 2p<p'
\eeqa
relative prime integers. Then, for this value of the coupling, the BD
model is identified with the perturbed minimal model with the action
(\ref{actionhat}). Taking the ratio of the VEV of the descendent field of
$\Phi_{lk}$
associated with the action
(\ref{actionhat}) and the VEV of the primary field itself, one obtains :
\beqa
\frac{<0_s|L_{-2}{\overline L}_{-2}\Phi_{lk}|0_s>}{<0_s|\Phi_{lk}|0_s>}
&=& -\Big[\frac{m\xi \Gamma(1+\frac{1+\xi}{3-3\xi})\Gamma(-\frac{2\xi}{3-3\xi})}
{\Gamma(\frac{1}{3}){\sqrt 3}2^{2/3+1/2}} \Big]^4
\frac{\gamma^2(1/3)}{\gamma(-\frac{4\xi}{3-3\xi})
\gamma(\frac{2+2\xi}{3-3\xi})}\nonumber\\
\nn &&\ \ \ \ \ \ \ \ \ \ \ \ \ \ \ \ \ \ \ \ \ \ \ \ \ \
 \times\ \ \ {\cal W}_{15}((\xi+1)l-\xi k)
\eeqa
with:
\beqa
{\cal W}_{15}(\eta)=\frac{1}{\xi^{2}(\xi+1)^{2}}
\times w(\eta;\ \xi+5,\ 4-4\xi,\ -1-5\xi,\ 1-\xi;\ 6-6\xi)\nonumber\ .
\eeqa

We want now to calculate the contribution of the VEV of the third level descendent fields ${\cal K}(\alpha)=
\frac{<(\partial\vph)^3({\overline\partial}\vph)^3e^{i\alpha\vph}>_{BD}}
{<e^{i\alpha\vph}>_{BD}}$ (\ref{Ka}) to (\ref{twop}).

In the (Gaussian) free field theory, the composite fields
$(\partial\vph)^3({\overline\partial}\vph)^3e^{i\alpha\vph}$ are
spinless with scale dimension:
\beqa D\equiv \Delta+{\overline \Delta}=2\alpha^2+6\ .\label{dimtr}
\eeqa
For $0<\beta^2<1$ the perturbation is relevant and, similarly to the second level, a finite number
of lower scale dimension counterterms are sufficient to cancel the
divergences arising in the VEVs of third level descendent fields. As before, we are looking for $(n|n')$ resonances with some primary fields (\ref{reson}). One can easily find
that a resonance can appear between the third level
descendent field
$(\partial\varphi)^3(\bar\partial\varphi)^3e^{i\alpha\varphi}$ and
the following primary fields:
\beqa (i) \ \ \ \ &&e^{i(\val-\beta)\vph}\ \ \ \ \mbox{i.e.}\ \
(n|n')=(1|4) \ \ \
\mbox{for}\ \ \ \alpha=\frac{1}{\beta}-\frac{\beta}{2}\ ;\label{nmt}\\
(ii) \ \ \ \ &&e^{i(\val+3\beta)\vph}\ \ \ \mbox{i.e.}\ \
(n|n')=(3|0) \ \ \
\mbox{for}\ \ \ \alpha=-\beta\ ;\nonumber\\
(iii) \ \ \ \ &&e^{i(\val-\frac{\beta}{2})\vph}\ \ \ \
\mbox{i.e.}\ \ (n|n')=(0|1) \ \ \
\mbox{for}\ \ \ \ \alpha=-\frac{2}{\beta}\ ;\nonumber\\
(iv) \ \ \ \ &&e^{i(\val-\frac{3\beta}{2})\vph}\ \ \ \
\mbox{i.e.}\ \  (n|n')=(0|3) \ \ \ \mbox{for}\ \ \
\alpha=\frac{\beta}{2}\ .\nonumber \eeqa
If we now look at the expression (\ref{twop}), we notice that the
contribution brought by the third level descendent field in
(\ref{Ka}), and that of any of the exponential fields in  $(i)$,
$(ii)$, $(iii)$ and $(iv)$, have the same power behavior in $r$
($r^{4\val_1\val_2 + 6}$) at short-distance for the corresponding
values of $\alpha$. The integrals which appear in these
contributions are, respectively:
 \beqa &&(i)\ \qquad {\cal
F}_{1,4}(\val_1\beta,\val_2\beta,\beta^2)\ ,\qquad \
\ \  (ii)\ \ \ j_3(\val_1\beta,\val_2\beta,\beta^2)\ ,\nonumber\\
&&(iii)\ \ \
j_1(-\frac{\val_1\beta}{2},-\frac{\val_2\beta}{2},\frac{\beta^2}{4})\
,\qquad \ (iv)\ \
j_3(-\frac{\val_1\beta}{2},-\frac{\val_2\beta}{2},\frac{\beta^2}{4})\
. \nonumber\eeqa

One can see that ${\cal K}(\val)$ (and similarly for the real
coupling case) exhibits the same poles in order that the divergent
contributions compensate each other. This last requirement leads
for instance to a set of relations for ${\cal K}(\alpha)$. The
third one reads:
\beqa \frac{\val_1^2\val_2^2(\val_1-\val_2)^2}{144}{\cal
R}es_{\alpha=-\frac{2}{\beta}}\ {\cal K}(\val) = \mu'\frac{{\cal
G}_{\val-\beta/2}}{{\cal G}_\val}|_{\val=-\frac{2}{\beta}}\ {\cal
R}es_{\alpha=-\frac{2}{\beta}} \
j_1(-\frac{\val_1\beta}{2},-\frac{\val_2\beta}{2},\frac{\beta^2}{4})\
,\label{residut} \eeqa
which is used to fix the $\val$-independent part (normalization)
of ${\cal K}(\val)$.

On the other hand, to determine the explicit form of the
$\val$-dependent part of ${\cal K}(\val)$, we use again the reflection
relations method. The calculations go along the same line as for the second level descendent.
Consequently, if we denote:
\beqa
K(a)=\frac{\langle(\partial\varphi)^3({\overline\partial}\varphi)^3e^{a\varphi}\rangle
_{BD}} {\langle e^{a\varphi }\rangle_{BD}}\ ,\label{Kreal} \eeqa
then we obtain the following two functional
relations:
\beqa
K(a)&=&\Big[\frac{(b+1/b-a)(b+2/b-a)(2b+1/b-a)}{a(a+1/b)(a+b)}\Big]^2K(Q-a)\ ,\label{reft}\\
K(-a)&=&\Big[\frac{(b/2+2/b-a)(b/2+4/b-a)(b+2/b-a)}{a(a+2/b)(a+b/2)}\Big]^2K(-Q'+a)\
.\nonumber \eeqa
Notice that these equations are invariant with respect to the
symmetry
 $b\rightarrow-\frac{2}{b}$ with $a\rightarrow -a$ in agreement with the
well-known self-duality of the BD-model. Assuming that $K(a)$ is a
meromorphic function in $a$, we find that the ``minimal'' solution
which follows from (\ref{residut}), (\ref{reft}) is:
\beqa
K(a)\!&=&\!-\frac{1}{a^2}\Big[\frac{m\Gamma(\frac{b^2}{h})\Gamma(\frac{2}{h})}{\Gamma(\frac{1}{3}){\sqrt
3}\ 2(Q+Q')^2}
\Big]^6\gamma\big(\frac{2ba+b^2+2}{h}\big)\gamma\big(\frac{-2ba-2}{h}\big)
\gamma\big(\frac{2ba-b^2+4}{h}\big)\times\nonumber\\
&&\times\gamma\big(\frac{-2ba-2b^2}{h}\big)
 \gamma\big(\frac{-2ba+2b^2-2}{h}\big)\gamma\big(\frac{2ba-4}{h}\big)
\gamma\big(\frac{-2ba+b^2+2}{h}\big)\gamma\big(\frac{2ba-b^2}{h}\big)\nn
\eeqa
where $h=6+3b^2$ is the ``deformed'' Coxeter number \cite{corr}.
Here we have used the exact relation between the
parameters $\mu$ and $\mu'$ and the
mass of the fundamental particle $m$ \cite{flzz}.

Notice that $K(a)$ is invariant under the duality transformation
${b\rightarrow -2/b}$ as expected, and contains all the expected
poles. Accepting this conjecture and taking $a=0$, we obtain for
instance:
\beqa\nn
\langle L_{-3}{\overline L}_{-3}{\mathbb
I}\rangle_{BD}=-\frac{m^2}{2^{10/3}}\frac{\Gamma^2(1+2/h)\Gamma^2(1+b^2/h)\Gamma^2(2/3)}
{\gamma(1/2+2/h)\gamma(1/2+b^2/h)\gamma(1/3+6/h)\gamma(1/3+3b^2/h)}f_{BD}^2
\eeqa
where $f_{BD}$ is the bulk free energy of the Bullough-Dodd model,
obtained in \cite{flzz}.

In the same way as above we can apply these results for the corresponding perturbed conformal field theories.
Let us consider, for example,
 the first case i.e. the $\Phi_{pert}\equiv\Phi_{12}$ perturbation,
  obtained for $\beta^2=p/p'$ \ with \ $ 1<p<p'$ relative prime
integers.
Using
the particle-breather identification \cite{flzz}
$m=2M\sin\big(\frac{\pi\xi}{3\xi+6}\big)$ \ and parameter
$a=i\big(\frac{l-1}{2\beta}-\frac{k-1}{2}\beta\big)$ in $K(a)$ it
is straightforward to get the VEV:
\beqa \frac{\langle0_s|L_{-3}{\overline
L}_{-3}\Phi_{lk}|0_s\rangle}{\langle0_s|\Phi_{lk}|0_s\rangle} &=&
-\Big[\frac{2^{2/3}\pi M\Gamma(\frac{2+2\xi}{3\xi+6})}{{\sqrt
3}\Gamma(\frac{1}{3})\Gamma(\frac{\xi}{3\xi+6})(1+\xi)}\Big]^6
\frac{1}{\xi^2(1+\xi)^2(3\xi+6)^2}\nonumber\\
&&\ \ \ \times\ \ \frac{\gamma(\frac{\eta-4\xi-3}{3\xi+6})
\gamma(\frac{-\eta-4\xi-3}{3\xi+6})\gamma(\frac{\eta+1+\xi}{3\xi+6})\gamma(\frac{-\eta+1+\xi}{3\xi+6})
}{\gamma(\frac{\eta+2\xi+3}{3\xi+6})
\gamma(\frac{-\eta+2\xi+3}{3\xi+6})\gamma(\frac{\eta-2\xi+1}{3\xi+6})\gamma(\frac{-\eta-2\xi+1}{3\xi+6})}
\ .\nn
\eeqa
The calculations for the other perturbations are straightforward so we will not report them here.

\subsection{ Hidden local, quasi-local and non-local symmetries in integrable systems}

As observed in \cite{ger,kuper}, the classical limit ($c\to
-\infty$) of CFT's is described by the second Hamiltonian structure of the
(usual) KdV which corresponds to $A_1^{(1)}$ in the Drinfeld-Sokolov
scheme \cite{dsoc}.The KdV variable  $u(x,t)$ is related to the mKdV variable
$v(x,t)$  by the Miura transformation
$u=-v^2 +v'$, which is the classical
counterpart of the Feigin-Fuchs transformation \cite{miura}.
In fact the mKdV equation is:
\be
\dt v=-\frac{3}{2}v^2v'-\frac{1}{4}v'''
\label{mkdv}
\ee
and the mKdV field $v=-\phi'$ is the derivative of the Darboux field $\phi$.
The equation (\ref{mkdv}) can be re-written  as a null curvature
condition $[ \dt - A_t , \dx - A_x ] = 0 $ for connections
belonging to the $A_1^{(1)}$  loop  algebra:
\ba
A_x &=& - v h + (e_0 + e_1),  \nonumber \\
A_t &=& \la^2(e_0 + e_1 - vh) -
\frac{1}{2}[(v^2-v')e_0 + (v^2+v')e_1] - \frac{1}{2}(\frac{v''}{2}-v^3)h
\label{lax}
\ea
where the generators $e_0,e_1,h$  are chosen in the
fundamental representation and canonical gradation
of the $A_1^{(1)}$ loop algebra:
\be
e_0=\la E=\left(\begin{array}{cc} 0 & \la \\
                                  0 & 0 \end{array}\right),\qquad
e_1=\la F=\left(\begin{array}{cc} 0 & 0 \\
                                  \la & 0 \end{array}\right)\virg
h  = H=\left(\begin{array}{cc} 1 & 0 \\
                               0 & -1 \end{array}\right) .
\label{gend}
\ee
Of special interest for us will be the so called
transfer matrix which performs the parallel transport
along the $x$-axis, and is thus the solution of the associated linear problem:
\be
\dx T(x;\la) = A_x(x;\la)T(x;\la)  .
\label{T1}
\ee
The formal solution of the previous equation is given by:
\be
T(x,\la) =e^{H\phi(x)}{\cal P}
\exp\lt\la \int_0^xdy
(e^{-2\phi(y)} E+ e^{2\phi(y)} F ) \rt =\left(\begin{array}{cc} A & B \\
                                                                C & D \end{array}\right) ,
\label{forsol}
\ee
where the expansions of the entries are:
\ba
A(x;\la) &=&e^{\phi(x)}+O(\la^2) \virg  B(x;\la)=\la e^{\phi(x)} \int_0^x dy e^{-2\phi(y)}+O(\la^3),
\nonumber \\
C(x;\la) &=& A(-\phi(x)) \virg D(x;\la)=B(-\phi(x)) .
\label{ABCD}
\ea
Note that the first terms of the expansion (\ref{ABCD}) are exactly the
classical limits of the two {\it elementary} vertex operators.
Besides, the expression (\ref{forsol}) defines $T(x,\la)$ as an entire
function of $\la$ with an essential singularity at $\la=\infty$ where it is governed
by the corresponding asymptotic expansion. The two expansions give rise to different algebraic
and geometric structures, as we shall see below.

Let us first consider the regular expansion.
In our case the formal solution (\ref{forsol}) can be expressed  as
an expansion in positive powers of $\la$ with an infinite
radius of convergence and non-local coefficients (similarly to what we did in Section 5.2 for the $A_2^2$ case):
\be
T(x;\la)=e^{H\phi(x)} \sum_{k=0}^{\infty}
\la^k \int_{x\geq x_1 \geq x_2 \geq ... \geq
x_k\geq 0}K(x_1)K(x_2)...K(x_k) dx_1 dx_2 ... dx_k
\label{regexp}
\ee
where $K(x)= e^{-2\phi(x)}E+e^{2\phi(x)}F$.
After calculating the expression (\ref{regexp}) for $x=L$  and taking
the
trace, we obtain the regular expansion for $\tau(\la)=tr{T(L;\la)}$ in terms of the non-local
conserved charges in involution (we slightly changed the notations here with respect to Section 5.2 where we had $L=2\pi$ and the trace was denoted as $T(\lambda)$). However, one may obtain a larger number
of non-local conserved charges not in
involution, i.e. so that the charges commute with local
hamiltonian of the mKdV (\ref{mkdv}) but not between themselves. This can
be carried out by means of the dressing techniques in the following way.
By assuming the regular expansion (\ref{regexp}), let us construct the
generic resolvent by dressing one of the generators $X=H,E,F$ (\ref{gend}):
\be
Z^X(x,\la)=(TXT^{-1})(x,\la)=\sum_{k=0}^\infty \la^k Z^X_k  .
\label{dres}
\ee
$Z^X(x,\la)$ is clearly a  resolvent for the operator ${\cL}=\dx -
A_x$   (\ref{lax}) since by construction it satisfies:
\be
[{\cL},Z^X(x;\la)]=0  .
\label{resdef}
\ee
The foregoing property of the resolvent assures that,
once we define the gauge connection of the dressing symmetries:
\be
\Theta^X_n(x;\la)=(\la^{-n}Z^X(x;\la))_-=\sum_{k=0}^{n-1} \la^{k-n} Z^X_k  ,
\label{theta}
\ee
the commutator $[{\cL},\Theta^X_n(x;\la)]$ is of degree zero in $\la$.
Therefore it is possible to construct the gauge transformation:
\ba\label{gautra}
\delta^X_n A_x &=& -\delta^X_n{\cL} = -[\Theta^X_n(x;\la),{\cL}] ,\\
\nonumber \delta^X_n A_t &=& -[\Theta^X_n(x;\la), \dt - A_t],
\ea
which preserves the zero curvature condition by construction. It will also be
a true symmetry of the model in case the last term in (\ref{gautra}) is
proportional to $H$:
\be
\delta^X_n A_x=H \delta^X_n  \phi' .
\label{var}
\ee
This depends, for X fixed, on whether n is even or
odd. Indeed, by directly substituting the regular
expansion (\ref{regexp}) in (\ref{dres}), one can obtain :
\ba
Z^H_{2m}(x)=a^H_{2m}(x)H  \virg   Z^H_{2m+1}(x)=b^H_{2m+1}(x)E+c^H_{2m+1}(x)F \nonumber\\
Z^E_{2n}(x)=b^E_{2n}(x)E+c^E_{2n}(x)F  \virg  Z^E_{2n+1}(x)=a^E_{2n+1}(x)H  \nonumber \\
Z^F_{2p}(x)=b^F_{2p}(x)E+c^F_{2p}(x)F \virg  Z^F_{2p+1}(x)=a^F_{2p+1}(x)H  .
\label{Z's}
\ea
The variation (\ref{gautra}) can be explicitly calculated as:
\be
\delta^X_n A_x = [Z^X_{n-1},E+F]
\label{expvar}
\ee
and hence it is clear that $Z^X_{n-1}$ cannot
contain any term proportional to $H$.
The conclusions are that:
\begin{itemize}
\item in the $\Theta^H_n$ case $n$, in
(\ref{gautra}), must be even,
\item in the  $\Theta^E_n$ and  $\Theta^F_n$ case $n$ must conversely be odd.
\end{itemize}
Besides, it is possible to show  by direct calculation  that these
infinitesimal transformation generators form a representation of  a (twisted) Borel subalgebra $A_1\otimes {\bold C}$, (of the loop algebra $A_1^{(1)}$):
\be
[\delta^X_m , \delta^Y_n] = \delta^{[X,Y]}_{m+n},\qquad X,Y = H,E,F  .
\label{delalg}
\ee
The first generators of this algebra are explicitly given by:
\ba
\delta^E_1 \phi'(x) &=&  e^{2\phi(x)}          \nonumber  \\
\delta^F_1 \phi'(x) &=&  - e^{-2\phi(x)}    \nonumber  \\
\delta^H_2 \phi'(x) &=& e^{2\phi(x)} \int_0^x dy e^{-2\phi(y)} +  e^{-2\phi(x)} \int_0^x dy
e^{2\phi(y)}
\label{che}
\ea
and the rest are derived from these by commutation. Note that they are essentially non-local (this is true also for the higher
ones).

At this point we want to make an important observation. Consider the KdV
variable $x$ as a {\it space direction} $x_-$ of some more general system (and
$\p_-\equiv \p_x$ as a space derivative). Introduce the {\it time} variable
$x_+$and the corresponding evolution defining:
\be
\p_+\equiv (\d^E_{-1}+\d^F_{-1}).
\label{timeevol}
\ee
It is then obvious from (\ref{che}) that the equation of motion for $\phi$
becomes:
\be
\p_+\p_-\phi=2\sinh(2\phi) \virg (or \spz 2sin(2\phi)\spz if\spz
\phi\rightarrow i\phi) \label{sgeq}
\ee
i.e. the sine-Gordon equation! We consider this observation very important
since it provides a {\it global} introduction of sine-Gordon dynamics in the
KdV  system.

Around the point $\la=\infty$ the system is governed by the asymptotic
expansion. It can be obtained through the procedure described in \cite{dsoc}.
Namely, the asymptotic expansion for a solution of (\ref{T1}) can
be written as:
\be
T(x;\la)=KG(x;\la)e^{-\int_0^x dy D(y)},
\label{asyexpd}
\ee
in terms of a constant matrix
$
K =\frac{\sqrt{2}}{2}\left(\begin{array}{cc}  1 & 1 \\
                                              1 & -1 \end{array}\right)$,
a diagonal matrix:
\be
D(x;\la)=d(x;\la)H , \spz
d(x;\la)=\sum_{k=-1}^{\infty}\la^{-k}d_k(x)
\label{lod}
\ee
and, finally, of the
off-diagonal matrices $G_j(x), j>0$:
\be\nn
G(x;\la)=\buno + \sum_{j=1}^{\infty}\la^{-j}G_j(x)
\ee
with entries $(G_j(x))_{12}=g_j(x)$ and $(G_j(x))_{21}=(-1)^{j+1}g_j(x)$. It can be shown that the latter satisfy certain recursion relations.
Note that the $d_{2n}(x)$ are
exactly the charge densities (of the mKdV equation) resulting from the
asymptotic expansion of $\tau(\la)=trT(\la)$.

It is likewise known \cite{dsoc} that the construction of the mKdV flows goes
through the definition of a  resolvent $Z(x;\la)$ defined through the
following property of its asymptotic expansion
\be
[{\cL},Z(x;\la)]=0,\spz
Z(x,\la)=\sum_{k=0}^\infty \la^{-k} Z_k, \spz Z_0=E+F .
\label{asyres}
\ee
From the definition (\ref{asyres}) it is
possible to derive the  resolvent $Z$, obtained by {\it
dressing} the generator $H$ with the asymptotic expansion of $T$ (\ref{asyexpd}):
\be
Z(x,\la)=(THT^{-1})(x,\la).
\label{asydred}
\ee
The modes of the $\la$-expansion are given by:
\be
Z_{2k}(x)=b_{2k}(x)E+c_{2k}(x)F \virg
Z_{2k+1}(x)=a_{2k+1}(x)H,
\label{solasyd}
\ee
where for example
\ba
a_1 &=& -v \virg a_3=\frac{1}{4}v^3 - \frac{1}{8} v'' \nonumber \\
b_2 &=& \frac{1}{4}v^2 + \frac{1}{4} v' \virg  b_4=-\frac{3}{16}v^4 + \frac{1}{8} v''v- \frac{1}{16} v'^2 - \frac{3}{8} v'v^2 + \frac{1}{16} v''', etc.     .
\label{ab}
\ea

As in the regular case, the system enjoys a gauge symmetry
of the form (\ref{gautra}) with the constraint (\ref{var}):
\be\nn
\delta_{2k+1}A_x =-[\theta_{2k+1}(x;\la),{\cL}],
\ee
where the $\theta_{2k+1}$ are the Lax
connections associated to $A_x$
\be\nn
\theta_{2k+1}(x;\la)=(\la^{2k+1}Z(x;\la))_+=\sum_{j=0}^{2k+1}
\la^{2k+1-j} Z_j(x) .
\ee
It happens that these
transformations coincide exactly with the commuting higher mKdV flows (or mKdV
hierarchy):
\be
\delta_{2k+1}\phi'(x)=\partial a_{2k+1}(x)
\label{mkdvh}
\ee
and are therefore local in contrast with the regular ones.
It turns out that the other entries of the resolvent $b_{2n}(x)$ are exactly
the conserved densities, namely:
\be
\d_{2k+1}\phi'(x)=\{I_{2k+1},\phi'(x)\} \virg I_{2k-1}=\int_0^L dx b_{2k}(x).
\label{bdens}
\ee
They differ from $d_{2k+1}$ (\ref{lod}) by a
total derivative. For example:
\ba
b_2 &=& -d_1+\half \phi'' ,\nn \\
b_4 &=& {3\over 4}d_3+\p({7\over 32}\phi''\phi'+{1\over 16}(\phi')^3+{1\over
16}\phi''') \spz etc.
\label{bd}
\ea
Let us note that it can be shown that these two kind of symmetries
(regular and asymptotic) commute with each other. In this sense the non-local
regular transformations provide a true symmetry of the KdV hierarchy.
One can construct also the flows deriving from $Z^E=TET^{-1}$ and $Z^F=TFT^{-1}$ and
no more commuting with the $\delta_{2k+1}$ of the hierarchy, but rather
closing with them a spectrum generating  algebra.

Finally, one can use the above constructions to propose an alternative description of the
spectrum of local fields. Namely, we use as basic objects the entries
of the resolvent $Z(x;\la)$ modulo the gauge transformations described
above. A number of constraints, or {\it classical null vectors}, appear in
this picture coming from the equation of motion
$\delta_{2k+1}Z=[\theta_{2k+1},Z]$ of the resolvent and the obvious
constraint $Z^2=\buno$ \cite{myne21}.

One can show that our approach is easily applicable to other integrable systems. We can
consider for example the case of the $A_2^{(2)}$-KdV equation. The reason is that it can be considered
as a different classical limit of the {\bf CFT's} as was discussed in the previous Section. It turns out
that all the constructions described above go perfectly well also in this case.

We want to show now that one can construct in a natural way more
general kinds of dressing-like symmetries.
 It is well known that the vector fields $l_m=\lambda^{m+1}\dla$ on the
circle realize the centerless Virasoro algebra:
\be\nn
[l_m,l_n]=(m-n)l_{m+n}.
\ee
A very natural dressing is represented by the resolvent:
\be
Z^V_{-m}=T_{reg}l_{-m}T^{-1}_{reg} , \spz m>0
\label{rvr}
\ee
where $T_{reg}$ indicates the regular expansion of the transfer matrix
(\ref{forsol}) and $m$ is a positive integer. Of course, this dressed
generator satisfies the usual property (\ref{resdef}) of being a resolvent:
\be
[{\cL},Z^V_{-m}(x;\la)]=0 .
\label{resdef1}
\ee
As in the previous cases, the property (\ref{resdef1}) allows us to
calculate the expansion modes $Z_n$ of:
\be\nn
Z^V_{-1}=T_{reg}l_{-1}T^{-1}_{reg}=\sum_{n=0}^{\infty}\la^n Z_n -\dla
\ee
and thus the expansion modes of the more general Virasoro resolvent
(\ref{rvr}). In the same way, (\ref{resdef1}) authorizes us to define a
gauge connection:
\be\nn
\theta^V_{-m}=(Z^V_{-m})_- =\sum_{n=0}^m\la^{n-1-m}Z_n -\dla
\ee
and the relative gauge transformation:
\be
\delta_{-m}^V A_x =-[\theta^V_{-m}(x;\la),{\cL}] .
\label{rvf}
\ee
Finally, we have to verify the consistency of this gauge transformation
requiring\\ $\delta_{-m}^V A_x=H \delta_{-m}^V  \phi'$ .
It is easy to see that this requirement imposes $m$ to be even.
Explicit examples of the first flows are:
\ba
\delta_{-2}^V \phi' &=&  e^{2\phi(x)} \int_0^x dy e^{-2\phi(y)} -
e^{-2\phi(x)} \int_0^x dy e^{2\phi(y)}=e^{2\phi(x)}B_1-e^{-2\phi(x)}C_1
\nonumber  \\
\delta_{-4}^V  \phi' &=&  e^{2\phi(x)}(3B_3(x)-A_2(x)B_1(x)) -
e^{-2\phi(x)}(3C_3(x)-D_2(x)C_1(x)) \nonumber  \\
\delta_{-6}^V  \phi' &=& e^{2\phi(x)} (5B_5(x)-3A_4(x)B_1(x)+A_2(x)B_3(x))
\nonumber  \\
&-& e^{-2\phi(x)}(5C_5(x)-3D_4(x)C_1(x)+D_2(x)C_3(x))
\label{frd's}
\ea
where $A_i,B_i,C_i,D_i$ stand for the coefficients in the $\l$-expansion of the matrix (\ref{forsol}).
We stress that these infinitesimal variations have a form very similar to
that of the regular dressing flows
((\ref{gautra}) with $X=H$)
$\delta_{-2r}^H$. Nevertheless, in spite of the commutativity
$[\delta_{-2r}^H,\delta_{-2s}^H]=0$
one can check by direct calculation that instead the flows (\ref{frd's})
obey Virasoro commutation relations:
$[\delta_{-2},\delta_{-4}]=\delta_{-6}$.
Actually, this is true also in the general case:
\be
[\delta_{-2m}^V ,\delta_{-2n}^V ]=(2n-2m)\delta_{-2m-2n}^V .
\label{rsa}
\ee
From (\ref{frd's}) the transformations of the classical {\it
primary fields}
$e^{\phi}$ follow. For example:
\ba
\delta_{-2}^V e^{\phi} &=& (D_2-A_2)e^\phi  \nonumber  \\
\delta_{-4}^V e^{\phi} &=&  [(3D_4-C_3B_1)-(3A_4-B_3C_1)]e^\phi .
\nonumber
\ea
It is understood of course that these fields are primary with respect to
the usual {\it space-time} Virasoro symmetry.

In the same way it is quite natural to generate a resolvent by dressing
the remaining vector fields $l_m=\lambda^{m+1}\dla$, $m\geq 0$:
\be
Z^V_{m}=T_{asy}l_{m}T^{-1}_{asy} ,\spz   m\geq 0
\label{avr}
\ee
through the asymptotic expansion of the transfer matrix (\ref{asyexpd})
$T_{asy}$. Now we have:
\be\nn
Z^V_{-1}=T_{asy}l_{-1}T^{-1}_{asy}=\sum_{n=0}^{\infty}\la^{-n} Z_n -\dla.
\ee
In general :
\be
Z_{2n}=\beta_{2n}E+\gamma_{2n}F ,\spz  Z_{2n+1}=\alpha_{2n+1}H \nonumber
\ee
where for example $\beta_0=x=\gamma_0$, $\alpha_1=2xg_1$,
$\beta_2=-xb_2-g_1+\int^x d_1$, $\gamma_2=-xc_2+g_1+\int^x d_1$ etc. .
In the same manner we define a gauge connection:
\be\nn
\theta^V_{m}=(Z^V_{m})_+=\sum_{n=0}^{m+1}\la^{m+1-n}Z_n -\dla
\ee
and the relative gauge transformation
$\delta_{m}^V A_x =-[\theta^V_{m}(x;\la),{\cL}]$.
The consistency condition of this gauge transformation,
$\delta_{m}^V A_x=H \delta_{m}^V  \phi'$,
impose $m$ to be even in this case too.
Actually, the first transformation:
\be
\d_0^V\phi'(x)=(x\p+1)\phi'(x)
\label{deltazero}
\ee
is exactly the scale transformation - it counts the dimension (or level).
The first non-trivial examples are:
\ba
\delta_{2}^V  \phi' &=& 2xa_3'-(\phi')^3+{3\over 4}\phi'''
+2a'_1\int^x_0 d_1, \nn \\
\delta_{4}^V  \phi' &=& 2xa_5'+(\phi')^5-{5\over 2}\phi'''(\phi')^2-{27\over
8} (\phi'')^2\phi'+{5\over 16}\phi^V +\nn \\
                    &+&2a_3'\int_0^x d_1 +6a_1'\int_0^x d_3 .
\label{deltatwo}
\ea
We note that these depend explicitly on $x$ and are quasi-local (they
contain some indefinite  integrals). For further reference we presented the
integrands in (\ref{deltatwo}) explicitly in terms of the entries of the basic
objects $T(x,\lambda)$ and $Z(x,\lambda)$, defined in (\ref{lod}),(\ref{solasyd}).
Furthermore, one
can find the transformation of the resolvent and therefore the transformation
of the conserved densities $\delta_{2k}b_{2n}(x)$. In particular the first
nontrivial transformations of the KdV variable $u=b_2$ read:
\ba
\delta_2^Vb_2 &=& \delta_2^V u= 2xb_4'+u''-2u^2-{1\over 2}u'\int^x_0u , \nn \\
\delta_4^Vb_2 &=& \delta_4^V u= 2xb_6'+2u^3+3uu''+{17\over 8}(u')^2+{3\over
8}u^{IV} + \nn \\
              &+& u'\int_0^x b_4 +b_4'\int_0^x u .
\label{deltau}
\ea

One can check by direct calculation that the first flows (\ref{deltatwo})
obey Virasoro commutation relations. Actually, in general one can show
that:
\ba
\delta_{2n}^VZ_{2m}^V&=& [\theta^V_{2n},Z^V_{2m}]-(2n-2m)Z^V_{2n+2m}
\nonumber \\
\delta_{2n}^V\theta_{2m}^V-\delta_{2m}^V\theta_{2n}^V&=&
[\theta^V_{2n},\theta^V_{2m}]
-(2n-2m)\theta^V_{2n+2m}\nn .
\ea
From these it is not difficult to see that the asymptotic flows also close
(half) the Virasoro algebra:
\be
[\delta_{2m}^V ,\delta_{2n}^V ]=(2m-2n)\delta_{2m+2n}^V ,\spz m,n\ge 0.
\label{asa}
\ee

An important question arises at this point: what are the commutation
relations between the asymptotic and regular transformations? This is a
very nontrivial question in view of the different character of the
corresponding vector fields - the asymptotic ones are quasilocal (they can
be made local by differentiating a certain number of times), the regular
instead are essentially non-local being expressed in terms of vertex
operators. We recall here that the (proper) regular dressing symmetries
(\ref{gautra}) commute with all the mKdV flows (\ref{mkdvh}).
We shall see that this is not the case here. In fact it is easy to compute
the most simple relations: $[\delta_0,\delta_{2n}]=-2n\delta_{2n}$, $n\in
{\bf Z}$ (i.e. $\delta_0$ counts the dimension or level). Using the
explicit formulae presented above one can also compute the first
nontrivial commutator: $[\delta_2,\delta_{-2}]=4\delta_0$. In fact, one
can show that in general \cite{myne23}:
\ba
\delta^V_mZ^V_{-n}&=& [\theta^V_m,Z^V_{-n}]-(m+n)Z^V_{m-n} \nonumber \\
\delta^V_{-m}Z^V_n&=& [\theta^V_{-m},Z^V_n]+(m+n)Z^V_{n-m}\nonumber  \\
\nn \delta^V_m\theta^V_{-n}-\delta^V_{-n}\theta^V_m&=&
[\theta^V_m,\theta^V_{-n}]-(m+n)\theta^V_{m-n}.
\ea
From these it is easy to deduce:
$[\delta^V_m,\delta^V_{-n}]=(m+n)\delta^V_{m-n}$, and therefore:
\be
[\delta^V_{2m},\delta^V_{2n}]=(2m-2n)\delta^V_{2m+2n},\qquad m,n\in {\bold Z}.
\label{vir}
\ee
We want to stress once more that this Virasoro symmetry is different from
the {\it space-time} one and is essentially non-local. The additional
symmetries coming from the regular dressing are very important for
applications. They complete the asymptotic ones forming an entire Virasoro
algebra and provide a possibility of a central extension.

With the aim of understanding the classical and quantum structure of the
mKdV system we present here the complete algebra of symmetries. The
Virasoro flows commute neither with the mKdV hierarchy (\ref{mkdvh}) nor
with the (proper) regular dressing flows (\ref{gautra}). In fact one can
show that \cite{myne23}:
\be
[\delta_{2k+1},\delta^V_{2m}]=(2k+1)\delta_{2k+1+2m} , \quad
[\delta^X_n,\delta^V_{2m}]= n \delta^X_{n+2m} .
\label{virdr}
\ee
Note that the indices of r.h.s. can become negative, for the first, or
positive, for the second of equations (\ref{virdr}). Explicit calculation
show that in these cases the commutator is exactly equal to zero. This
fact confirms the self-consistensy of the construction.

\subsection{Hidden Virasoro symmetry of (soliton solutions of) the sine-Gordon theory}

We would like now to restrict the Virasoro symmetry to the soliton solutions
of the (m)KdV theory. One can expect that in this case it simplifies
considerably.

We start with a brief description of the well known soliton solutions of
(m)KdV. They are best expressed in terms of the so-called {\it
tau-function}. In the case of N-soliton solution of (m)KdV it has the form:
\be
\tau(X_1,...,X_N| B_1,...,B_N)=\det(1+V)
\label{tau}
\ee
where $V$ is a matrix:
\be\nn
V_{ij}=2{B_iX_i(x)\over B_i+B_j} \virg i,j=1,...,N.
\ee
The m-KdV field is then expressed as:
\be
e^\phi={\tau_-\over \tau_+} ,
\label{phiintau}
\ee
where:
\be
\tau_\pm(x)=\tau(\pm X(x)|B)
\label{taupm}
\ee
and $X_i(x)$ is simply given by:
\be\nn
X_i(x)=X_i\exp(2B_ix).
\ee
The variables $B_i$ and $X_i$ are the parameters describing the solitons:
$\beta_i=\log B_i$ are the so-called rapidities and $X_i$ are related to the
positions. The integrals of motion, restricted to the N-soliton solutions have
the form:
\be
I_{2n+1}=\sum_{i=1}^N B_i^{2n+1} \virg n\ge0 .
\label{inmotb}
\ee
It is well known that (m)KdV admits a non-degenerate symplectic structure. One
can find the corresponding Poisson brackets between the basic variables $B_i$
and $X_i$ \cite{baber}. The (m)KdV flows are then generated by (\ref{inmotb})
via:
\be
\delta_{2n+1}*=\{\sum_{i=1}^N B^{2n+1},*\} \virg n\ge0 .
\label{solkdv}
\ee

Our final goal is the quantization of solitons and of the Virasoro symmetry. It
was argued in \cite{bbs2} that this is best performed in another set of
variables $\{A_i,B_i\}$. The latter are the soliton limit of certain variables
describing the more general quasi-periodic finite-zone solutions of (m)KdV and are known as analytical variables.

Explicitly, the change of variables is given by:
\be\nn
X_j\prod_{k\ne j}{B_j-B_k\over B_j+B_k}=\prod_{k=1}^N{B_j-A_k\over B_j+A_k}
\virg j=1,...,N .
\ee
 The non-vanishing Poisson
brackets expressed in terms of these new variables
take the form:
\be
\{A_i,B_j\}={\prod_{k\ne i}(B_j^2-A_k^2)\prod_{k\ne j}(A_i^2-B_k^2)\over
\prod_{k\ne i}(A_i^2-A_k^2)\prod_{k\ne j}(B_j^2-B_k^2)}(A_i^2-B_j^2).
\label{poisson}
\ee
The corresponding tau-functions have also a very compact form in terms of the
analytical variables:
\ba
\tau_+ &=& 2^N\prod_{j=1}^N B_j\{{\prod_{i<j}(A_i+A_j)\prod_{i<j}(B_i+B_j)\over
\prod_{i,j}(B_i+A_j)}\} \nonumber \\
\tau_- &=& 2^N\prod_{j=1}^N A_j\{{\prod_{i<j}(A_i+A_j)\prod_{i<j}(B_i+B_j)\over
\prod_{i,j}(B_i+A_j)}\}.
\label{anlitictau}
\ea
Therefore, from the explicit form of the m-KdV field in terms of the
tau-functions (\ref{phiintau}) we obtain the following very simple expression:
\be
e^\phi\equiv{\tau_-\over\tau_+}=\prod_{j=1}^N{A_j\over B_j}.
\label{analiticphi}
\ee
The equation of motion of the $A_i$ variable is given by:
\be
\p_xA_i\equiv \d_1A_i=\{I_1,A_i\}=\prod_{j=1}^N(A_i^2-B_j^2)\prod_{j\ne
i}{1\over(A_i^2-A_j^2)}.
\label{eqmot}
\ee
One can verify that, as a consequence, the usual KdV variable $u$ is expressed
as:
\be
b_2\equiv u=\half (\phi')^2+\half \phi''=\sum_{j=1}^NA_j^2-\sum_{j=1}^NB_j^2.
\label{analitcu}
\ee
One can restrict also the higher KdV flows to the soliton solutions. For
example it is clear from (\ref{solkdv}) that:
\be\nn
\d_{2n+1}B_i=0 \virg n\ge 0 .
\ee
The variation of the $A_i$ variables can be easily computed as :
\be
\d_{2n+1}A_i=\{I_{2n+1},A_i\} \virg n\ge0
\ee
using the Poisson brackets (\ref{poisson}).

Now, we want to restrict the Virasoro symmetry of (m)KdV constructed in the previous Section to
the case of soliton solutions.  We shall be only interested in
the positive part of the latter. The transformation of the rapidities can be
easily deduced as a soliton limit of the Virasoro action on the finite-zone
solutions described in \cite{grin}:
\be
\d_{2n}B_i=B_i^{2n+1} \virg n\ge 0 .
\label{btransf}
\ee
What remains is to obtain the transformations of the $A_i$ variables. We found
it quite difficult to deduce them as a soliton limit of the corresponding
transformations of  \cite{grin}. Instead, we propose here another approach.
Namely, we use the transformation of the fields $\d_{2n}\phi$, $\d_{2n}\phi'$,
$\d_{2n}u$ {\it etc.} which we found before, restricted to the soliton
solutions using (\ref{analiticphi}),(\ref{analitcu}). The problem is
simplified by the fact that the Virasoro algebra is freely generated ,
i.e. we need to compute only the $\d_0$, $\d_2$ and $\d_4$ transformations, the
remaining ones are then obtained by commutation. In practice, we perform the
computation for the first few cases of $N=1,2,3$ solitons and then proceed by
induction.

Let us make an important observation. As we have stressed, the transformation
of the basic objects in the field theory of (m)KdV are quasi-local -- they
contain certain indefinite integrals. It happens that the corresponding
integrands become total derivatives when restricted to the soliton solutions.
For example:
\ba
b_2 &\equiv &u=\p_x\sum_{i=1}^NA_i(x) , \nonumber \\
b_4 &=& \p_x\sum_{i=1}^N A_i^3-\half u'\equiv \p_x[\sum_{i=1}^N(A_i^3-\half
\p_xA_i)].
\label{totderiv}
\ea
Therefore the Virasoro transformations become {\it local} when restricted to
the soliton solutions! The calculation is straightforward but quite tedious
so we present here only the final result:
\ba
\d_0 A_i &=& (x\p_x+1)A_i , \nonumber \\
\d_2 A_i &=&{1\over 3}x\d_3A_i+A_i^3-(\sum_{j=1}^N A_j)\p_xA_i , \nonumber \\
\d_4 A_i &=&{1\over 5}x\d_5A_i+A_i^5-\{\sum_{j\ne
i}A_i(A_i^2-A_j^2)+\sum_{j=1}^NA_j \sum_{k=1}^N B_k^2\}\p_xA_i ,
\label{result}
\ea
where the KdV flows read explicitly:
\ba
{1\over 3}\d_3A_i &=& (\sum_{j=1}^N B_j^2-\sum_{k\ne i}A_k^2)\p_xA_i ,
\nonumber \\
{1\over 5}\d_5A_i &=& (\sum_{j=1}^N B_j^4-\sum_{k\ne i}A_k^4)\p_xA_i -
\sum_{j\ne i}(A_i^2-A_j^2)\p_xA_i\p_xA_j.
\label{kdvonsol}
\ea
As we already mentioned, the remaining transformations can be obtained by
commutation, for example:
\be\nn
2\d_6A_i=[\d_4,\d_2]A_i \virg etc.
\ee

Now we pass to the most important part of this Section. We would like to extend
the construction presented above in (m)KdV theory to the case of sine-Gordon.
For this purpose one has to find a way of extending the mKdV dynamics up to the
sine-Gordon one. It is to some extent known how this can be done in the case
of the soliton solutions \cite{bbs2}. The idea is close to what we proposed
in the previous Section directly in the field theory of (m)KdV. Namely, let us consider the KdV
variable $x$ as a {\it space} variable of some more general system and call it
$x_-$ ( and $\p_-\equiv \p_x$ correspondingly ). We would like to introduce a
new {\it time} variable $x_+$ and the corresponding time dynamics. In the case
of the N - soliton solutions the latter is generated by the Hamiltonian:
\be\nn
I_{-1}=\sum_{i=1}^NB_i^{-1}
\ee
( essentially the inverse power of the momentum ) so that the {\it time} flow
is given by:
\be
\p_+*=\d_{-1}*=\{I_{-1},*\}
\label{timeflow}
\ee
using again the Poisson brackets (\ref{poisson}). In particular:
\be
\p_+A_i=\prod_{j=1}^N{A_i^2-B_j^2)\over B_j^2}\prod_{j\ne i}{A_j^2\over
(A_i^2-A_j^2)}.
\label{plusa}
\ee
One can check, using (\ref{analiticphi}, \ref{eqmot}), that with this
definition the resulting equation for the field $\phi$ is:
\be
\p_+\p_-\phi=2\sinh (2\phi)
\label{sinhg}
\ee
or under the change $\phi\rightarrow i\phi$:
\be
\p_+\p_-\phi=2\sin (2\phi)
\label{sing}
\ee
i.e. the sine-Gordon equation.
In a similar manner one can introduce
the rest of the sine-Gordon Hamiltonians:
\be
I_{-2n-1}=\sum_{i=1}^NB^{-2n-1}_i \virg n\ge 0 .
\label{negcharges}
\ee
They generate the ``negative KdV flows'' via the Poisson brackets
(\ref{poisson}):
\ba
\d_{-2n-1}B_i &=& 0 , \nonumber \\
\d_{-2n-1}A_i &=& \{I_{-2n-1},A_i\} \virg n\ge 0 .
\label{negkdv}
\ea

Now we arrive at the {\bf main conjecture} of this Section. Having in mind the
symmetric role the derivatives $\p_-$ and $\p_+$ are playing in the sine-Gordon
equation {\it we would like to suppose that one can obtain another half
Virasoro algebra by using the same construction as above but with $\p_-$
interchanged with $\p_+$}!

So let us define as before:
\be
\d_{-2n}B_i=-B_i^{-2n+1} \virg n\ge 0
\label{negb}
\ee
(note the additional -- sign in the r.h.s. which is needed for
the self-consistency of the construction). Following our conjecture we
construct the negative flows of the $A_i$ variable in the same way as before
but with the change $\p_-\rightarrow\p_+$. We have for example:
\ba
\d_{-2}\phi &=& x_+(2a_3^+)+b_2^+-2a_1^+\int^{x_+}_0b_2^+ , \nn \\
\d_{-2}b_2^- &\equiv& \d_{-2}u ={1\over 3}x_+\d_{-3}u
+(\p_+\phi-\int^{x_+}_0b_2^+)\p_-e^{2\phi} ,
\label{negu}
\ea
where $\d_{-3}u\equiv \{I_{-3},u\}$ {\it etc.} In (\ref{negu}) the + subscript
means that we take the same objects as defined in (\ref{lod}),
(\ref{solasyd}) but with $\p_-$ changed by $\p_+$. For example:
\ba
b_2^+ &=& \half (\p_+\phi)^2 +\half \p_+^2\phi ,\nonumber \\
a_3^+ &=& -{1\over 4}(\p_+\phi)^3 +{1\over 8} \p_+^3\phi \spz etc.
\label{pluses}
\ea
At this point we
want to make an important remark. Very non-trivially, it happens again that the
integrands in the expressions (\ref{negu}) and similar become total derivatives
when restricted to the N-soliton solutions. So that again the (negative)
Virasoro symmetry is {\it local} in the case of solitons! We present below the
first examples of this phenomenon:
\ba
b_2^+ &=& \p_+\{\sum_{i,j=1}^N{A_iA_j\over
B_i^2B_j^2}\sum_{i=1}^NA_i-\p_-\sum_{i,j=1}^N{A_iA_j\over B_i^2B_j^2}\},
\nonumber \\
b_4^+ &=& \p_+\{\sum_{i,j=1}^N{A_iA_j\over
B_i^4B_j^4}\sum_{i=1}^NA_i^3+ b_2^-\p_-\sum_{i,j=1}^N{A_iA_j\over
B_i^4B_j^4}- \nonumber \\
&-& \p_-b_2^-\sum_{i,j=1}^N{A_iA_j\over
B_i^4B_j^4}\} \spz etc.
\label{totderplus}
\ea
We then proceed as in the case of the positive Virasoro flows, i.e. we restrict
the transformations of the fields thus obtained to the soliton solutions.
As we explained, it is enough to find only the first transformations
$\d_{-2}A_i$ and $\d_{-4}A_i$ and the remaining ones are found by
commutation. Following our approach we do the computation explicitly in the
case of $N=1,2,3$ solitons and then proceed by induction. Here we give
the final results only:
\ba
\d_{-2}A_i &=& \frac{1}{3} x_+\d_{-3}A_i
-A_i^{-1}-(\sum_{j=1}^NA_j^{-1})\p_+A_i, \nn \\
\d_{-4}A_i &=& \frac{1}{5} x_+\d_{-5}A_i
-A_i^{-3}- \nn \\
&-&\{\sum_{j\neq i}^N{1\over A_i}({1\over
 A_i^2}-{1\over A_j^2})+\sum_{j=1}^N{1\over A_j}\sum_{k=1}^N{1\over
 B_k^2}\}\p_+A_i ,
\ea
where as before $\d_{-3}A_i=\{\sum_{j=1}^NB_j^{-3},A_i\}$ etc. As stated
above, we then can compute $2\d_{-6}A_i=[\d_{-2},\d_{-4}]A_i$ etc.

Now, we come to the important problem of the commutation relations between the
two half Virasoro algebras so constructed. This is a non-trivial question in
view of the different way we obtained them. In fact, it is clear that, by
construction, the positive (negative) Virasoro flows commute with the
corresponding $\p_-$ ( $\p_+$ ) derivatives:
\ba
[\d_{2n},\partial_-]A_i &=& 0 , \nn \\
\vspace{0.5cm}
[\d_{-2n},\partial_+]A_i &=& 0 \virg n\ge 0 .
\label{commut}
\ea
It is easy to see that this is not true for the ``cross commutators''.
Actually, one finds in this case:
\ba
[\d_{2n},\p_+]A_i &=& -\d_{2n-1}A_i , \nn \\
\vspace{0.5cm}
[\d_{-2n},\p_-]A_i &=& -\d_{-2n+1}A_i \virg n\ge 0 .
\label{noncommut}
\ea
It is clear that we are interested in a {\it true symmetry} of the sine-Gordon
theory. We must therefore obtain transformations that commute with the $\p_-$
and $\p_+$ flows and as a consequence with the corresponding Hamiltonians. It
is obvious from (\ref{commut}),(\ref{noncommut}) that this is acheeved by a
simple modification of the flows, i.e. let us define:
\ba
\d_{2n}' &=& \d_{2n}-x_+\d_{2n-1} , \nn \\
\d_{-2n}' &=& \d_{-2n}-x_-\d_{-2n+1} \virg n\ge 0 .\nn
\ea
Then, for the modified transformation we obtain:
\be\nn
[\d_{2n}',\p_{\pm}]A_i=0 \virg n\in {\Bbb Z}.
\ee
Finally, one can show that, with this modification, the commutation relations
between the positive and negative parts of the transformations close exactly
the whole Virasoro algebra:
\be
[\d_{2n}',\d_{2m}']A_i=(2n-2m)\d_{2n+2m}' A_i \virg n,m\in {\Bbb Z}.\nn
\ee

\section{Contributions}

\begin{enumerate}

\item{ We present an explicit construction of the Ramond sector  of the superconformal minimal models in terms of the Coulomb gas representation. The basic ingredients are written in terms of the Ising model variables (the order-disorder parameter fields and a free Majorana fermion) and a free scalar field. This allows the explicit construction of the fusion rules in all sectors. We compute also the four-point functions and the structure constants of the simplest Ramond fields.}

\item{ We compute the genus $g=2$ partition function for the $N=1$ superconformal minimal models on $Z_2$ hyperelliptic supersurfaces. The latter are mapped onto the minimal models of the $D_4^{p=2}$ parafermionic algebra on the branched sphere. The partition functions are written in terms of the multi-point Ising correlation function on the ordinary hyperelliptic surface and the $Z_2$ orbifold correlation functions.}

\item{ We describe the renormalization group flow of the $N=1$ superconformal minimal models perturbed by the least relevant field. For that purpose we compute the conformal blocks and the corresponding four-point functions of certain fields in NS and R sectors in the leading order. The anomalous dimensions and the mixing matrix of these fields are obtained. It turns out that the latter is finite and coincides exactly with that found in the non-supersymmetric case.}

\item{ We compute the exact three-point correlation functions of the NS and R fields in the $N=1$ supersymmetric Liouville field theory. They are expressed in terms of some generalized special function. Using the properties of the NS and R fields we obtain also the reflection amplitudes (or two-point functions) for the supersymmetric case.}

\item{ We compute the exact one-point function of the $N=1$ super-Liouville field theory with appropriate boundary conditions. Exact results are derived both for the theory defined on a pseudosphere with discrete (NS) boundary conditions and for the theory with explicit boundary action which preserves the superconformal symmetry. We also show that these one-point functions can be related to a generalized Cardy conditions along with corresponding modular S-matrices.}

\item{ We present the Coulomb gas representation of the $N=2$ superconformal minimal models. The basic ingredient of the construction is the system of two scalar fields and two free fermionic ones. We show that the dynamics of the models is generated by two different kinds of screening operators - one based on chiral (antichiral) superfields and the other on a scalar superfield. The Ramond and twisted primary fields are represented by vertex operators involving the lowest dimensional spin  (for fermions) and twisted (for scalars) fields.}

\item{ We use the parafermionic construction of the $N=2$ auperconformal algebra to derive the fusion rules, four-point functions and structure constants in all sectors of the $N=2$ minimal models. This is used to reveal the origin of the $Z_{p+2}$ symmetry of the p-th minimal model. We show that it is generated by specific $N=2$ superfields which, together with the super-stress tensor, close an $N=2$ super-parafemionic algebra.}

\item{ We find the general form and the exact Yukawa coupling constants of the low-energy effective superpotential for the three-generation Gepner's tensor product model.}

\item{ We propose a dual action for the $N=2$ super-Liouville field theory based on a scalar superfield. We claim that it realizes the strong-weak coupling duality ($b\to 1/b$) of the theory. We compute the reflection amplitudes (or two-point functions) of the NS and R fields based on the conjectured dual action and show that the results are consistent with the known results.}

\item{ We find the conformal boundary conditions and the corresponding one-point functions of the $N=2$ super-Liouville theory. This is done using the conformal and the modular bootstrap methods. We find both continuous ( FZZT branes) and discrete (ZZ branes) boundary conditions.}

\item{ We present an infinite set of higher equations of motion in $N=2$ super-Liouville field theory. They are in one to one correspondence with the degenerate representations and are enumerated by the $U(1)$ charge and by a pair of positive integers. We check that in the classical limit these equations hold as relations among the classical fields.}

\item{ We show that the higher level $\hat{su}(2)$ coset models can be represented as projected tensor products of lower level models or, finally, as products of Virasoro models. We construct the monodromy invariant correlation functions for arbitrary level fields and calculate some of the structure constants.}

\item{ We describe a RG flow in a general $\hat{su}(2)$ coset model perturbed by the least relevant field. Using our (projected) tensor product construction we obtain the structure constants and the four-point functions in the leading order. This allows us to compute the mixing coefficients among the fields in the UV and the IR theory up to the second order in the perturbation theory. It turns out that they are finite and exactly the same for all levels and, in that sense, universal.}

\item{ An RG flow in a general $\hat{su}(2)$ coset model perturbed by the least relevant field is considered using the (non-perturbative) RG domain wall construction. The mixing matrix between the UV and the IR fields in this construction is expressed in terms of one-point functions of these fields in the presence of a special boundary condition. We compute these one-point functions and show that the result agrees with the perturbative calculation up to second order.}


\item{ We propose an alternative description of the two-dimensional conformal field theory in terms of quantum inverse scattering. It is based on the generalized KdV system attached to $A_2^{(2)}$, yielding the classical limit of Virasoro as Poisson bracket structure. We classify the primary operators of the minimal models that commute with all the integrals of motion, and are therefore candidates to perturb the model by keeping the conservation laws. For our $A_2^{(2)}$ structure these happen to be the fields $\phi_{1,2},\phi_{2,1},\phi_{1,5}$.}

\item{ We calculate the exact vacuum expectation values (VEV's) of the second and third order level descendent fields in the Boulough-Dodd model. By performing quantum group restrictions we obtain the VEV's of the corresponding descendants of primary fields in the $\phi_{1,2},\phi_{2,1},\phi_{1,5}$ perturbed minimal models.}

\item{ We propose an alternative description of the spectrum of local fields in the classical limit of the integrable quantum field theories. It is essentially a variation of the inverse scattering method and is based on the so called dressing symmetry transformations. Our approach provides a systematic way of deriving the null-vectors that appear in this construction.}

\item{ We generalize the dressing symmetry construction in the mKdV hierarchy. This leads to non-local vector fields (expressed in terms of vertex operators) closing a Virasoro algebra. We argue that this algebra should play an important role in the study of the two-dimensional integrable field theories and in particular should be related to the deformed Virasoro algebra when the construction is perturbed out of the critical theory.}

\item{ We present a construction of a Virasoro symmetry in the sine-Gordon theory. It is a dynamical one and is not related to the space-time Virasoro symmetry of 2D CFT. We are mainly concerned with the corresponding N-soliton solutions. We present explicit expressions for the infinitesimal transformations and show that they are local in this case.}

\end{enumerate}


\newpage
\section*{Acknowledgements}

\no This work is supported in part by the NSFB grant DFNI T02/6 and the NSFB grant DNTS France 01/6.

\section*{Appendices}
\def\theequation{A.\arabic{equation}}
\setcounter{equation}{0}
\begin{appendix}
\section{Structure constants in general $\hat{su}(2)$ coset models}

In this Appendix we present the solution of the equations for the structure constants written in Section 4.2.
Here is the list of the structure constants we need:
\bea\label{cnn}
\CC_{(33)(nn)}^{(nn)}(l,p) &=&{\CG_n(p+l-1)\over\CG_n(p-1)},\\
\nn \CC_{(33)(nn)}^{(n+2n+2)}(l,p) &=&{\tilde\CG_n(p+l-1)\over\tilde\CG_n(p-1)},\\
\nn \CC_{(33)(nn)}^{(nn+2)}(l,p) &=&\sqrt{{l\ov (p-n-1)(p+l-n-1)}}{\tilde\CG_n(p+l-1)\over\CG_n(p-1)},\\
\nn \CC_{(33)(nn+2)}^{(n+2n+2)}(l,p) &=&-2\sqrt{{l\ov (p-n-1)(p+l-n-1)}}{\CG_{n+2}(p+l-1)\over\tilde\CG_n(p-1)}
\eea
\bea\label{resn}
\CC_{(13)(nn)}^{(nn)}(l,p) &=&-(n-1)\sqrt{{l\ov (p+l-2)(p-2)}} \CG_n(p+l-1),\\
\nn \CC_{(13)(nn)}^{(nn+2)}(l,p) &=&\sqrt{{(p+l-2)(p-n-1)\ov (p+l-n-1)(p-2)}} \tilde\CG_n(p+l-1),\\
\nn \CC_{(13)(nn+2)}^{(nn+2)}(l,p) &=&\(-l(n+1)+{2(p+l-2)(p-n-1)\ov p+l-n-1}\){\CG_{n+2}(p+l-1)\ov \sqrt{l(p+l-2)(p-2)}},\\
\nn \CC_{(33)(nn+2)}^{(nn+2)}(l,p) &=&(1-{2l\ov (p-n-1)(p+l-n-1)}){\CG_{-n+2}(p+l-1)\over\CG_{-n}(p-1)}
\eea
where we introduced the functions:
\bea\label{gamn}
\CG_n(p)&=&\[\g^3({p\ov p+1})\g^2({2\ov p+1})\g^2({n-1\ov p+1})\g^2({p-n\ov p+1})\g({3\ov p+1})\]^{1\ov 4},\\
\nn \tilde\CG_n(p)&=&\[\g({p\ov p+1})\g({n\ov p+1})\g({p-n-1\ov p+1})\g({3\ov p+1})\]^{1\ov 4}.
\eea
We want to stress that the "structure constants" thus obtained are actually square roots of the true structure constants $C$. The reason is that our construction makes use of "chiral" one-dimensional fields instead of the real two-dimensional ones (see Section 4.1).
Therefore the true structure constants are squares of those in (\ref{cnn}) and (\ref{resn}).

The coefficients in the construction (\ref{defn}) are given by:
$$
x=\sqrt{{(l-1)(p-n-1)\ov l(p-n)}} \qquad y=\sqrt{{p+l-n-1\ov l(p-n)}}.
$$

In exactly the same way one obtains the structure constants (and the coefficients $\tilde x$, $\tilde y$) involving the field $\tilde\phi_{n,n-2}(l,p)$. It turns out that they are obtained from the corresponding ones for $\tilde\phi_{n,n+2}(l,p)$ by simply changing $n\rightarrow -n$. This anticipates our observation in the main text that the two-point functions involving the field $\tilde\phi_{n,n-2}(l,p)$ are obtained from those of $\tilde\phi_{n,n+2}(l,p)$ by the same substitution.

Finally $\CC_{(13)(nn+2)}^{(nn-2)}(l,p)=0$ as can be seen by examining recursively the OPEs and fusion rules of the fields.

\def\theequation{B.\arabic{equation}}
\setcounter{equation}{0}

\section{Correlation function $<\tphi(x)\tphi(0)\tphi(1)\tphi(\infty)>$}

In this Appendix we present the calculation of the correlation function of the field $\tphi(x)$ defined in Section 4.2:
\bea\label{fotri}
&&<\tphi(x)\tphi(0)\tphi(1)\tphi(\infty)>=\\
\nn &&=<\prod_{i=1}^4\left(a(l,p)\phi_{1,1}(1,p)\tilde\phi_{1,3}(l-1,p+1)+b(l,p)\phi_{1,3}(1,p)\phi_{3,3}(l-1,p+1)\right)(x_i)>.
\eea
It defines the $\b$-function and the fixed point up to a second order of the perturbation theory. As explained in Section 4 the conformal blocks corresponding to (\ref{fotri}) are linear combinations of products of conformal blocks at levels $1$ and $l-1$ (\ref{cbprod}). There are in general 16 terms in (\ref{fotri}). Some of them are absent because of the fusion rules in each intermediate channel.
Here there are three channels: identity $\phi_{1,1}$, the field $\tilde\phi_{1,3}$ itself and $\tilde\phi_{1,5}$ which was defined in (\ref{five}). We present the calculation of the corresponding conformal blocks separately.
Our strategy here is to compute the conformal blocks up to a sufficiently high order and to make a guess. For $l=1$ this was done in \cite{pogsc2}. For $l-1$ we proceed recursively and use the fact that we know the result for $l=2,3,5$. The calculations are simplified significantly by the fact that we need the result in the leading order in $\e\rightarrow 0$.

\vskip.5cm
\no\rb{\bf Channel $\phi_{1,1}$}

The possible internal channels in the product (\ref{cbprod}) are $r,t=1,n$ and $t,s=n,1$ with $n=1,3,5,...$ (odd integer) corresponding to descendants at higher level as in (\ref{exp}). We examine the various terms that enter the sum (\ref{fotri}) and call for simplicity the corresponding conformal block at level $l$ $F_l$ omitting the indexes.
We do not present here the detailed calculations which are straightforward but quite tedious \cite{myne15}. As a result, we get a recursive equation for the conformal block at level $l$:
\be\label{reco}
F_l=a^4F_{l-1}+b^4F_1+2a^2b^2+2a^2b^2 x^2 \CC_{(13)(33)}^{(31)}(l-1)\left(1+{1\ov (1-x)^2}\right)
\ee
(note that we dropped the overall factor $x^{-2}$ for the time being).
The values of the coefficients in (\ref{reco}) in the leading order are:
$$
a\sim\sqrt{{l-1\ov l}},\qquad b\sim\sqrt{{1\ov l}},\qquad \CC_{(13)(33)}^{(31)}\sim {1\ov 3}.
$$
Also:
$$
F_1={1\over (1-x)^2}(1-2x+3x^2-2x^3+1/3x^4)
$$
as computed in \cite{pogsc2}. Introducing the useful notation:
$$
\tilde F_l=(1-x)^2F_l
$$
the recursion equation (\ref{reco}) becomes:
\be\label{rect}
l^2\tilde F_l=(l-1)^2\tilde F_{l-1}+\tilde F_1+2(l-1)f(x)
\ee
where we defined:
\be\label{fxx}
f(x)=(1-x)^2+{x^2\ov 3}(1+(1-x)^2).
\ee
The solution of this equation is given by:
$$
\tilde F_l={1\ov l}\tilde F_1+{l-1\ov l}f(x).
$$

Inserting $f(x)$ and returning to the initial notations (and restoring the overall $x^{-2}$) we get the final result for the conformal block:
\bea\label{cbot}
&&<\tphi_{1,3}(x)\tphi_{1,3}(0)|_{11}\tphi_{1,3}(1)\tphi_{1,3}(\infty)>=\\
\nn &=&{1\ov x^2(1-x)^2}\left[1-2x+({5\ov 3}+{4\ov 3l})x^2-({2\ov 3}+{4\ov 3l})x^3+{1\ov 3}x^4\right].
\eea
This result is in perfect agreement with $l=1$ \cite{pogsc2} and $l=2$ (Section 2.2).

\vskip.5cm
\no\rb{\bf Channel $\tphi_{1,5}$}

The field $\tphi_{1,5}$ was defined in (\ref{five}) and has a dimension close to 2. Therefore the possible internal channels in the product (\ref{cbprod}) are $r,t=1,n$ and $t,s=n,5$ with $n=1,3,5,...$.
Following the same logic as before we arrive at a solution similar to the one we obtained above:
$$
\tilde F_l={1\ov l}x^2\tilde F_1+{l-1\ov l}f(x)
$$
where now $\tilde F_l=(1-x)^2\CC_{(13)(13)}^{(15)}(l)F_{l}$ and:
$$
f(x)={\sqrt{5}\ov 3}(1+(1-x)^2)={\sqrt{5}\ov 3}(2-2x+x^2).
$$
Now we use the fact that we know the conformal block $\tilde F_1={\sqrt{5}\ov 3}$ from \cite{pogsc2}. Restoring the initial notations we obtain for the conformal block with internal channel $\tphi_{1,5}$ (note that here the overall power of $x$ is simply $x^0=1$):
\be\label{cbof}
<\tphi_{1,3}(x)\tphi_{1,3}(0)|_{15}\tphi_{1,3}(1)\tphi_{1,3}(\infty)>={1\ov (1-x)^2}\left[1-x+{l\ov 2(l-1)}x^2\right].
\ee

\vskip.5cm
\no\rb{\bf Channel $\tphi_{1,3}$}

One can proceed in the same way as for the previous channels. It turns out however that in this case some of the conformal blocks that enter the sum (\ref{fotri}) are divergent as $p\rightarrow\infty$. These divergences  are exactly compensated by the zeros of the corresponding structure constants in (\ref{cbprod}). Since the analysis similar to the above channels is more complicated here we adopt another strategy. Namely, we use the crossing symmetry of the correlation function (\ref{fotri}).  We ask that it is invariant under the transformation $x\to 1/x$ and use the explicit form of the remaining conformal blocks that we obtained above. This leads to linear equations for the coefficients in the $x$-expansion of the desired conformal block. The result is:
\be\label{cbthr}
<\tphi_{1,3}(x)\tphi_{1,3}(0)|_{13}\tphi_{1,3}(1)\tphi_{1,3}(\infty)>={1\ov x(1-x)^2}\left[1-{3\ov 2}x+{l+1\ov 2}x^2-{l\ov 4}x^3\right].
\ee

Combining altogether we finally obtain the 2D correlation function:
\bea\nn
<\tphi(x)\tphi(0)\tphi(1)\tphi(\infty)>&=&\left|{1\ov x^2(1-x)^2}\left[1-2x+({5\ov 3}+{4\ov 3l})x^2-({2\ov 3}+{4\ov 3l})x^3+{1\ov 3}x^4\right]\right|^2+\\
\label{fper}&+&{16\ov 3l^2}\left|{1\ov x(1-x)^2}\left[1-{3\ov 2}x+{l+1\ov 2}x^2-{l\ov 4}x^3\right]\right|^2+\\
\nn &+&{5\ov 9}\left({2(l-1)\ov l}\right)^2\left|{1\ov (1-x)^2}\left[1-x+{l\ov 2(l-1)}x^2\right]\right|^2.
\eea

We used this function in Section 4.2 for the computation of the $\b$-function and the fixed point.

\def\theequation{C.\arabic{equation}}
\setcounter{equation}{0}

\section{Other correlation functions}

In this Appendix we present the calculation of the other correlation functions we used in Section 4.2 to describe the mixing of the fields.

First, we notice that the computation of the function $<\tphi(x)\tphi(0)\tphi_{n,n+2}(1)\tphi_{n,n+2}(\infty)>$ goes in the same way as that of the function of the perturbing field itself, the latter being just a special case $n=1$. There are again the same three internal channels. It turns out that the corresponding conformal blocks are exactly the same, in agreement with $l=1$ and $l=2$ cases. The difference is only in the structure constants. Omitting the details we present the final result:
\bea\label{crfn}
&<&\tphi(x)\tphi(0)\tphi_{n,n+2}(1)\tphi_{n,n+2}(\infty)>=\\
\nn &=&\left|{1\ov x^2(1-x)^2}\left[1-2x+({5\ov 3}+{4\ov 3l})x^2-({2\ov 3}+{4\ov 3l})x^3+{1\ov 3}x^4\right]\right|^2+\\
\nn &+&{8\ov 3l^2}{n+3\ov n+1}\left|{1\ov x(1-x)^2}\left[1-{3\ov 2}x+{l+1\ov 2}x^2-{l\ov 4}x^3\right]\right|^2+\\
\nn &+&\left({2(l-1)\ov l}\right)^2{(n+3)(n+4)\ov 18 n(n+1)}\left|{1\ov (1-x)^2}\left(1-x+{l\ov 2(l-1)}x^2\right)\right|^2.
\eea

\vskip.5cm
\no\rb{\bf Function $<\tphi(x)\tphi(0)\tphi_{n,n+2}(1)\tphi_{n,n-2}(\infty)>$}

The only internal channel in this function corresponds to the field $\tphi_{1,5}$.  Denoting again $\hat F_l=C_l F_l$ we obtain explicitly the recursion equation:
$$
\hat F_l={(l-1)^2\ov l^2}\hat F_{l-1}+{1\ov l^2}x^2\hat F_1+2{(l-1)\ov l^2}\CC f(x)
$$
where:
$$
f(x)=1+{1\ov (1-x)^2},\qquad \CC=\sqrt{{1\ov 3n}\sqrt{n^2-4}}
$$
($\CC$ turns to coincide exactly with $\CC_1$). The solution of the equation then in terms of $\tilde F_l=(1-x)^2\hat F_l$ is:
$$
\tilde F=\CC\left( {2(l-1)\ov l}- {2(l-1)\ov l}x+x^2\right).
$$
Returning back to the original notations we can write finally the result for the correlation function:
\bea\label{cfpm}
&<&\tphi(x)\tphi(0)\tphi_{n,n+2}(1)\tphi_{n,n-2}(\infty)>=\\
\nn &=&{1\ov 3n}\sqrt{n^2-4}\left( {2(l-1)\ov l}\right)^2
\left|{1\ov (1-x)^2}\left(1-x+{l\ov 2(l-1)}x^2\right)\right|^2.
\eea

\vskip.5cm
\no\rb{\bf Function $<\tphi(x)\tphi(0)\phi_{n,n}(1)\tphi_{n,n+2}(\infty)>$}

There is only one relevant intermediate internal channel in the leading order in this function corresponding to $\tphi_{1,3}$. Inserting the values of the structure constants from Appendix A in the leading order we get for this correlation function:
\be\label{cfnpl}
<\tphi(x)\tphi(0)\tphi_{n,n}(1)\tphi_{n,n+2}(\infty)>={4\ov 3l}\sqrt{{n+2\ov n}}|x|^{-2}.
\ee

\vskip.5cm
\no\rb{\bf Function $<\tphi(x)\tphi(0)\phi_{n,n}(1)\phi_{n,n}(\infty)>$}

It was mentioned in Section 4.2 that this correlation function is exactly equal to those of $l=1$ and $l=2$. Here we want to explain in more details what is the reason for that. Note that, as mentioned in \cite{pogsc2}, we have to keep terms up to order $\e^2$ in this correlation function. Since the correlation function is quadratic in the conformal blocks we keep in the latter only terms up to order $\e$.

Since $\phi_{n,n}$ is just a primary field only four terms appear in this correlation function.
There are two relevant contributions in the intermediate channels corresponding to $\phi_{1,1}$ and $\tphi_{1,3}$.

Let us consider first the contribution from $\phi_{1,1}$.
There are two terms proportional to square roots of products of the constants $\CC_{(nn)(nn)}^{(13)}\CC_{(nn)(nn)}^{(31)}$  which are of order $\e^4$. As explained above we drop them.  Inserting the values of the structure constants in the leading order in the remaining two terms gives:
$$
F_l=a^2F_{l-1}+b^2F_1={l-1\ov l}F_{l-1}+{1\ov l}F_1.
$$
This equation is easily solved recursively:
$$
F_l=F_1
$$
(where $F_1$ is that of \cite{pogsc2}).

Similarly, in the channel corresponding to $\tphi_{1,3}$ there remain two terms, the other being of order $\e^2$
so we drop them. Then the equation reads:
\bea\nn
\hat F_l&=&a^2 F_{l-1}\sqrt{\CC_{(13)(13)}^{(13)}(l-1)\CC_{(nn)(nn)}^{(13)}(l-1)}+\\
\nn &+&b^2 F_{1}\sqrt{\CC_{(13)(13)}^{(13)}(1)\CC_{(nn)(nn)}^{(13)}(1)\CC_{(33)(33)}^{(33)}(l-1)\CC_{(33)(nn)}^{(nn)}(l-1)}=\\
\nn &=&\sqrt{{2(n^2-1)\ov 3p^2}}\left(a^2F_{l-1}+b^2F_1\right)
\eea
and we inserted the values of the structure constants. We see that the overall constant do not depend on $l$ so that this equation is very similar to the previous one and again the solution is:
$$
\hat F_l=\hat F_1=\sqrt{{2(n^2-1)\ov 3p^2}}F_1.
$$

As a result the correlation function is the same for all $l$ and reads (up to order $\e^2$):
\bea\label{cfnn}
&<&\tphi(x)\tphi(0)\tphi_{n,n}(1)\tphi_{n,n}(\infty)>=\left|F_1(1,1)\right|^2+{2(n^2-1)\ov 3p^2}\left|F_1(1,3)\right|^2=\\
\nn &=&|x|^{-4}+{(n^2-1)\e^2\ov 12}|x|^{-4}\left({x^2\ov 2(1-x)}+{\bar x^2\ov 2(1-\bar x)}+(\log (1-x)+\log (1-\bar x))^2\right).
\eea


\end{appendix}

\end{document}